\documentclass[aps,prl,reprint,a4paper,superscriptaddress]{revtex4-2}

\usepackage[english]{babel}

\usepackage{amsmath}
\usepackage{amssymb}
\usepackage{amsfonts}
\usepackage{amsthm}
\usepackage{mathtools} 
\usepackage{mathrsfs}
\usepackage{color,verbatim,graphicx}
\usepackage{url}
\usepackage{multirow}
\usepackage[hidelinks]{hyperref}
\usepackage{cleveref}
\usepackage{diagbox}
\usepackage{booktabs}
\usepackage[detect-weight=true,detect-family=true]{siunitx}
\usepackage{csquotes}
\usepackage{microtype}
\usepackage{multirow}

\usepackage{tikz}
\usetikzlibrary{backgrounds}
\usetikzlibrary{backgrounds,calc,shapes,arrows}
\usepackage{pgfplots}
\usepgfplotslibrary{fillbetween}
\usepackage[normalem]{ulem}

\usepackage{xcolor}
\definecolor{mlblue}{rgb}{0,0.4470,0.7410}
\definecolor{mlred}{rgb}{0.8500,0.3250,0.0980}
\definecolor{mlorange}{rgb}{0.9290,0.6940,0.1250}
\definecolor{mlviolet}{rgb}{0.4940,0.1840,0.5560}
\definecolor{mlgreen}{rgb}{0.4660,0.6740,0.1880}
\definecolor{mldarkred}{rgb}{0.6350,0.0780,0.1840}

\newcommand{\bX}{\mathbf{X}}
\newcommand{\bY}{\mathbf{Y}}
\newcommand{\bA}{\mathbf{A}}

\newcommand{\bB}{\mathbf{B}}

\newcommand{\bM}{\mathbf{M}}

\newcommand{\bK}{\mathbf{K}}

\newcommand{\thresh}{{\mathrm{thr}}} 

\newcommand{\esnd}{\varepsilon_\mathrm{snd}}

\newcommand{\srplus}{\textsuperscript{88}Sr\textsuperscript{+}}

\setcitestyle{super}

\usepackage[caption=false]{subfig}
\makeatletter
\renewcommand*{\fnum@figure}{{\normalfont\bfseries \figurename~\thefigure}}
\renewcommand*{\fnum@table}{{\normalfont\bfseries \tablename~\thetable}}
\renewcommand*{\@caption@fignum@sep}{\textbf{:~}}
\makeatother

\usepackage{xcolor}
\definecolor{lasallegreen}{rgb}{0.03, 0.47, 0.19}
\definecolor{darkblue}{rgb}{0.0, 0.0, 0.745}

\begin{document}
\renewcommand{\figurename}{Fig.}
\crefname{figure}{Fig.}{Figs.}

\title{Experimental quantum key distribution certified by Bell's theorem}
\author{D.~P.~Nadlinger}
\author{P.~Drmota}
\author{B.~C.~Nichol}
\author{G.~Araneda}
\author{D.~Main}
\author{R.~Srinivas}
\author{D.~M.~Lucas}
\author{C.~J.~Ballance}
\affiliation{Department of Physics, University of Oxford, Clarendon Laboratory, Parks Road, Oxford OX1 3PU, U.K.}

\author{K.~Ivanov}
\affiliation{School of Computer and Communication Sciences, EPFL, 1015 Lausanne, Switzerland}

\author{E.~Y-Z.~Tan}
\affiliation{Institute for Theoretical Physics, ETH Zürich, 8093 Zürich, Switzerland}

\author{P.~Sekatski}
\affiliation{Department of Applied Physics, University of Geneva, Rue de l’École-de-Médecine, 1211 Geneva, Switzerland}

\author{R.~L.~Urbanke}
\affiliation{School of Computer and Communication Sciences, EPFL, 1015 Lausanne, Switzerland}

\author{R.~Renner}
\affiliation{Institute for Theoretical Physics, ETH Zürich, 8093 Zürich, Switzerland}

\author{N.~Sangouard}
\affiliation{Universit\'e Paris-Saclay, CEA, CNRS, Institut de physique th\'eorique, 91191, Gif-sur-Yvette, France}

\author{J-D.~Bancal}
\affiliation{Universit\'e Paris-Saclay, CEA, CNRS, Institut de physique th\'eorique, 91191, Gif-sur-Yvette, France}

\date{September 29, 2021; this revision as accepted to Nature: May 10, 2022}

\begin{abstract}\noindent%
Cryptographic key exchange protocols traditionally rely on computational conjectures such as the hardness of prime factorisation\cite{AdlemanCotA1978methodobtainingdigital} to provide security against eavesdropping attacks. Remarkably, quantum key distribution protocols like the one proposed by Bennett and Brassard\cite{Bennett84} provide information-theoretic security against such attacks, a much stronger form of security unreachable by classical means. However, quantum protocols realised so far are subject to a new class of attacks exploiting a mismatch between the quantum states or measurements implemented and their theoretical modelling, as demonstrated in numerous experiments\cite{Zhao08, Lydersen10, Gerhardt11, Weier11}. Here, we present the experimental realisation of a complete quantum key distribution protocol immune to these vulnerabilities, following Ekert's pioneering proposal\cite{Ekert91} to use entanglement to bound an adversary's information from Bell's theorem\cite{Bell64}. By combining theoretical developments with an improved optical fibre link generating entanglement between two trapped-ion qubits, we obtain \num{95628} key bits with device-independent security\cite{Mayers04,Acin07,Vazirani14,ArnonFriedman18} from \num{1.5} million Bell pairs created during eight hours of run time. We take steps to ensure that information on the measurement results is inaccessible to an eavesdropper. These measurements are performed without space-like separation. Our result shows that provably secure cryptography under general assumptions is possible with real-world devices, and paves the way for further quantum information applications based on the device-independence principle.
\end{abstract}

\maketitle

\noindent%
Private communication over shared network infrastructure is of fundamental importance to the modern world. Classically, shared secrets cannot be created with information-theoretic security; real-world key exchange protocols rely on unproven conjectures regarding the computational intractability of certain operations. Quantum theory, however, promises that measurements on two correlated quantum systems can yield identical outcomes that are fundamentally unpredictable to any third party. This possibility of generating secret correlated outcomes at a distance forms the basic idea of quantum key distribution (QKD)\cite{Gisin02, Scarani09, Lo14}. Importantly, the security guarantees provided by QKD are unique in that they do not rely on the assumption that the adversary has limited computational power. Instead, the only required assumptions are that: (i) quantum theory is correct, (ii) the parties can isolate their systems to prevent information leaking to an adversary, (iii) they can privately choose random classical inputs to their quantum devices, and (iv) they can process classical information on trusted computers.

Existing QKD systems rely on an additional assumption that is hard to satisfy in practice: they require the quantum states and measurements used to distribute the key to be accurately characterised (e.g.~including their dimension)\cite{Scarani09, Lo14, Acin06}. In other words, the quantum devices are assumed to be trusted and to maintain perfect calibration. Deviations from the expected behaviour can be difficult to detect, which has been exploited in a number of demonstrations where real-world QKD systems were compromised\cite{Zhao08, Lydersen10, Gerhardt11, Weier11}. So-called measurement-device-independent QKD protocols allow untrusted measurements to be used as part of the system but still require well-characterised sources\cite{Braunstein12, Lo12, Rubenok13, Ferreira13, Liu13}.

\begin{figure*}
    \includegraphics{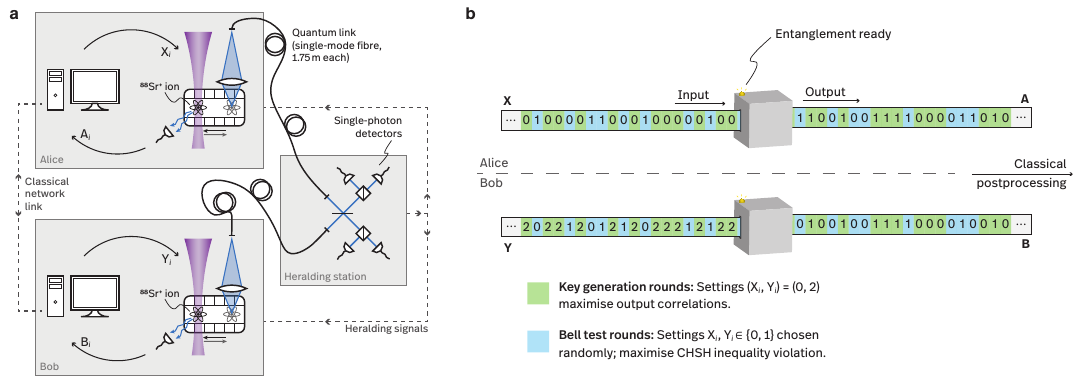}
    \caption{
        \textbf{DIQKD with trapped ions. a,} Alice and Bob each operate independent ion-trap nodes, here separated by about \SI{2}{\metre}, where one \srplus{} qubit is confined in a vacuum chamber. To establish entanglement the ions are simultaneously excited to an electronic state which spontaneously decays, during which a single photon each is emitted whose polarisation is entangled with the ion's internal state. These photons are collected into optical fibres using free-space optics and sent to a central station, where a probabilistic Bell-basis measurement is performed. This projects the ions into a maximally entangled state, heralded by the coincident detection of a pair of photons. On success, the ions are each transported to a different location in the trap to disconnect the photonic link such that local projective measurements can be applied without either the measurement basis settings $X_i$, $Y_i$ or the outcomes $A_i, B_i \in \{0,1\}$ leaking into the environment. The state is then reset and the link reconnected, and the process is repeated for a large number of rounds.
        \textbf{b,} In the device-independent security proof, no further assumptions are made about the internal workings of the quantum devices; they are modelled as \enquote{black boxes} with classical inputs and outputs. By randomly alternating between measurement settings realising a Bell test and settings which lead to highly correlated outputs, the output bit strings can be certified to originate from appropriate measurements on a quantum state that is close to being maximally entangled, ensuring secrecy.
    }
    \label{fig:schematic}
\end{figure*}

Device-independent QKD~(DIQKD) protocols\cite{Mayers04, Acin07, Vazirani14, ArnonFriedman18} make no additional assumptions about the physical apparatus. According to Bell's theorem, one can guarantee that two systems produce outcomes sharing exclusive correlations  -- preventing a third party from knowing these results -- without assuming how these outcomes are produced\cite{Brunner14}. This remarkable fact can be used to construct key distribution protocols with security guarantees independent of any assumption about the states measured and measurements performed. Rather, the underlying state and measurements are certified in the process~\cite{Mayers04}. Imperfections that might lead to key leakage in conventional QKD due to an inappropriate description of the internal quantum state or measurement instead just result in the protocol aborting. However, this enhanced security comes at the cost of far more stringent experimental requirements.
The certifiable amount of private information directly depends on the size of a Bell inequality violation, necessitating a platform capable of distributing and measuring high-quality entangled states while closing the detection loophole.
To successfully extract a shared key using state-of-the-art devices, a tight theoretical analysis and an efficient classical post-processing pipeline are needed, in particular regarding finite-size effects resulting from practical limits on the number of measurements.
Despite significant theoretical progress\cite{Barrett05,Acin07,Masanes09,Reichardt13,Vazirani14,ArnonFriedman18,Ho20,Schwonnek21,Woodhead21,Sekatski21,Brown21,Masini21} a practical demonstration of these protocols has remained out of reach.

Here we report the first experimental distribution of a key with device-independent security guarantees. Using a hybrid system where heralded entanglement between stationary trapped ion qubits is established via flying photonic qubits (see \cref{fig:schematic}), we are able to generate high-quality entangled states between two ions separated by about \SI{2}{\metre}, resulting in a record-high detection-loophole-free Bell inequality violation with isolated systems. We propose a concrete, practical DIQKD protocol, design a universal capacity-approaching error-correction code optimised for this context, use extractors and authentication schemes that are frugal in their randomness consumption, and provide a detailed finite-statistics security proof.
Combining these theoretical and experimental advances, we successfully obtain a key whose length is more than two orders of magnitude greater than the amount of private shared randomness consumed by the protocol.

\begin{figure*}
    \includegraphics{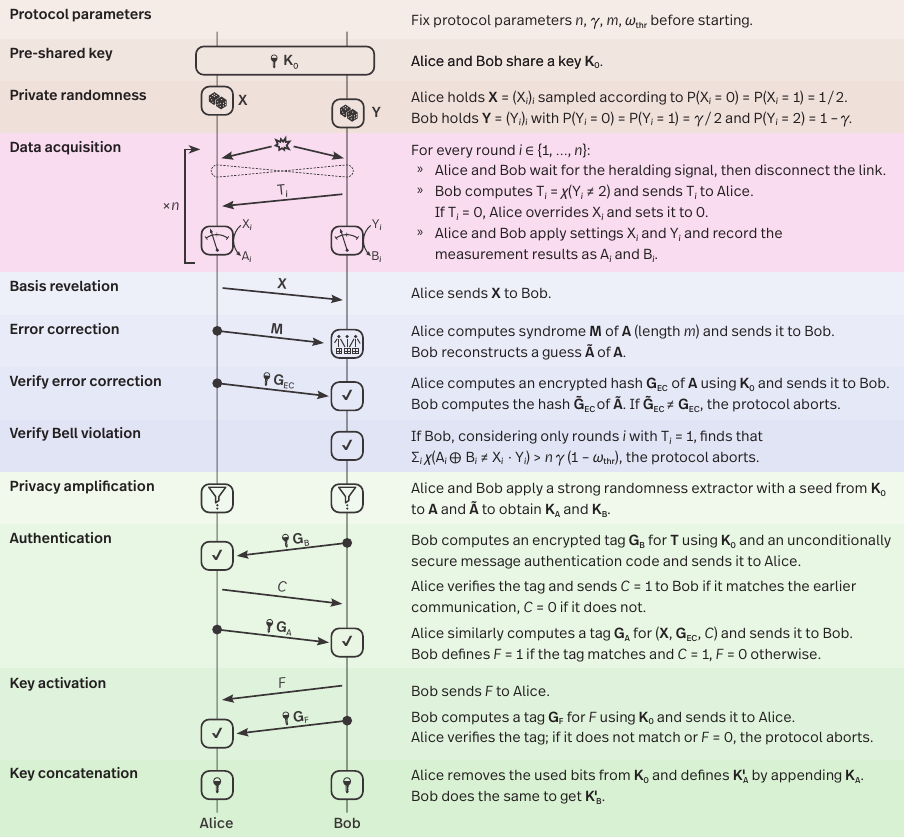}
    \caption{
        \textbf{DIQKD protocol structure.} Before the protocol starts, the number of rounds $n$, the probability $\gamma$ of each round being chosen to be Bell test round, an acceptance threshold $\omega_\thresh$ for the CHSH winning probability, and the length $m$ of the error correction syndrome are fixed. An initial key $\mathbf{K_0}$, mostly reusable, is required to seed the privacy amplification and authentication algorithms, and as a one-time pad to encrypt a few short messages (indicated using a key symbol).
        Arrows indicate the classical messages exchanged between the parties, bold letters strings consisting of multiple bits, $\chi$ the indicator function with $\chi(P) = 1$ if $P$ is true and $0$ otherwise.
    }
    \label{fig:protocol}
\end{figure*}

Our DIQKD protocol is illustrated in \cref{fig:protocol}. All parameters are chosen before the start of the protocol. The data acquisition phase of our protocol consists of $n$ sequential rounds. At the start of each, Alice and Bob wait for a valid heralding signal from the central heralding station indicating the creation of remote entanglement. For Bell test rounds, Alice and Bob randomly select inputs $X_i,Y_i \in \{0,1\}$ for their measurements, which are implemented so that the outcomes $A_i, B_i \in \{0,1\}$ maximise the probability of winning the Clauser-Horne-Shimony-Holt (CHSH) game\cite{CHSH69} $A_i \oplus B_i = X_i \cdot Y_i$. A high winning probability $\omega$, customarily expressed in terms of the CHSH score $S=4\,(2 \omega - 1)$, tightly bounds the information any adversary can have about the outcomes. For so-called key generation rounds, the inputs are fixed to $X_i=0$ and $Y_i=2$, maximising output correlations as quantified by a low quantum bit error rate $Q=P(A_i\neq B_i|X_i=0,Y_i=2)$. Bob randomly chooses between the round types after the heralding signal is received and the links are disconnected, choosing Bell test rounds with probability $\gamma$, and communicates this choice to Alice, which avoids sifting (discarding of rounds with mismatched measurement bases).
The parties keep private records of their measurement settings $\bX = X_1 X_2 \ldots$ and $\bY = Y_1 Y_2\ldots$ as well as the outcomes $\bA = A_1 A_2 \ldots$ and $\bB = B_1 B_2 \ldots$.
After the $n$ measurement rounds are complete, Alice and Bob need to verify the CHSH score and extract a shared key from the noisy output correlations.
Alice openly sends her inputs $\bX$ to Bob, along with a shorter error-correction syndrome string $\bM$ which allows Bob to reconstruct a copy $\tilde\bA$ of Alice's outcomes $\bA$. Now holding $\bX\bY\tilde\bA\bB$, Bob is able to verify whether the CHSH score achieved during the Bell test rounds exceeds a pre-agreed threshold. If this it not the case, or if the reconstruction of $\tilde\bA$ fails, the security might be compromised so the protocol aborts. Otherwise, the parties locally process the outcomes $\bA$ and $\tilde\bA$ to extract identical final keys, with a guaranteed, arbitrarily low, bound on the information leaked to any adversary.

A crucial step in the security analysis is to infer an upper bound on the knowledge an adversary can have about Alice's outcomes. In the device-independent setting, where quantum states and measurements are a priori uncharacterised and where we allow memory effects to correlate successive measurement rounds, obtaining a precise bound is nontrivial. A recent breakthrough in this respect is the entropy accumulation theorem (EAT)\cite{Dupuis20,ArnonFriedman18}. It shows that in an $n$-round protocol characterised by a CHSH score $S$, the amount of randomness generated by both parties, with respect to an adversary limited by quantum theory, is lower-bounded by $n\, h(S)$ up to a correction of order $\sqrt{n}$, where $h(S)$ is the worst-case von Neumann entropy of the outcomes for an individual round of the protocol compatible with the score $S$. Using this result, we derive a bound on Alice's private randomness from the CHSH winning probability threshold $\omega_\thresh$ chosen before start of the protocol\cite{ArnonFriedman18}. We use a recent version of the EAT\cite{Liu21} for its improved statistical performance, proving security against the most general types of attacks (as detailed in the Supplementary Material).

The length of the final key is given by the difference between the amount of entropy certified by the EAT and the amount of information revealed through the classical messages exchanged by the parties as part of the protocol.
To obtain a positive key rate it is therefore crucial to reconcile Alice and Bob's outcomes with an error correction syndrome that is as short as possible. Asymptotically, the cost of reconciliation per round is given by $\operatorname{H}(\bA|\bX\bY\bB)$, the entropy of Alice's outcomes conditioned on the knowledge of inputs and Bob's outcomes. Reconciliation in presence of finite statistics however comes at a higher price and the best algorithms are often not realisable in practice. In fact, all finite-statistics DIQKD analyses so far have assumed computationally intractable error correction schemes with decodings involving  hash pre-image computations~\cite{Vazirani14,ArnonFriedman18,Murta19,Tan20}.
We address this challenge using a custom error correction solution based on spatially-coupled low-density parity-check (SC-LDPC) encoding. Our scheme admits an efficient decoding algorithm and requires less overhead than some previously considered impractical codes (see Supplementary Material).

\begin{figure}
    \includegraphics{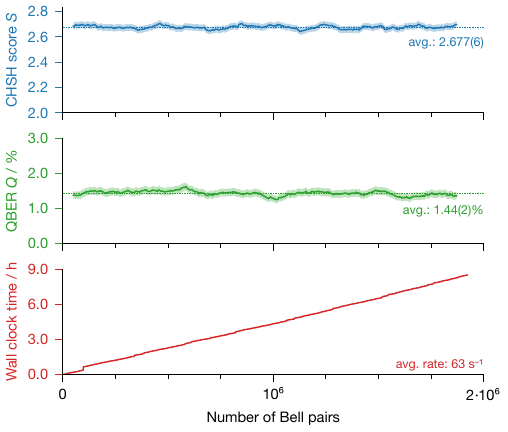}
    \caption{
        \textbf{Quantum link performance.}
        A separate characterisation run of \num{1920000} total Bell pairs shows stable link performance.
        Inputs and outputs from both nodes were collected to compute a moving average for CHSH score and quantum bit error rate (window length: \num{100000} rounds; test round fraction: $\gamma = 1 / 2$). Shaded bands indicate \SI{95}{\percent} confidence intervals from binomial statistics in each window.
        The bottom panel shows the acquisition timestamp for each Bell pair during the \SI{8.5}{\hour} experiment duration. The vertical steps, where time passed without heralds, correspond to ion loss and recalibration events.
    }
    \label{fig:link-performance}
\end{figure}

In our experiment, Alice and Bob each control a trapped-ion quantum network node\cite{moehringEntanglementSingleatomQuantum2007,stephensonHighRateHighFidelityEntanglement2020} holding a single \srplus{} ion qubit each. Dissipative and coherent operations are implemented using resonant laser fields, which each node derives from a shared set of laser sources. Entanglement between the ion and the polarisation of a spontaneously emitted \SI{422}{\nano\metre} photon is locally created at each node by excitation to a short-lived state with two decay paths to the Zeeman-split qubit states in the ground level. The photons are coupled into single-mode optical fibres connecting the nodes to a central heralding station, where a linear optics setup and four avalanche photodetectors implement a Bell-basis measurement (see \cref{fig:schematic}a and Supplementary Material). The overall detection efficiency for each spontaneous decay photon is $\approx \SI{2}{\percent}$, necessitating many attempts to observe a two-photon coincidence heralding the successful creation of entanglement between Alice's and Bob's ion qubits.
We perform state tomography to characterise the ion-ion entanglement, measuring a fidelity to the closest maximally entangled state of \SI{96.0(1)}{\percent}, which exceeds previously reported measurements for remote entanglement between matter qubits\cite{maunzHeraldedQuantumGate2009,lettnerRemoteEntanglementSingle2011,Ritter2012,Hensen2015,rosenfeldEventReadyBellTest2017,stephensonHighRateHighFidelityEntanglement2020}.

\begin{figure}
    \includegraphics{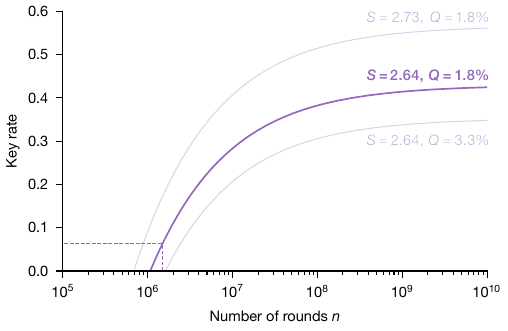}
    \caption{\textbf{Finite-statistics key rate.} Certified key rate as a function of the number of rounds for $S=2.64$, $Q=1.8\%$, $\gamma=13/256$ and $\esnd=10^{-10}$. We operate the DIQKD protocol at the point $n=1.5\times 10^6$ on this curve, corresponding to a threshold winning probability of $\omega_\thresh \approx 0.825538$ and a key rate of $\num{95884} / n \approx 6.39\%$. Two key rate curves for alternate parameter choices illustrate its sensitivity to the link performance.}
\end{figure}

To implement the \enquote{black box} primitives for DIQKD, each node first disconnects the qubit from the optical link following a herald event by shuttling it away from the focus of the fibre-coupling objective lens, which ensures the basis choices and measurement results do not leak to a third party (see Methods). Bob decides between Bell test and key generation rounds using private randomness obtained from a commercial quantum random number generator, sending the result to Alice in real time. In case of a Bell test round, both nodes then similarly use a private random string to choose between the coherent operations defining the measurement bases, i.e.~the inputs $X_i$ and $Y_i$.
As each DIQKD round only starts after a heralding success, this approach is intrinsically detection-loophole-free; losses in the optical links and limited detector quantum efficiencies, which pose a challenge for DIQKD on photonic platforms\cite{Murta19}, affect the heralding rate, but not $S$ or $Q$.
The measurements of one party are not space-like separated from the choice of measurements of the other party.
We characterise the link performance by measuring the probabilities $P(A_i B_i | X_i Y_i)$ in a separate experiment. A representative dataset, shown in \cref{fig:link-performance}, results in a CHSH score of $S = \num{2.677(6)}$ and a quantum bit error rate of $Q = \SI{1.44(2)}{\percent}$.

We now implement the complete protocol as described in \cref{fig:protocol}, using Trevisan's quantum-proof construction as a strong randomness extractor and short tags based on a strongly universal hash family for classical message authentication. The two nodes are operated completely independently: they have independent control systems, and the input and output data is only exchanged over a classical network link as prescribed by the protocol (see Methods). In order to guarantee a high success probability of the protocol, we conservatively choose $n = \num{1.5e6}$, $\gamma \approx \SI{5.1}{\percent}$, $\omega_\thresh \approx \num{0.825538}$,
and an error correction code with syndrome length $m = \num{296517}$ based on the previous link characterisation (see Supplementary Material for more details).
We acquire data over \SI{7.9}{\hour} of wall-clock time (with one pause of \SI{4.4}{\hour} due to a laser failure), and successfully extract a secret key of \num{95884} bits with soundness error $\esnd=10^{-10}$, with a post-processing time of \SI{5}{\minute} (see \cref{tab:parameters}). This greatly exceeds the \num{256}~bits of private randomness consumed during the protocol, thus marking the first demonstration of a fully device-independent QKD.
\setlength{\extrarowheight}{1pt}
\begin{table}
\begin{tabular}{c||c|c|c}
Protocol step & Consumed & Used (reusable) & Generated \\
\hline\hline
Validation of & \multirow{2}{*}{64} & \multirow{4}{*}{\num{1280}} & \multirow{2}{*}{0}\\
error correction &  &  & \\
\cline{1-2}\cline{4-4}
Authentication & 128 &  & 0\\
\cline{1-2}\cline{4-4}
Key activation & 64 & & 0\\
\hline
Privacy & \multirow{2}{*}{0} & \multirow{2}{*}{\num{1201886}} & \multirow{2}{*}{\num{95884}} \\
amplification & & & \\
\hline\hline
Total & \num{256} & \num{1203166} & \num{95884}
\end{tabular}
\caption{\textbf{Secret key balance sheet.} The experiment requires an initial \num{1203422}-bits long key $\bK_0$ shared between Alice and Bob. \num{256} of these bits are consumed by the protocol while generating \num{95884} new secret shared bits. This results in the creation of a longer \num{1299050}-bits long shared secret key $\bK_1=\bK_A'$ between Alice and Bob, effectively extending the initial key by \num{95628} bits.}
\label{tab:parameters}
\end{table}

The resulting key is secure under very general assumptions, including attacks where the adversary can act arbitrarily on the quantum side information from all rounds. Moreover, the measurement devices can have memory and operate according to the results of previous rounds, and the security of the protocol is composable with other cryptographic protocols using different devices\cite{Portmann21,Barrett13}.
Out of the remaining assumptions (i)-(iv), isolation is arguably the most critical one from a practical perspective. Space-like separation between measurement events could exclude some types of leakage but does not guarantee isolation\cite{Pironio10,Pironio09,Murta19,Pirandola20} (see Methods). We envision that a stronger confidence in the isolation of the nodes will be possible when all quantum systems can be distributed and stored in advance~\cite{Heshami16}, thereby avoiding the necessity for quantum communication during the protocol.
The present apparatus could potentially be adapted to span building-scale local-area networks, with optical fibre losses at the \SI{422}{\nano\metre} wavelength reducing the rate by approximately a factor of two every hundred metres. Key distribution over kilometres is in principle possible with downconversion to the telecom band\cite{wrightTwoWayPhotonicInterface2018}, or through free-space optics. Improved key-generation rates might be achievable through improvements to the photon collection efficiency, for instance through the use of optical cavities\cite{schuppInterfaceTrappedIonQubits2021}. Combining multiple instances of telecom and cavity-enhanced links into a quantum repeater architecture\cite{Sangouard11} might enable fully quantum-secure long distance communication.

\section*{Methods}

\textbf{Loopholes.}
In the context of Bell experiments seeking to disprove local causality, several \enquote{loopholes} have been discussed\cite{Brunner14}, referring to requirements in the theoretical analysis not met in experimental realisations (consequently permitting a locally causal explanation for the observations even in the presence of a Bell inequality violation). One of them, known as the \enquote{locality loophole}, is the requirement that no information about the input of one party be known by the device of the other party before it produces an outcome. In a fundamental test of local causality, information transfer of any kind must be excluded, even that potentially described by an as-yet-unknown theory. Under the assumption that measurement choices are made locally in a truly random manner so that no information about them is available in advance, it is generally admitted that this loophole can be closed through space-like separation by requiring that the measurements of both parties are space-like separated from the events defining the other party's setting choice. The condition that the inputs of one party be unknown to the other party upon measurement is also necessary for DIQKD (see Supplementary Material for more details). Space-like separation between the parties' measurement events is advantageous for this and can possibly already eliminate specific attacks by itself. However, the cryptographic context is more demanding. Not only must the input choices be unknown at other locations but also, even more obviously, the output results. Moreover, no cryptographic security can be achieved if information is allowed to travel to another party (e.g.~the adversary), even if the speed of that transfer is limited to that of light. Therefore, space-like separation is not sufficient for DIQKD and instead, the parties must be well isolated\cite{Pironio09,Pirandola20}. Isolation of each node, i.e.~control over the flow of information from and to the outside, is thus an unavoidable assumption in any key exchange protocol and must be achieved through separate means. The locality condition being a consequence of the isolation assumption, no further conditions such as space-like separation are needed: if local information can be kept locally, then it is not available to a distant device. The isolation assumption has however no consequence on the \enquote{detection loophole}, implying that a similar treatment of this loophole is needed in the context of DIQKD as for conventional Bell tests.

\textbf{Link disconnection and isolation.}
By design, the trapped-ion quantum bits are well-coupled to the optical fibre linking the nodes during the entanglement generation phase.
To meet the isolation assumption, the ions in each node are shuttled \SI{3}{\micro\metre} from the focus of the imaging system after a successful herald by modifying the trapping potential. This way, the coupling from Alice's and Bob's qubits to the outside of their laboratories is reduced by a factor of $>10^4$ before the basis choices are applied and the states are read out. Similarly, the qubit state left after readout is scrambled by applying deshelving and cooling lasers for \SI{100}{\micro\second} (${>10}$ time constants) before reconnecting the link. The information that could possibly leak through the remaining weak coupling is insufficient to compromise our conclusion of a positive key rate. If desired, one could arbitrarily reduce the amount of leaked information by shuttling the ion a further distance from the focus (${>\SI{1}{\milli\metre}}$ is accessible in the trap chips used here), or by adding a macroscopic mechanism to disconnect the optical fibre, e.g.~a fast micro-electromechanical fibre switch.
The laser control fields for both nodes were derived from the same set of sources. An independent set of laser sources could be used instead, as the only synchronisation requirements are for the qubit frequencies to be matched and for the single-photon wave packets at the heralding station to be well-aligned compared to their \SI{7}{\nano\second} duration.

\textbf{Data handling.}
We make extensive efforts to perform our experiments in a fashion representative of a real deployment and compatible with the security and isolation assumptions throughout. Each node is driven by an independent instance of the Advanced Real-Time Infrastructure for Quantum physics (ARTIQ) open-source real-time control system~\cite{ARTIQ}; the combination of personal computer (PC) and Field Programmable Gate Arrays (FPGA) makes up the trusted local classical computation hardware as per the assumptions. The random bit strings for the CHSH basis and round type choices are derived from a commercial quantum random number generator and stored in computer memory before the start of the experiment. No post-selection takes place; a heralding event always results in a bit added to the final $\bA$ and $\bB$ strings. The outcome strings, along with other private data, neither leave the respective node PCs, nor are they manually inspected. A classical network link connects the two nodes, used to exchange the public messages required by the protocol (e.g.~the error correction syndrome data), and to coordinate the overall experiment schedule (e.g.~for periodic calibrations, or to remedy ion loss events).

\section*{Data availability}
The datasets generated during the current study are available from DPN and CJB on reasonable request.

\section*{Code availability}
The custom code generated during the current study is available from the corresponding authors on reasonable request.

\bibliography{references}

\begin{thebibliography}{53}%
\makeatletter
\providecommand \@ifxundefined [1]{%
 \@ifx{#1\undefined}
}%
\providecommand \@ifnum [1]{%
 \ifnum #1\expandafter \@firstoftwo
 \else \expandafter \@secondoftwo
 \fi
}%
\providecommand \@ifx [1]{%
 \ifx #1\expandafter \@firstoftwo
 \else \expandafter \@secondoftwo
 \fi
}%
\providecommand \natexlab [1]{#1}%
\providecommand \enquote  [1]{``#1''}%
\providecommand \bibnamefont  [1]{#1}%
\providecommand \bibfnamefont [1]{#1}%
\providecommand \citenamefont [1]{#1}%
\providecommand \href@noop [0]{\@secondoftwo}%
\providecommand \href [0]{\begingroup \@sanitize@url \@href}%
\providecommand \@href[1]{\@@startlink{#1}\@@href}%
\providecommand \@@href[1]{\endgroup#1\@@endlink}%
\providecommand \@sanitize@url [0]{\catcode `\\12\catcode `\$12\catcode
  `\&12\catcode `\#12\catcode `\^12\catcode `\_12\catcode `\%12\relax}%
\providecommand \@@startlink[1]{}%
\providecommand \@@endlink[0]{}%
\providecommand \url  [0]{\begingroup\@sanitize@url \@url }%
\providecommand \@url [1]{\endgroup\@href {#1}{\urlprefix }}%
\providecommand \urlprefix  [0]{URL }%
\providecommand \Eprint [0]{\href }%
\providecommand \doibase [0]{https://doi.org/}%
\providecommand \selectlanguage [0]{\@gobble}%
\providecommand \bibinfo  [0]{\@secondoftwo}%
\providecommand \bibfield  [0]{\@secondoftwo}%
\providecommand \translation [1]{[#1]}%
\providecommand \BibitemOpen [0]{}%
\providecommand \bibitemStop [0]{}%
\providecommand \bibitemNoStop [0]{.\EOS\space}%
\providecommand \EOS [0]{\spacefactor3000\relax}%
\providecommand \BibitemShut  [1]{\csname bibitem#1\endcsname}%
\let\auto@bib@innerbib\@empty
\bibitem [{\citenamefont {Rivest}\ \emph {et~al.}(1978)\citenamefont {Rivest},
  \citenamefont {Shamir},\ and\ \citenamefont
  {Adleman}}]{AdlemanCotA1978methodobtainingdigital}%
  \BibitemOpen
  \bibfield  {author} {\bibinfo {author} {\bibfnamefont {R.~L.}\ \bibnamefont
  {Rivest}}, \bibinfo {author} {\bibfnamefont {A.}~\bibnamefont {Shamir}},\
  and\ \bibinfo {author} {\bibfnamefont {L.}~\bibnamefont {Adleman}},\
  }\bibfield  {title} {\bibinfo {title} {A method for obtaining digital
  signatures and public-key cryptosystems},\ }\href
  {https://doi.org/10.1145/359340.359342} {\bibfield  {journal} {\bibinfo
  {journal} {Commun. ACM}\ }\textbf {\bibinfo {volume} {21}},\ \bibinfo {pages}
  {120} (\bibinfo {year} {1978})}\BibitemShut {NoStop}%
\bibitem [{\citenamefont {Bennett}\ and\ \citenamefont
  {Brassard}(1984)}]{Bennett84}%
  \BibitemOpen
  \bibfield  {author} {\bibinfo {author} {\bibfnamefont {C.~H.}\ \bibnamefont
  {Bennett}}\ and\ \bibinfo {author} {\bibfnamefont {G.}~\bibnamefont
  {Brassard}},\ }\bibfield  {title} {\bibinfo {title} {Quantum cryptography:
  public key distribution and coin tossing.},\ }\href@noop {} {\bibfield
  {journal} {\bibinfo  {journal} {Theor. Comput. Sci.}\ }\textbf {\bibinfo
  {volume} {560}},\ \bibinfo {pages} {7} (\bibinfo {year} {1984})}\BibitemShut
  {NoStop}%
\bibitem [{\citenamefont {Zhao}\ \emph {et~al.}(2008)\citenamefont {Zhao},
  \citenamefont {Fung}, \citenamefont {Qi}, \citenamefont {Chen},\ and\
  \citenamefont {Lo}}]{Zhao08}%
  \BibitemOpen
  \bibfield  {author} {\bibinfo {author} {\bibfnamefont {Y.}~\bibnamefont
  {Zhao}}, \bibinfo {author} {\bibfnamefont {C.-H.~F.}\ \bibnamefont {Fung}},
  \bibinfo {author} {\bibfnamefont {B.}~\bibnamefont {Qi}}, \bibinfo {author}
  {\bibfnamefont {C.}~\bibnamefont {Chen}},\ and\ \bibinfo {author}
  {\bibfnamefont {H.-K.}\ \bibnamefont {Lo}},\ }\bibfield  {title} {\bibinfo
  {title} {{Quantum hacking: Experimental demonstration of time-shift attack
  against practical quantum-key-distribution systems}},\ }\href
  {https://doi.org/10.1103/PhysRevA.78.042333} {\bibfield  {journal} {\bibinfo
  {journal} {Phys. Rev. A}\ }\textbf {\bibinfo {volume} {78}},\ \bibinfo
  {pages} {042333} (\bibinfo {year} {2008})}\BibitemShut {NoStop}%
\bibitem [{\citenamefont {Lydersen}\ \emph {et~al.}(2010)\citenamefont
  {Lydersen}, \citenamefont {Wiechers}, \citenamefont {Wittmann}, \citenamefont
  {Elser}, \citenamefont {Skaar},\ and\ \citenamefont {Makarov}}]{Lydersen10}%
  \BibitemOpen
  \bibfield  {author} {\bibinfo {author} {\bibfnamefont {L.}~\bibnamefont
  {Lydersen}}, \bibinfo {author} {\bibfnamefont {C.}~\bibnamefont {Wiechers}},
  \bibinfo {author} {\bibfnamefont {C.}~\bibnamefont {Wittmann}}, \bibinfo
  {author} {\bibfnamefont {D.}~\bibnamefont {Elser}}, \bibinfo {author}
  {\bibfnamefont {J.}~\bibnamefont {Skaar}},\ and\ \bibinfo {author}
  {\bibfnamefont {V.}~\bibnamefont {Makarov}},\ }\bibfield  {title} {\bibinfo
  {title} {{Hacking commercial quantum cryptography systems by tailored bright
  illumination}},\ }\href {https://doi.org/10.1038/nphoton.2010.214} {\bibfield
   {journal} {\bibinfo  {journal} {Nat. Photonics}\ }\textbf {\bibinfo {volume}
  {4}},\ \bibinfo {pages} {686} (\bibinfo {year} {2010})}\BibitemShut {NoStop}%
\bibitem [{\citenamefont {Gerhardt}\ \emph {et~al.}(2011)\citenamefont
  {Gerhardt}, \citenamefont {Liu}, \citenamefont {Lamas-Linares}, \citenamefont
  {Skaar}, \citenamefont {Kurtsiefer},\ and\ \citenamefont
  {Makarov}}]{Gerhardt11}%
  \BibitemOpen
  \bibfield  {author} {\bibinfo {author} {\bibfnamefont {I.}~\bibnamefont
  {Gerhardt}}, \bibinfo {author} {\bibfnamefont {Q.}~\bibnamefont {Liu}},
  \bibinfo {author} {\bibfnamefont {A.}~\bibnamefont {Lamas-Linares}}, \bibinfo
  {author} {\bibfnamefont {J.}~\bibnamefont {Skaar}}, \bibinfo {author}
  {\bibfnamefont {C.}~\bibnamefont {Kurtsiefer}},\ and\ \bibinfo {author}
  {\bibfnamefont {V.}~\bibnamefont {Makarov}},\ }\bibfield  {title} {\bibinfo
  {title} {Full-field implementation of a perfect eavesdropper on a quantum
  cryptography system},\ }\bibfield  {journal} {\bibinfo  {journal} {Nat.
  Commun.}\ }\textbf {\bibinfo {volume} {2}},\ \href
  {https://doi.org/10.1038/ncomms1348} {10.1038/ncomms1348} (\bibinfo {year}
  {2011})\BibitemShut {NoStop}%
\bibitem [{\citenamefont {Weier}\ \emph {et~al.}(2011)\citenamefont {Weier},
  \citenamefont {Krauss}, \citenamefont {Rau}, \citenamefont {F{\"u}rst},
  \citenamefont {Nauerth},\ and\ \citenamefont {Weinfurter}}]{Weier11}%
  \BibitemOpen
  \bibfield  {author} {\bibinfo {author} {\bibfnamefont {H.}~\bibnamefont
  {Weier}}, \bibinfo {author} {\bibfnamefont {H.}~\bibnamefont {Krauss}},
  \bibinfo {author} {\bibfnamefont {M.}~\bibnamefont {Rau}}, \bibinfo {author}
  {\bibfnamefont {M.}~\bibnamefont {F{\"u}rst}}, \bibinfo {author}
  {\bibfnamefont {S.}~\bibnamefont {Nauerth}},\ and\ \bibinfo {author}
  {\bibfnamefont {H.}~\bibnamefont {Weinfurter}},\ }\bibfield  {title}
  {\bibinfo {title} {Quantum eavesdropping without interception: an attack
  exploiting the dead time of single-photon detectors},\ }\href@noop {}
  {\bibfield  {journal} {\bibinfo  {journal} {New J. Phys.}\ }\textbf {\bibinfo
  {volume} {13}},\ \bibinfo {pages} {073024} (\bibinfo {year}
  {2011})}\BibitemShut {NoStop}%
\bibitem [{\citenamefont {Ekert}(1991)}]{Ekert91}%
  \BibitemOpen
  \bibfield  {author} {\bibinfo {author} {\bibfnamefont {A.~K.}\ \bibnamefont
  {Ekert}},\ }\bibfield  {title} {\bibinfo {title} {{Quantum cryptography based
  on Bell’s theorem}},\ }\href@noop {} {\bibfield  {journal} {\bibinfo
  {journal} {Phys. Rev. Lett.}\ }\textbf {\bibinfo {volume} {67}},\ \bibinfo
  {pages} {661} (\bibinfo {year} {1991})}\BibitemShut {NoStop}%
\bibitem [{\citenamefont {Bell}(1964)}]{Bell64}%
  \BibitemOpen
  \bibfield  {author} {\bibinfo {author} {\bibfnamefont {J.~S.}\ \bibnamefont
  {Bell}},\ }\bibfield  {title} {\bibinfo {title} {{On the Einstein Podolsky
  Rosen paradox}},\ }\href
  {https://doi.org/10.1103/PhysicsPhysiqueFizika.1.195} {\bibfield  {journal}
  {\bibinfo  {journal} {Phys. Phys. Fiz}\ }\textbf {\bibinfo {volume} {1}},\
  \bibinfo {pages} {195} (\bibinfo {year} {1964})}\BibitemShut {NoStop}%
\bibitem [{\citenamefont {Mayers}\ and\ \citenamefont {Yao}(2004)}]{Mayers04}%
  \BibitemOpen
  \bibfield  {author} {\bibinfo {author} {\bibfnamefont {D.}~\bibnamefont
  {Mayers}}\ and\ \bibinfo {author} {\bibfnamefont {A.}~\bibnamefont {Yao}},\
  }\bibfield  {title} {\bibinfo {title} {{Self Testing Quantum Apparatus}},\
  }\href@noop {} {\bibfield  {journal} {\bibinfo  {journal} {Quantum Info.
  Comput.}\ }\textbf {\bibinfo {volume} {4}},\ \bibinfo {pages} {273–286}
  (\bibinfo {year} {2004})}\BibitemShut {NoStop}%
\bibitem [{\citenamefont {Ac\'{\i}n}\ \emph {et~al.}(2007)\citenamefont
  {Ac\'{\i}n}, \citenamefont {Brunner}, \citenamefont {Gisin}, \citenamefont
  {Massar}, \citenamefont {Pironio},\ and\ \citenamefont {Scarani}}]{Acin07}%
  \BibitemOpen
  \bibfield  {author} {\bibinfo {author} {\bibfnamefont {A.}~\bibnamefont
  {Ac\'{\i}n}}, \bibinfo {author} {\bibfnamefont {N.}~\bibnamefont {Brunner}},
  \bibinfo {author} {\bibfnamefont {N.}~\bibnamefont {Gisin}}, \bibinfo
  {author} {\bibfnamefont {S.}~\bibnamefont {Massar}}, \bibinfo {author}
  {\bibfnamefont {S.}~\bibnamefont {Pironio}},\ and\ \bibinfo {author}
  {\bibfnamefont {V.}~\bibnamefont {Scarani}},\ }\bibfield  {title} {\bibinfo
  {title} {{Device-Independent Security of Quantum Cryptography against
  Collective Attacks}},\ }\href {https://doi.org/10.1103/PhysRevLett.98.230501}
  {\bibfield  {journal} {\bibinfo  {journal} {Phys. Rev. Lett.}\ }\textbf
  {\bibinfo {volume} {98}},\ \bibinfo {pages} {230501} (\bibinfo {year}
  {2007})}\BibitemShut {NoStop}%
\bibitem [{\citenamefont {Vazirani}\ and\ \citenamefont
  {Vidick}(2014)}]{Vazirani14}%
  \BibitemOpen
  \bibfield  {author} {\bibinfo {author} {\bibfnamefont {U.}~\bibnamefont
  {Vazirani}}\ and\ \bibinfo {author} {\bibfnamefont {T.}~\bibnamefont
  {Vidick}},\ }\bibfield  {title} {\bibinfo {title} {{Fully Device-Independent
  Quantum Key Distribution}},\ }\href
  {https://doi.org/10.1103/PhysRevLett.113.140501} {\bibfield  {journal}
  {\bibinfo  {journal} {Phys. Rev. Lett.}\ }\textbf {\bibinfo {volume} {113}},\
  \bibinfo {pages} {140501} (\bibinfo {year} {2014})}\BibitemShut {NoStop}%
\bibitem [{\citenamefont {Arnon-Friedman}\ \emph {et~al.}(2018)\citenamefont
  {Arnon-Friedman}, \citenamefont {Dupuis}, \citenamefont {Fawzi},
  \citenamefont {Renner},\ and\ \citenamefont {Vidick}}]{ArnonFriedman18}%
  \BibitemOpen
  \bibfield  {author} {\bibinfo {author} {\bibfnamefont {R.}~\bibnamefont
  {Arnon-Friedman}}, \bibinfo {author} {\bibfnamefont {F.}~\bibnamefont
  {Dupuis}}, \bibinfo {author} {\bibfnamefont {O.}~\bibnamefont {Fawzi}},
  \bibinfo {author} {\bibfnamefont {R.}~\bibnamefont {Renner}},\ and\ \bibinfo
  {author} {\bibfnamefont {T.}~\bibnamefont {Vidick}},\ }\bibfield  {title}
  {\bibinfo {title} {Practical device-independent quantum cryptography via
  entropy accumulation},\ }\href@noop {} {\bibfield  {journal} {\bibinfo
  {journal} {Nat. Commun.}\ }\textbf {\bibinfo {volume} {9}},\ \bibinfo {pages}
  {459} (\bibinfo {year} {2018})}\BibitemShut {NoStop}%
\bibitem [{\citenamefont {Gisin}\ \emph {et~al.}(2002)\citenamefont {Gisin},
  \citenamefont {Ribordy}, \citenamefont {Tittel},\ and\ \citenamefont
  {Zbinden}}]{Gisin02}%
  \BibitemOpen
  \bibfield  {author} {\bibinfo {author} {\bibfnamefont {N.}~\bibnamefont
  {Gisin}}, \bibinfo {author} {\bibfnamefont {G.}~\bibnamefont {Ribordy}},
  \bibinfo {author} {\bibfnamefont {W.}~\bibnamefont {Tittel}},\ and\ \bibinfo
  {author} {\bibfnamefont {H.}~\bibnamefont {Zbinden}},\ }\bibfield  {title}
  {\bibinfo {title} {{Quantum cryptography}},\ }\href
  {https://doi.org/10.1103/RevModPhys.74.145} {\bibfield  {journal} {\bibinfo
  {journal} {Rev. Mod. Phys.}\ }\textbf {\bibinfo {volume} {74}},\ \bibinfo
  {pages} {145} (\bibinfo {year} {2002})}\BibitemShut {NoStop}%
\bibitem [{\citenamefont {Scarani}\ \emph {et~al.}(2009)\citenamefont
  {Scarani}, \citenamefont {Bechmann-Pasquinucci}, \citenamefont {Cerf},
  \citenamefont {Du\ifmmode~\check{s}\else \v{s}\fi{}ek}, \citenamefont
  {L\"utkenhaus},\ and\ \citenamefont {Peev}}]{Scarani09}%
  \BibitemOpen
  \bibfield  {author} {\bibinfo {author} {\bibfnamefont {V.}~\bibnamefont
  {Scarani}}, \bibinfo {author} {\bibfnamefont {H.}~\bibnamefont
  {Bechmann-Pasquinucci}}, \bibinfo {author} {\bibfnamefont {N.~J.}\
  \bibnamefont {Cerf}}, \bibinfo {author} {\bibfnamefont {M.}~\bibnamefont
  {Du\ifmmode~\check{s}\else \v{s}\fi{}ek}}, \bibinfo {author} {\bibfnamefont
  {N.}~\bibnamefont {L\"utkenhaus}},\ and\ \bibinfo {author} {\bibfnamefont
  {M.}~\bibnamefont {Peev}},\ }\bibfield  {title} {\bibinfo {title} {{The
  security of practical quantum key distribution}},\ }\href
  {https://doi.org/10.1103/RevModPhys.81.1301} {\bibfield  {journal} {\bibinfo
  {journal} {Rev. Mod. Phys.}\ }\textbf {\bibinfo {volume} {81}},\ \bibinfo
  {pages} {1301} (\bibinfo {year} {2009})}\BibitemShut {NoStop}%
\bibitem [{\citenamefont {Lo}\ \emph {et~al.}(2014)\citenamefont {Lo},
  \citenamefont {Curty},\ and\ \citenamefont {Tamaki}}]{Lo14}%
  \BibitemOpen
  \bibfield  {author} {\bibinfo {author} {\bibfnamefont {H.-K.}\ \bibnamefont
  {Lo}}, \bibinfo {author} {\bibfnamefont {M.}~\bibnamefont {Curty}},\ and\
  \bibinfo {author} {\bibfnamefont {K.}~\bibnamefont {Tamaki}},\ }\bibfield
  {title} {\bibinfo {title} {{Secure quantum key distribution}},\ }\href
  {https://doi.org/10.1038/nphoton.2014.149} {\bibfield  {journal} {\bibinfo
  {journal} {Nat. Photonics}\ }\textbf {\bibinfo {volume} {8}},\ \bibinfo
  {pages} {595–604} (\bibinfo {year} {2014})}\BibitemShut {NoStop}%
\bibitem [{\citenamefont {Ac\'{\i}n}\ \emph {et~al.}(2006)\citenamefont
  {Ac\'{\i}n}, \citenamefont {Gisin},\ and\ \citenamefont {Masanes}}]{Acin06}%
  \BibitemOpen
  \bibfield  {author} {\bibinfo {author} {\bibfnamefont {A.}~\bibnamefont
  {Ac\'{\i}n}}, \bibinfo {author} {\bibfnamefont {N.}~\bibnamefont {Gisin}},\
  and\ \bibinfo {author} {\bibfnamefont {L.}~\bibnamefont {Masanes}},\
  }\bibfield  {title} {\bibinfo {title} {{From Bell's Theorem to Secure Quantum
  Key Distribution}},\ }\href {https://doi.org/10.1103/PhysRevLett.97.120405}
  {\bibfield  {journal} {\bibinfo  {journal} {Phys. Rev. Lett.}\ }\textbf
  {\bibinfo {volume} {97}},\ \bibinfo {pages} {120405} (\bibinfo {year}
  {2006})}\BibitemShut {NoStop}%
\bibitem [{\citenamefont {Braunstein}\ and\ \citenamefont
  {Pirandola}(2012)}]{Braunstein12}%
  \BibitemOpen
  \bibfield  {author} {\bibinfo {author} {\bibfnamefont {S.~L.}\ \bibnamefont
  {Braunstein}}\ and\ \bibinfo {author} {\bibfnamefont {S.}~\bibnamefont
  {Pirandola}},\ }\bibfield  {title} {\bibinfo {title} {{Side-Channel-Free
  Quantum Key Distribution}},\ }\href
  {https://doi.org/10.1103/PhysRevLett.108.130502} {\bibfield  {journal}
  {\bibinfo  {journal} {Phys. Rev. Lett.}\ }\textbf {\bibinfo {volume} {108}},\
  \bibinfo {pages} {130502} (\bibinfo {year} {2012})}\BibitemShut {NoStop}%
\bibitem [{\citenamefont {Lo}\ \emph {et~al.}(2012)\citenamefont {Lo},
  \citenamefont {Curty},\ and\ \citenamefont {Qi}}]{Lo12}%
  \BibitemOpen
  \bibfield  {author} {\bibinfo {author} {\bibfnamefont {H.-K.}\ \bibnamefont
  {Lo}}, \bibinfo {author} {\bibfnamefont {M.}~\bibnamefont {Curty}},\ and\
  \bibinfo {author} {\bibfnamefont {B.}~\bibnamefont {Qi}},\ }\bibfield
  {title} {\bibinfo {title} {{Measurement-Device-Independent Quantum Key
  Distribution}},\ }\href {https://doi.org/10.1103/PhysRevLett.108.130503}
  {\bibfield  {journal} {\bibinfo  {journal} {Phys. Rev. Lett.}\ }\textbf
  {\bibinfo {volume} {108}},\ \bibinfo {pages} {130503} (\bibinfo {year}
  {2012})}\BibitemShut {NoStop}%
\bibitem [{\citenamefont {Rubenok}\ \emph {et~al.}(2013)\citenamefont
  {Rubenok}, \citenamefont {Slater}, \citenamefont {Chan}, \citenamefont
  {Lucio-Martinez},\ and\ \citenamefont {Tittel}}]{Rubenok13}%
  \BibitemOpen
  \bibfield  {author} {\bibinfo {author} {\bibfnamefont {A.}~\bibnamefont
  {Rubenok}}, \bibinfo {author} {\bibfnamefont {J.~A.}\ \bibnamefont {Slater}},
  \bibinfo {author} {\bibfnamefont {P.}~\bibnamefont {Chan}}, \bibinfo {author}
  {\bibfnamefont {I.}~\bibnamefont {Lucio-Martinez}},\ and\ \bibinfo {author}
  {\bibfnamefont {W.}~\bibnamefont {Tittel}},\ }\bibfield  {title} {\bibinfo
  {title} {{Real-World Two-Photon Interference and Proof-of-Principle Quantum
  Key Distribution Immune to Detector Attacks}},\ }\href
  {https://doi.org/10.1103/PhysRevLett.111.130501} {\bibfield  {journal}
  {\bibinfo  {journal} {Phys. Rev. Lett.}\ }\textbf {\bibinfo {volume} {111}},\
  \bibinfo {pages} {130501} (\bibinfo {year} {2013})}\BibitemShut {NoStop}%
\bibitem [{\citenamefont {Ferreira~da Silva}\ \emph {et~al.}(2013)\citenamefont
  {Ferreira~da Silva}, \citenamefont {Vitoreti}, \citenamefont {Xavier},
  \citenamefont {do~Amaral}, \citenamefont {Tempor\~ao},\ and\ \citenamefont
  {von~der Weid}}]{Ferreira13}%
  \BibitemOpen
  \bibfield  {author} {\bibinfo {author} {\bibfnamefont {T.}~\bibnamefont
  {Ferreira~da Silva}}, \bibinfo {author} {\bibfnamefont {D.}~\bibnamefont
  {Vitoreti}}, \bibinfo {author} {\bibfnamefont {G.~B.}\ \bibnamefont
  {Xavier}}, \bibinfo {author} {\bibfnamefont {G.~C.}\ \bibnamefont
  {do~Amaral}}, \bibinfo {author} {\bibfnamefont {G.~P.}\ \bibnamefont
  {Tempor\~ao}},\ and\ \bibinfo {author} {\bibfnamefont {J.~P.}\ \bibnamefont
  {von~der Weid}},\ }\bibfield  {title} {\bibinfo {title} {{Proof-of-principle
  demonstration of measurement-device-independent quantum key distribution
  using polarization qubits}},\ }\href
  {https://doi.org/10.1103/PhysRevA.88.052303} {\bibfield  {journal} {\bibinfo
  {journal} {Phys. Rev. A}\ }\textbf {\bibinfo {volume} {88}},\ \bibinfo
  {pages} {052303} (\bibinfo {year} {2013})}\BibitemShut {NoStop}%
\bibitem [{\citenamefont {Liu}\ \emph {et~al.}(2013)\citenamefont {Liu},
  \citenamefont {Chen}, \citenamefont {Wang}, \citenamefont {Liang},
  \citenamefont {Shentu}, \citenamefont {Wang}, \citenamefont {Cui},
  \citenamefont {Yin}, \citenamefont {Liu}, \citenamefont {Li}, \citenamefont
  {Ma}, \citenamefont {Pelc}, \citenamefont {Fejer}, \citenamefont {Peng},
  \citenamefont {Zhang},\ and\ \citenamefont {Pan}}]{Liu13}%
  \BibitemOpen
  \bibfield  {author} {\bibinfo {author} {\bibfnamefont {Y.}~\bibnamefont
  {Liu}}, \bibinfo {author} {\bibfnamefont {T.-Y.}\ \bibnamefont {Chen}},
  \bibinfo {author} {\bibfnamefont {L.-J.}\ \bibnamefont {Wang}}, \bibinfo
  {author} {\bibfnamefont {H.}~\bibnamefont {Liang}}, \bibinfo {author}
  {\bibfnamefont {G.-L.}\ \bibnamefont {Shentu}}, \bibinfo {author}
  {\bibfnamefont {J.}~\bibnamefont {Wang}}, \bibinfo {author} {\bibfnamefont
  {K.}~\bibnamefont {Cui}}, \bibinfo {author} {\bibfnamefont {H.-L.}\
  \bibnamefont {Yin}}, \bibinfo {author} {\bibfnamefont {N.-L.}\ \bibnamefont
  {Liu}}, \bibinfo {author} {\bibfnamefont {L.}~\bibnamefont {Li}}, \bibinfo
  {author} {\bibfnamefont {X.}~\bibnamefont {Ma}}, \bibinfo {author}
  {\bibfnamefont {J.~S.}\ \bibnamefont {Pelc}}, \bibinfo {author}
  {\bibfnamefont {M.~M.}\ \bibnamefont {Fejer}}, \bibinfo {author}
  {\bibfnamefont {C.-Z.}\ \bibnamefont {Peng}}, \bibinfo {author}
  {\bibfnamefont {Q.}~\bibnamefont {Zhang}},\ and\ \bibinfo {author}
  {\bibfnamefont {J.-W.}\ \bibnamefont {Pan}},\ }\bibfield  {title} {\bibinfo
  {title} {{Experimental Measurement-Device-Independent Quantum Key
  Distribution}},\ }\href {https://doi.org/10.1103/PhysRevLett.111.130502}
  {\bibfield  {journal} {\bibinfo  {journal} {Phys. Rev. Lett.}\ }\textbf
  {\bibinfo {volume} {111}},\ \bibinfo {pages} {130502} (\bibinfo {year}
  {2013})}\BibitemShut {NoStop}%
\bibitem [{\citenamefont {Brunner}\ \emph {et~al.}(2014)\citenamefont
  {Brunner}, \citenamefont {Cavalcanti}, \citenamefont {Pironio}, \citenamefont
  {Scarani},\ and\ \citenamefont {Wehner}}]{Brunner14}%
  \BibitemOpen
  \bibfield  {author} {\bibinfo {author} {\bibfnamefont {N.}~\bibnamefont
  {Brunner}}, \bibinfo {author} {\bibfnamefont {D.}~\bibnamefont {Cavalcanti}},
  \bibinfo {author} {\bibfnamefont {S.}~\bibnamefont {Pironio}}, \bibinfo
  {author} {\bibfnamefont {V.}~\bibnamefont {Scarani}},\ and\ \bibinfo {author}
  {\bibfnamefont {S.}~\bibnamefont {Wehner}},\ }\bibfield  {title} {\bibinfo
  {title} {{Bell nonlocality}},\ }\href
  {https://doi.org/10.1103/RevModPhys.86.419} {\bibfield  {journal} {\bibinfo
  {journal} {Rev. Mod. Phys.}\ }\textbf {\bibinfo {volume} {86}},\ \bibinfo
  {pages} {419} (\bibinfo {year} {2014})}\BibitemShut {NoStop}%
\bibitem [{\citenamefont {Barrett}\ \emph {et~al.}(2005)\citenamefont
  {Barrett}, \citenamefont {Hardy},\ and\ \citenamefont {Kent}}]{Barrett05}%
  \BibitemOpen
  \bibfield  {author} {\bibinfo {author} {\bibfnamefont {J.}~\bibnamefont
  {Barrett}}, \bibinfo {author} {\bibfnamefont {L.}~\bibnamefont {Hardy}},\
  and\ \bibinfo {author} {\bibfnamefont {A.}~\bibnamefont {Kent}},\ }\bibfield
  {title} {\bibinfo {title} {{No Signaling and Quantum Key Distribution}},\
  }\href {https://doi.org/10.1103/PhysRevLett.95.010503} {\bibfield  {journal}
  {\bibinfo  {journal} {Phys. Rev. Lett.}\ }\textbf {\bibinfo {volume} {95}},\
  \bibinfo {pages} {010503} (\bibinfo {year} {2005})}\BibitemShut {NoStop}%
\bibitem [{\citenamefont {Masanes}(2009)}]{Masanes09}%
  \BibitemOpen
  \bibfield  {author} {\bibinfo {author} {\bibfnamefont {L.}~\bibnamefont
  {Masanes}},\ }\bibfield  {title} {\bibinfo {title} {{Universally Composable
  Privacy Amplification from Causality Constraints}},\ }\href
  {https://doi.org/10.1103/PhysRevLett.102.140501} {\bibfield  {journal}
  {\bibinfo  {journal} {Phys. Rev. Lett.}\ }\textbf {\bibinfo {volume} {102}},\
  \bibinfo {pages} {140501} (\bibinfo {year} {2009})}\BibitemShut {NoStop}%
\bibitem [{\citenamefont {Reichardt}\ \emph {et~al.}(2013)\citenamefont
  {Reichardt}, \citenamefont {Unger},\ and\ \citenamefont
  {Vazirani}}]{Reichardt13}%
  \BibitemOpen
  \bibfield  {author} {\bibinfo {author} {\bibfnamefont {B.~W.}\ \bibnamefont
  {Reichardt}}, \bibinfo {author} {\bibfnamefont {F.}~\bibnamefont {Unger}},\
  and\ \bibinfo {author} {\bibfnamefont {U.}~\bibnamefont {Vazirani}},\
  }\bibfield  {title} {\bibinfo {title} {{Classical command of quantum
  systems}},\ }\href@noop {} {\bibfield  {journal} {\bibinfo  {journal}
  {Nature}\ }\textbf {\bibinfo {volume} {496}},\ \bibinfo {pages} {456}
  (\bibinfo {year} {2013})}\BibitemShut {NoStop}%
\bibitem [{\citenamefont {Ho}\ \emph {et~al.}(2020)\citenamefont {Ho},
  \citenamefont {Sekatski}, \citenamefont {Tan}, \citenamefont {Renner},
  \citenamefont {Bancal},\ and\ \citenamefont {Sangouard}}]{Ho20}%
  \BibitemOpen
  \bibfield  {author} {\bibinfo {author} {\bibfnamefont {M.}~\bibnamefont
  {Ho}}, \bibinfo {author} {\bibfnamefont {P.}~\bibnamefont {Sekatski}},
  \bibinfo {author} {\bibfnamefont {E.~Y.-Z.}\ \bibnamefont {Tan}}, \bibinfo
  {author} {\bibfnamefont {R.}~\bibnamefont {Renner}}, \bibinfo {author}
  {\bibfnamefont {J.-D.}\ \bibnamefont {Bancal}},\ and\ \bibinfo {author}
  {\bibfnamefont {N.}~\bibnamefont {Sangouard}},\ }\bibfield  {title} {\bibinfo
  {title} {{Noisy Preprocessing Facilitates a Photonic Realization of
  Device-Independent Quantum Key Distribution}},\ }\href
  {https://doi.org/10.1103/PhysRevLett.124.230502} {\bibfield  {journal}
  {\bibinfo  {journal} {Phys. Rev. Lett.}\ }\textbf {\bibinfo {volume} {124}},\
  \bibinfo {pages} {230502} (\bibinfo {year} {2020})}\BibitemShut {NoStop}%
\bibitem [{\citenamefont {Schwonnek}\ \emph {et~al.}(2021)\citenamefont
  {Schwonnek}, \citenamefont {Goh}, \citenamefont {Primaatmaja}, \citenamefont
  {Tan}, \citenamefont {Wolf}, \citenamefont {Scarani},\ and\ \citenamefont
  {Lim}}]{Schwonnek21}%
  \BibitemOpen
  \bibfield  {author} {\bibinfo {author} {\bibfnamefont {R.}~\bibnamefont
  {Schwonnek}}, \bibinfo {author} {\bibfnamefont {K.~T.}\ \bibnamefont {Goh}},
  \bibinfo {author} {\bibfnamefont {I.~W.}\ \bibnamefont {Primaatmaja}},
  \bibinfo {author} {\bibfnamefont {E.~Y.-Z.}\ \bibnamefont {Tan}}, \bibinfo
  {author} {\bibfnamefont {R.}~\bibnamefont {Wolf}}, \bibinfo {author}
  {\bibfnamefont {V.}~\bibnamefont {Scarani}},\ and\ \bibinfo {author}
  {\bibfnamefont {C.~C.-W.}\ \bibnamefont {Lim}},\ }\bibfield  {title}
  {\bibinfo {title} {{Device-independent quantum key distribution with random
  key basis}},\ }\href {https://doi.org/10.1038/s41467-021-23147-3} {\bibfield
  {journal} {\bibinfo  {journal} {Nat. Commun.}\ }\textbf {\bibinfo {volume}
  {12}},\ \bibinfo {pages} {2880} (\bibinfo {year} {2021})}\BibitemShut
  {NoStop}%
\bibitem [{\citenamefont {Woodhead}\ \emph {et~al.}(2021)\citenamefont
  {Woodhead}, \citenamefont {Acín},\ and\ \citenamefont
  {Pironio}}]{Woodhead21}%
  \BibitemOpen
  \bibfield  {author} {\bibinfo {author} {\bibfnamefont {E.}~\bibnamefont
  {Woodhead}}, \bibinfo {author} {\bibfnamefont {A.}~\bibnamefont {Acín}},\
  and\ \bibinfo {author} {\bibfnamefont {S.}~\bibnamefont {Pironio}},\
  }\bibfield  {title} {\bibinfo {title} {{Device-independent quantum key
  distribution with asymmetric CHSH inequalities}},\ }\href
  {https://doi.org/10.22331/q-2021-04-26-443} {\bibfield  {journal} {\bibinfo
  {journal} {Quantum}\ }\textbf {\bibinfo {volume} {5}},\ \bibinfo {pages}
  {443} (\bibinfo {year} {2021})}\BibitemShut {NoStop}%
\bibitem [{\citenamefont {Sekatski}\ \emph {et~al.}(2021)\citenamefont
  {Sekatski}, \citenamefont {Bancal}, \citenamefont {Valcarce}, \citenamefont
  {Tan}, \citenamefont {Renner},\ and\ \citenamefont {Sangouard}}]{Sekatski21}%
  \BibitemOpen
  \bibfield  {author} {\bibinfo {author} {\bibfnamefont {P.}~\bibnamefont
  {Sekatski}}, \bibinfo {author} {\bibfnamefont {J.-D.}\ \bibnamefont
  {Bancal}}, \bibinfo {author} {\bibfnamefont {X.}~\bibnamefont {Valcarce}},
  \bibinfo {author} {\bibfnamefont {E.-Z.}\ \bibnamefont {Tan}}, \bibinfo
  {author} {\bibfnamefont {R.}~\bibnamefont {Renner}},\ and\ \bibinfo {author}
  {\bibfnamefont {N.}~\bibnamefont {Sangouard}},\ }\bibfield  {title} {\bibinfo
  {title} {{Device-independent quantum key distribution from generalized CHSH
  inequalities}},\ }\href {https://doi.org/10.22331/q-2021-04-26-444}
  {\bibfield  {journal} {\bibinfo  {journal} {Quantum}\ }\textbf {\bibinfo
  {volume} {5}},\ \bibinfo {pages} {444} (\bibinfo {year} {2021})}\BibitemShut
  {NoStop}%
\bibitem [{\citenamefont {Brown}\ \emph {et~al.}(2021)\citenamefont {Brown},
  \citenamefont {Fawzi},\ and\ \citenamefont {Fawzi}}]{Brown21}%
  \BibitemOpen
  \bibfield  {author} {\bibinfo {author} {\bibfnamefont {P.}~\bibnamefont
  {Brown}}, \bibinfo {author} {\bibfnamefont {H.}~\bibnamefont {Fawzi}},\ and\
  \bibinfo {author} {\bibfnamefont {O.}~\bibnamefont {Fawzi}},\ }\bibfield
  {title} {\bibinfo {title} {{Device-independent lower bounds on the
  conditional von Neumann entropy}},\ }\href@noop {} {\bibfield  {journal}
  {\bibinfo  {journal} {arXiv:2106.13692}\ } (\bibinfo {year}
  {2021})}\BibitemShut {NoStop}%
\bibitem [{\citenamefont {Masini}\ \emph {et~al.}(2021)\citenamefont {Masini},
  \citenamefont {Pironio},\ and\ \citenamefont {Woodhead}}]{Masini21}%
  \BibitemOpen
  \bibfield  {author} {\bibinfo {author} {\bibfnamefont {M.}~\bibnamefont
  {Masini}}, \bibinfo {author} {\bibfnamefont {S.}~\bibnamefont {Pironio}},\
  and\ \bibinfo {author} {\bibfnamefont {E.}~\bibnamefont {Woodhead}},\
  }\bibfield  {title} {\bibinfo {title} {{Simple and practical DIQKD security
  analysis via BB84-type uncertainty relations and Pauli correlation
  constraints}},\ }\href@noop {} {\bibfield  {journal} {\bibinfo  {journal}
  {arXiv:2107.08894}\ } (\bibinfo {year} {2021})}\BibitemShut {NoStop}%
\bibitem [{\citenamefont {Clauser}\ \emph {et~al.}(1969)\citenamefont
  {Clauser}, \citenamefont {Horne}, \citenamefont {Shimony},\ and\
  \citenamefont {Holt}}]{CHSH69}%
  \BibitemOpen
  \bibfield  {author} {\bibinfo {author} {\bibfnamefont {J.~F.}\ \bibnamefont
  {Clauser}}, \bibinfo {author} {\bibfnamefont {M.~A.}\ \bibnamefont {Horne}},
  \bibinfo {author} {\bibfnamefont {A.}~\bibnamefont {Shimony}},\ and\ \bibinfo
  {author} {\bibfnamefont {R.~A.}\ \bibnamefont {Holt}},\ }\bibfield  {title}
  {\bibinfo {title} {Proposed experiment to test local hidden-variable
  theories},\ }\href@noop {} {\bibfield  {journal} {\bibinfo  {journal} {Phys.
  Rev. Lett.}\ }\textbf {\bibinfo {volume} {23}},\ \bibinfo {pages} {880}
  (\bibinfo {year} {1969})}\BibitemShut {NoStop}%
\bibitem [{\citenamefont {Dupuis}\ \emph {et~al.}(2020)\citenamefont {Dupuis},
  \citenamefont {Fawzi},\ and\ \citenamefont {Renner}}]{Dupuis20}%
  \BibitemOpen
  \bibfield  {author} {\bibinfo {author} {\bibfnamefont {F.}~\bibnamefont
  {Dupuis}}, \bibinfo {author} {\bibfnamefont {O.}~\bibnamefont {Fawzi}},\ and\
  \bibinfo {author} {\bibfnamefont {R.}~\bibnamefont {Renner}},\ }\bibfield
  {title} {\bibinfo {title} {{Entropy Accumulation}},\ }\href@noop {}
  {\bibfield  {journal} {\bibinfo  {journal} {Commun. Math. Phys.}\ }\textbf
  {\bibinfo {volume} {379}},\ \bibinfo {pages} {867} (\bibinfo {year}
  {2020})}\BibitemShut {NoStop}%
\bibitem [{\citenamefont {Liu}\ \emph {et~al.}(2021)\citenamefont {Liu},
  \citenamefont {Li}, \citenamefont {Ragy}, \citenamefont {Zhao}, \citenamefont
  {Bai}, \citenamefont {Liu}, \citenamefont {Brown}, \citenamefont {Zhang},
  \citenamefont {Colbeck}, \citenamefont {Fan}, \citenamefont {Zhang},\ and\
  \citenamefont {Pan}}]{Liu21}%
  \BibitemOpen
  \bibfield  {author} {\bibinfo {author} {\bibfnamefont {W.-Z.}\ \bibnamefont
  {Liu}}, \bibinfo {author} {\bibfnamefont {M.-H.}\ \bibnamefont {Li}},
  \bibinfo {author} {\bibfnamefont {S.}~\bibnamefont {Ragy}}, \bibinfo {author}
  {\bibfnamefont {S.-R.}\ \bibnamefont {Zhao}}, \bibinfo {author}
  {\bibfnamefont {B.}~\bibnamefont {Bai}}, \bibinfo {author} {\bibfnamefont
  {Y.}~\bibnamefont {Liu}}, \bibinfo {author} {\bibfnamefont {P.~J.}\
  \bibnamefont {Brown}}, \bibinfo {author} {\bibfnamefont {J.}~\bibnamefont
  {Zhang}}, \bibinfo {author} {\bibfnamefont {R.}~\bibnamefont {Colbeck}},
  \bibinfo {author} {\bibfnamefont {J.}~\bibnamefont {Fan}}, \bibinfo {author}
  {\bibfnamefont {Q.}~\bibnamefont {Zhang}},\ and\ \bibinfo {author}
  {\bibfnamefont {J.-W.}\ \bibnamefont {Pan}},\ }\bibfield  {title} {\bibinfo
  {title} {{Device-independent randomness expansion against quantum side
  information}},\ }\href@noop {} {\bibfield  {journal} {\bibinfo  {journal}
  {Nat. Phys.}\ }\textbf {\bibinfo {volume} {17}},\ \bibinfo {pages} {448}
  (\bibinfo {year} {2021})}\BibitemShut {NoStop}%
\bibitem [{\citenamefont {Murta}\ \emph {et~al.}(2019)\citenamefont {Murta},
  \citenamefont {van Dam}, \citenamefont {Ribeiro}, \citenamefont {Hanson},\
  and\ \citenamefont {Wehner}}]{Murta19}%
  \BibitemOpen
  \bibfield  {author} {\bibinfo {author} {\bibfnamefont {G.}~\bibnamefont
  {Murta}}, \bibinfo {author} {\bibfnamefont {S.~B.}\ \bibnamefont {van Dam}},
  \bibinfo {author} {\bibfnamefont {J.}~\bibnamefont {Ribeiro}}, \bibinfo
  {author} {\bibfnamefont {R.}~\bibnamefont {Hanson}},\ and\ \bibinfo {author}
  {\bibfnamefont {S.}~\bibnamefont {Wehner}},\ }\bibfield  {title} {\bibinfo
  {title} {{Towards a realization of device-independent quantum key
  distribution}},\ }\href@noop {} {\bibfield  {journal} {\bibinfo  {journal}
  {Quantum Sci. Technol.}\ }\textbf {\bibinfo {volume} {4}},\ \bibinfo {pages}
  {035011} (\bibinfo {year} {2019})}\BibitemShut {NoStop}%
\bibitem [{\citenamefont {Tan}\ \emph {et~al.}(2020)\citenamefont {Tan},
  \citenamefont {Sekatski}, \citenamefont {Bancal}, \citenamefont {Schwonnek},
  \citenamefont {Renner}, \citenamefont {Sangouard},\ and\ \citenamefont
  {Lim}}]{Tan20}%
  \BibitemOpen
  \bibfield  {author} {\bibinfo {author} {\bibfnamefont {E.~Y.-Z.}\
  \bibnamefont {Tan}}, \bibinfo {author} {\bibfnamefont {P.}~\bibnamefont
  {Sekatski}}, \bibinfo {author} {\bibfnamefont {J.-D.}\ \bibnamefont
  {Bancal}}, \bibinfo {author} {\bibfnamefont {R.}~\bibnamefont {Schwonnek}},
  \bibinfo {author} {\bibfnamefont {R.}~\bibnamefont {Renner}}, \bibinfo
  {author} {\bibfnamefont {N.}~\bibnamefont {Sangouard}},\ and\ \bibinfo
  {author} {\bibfnamefont {C.~C.-W.}\ \bibnamefont {Lim}},\ }\bibfield  {title}
  {\bibinfo {title} {{Improved DIQKD protocols with finite-size analysis}},\
  }\href@noop {} {\bibfield  {journal} {\bibinfo  {journal} {arXiv:2012.08714}\
  } (\bibinfo {year} {2020})}\BibitemShut {NoStop}%
\bibitem [{\citenamefont {Moehring}\ \emph {et~al.}(2007)\citenamefont
  {Moehring}, \citenamefont {Maunz}, \citenamefont {Olmschenk}, \citenamefont
  {Younge}, \citenamefont {Matsukevich}, \citenamefont {Duan},\ and\
  \citenamefont {Monroe}}]{moehringEntanglementSingleatomQuantum2007}%
  \BibitemOpen
  \bibfield  {author} {\bibinfo {author} {\bibfnamefont {D.~L.}\ \bibnamefont
  {Moehring}}, \bibinfo {author} {\bibfnamefont {P.}~\bibnamefont {Maunz}},
  \bibinfo {author} {\bibfnamefont {S.}~\bibnamefont {Olmschenk}}, \bibinfo
  {author} {\bibfnamefont {K.~C.}\ \bibnamefont {Younge}}, \bibinfo {author}
  {\bibfnamefont {D.~N.}\ \bibnamefont {Matsukevich}}, \bibinfo {author}
  {\bibfnamefont {L.-M.}\ \bibnamefont {Duan}},\ and\ \bibinfo {author}
  {\bibfnamefont {C.}~\bibnamefont {Monroe}},\ }\bibfield  {title} {\bibinfo
  {title} {{Entanglement of Single-Atom Quantum Bits at a Distance}},\ }\href
  {https://doi.org/10.1038/nature06118} {\bibfield  {journal} {\bibinfo
  {journal} {Nature}\ }\textbf {\bibinfo {volume} {449}},\ \bibinfo {pages}
  {68} (\bibinfo {year} {2007})}\BibitemShut {NoStop}%
\bibitem [{\citenamefont {Stephenson}\ \emph {et~al.}(2020)\citenamefont
  {Stephenson}, \citenamefont {Nadlinger}, \citenamefont {Nichol},
  \citenamefont {An}, \citenamefont {Drmota}, \citenamefont {Ballance},
  \citenamefont {Thirumalai}, \citenamefont {Goodwin}, \citenamefont {Lucas},\
  and\ \citenamefont
  {Ballance}}]{stephensonHighRateHighFidelityEntanglement2020}%
  \BibitemOpen
  \bibfield  {author} {\bibinfo {author} {\bibfnamefont {L.~J.}\ \bibnamefont
  {Stephenson}}, \bibinfo {author} {\bibfnamefont {D.~P.}\ \bibnamefont
  {Nadlinger}}, \bibinfo {author} {\bibfnamefont {B.~C.}\ \bibnamefont
  {Nichol}}, \bibinfo {author} {\bibfnamefont {S.}~\bibnamefont {An}}, \bibinfo
  {author} {\bibfnamefont {P.}~\bibnamefont {Drmota}}, \bibinfo {author}
  {\bibfnamefont {T.~G.}\ \bibnamefont {Ballance}}, \bibinfo {author}
  {\bibfnamefont {K.}~\bibnamefont {Thirumalai}}, \bibinfo {author}
  {\bibfnamefont {J.~F.}\ \bibnamefont {Goodwin}}, \bibinfo {author}
  {\bibfnamefont {D.~M.}\ \bibnamefont {Lucas}},\ and\ \bibinfo {author}
  {\bibfnamefont {C.~J.}\ \bibnamefont {Ballance}},\ }\bibfield  {title}
  {\bibinfo {title} {{High-{{Rate}}, {{High}}-{{Fidelity Entanglement}} of
  {{Qubits Across}} an {{Elementary Quantum Network}}}},\ }\href
  {https://doi.org/10.1103/PhysRevLett.124.110501} {\bibfield  {journal}
  {\bibinfo  {journal} {Phys. Rev. Lett.}\ }\textbf {\bibinfo {volume} {124}},\
  \bibinfo {pages} {110501} (\bibinfo {year} {2020})}\BibitemShut {NoStop}%
\bibitem [{\citenamefont {Maunz}\ \emph {et~al.}(2009)\citenamefont {Maunz},
  \citenamefont {Olmschenk}, \citenamefont {Hayes}, \citenamefont
  {Matsukevich}, \citenamefont {Duan},\ and\ \citenamefont
  {Monroe}}]{maunzHeraldedQuantumGate2009}%
  \BibitemOpen
  \bibfield  {author} {\bibinfo {author} {\bibfnamefont {P.}~\bibnamefont
  {Maunz}}, \bibinfo {author} {\bibfnamefont {S.}~\bibnamefont {Olmschenk}},
  \bibinfo {author} {\bibfnamefont {D.}~\bibnamefont {Hayes}}, \bibinfo
  {author} {\bibfnamefont {D.~N.}\ \bibnamefont {Matsukevich}}, \bibinfo
  {author} {\bibfnamefont {L.-M.}\ \bibnamefont {Duan}},\ and\ \bibinfo
  {author} {\bibfnamefont {C.}~\bibnamefont {Monroe}},\ }\bibfield  {title}
  {\bibinfo {title} {Heralded {{Quantum Gate}} between {{Remote Quantum
  Memories}}},\ }\href {https://doi.org/10.1103/PhysRevLett.102.250502}
  {\bibfield  {journal} {\bibinfo  {journal} {Phys. Rev. Lett.}\ }\textbf
  {\bibinfo {volume} {102}},\ \bibinfo {pages} {250502} (\bibinfo {year}
  {2009})}\BibitemShut {NoStop}%
\bibitem [{\citenamefont {Lettner}\ \emph {et~al.}(2011)\citenamefont
  {Lettner}, \citenamefont {M{\"u}cke}, \citenamefont {Riedl}, \citenamefont
  {Vo}, \citenamefont {Hahn}, \citenamefont {Baur}, \citenamefont {Bochmann},
  \citenamefont {Ritter}, \citenamefont {D{\"u}rr},\ and\ \citenamefont
  {Rempe}}]{lettnerRemoteEntanglementSingle2011}%
  \BibitemOpen
  \bibfield  {author} {\bibinfo {author} {\bibfnamefont {M.}~\bibnamefont
  {Lettner}}, \bibinfo {author} {\bibfnamefont {M.}~\bibnamefont {M{\"u}cke}},
  \bibinfo {author} {\bibfnamefont {S.}~\bibnamefont {Riedl}}, \bibinfo
  {author} {\bibfnamefont {C.}~\bibnamefont {Vo}}, \bibinfo {author}
  {\bibfnamefont {C.}~\bibnamefont {Hahn}}, \bibinfo {author} {\bibfnamefont
  {S.}~\bibnamefont {Baur}}, \bibinfo {author} {\bibfnamefont {J.}~\bibnamefont
  {Bochmann}}, \bibinfo {author} {\bibfnamefont {S.}~\bibnamefont {Ritter}},
  \bibinfo {author} {\bibfnamefont {S.}~\bibnamefont {D{\"u}rr}},\ and\
  \bibinfo {author} {\bibfnamefont {G.}~\bibnamefont {Rempe}},\ }\bibfield
  {title} {\bibinfo {title} {Remote {{Entanglement}} between a {{Single Atom}}
  and a {{Bose}}-{{Einstein Condensate}}},\ }\href
  {https://doi.org/10.1103/PhysRevLett.106.210503} {\bibfield  {journal}
  {\bibinfo  {journal} {Phys. Rev. Lett.}\ }\textbf {\bibinfo {volume} {106}},\
  \bibinfo {pages} {210503} (\bibinfo {year} {2011})}\BibitemShut {NoStop}%
\bibitem [{\citenamefont {Ritter}\ \emph {et~al.}(2012)\citenamefont {Ritter},
  \citenamefont {N{\"o}lleke}, \citenamefont {Hahn}, \citenamefont {Reiserer},
  \citenamefont {Neuzner}, \citenamefont {Uphoff}, \citenamefont {M{\"u}cke},
  \citenamefont {Figueroa}, \citenamefont {Bochmann},\ and\ \citenamefont
  {Rempe}}]{Ritter2012}%
  \BibitemOpen
  \bibfield  {author} {\bibinfo {author} {\bibfnamefont {S.}~\bibnamefont
  {Ritter}}, \bibinfo {author} {\bibfnamefont {C.}~\bibnamefont {N{\"o}lleke}},
  \bibinfo {author} {\bibfnamefont {C.}~\bibnamefont {Hahn}}, \bibinfo {author}
  {\bibfnamefont {A.}~\bibnamefont {Reiserer}}, \bibinfo {author}
  {\bibfnamefont {A.}~\bibnamefont {Neuzner}}, \bibinfo {author} {\bibfnamefont
  {M.}~\bibnamefont {Uphoff}}, \bibinfo {author} {\bibfnamefont
  {M.}~\bibnamefont {M{\"u}cke}}, \bibinfo {author} {\bibfnamefont
  {E.}~\bibnamefont {Figueroa}}, \bibinfo {author} {\bibfnamefont
  {J.}~\bibnamefont {Bochmann}},\ and\ \bibinfo {author} {\bibfnamefont
  {G.}~\bibnamefont {Rempe}},\ }\bibfield  {title} {\bibinfo {title} {An
  elementary quantum network of single atoms in optical cavities},\ }\href
  {https://doi.org/10.1038/nature11023} {\bibfield  {journal} {\bibinfo
  {journal} {Nature}\ }\textbf {\bibinfo {volume} {484}},\ \bibinfo {pages}
  {195} (\bibinfo {year} {2012})}\BibitemShut {NoStop}%
\bibitem [{\citenamefont {Hensen}\ \emph {et~al.}(2015)\citenamefont {Hensen},
  \citenamefont {Bernien}, \citenamefont {Dr{\'e}au}, \citenamefont {Reiserer},
  \citenamefont {Kalb}, \citenamefont {Blok}, \citenamefont {Ruitenberg},
  \citenamefont {Vermeulen}, \citenamefont {Schouten}, \citenamefont
  {Abell{\'a}n}, \citenamefont {Amaya}, \citenamefont {Pruneri}, \citenamefont
  {Mitchell}, \citenamefont {Markham}, \citenamefont {Twitchen}, \citenamefont
  {Elkouss}, \citenamefont {Wehner}, \citenamefont {Taminiau},\ and\
  \citenamefont {Hanson}}]{Hensen2015}%
  \BibitemOpen
  \bibfield  {author} {\bibinfo {author} {\bibfnamefont {B.}~\bibnamefont
  {Hensen}}, \bibinfo {author} {\bibfnamefont {H.}~\bibnamefont {Bernien}},
  \bibinfo {author} {\bibfnamefont {A.~E.}\ \bibnamefont {Dr{\'e}au}}, \bibinfo
  {author} {\bibfnamefont {A.}~\bibnamefont {Reiserer}}, \bibinfo {author}
  {\bibfnamefont {N.}~\bibnamefont {Kalb}}, \bibinfo {author} {\bibfnamefont
  {M.~S.}\ \bibnamefont {Blok}}, \bibinfo {author} {\bibfnamefont
  {J.}~\bibnamefont {Ruitenberg}}, \bibinfo {author} {\bibfnamefont {R.~F.~L.}\
  \bibnamefont {Vermeulen}}, \bibinfo {author} {\bibfnamefont {R.~N.}\
  \bibnamefont {Schouten}}, \bibinfo {author} {\bibfnamefont {C.}~\bibnamefont
  {Abell{\'a}n}}, \bibinfo {author} {\bibfnamefont {W.}~\bibnamefont {Amaya}},
  \bibinfo {author} {\bibfnamefont {V.}~\bibnamefont {Pruneri}}, \bibinfo
  {author} {\bibfnamefont {M.~W.}\ \bibnamefont {Mitchell}}, \bibinfo {author}
  {\bibfnamefont {M.}~\bibnamefont {Markham}}, \bibinfo {author} {\bibfnamefont
  {D.~J.}\ \bibnamefont {Twitchen}}, \bibinfo {author} {\bibfnamefont
  {D.}~\bibnamefont {Elkouss}}, \bibinfo {author} {\bibfnamefont
  {S.}~\bibnamefont {Wehner}}, \bibinfo {author} {\bibfnamefont {T.~H.}\
  \bibnamefont {Taminiau}},\ and\ \bibinfo {author} {\bibfnamefont
  {R.}~\bibnamefont {Hanson}},\ }\bibfield  {title} {\bibinfo {title}
  {Loophole-free {{Bell}} inequality violation using electron spins separated
  by 1.3 kilometres},\ }\href {https://doi.org/10.1038/nature15759} {\bibfield
  {journal} {\bibinfo  {journal} {Nature}\ }\textbf {\bibinfo {volume} {526}},\
  \bibinfo {pages} {682} (\bibinfo {year} {2015})}\BibitemShut {NoStop}%
\bibitem [{\citenamefont {Rosenfeld}\ \emph {et~al.}(2017)\citenamefont
  {Rosenfeld}, \citenamefont {Burchardt}, \citenamefont {Garthoff},
  \citenamefont {Redeker}, \citenamefont {Ortegel}, \citenamefont {Rau},\ and\
  \citenamefont {Weinfurter}}]{rosenfeldEventReadyBellTest2017}%
  \BibitemOpen
  \bibfield  {author} {\bibinfo {author} {\bibfnamefont {W.}~\bibnamefont
  {Rosenfeld}}, \bibinfo {author} {\bibfnamefont {D.}~\bibnamefont
  {Burchardt}}, \bibinfo {author} {\bibfnamefont {R.}~\bibnamefont {Garthoff}},
  \bibinfo {author} {\bibfnamefont {K.}~\bibnamefont {Redeker}}, \bibinfo
  {author} {\bibfnamefont {N.}~\bibnamefont {Ortegel}}, \bibinfo {author}
  {\bibfnamefont {M.}~\bibnamefont {Rau}},\ and\ \bibinfo {author}
  {\bibfnamefont {H.}~\bibnamefont {Weinfurter}},\ }\bibfield  {title}
  {\bibinfo {title} {Event-{{Ready Bell Test Using Entangled Atoms
  Simultaneously Closing Detection}} and {{Locality Loopholes}}},\ }\href
  {https://doi.org/10.1103/PhysRevLett.119.010402} {\bibfield  {journal}
  {\bibinfo  {journal} {Phys. Rev. Lett.}\ }\textbf {\bibinfo {volume} {119}},\
  \bibinfo {pages} {010402} (\bibinfo {year} {2017})}\BibitemShut {NoStop}%
\bibitem [{\citenamefont {Portmann}\ and\ \citenamefont
  {Renner}(2021)}]{Portmann21}%
  \BibitemOpen
  \bibfield  {author} {\bibinfo {author} {\bibfnamefont {C.}~\bibnamefont
  {Portmann}}\ and\ \bibinfo {author} {\bibfnamefont {R.}~\bibnamefont
  {Renner}},\ }\bibfield  {title} {\bibinfo {title} {{Security in Quantum
  Cryptography}},\ }\href {http://arxiv.org/abs/2102.00021} {\bibfield
  {journal} {\bibinfo  {journal} {arXiv:2102.00021}\ } (\bibinfo {year}
  {2021})}\BibitemShut {NoStop}%
\bibitem [{\citenamefont {Barrett}\ \emph {et~al.}(2013)\citenamefont
  {Barrett}, \citenamefont {Colbeck},\ and\ \citenamefont {Kent}}]{Barrett13}%
  \BibitemOpen
  \bibfield  {author} {\bibinfo {author} {\bibfnamefont {J.}~\bibnamefont
  {Barrett}}, \bibinfo {author} {\bibfnamefont {R.}~\bibnamefont {Colbeck}},\
  and\ \bibinfo {author} {\bibfnamefont {A.}~\bibnamefont {Kent}},\ }\bibfield
  {title} {\bibinfo {title} {{Memory Attacks on Device-Independent Quantum
  Cryptography}},\ }\href {https://doi.org/10.1103/PhysRevLett.110.010503}
  {\bibfield  {journal} {\bibinfo  {journal} {Phys. Rev. Lett.}\ }\textbf
  {\bibinfo {volume} {110}},\ \bibinfo {pages} {010503} (\bibinfo {year}
  {2013})}\BibitemShut {NoStop}%
\bibitem [{\citenamefont {Pironio}\ \emph {et~al.}(2010)\citenamefont
  {Pironio}, \citenamefont {Acín}, \citenamefont {Massar}, \citenamefont
  {Boyer de~la Giroday}, \citenamefont {Matsukevich}, \citenamefont {Maunz},
  \citenamefont {Olmschenk}, \citenamefont {Hayes}, \citenamefont {Luo},
  \citenamefont {Manning},\ and\ \citenamefont {Monroe}}]{Pironio10}%
  \BibitemOpen
  \bibfield  {author} {\bibinfo {author} {\bibfnamefont {S.}~\bibnamefont
  {Pironio}}, \bibinfo {author} {\bibfnamefont {A.}~\bibnamefont {Acín}},
  \bibinfo {author} {\bibfnamefont {S.}~\bibnamefont {Massar}}, \bibinfo
  {author} {\bibfnamefont {A.}~\bibnamefont {Boyer de~la Giroday}}, \bibinfo
  {author} {\bibfnamefont {D.~N.}\ \bibnamefont {Matsukevich}}, \bibinfo
  {author} {\bibfnamefont {P.}~\bibnamefont {Maunz}}, \bibinfo {author}
  {\bibfnamefont {S.}~\bibnamefont {Olmschenk}}, \bibinfo {author}
  {\bibfnamefont {D.}~\bibnamefont {Hayes}}, \bibinfo {author} {\bibfnamefont
  {L.}~\bibnamefont {Luo}}, \bibinfo {author} {\bibfnamefont {T.~A.}\
  \bibnamefont {Manning}},\ and\ \bibinfo {author} {\bibfnamefont
  {C.}~\bibnamefont {Monroe}},\ }\bibfield  {title} {\bibinfo {title} {{Random
  numbers certified by Bell’s theorem}},\ }\href
  {https://doi.org/10.1038/nature09008} {\bibfield  {journal} {\bibinfo
  {journal} {Nature}\ }\textbf {\bibinfo {volume} {464}},\ \bibinfo {pages}
  {1021} (\bibinfo {year} {2010})}\BibitemShut {NoStop}%
\bibitem [{\citenamefont {Pironio}\ \emph {et~al.}(2009)\citenamefont
  {Pironio}, \citenamefont {Acin}, \citenamefont {Brunner}, \citenamefont
  {Gisin}, \citenamefont {Massar},\ and\ \citenamefont {Scarani}}]{Pironio09}%
  \BibitemOpen
  \bibfield  {author} {\bibinfo {author} {\bibfnamefont {S.}~\bibnamefont
  {Pironio}}, \bibinfo {author} {\bibfnamefont {A.}~\bibnamefont {Acin}},
  \bibinfo {author} {\bibfnamefont {N.}~\bibnamefont {Brunner}}, \bibinfo
  {author} {\bibfnamefont {N.}~\bibnamefont {Gisin}}, \bibinfo {author}
  {\bibfnamefont {S.}~\bibnamefont {Massar}},\ and\ \bibinfo {author}
  {\bibfnamefont {V.}~\bibnamefont {Scarani}},\ }\bibfield  {title} {\bibinfo
  {title} {Device-independent quantum key distribution secure against
  collective attacks},\ }\href@noop {} {\bibfield  {journal} {\bibinfo
  {journal} {New J. Phys.}\ }\textbf {\bibinfo {volume} {11}},\ \bibinfo
  {pages} {045021} (\bibinfo {year} {2009})}\BibitemShut {NoStop}%
\bibitem [{\citenamefont {Pirandola}\ \emph {et~al.}(2020)\citenamefont
  {Pirandola}, \citenamefont {Andersen}, \citenamefont {Banchi}, \citenamefont
  {Berta}, \citenamefont {Bunandar}, \citenamefont {Colbeck}, \citenamefont
  {Englund}, \citenamefont {Gehring}, \citenamefont {Lupo}, \citenamefont
  {Ottaviani}, \citenamefont {Pereira}, \citenamefont {Razavi}, \citenamefont
  {Shaari}, \citenamefont {Tomamichel}, \citenamefont {Usenko}, \citenamefont
  {Vallone}, \citenamefont {Villoresi},\ and\ \citenamefont
  {Wallden}}]{Pirandola20}%
  \BibitemOpen
  \bibfield  {author} {\bibinfo {author} {\bibfnamefont {S.}~\bibnamefont
  {Pirandola}}, \bibinfo {author} {\bibfnamefont {U.~L.}\ \bibnamefont
  {Andersen}}, \bibinfo {author} {\bibfnamefont {L.}~\bibnamefont {Banchi}},
  \bibinfo {author} {\bibfnamefont {M.}~\bibnamefont {Berta}}, \bibinfo
  {author} {\bibfnamefont {D.}~\bibnamefont {Bunandar}}, \bibinfo {author}
  {\bibfnamefont {R.}~\bibnamefont {Colbeck}}, \bibinfo {author} {\bibfnamefont
  {D.}~\bibnamefont {Englund}}, \bibinfo {author} {\bibfnamefont
  {T.}~\bibnamefont {Gehring}}, \bibinfo {author} {\bibfnamefont
  {C.}~\bibnamefont {Lupo}}, \bibinfo {author} {\bibfnamefont {C.}~\bibnamefont
  {Ottaviani}}, \bibinfo {author} {\bibfnamefont {J.~L.}\ \bibnamefont
  {Pereira}}, \bibinfo {author} {\bibfnamefont {M.}~\bibnamefont {Razavi}},
  \bibinfo {author} {\bibfnamefont {J.~S.}\ \bibnamefont {Shaari}}, \bibinfo
  {author} {\bibfnamefont {M.}~\bibnamefont {Tomamichel}}, \bibinfo {author}
  {\bibfnamefont {V.~C.}\ \bibnamefont {Usenko}}, \bibinfo {author}
  {\bibfnamefont {G.}~\bibnamefont {Vallone}}, \bibinfo {author} {\bibfnamefont
  {P.}~\bibnamefont {Villoresi}},\ and\ \bibinfo {author} {\bibfnamefont
  {P.}~\bibnamefont {Wallden}},\ }\bibfield  {title} {\bibinfo {title}
  {Advances in quantum cryptography},\ }\href
  {https://doi.org/10.1364/AOP.361502} {\bibfield  {journal} {\bibinfo
  {journal} {Adv. Opt. Photon.}\ }\textbf {\bibinfo {volume} {12}},\ \bibinfo
  {pages} {1012} (\bibinfo {year} {2020})}\BibitemShut {NoStop}%
\bibitem [{\citenamefont {Heshami}\ \emph {et~al.}(2016)\citenamefont
  {Heshami}, \citenamefont {England}, \citenamefont {Humphreys}, \citenamefont
  {Bustard}, \citenamefont {Acosta}, \citenamefont {Nunn},\ and\ \citenamefont
  {Sussman}}]{Heshami16}%
  \BibitemOpen
  \bibfield  {author} {\bibinfo {author} {\bibfnamefont {K.}~\bibnamefont
  {Heshami}}, \bibinfo {author} {\bibfnamefont {D.~G.}\ \bibnamefont
  {England}}, \bibinfo {author} {\bibfnamefont {P.~C.}\ \bibnamefont
  {Humphreys}}, \bibinfo {author} {\bibfnamefont {P.~J.}\ \bibnamefont
  {Bustard}}, \bibinfo {author} {\bibfnamefont {V.~M.}\ \bibnamefont {Acosta}},
  \bibinfo {author} {\bibfnamefont {J.}~\bibnamefont {Nunn}},\ and\ \bibinfo
  {author} {\bibfnamefont {B.~J.}\ \bibnamefont {Sussman}},\ }\bibfield
  {title} {\bibinfo {title} {Quantum memories: emerging applications and recent
  advances},\ }\href {https://doi.org/10.1080/09500340.2016.1148212} {\bibfield
   {journal} {\bibinfo  {journal} {Journal of Modern Optics}\ }\textbf
  {\bibinfo {volume} {63}},\ \bibinfo {pages} {2005} (\bibinfo {year}
  {2016})}\BibitemShut {NoStop}%
\bibitem [{\citenamefont {Wright}\ \emph {et~al.}(2018)\citenamefont {Wright},
  \citenamefont {{Francis-Jones}}, \citenamefont {Gawith}, \citenamefont
  {Becker}, \citenamefont {Ledingham}, \citenamefont {Smith}, \citenamefont
  {Nunn}, \citenamefont {Mosley}, \citenamefont {Brecht},\ and\ \citenamefont
  {Walmsley}}]{wrightTwoWayPhotonicInterface2018}%
  \BibitemOpen
  \bibfield  {author} {\bibinfo {author} {\bibfnamefont {T.~A.}\ \bibnamefont
  {Wright}}, \bibinfo {author} {\bibfnamefont {R.~J.~A.}\ \bibnamefont
  {{Francis-Jones}}}, \bibinfo {author} {\bibfnamefont {C.~B.~E.}\ \bibnamefont
  {Gawith}}, \bibinfo {author} {\bibfnamefont {J.~N.}\ \bibnamefont {Becker}},
  \bibinfo {author} {\bibfnamefont {P.~M.}\ \bibnamefont {Ledingham}}, \bibinfo
  {author} {\bibfnamefont {P.~G.~R.}\ \bibnamefont {Smith}}, \bibinfo {author}
  {\bibfnamefont {J.}~\bibnamefont {Nunn}}, \bibinfo {author} {\bibfnamefont
  {P.~J.}\ \bibnamefont {Mosley}}, \bibinfo {author} {\bibfnamefont
  {B.}~\bibnamefont {Brecht}},\ and\ \bibinfo {author} {\bibfnamefont {I.~A.}\
  \bibnamefont {Walmsley}},\ }\bibfield  {title} {\bibinfo {title} {Two-way
  photonic interface for linking the {Sr+} transition at 422 nm to the
  telecommunication {C} band},\ }\href
  {https://doi.org/10.1103/PhysRevApplied.10.044012} {\bibfield  {journal}
  {\bibinfo  {journal} {Phys. Rev. Appl.}\ }\textbf {\bibinfo {volume} {10}},\
  \bibinfo {pages} {044012} (\bibinfo {year} {2018})}\BibitemShut {NoStop}%
\bibitem [{\citenamefont {Schupp}\ \emph {et~al.}(2021)\citenamefont {Schupp},
  \citenamefont {Krcmarsky}, \citenamefont {Krutyanskiy}, \citenamefont
  {Meraner}, \citenamefont {Northup},\ and\ \citenamefont
  {Lanyon}}]{schuppInterfaceTrappedIonQubits2021}%
  \BibitemOpen
  \bibfield  {author} {\bibinfo {author} {\bibfnamefont {J.}~\bibnamefont
  {Schupp}}, \bibinfo {author} {\bibfnamefont {V.}~\bibnamefont {Krcmarsky}},
  \bibinfo {author} {\bibfnamefont {V.}~\bibnamefont {Krutyanskiy}}, \bibinfo
  {author} {\bibfnamefont {M.}~\bibnamefont {Meraner}}, \bibinfo {author}
  {\bibfnamefont {T.}~\bibnamefont {Northup}},\ and\ \bibinfo {author}
  {\bibfnamefont {B.}~\bibnamefont {Lanyon}},\ }\bibfield  {title} {\bibinfo
  {title} {Interface between {{Trapped}}-{{Ion Qubits}} and {{Traveling
  Photons}} with {{Close}}-to-{{Optimal Efficiency}}},\ }\href
  {https://doi.org/10.1103/PRXQuantum.2.020331} {\bibfield  {journal} {\bibinfo
   {journal} {PRX Quantum}\ }\textbf {\bibinfo {volume} {2}},\ \bibinfo {pages}
  {020331} (\bibinfo {year} {2021})}\BibitemShut {NoStop}%
\bibitem [{\citenamefont {Sangouard}\ \emph {et~al.}(2011)\citenamefont
  {Sangouard}, \citenamefont {Simon}, \citenamefont {de~Riedmatten},\ and\
  \citenamefont {Gisin}}]{Sangouard11}%
  \BibitemOpen
  \bibfield  {author} {\bibinfo {author} {\bibfnamefont {N.}~\bibnamefont
  {Sangouard}}, \bibinfo {author} {\bibfnamefont {C.}~\bibnamefont {Simon}},
  \bibinfo {author} {\bibfnamefont {H.}~\bibnamefont {de~Riedmatten}},\ and\
  \bibinfo {author} {\bibfnamefont {N.}~\bibnamefont {Gisin}},\ }\bibfield
  {title} {\bibinfo {title} {{Quantum repeaters based on atomic ensembles and
  linear optics}},\ }\href {https://doi.org/10.1103/RevModPhys.83.33}
  {\bibfield  {journal} {\bibinfo  {journal} {Rev. Mod. Phys.}\ }\textbf
  {\bibinfo {volume} {83}},\ \bibinfo {pages} {33} (\bibinfo {year}
  {2011})}\BibitemShut {NoStop}%
\bibitem [{\citenamefont {Bourdeauducq}\ \emph {et~al.}(2021)\citenamefont
  {Bourdeauducq} \emph {et~al.}}]{ARTIQ}%
  \BibitemOpen
  \bibfield  {author} {\bibinfo {author} {\bibfnamefont {S.}~\bibnamefont
  {Bourdeauducq}} \emph {et~al.},\ }\href
  {https://doi.org/10.5281/zenodo.1492176} {\bibinfo {title} {{m-labs/artiq:
  6.0 (Version 6.0)}}} (\bibinfo {year} {2021})\BibitemShut {NoStop}%
\end{thebibliography}%


\begin{thebibliography}{77}%
\makeatletter
\providecommand \@ifxundefined [1]{%
 \@ifx{#1\undefined}
}%
\providecommand \@ifnum [1]{%
 \ifnum #1\expandafter \@firstoftwo
 \else \expandafter \@secondoftwo
 \fi
}%
\providecommand \@ifx [1]{%
 \ifx #1\expandafter \@firstoftwo
 \else \expandafter \@secondoftwo
 \fi
}%
\providecommand \natexlab [1]{#1}%
\providecommand \enquote  [1]{``#1''}%
\providecommand \bibnamefont  [1]{#1}%
\providecommand \bibfnamefont [1]{#1}%
\providecommand \citenamefont [1]{#1}%
\providecommand \href@noop [0]{\@secondoftwo}%
\providecommand \href [0]{\begingroup \@sanitize@url \@href}%
\providecommand \@href[1]{\@@startlink{#1}\@@href}%
\providecommand \@@href[1]{\endgroup#1\@@endlink}%
\providecommand \@sanitize@url [0]{\catcode `\\12\catcode `\$12\catcode
  `\&12\catcode `\#12\catcode `\^12\catcode `\_12\catcode `\%12\relax}%
\providecommand \@@startlink[1]{}%
\providecommand \@@endlink[0]{}%
\providecommand \url  [0]{\begingroup\@sanitize@url \@url }%
\providecommand \@url [1]{\endgroup\@href {#1}{\urlprefix }}%
\providecommand \urlprefix  [0]{URL }%
\providecommand \Eprint [0]{\href }%
\providecommand \doibase [0]{https://doi.org/}%
\providecommand \selectlanguage [0]{\@gobble}%
\providecommand \bibinfo  [0]{\@secondoftwo}%
\providecommand \bibfield  [0]{\@secondoftwo}%
\providecommand \translation [1]{[#1]}%
\providecommand \BibitemOpen [0]{}%
\providecommand \bibitemStop [0]{}%
\providecommand \bibitemNoStop [0]{.\EOS\space}%
\providecommand \EOS [0]{\spacefactor3000\relax}%
\providecommand \BibitemShut  [1]{\csname bibitem#1\endcsname}%
\let\auto@bib@innerbib\@empty
\bibitem [{\citenamefont {Stephenson}\ \emph {et~al.}(2020)\citenamefont
  {Stephenson}, \citenamefont {Nadlinger}, \citenamefont {Nichol},
  \citenamefont {An}, \citenamefont {Drmota}, \citenamefont {Ballance},
  \citenamefont {Thirumalai}, \citenamefont {Goodwin}, \citenamefont {Lucas},\
  and\ \citenamefont
  {Ballance}}]{stephensonHighRateHighFidelityEntanglement2020}%
  \BibitemOpen
  \bibfield  {author} {\bibinfo {author} {\bibfnamefont {L.~J.}\ \bibnamefont
  {Stephenson}}, \bibinfo {author} {\bibfnamefont {D.~P.}\ \bibnamefont
  {Nadlinger}}, \bibinfo {author} {\bibfnamefont {B.~C.}\ \bibnamefont
  {Nichol}}, \bibinfo {author} {\bibfnamefont {S.}~\bibnamefont {An}}, \bibinfo
  {author} {\bibfnamefont {P.}~\bibnamefont {Drmota}}, \bibinfo {author}
  {\bibfnamefont {T.~G.}\ \bibnamefont {Ballance}}, \bibinfo {author}
  {\bibfnamefont {K.}~\bibnamefont {Thirumalai}}, \bibinfo {author}
  {\bibfnamefont {J.~F.}\ \bibnamefont {Goodwin}}, \bibinfo {author}
  {\bibfnamefont {D.~M.}\ \bibnamefont {Lucas}},\ and\ \bibinfo {author}
  {\bibfnamefont {C.~J.}\ \bibnamefont {Ballance}},\ }\bibfield  {title}
  {\bibinfo {title} {{High-{{Rate}}, {{High}}-{{Fidelity Entanglement}} of
  {{Qubits Across}} an {{Elementary Quantum Network}}}},\ }\href
  {https://doi.org/10.1103/PhysRevLett.124.110501} {\bibfield  {journal}
  {\bibinfo  {journal} {Phys. Rev. Lett.}\ }\textbf {\bibinfo {volume} {124}},\
  \bibinfo {pages} {110501} (\bibinfo {year} {2020})}\BibitemShut {NoStop}%
\bibitem [{\citenamefont {Maunz}(2016)}]{maunzHighOpticalAccess2016}%
  \BibitemOpen
  \bibfield  {author} {\bibinfo {author} {\bibfnamefont {P.~L.~W.}\
  \bibnamefont {Maunz}},\ }\href@noop {} {\emph {\bibinfo {title} {{High
  {{Optical Access Trap}} 2.0}}}},\ \bibinfo {type} {Technical {{Report}}}\
  \bibinfo {number} {SAND-2016-0796R}\ (\bibinfo  {institution} {{Sandia
  National Laboratories}},\ \bibinfo {address} {{Albuquerque, New Mexico}},\
  \bibinfo {year} {2016})\BibitemShut {NoStop}%
\bibitem [{\citenamefont {Reh{\'{a}}{\v{c}}ek}\ \emph
  {et~al.}(2007)\citenamefont {Reh{\'{a}}{\v{c}}ek}, \citenamefont {Hradil},
  \citenamefont {Knill},\ and\ \citenamefont {Lvovsky}}]{Rehacek2007}%
  \BibitemOpen
  \bibfield  {author} {\bibinfo {author} {\bibfnamefont {J.}~\bibnamefont
  {Reh{\'{a}}{\v{c}}ek}}, \bibinfo {author} {\bibfnamefont {Z.}~\bibnamefont
  {Hradil}}, \bibinfo {author} {\bibfnamefont {E.}~\bibnamefont {Knill}},\ and\
  \bibinfo {author} {\bibfnamefont {A.~I.}\ \bibnamefont {Lvovsky}},\
  }\bibfield  {title} {\bibinfo {title} {{Diluted Maximum-Likelihood Algorithm
  for Quantum Tomography}},\ }\href
  {https://doi.org/10.1103/PhysRevA.75.042108} {\bibfield  {journal} {\bibinfo
  {journal} {Phys. Rev. A}\ }\textbf {\bibinfo {volume} {75}},\ \bibinfo
  {pages} {042108} (\bibinfo {year} {2007})}\BibitemShut {NoStop}%
\bibitem [{\citenamefont {Brown}()}]{DieharderCode}%
  \BibitemOpen
  \bibfield  {author} {\bibinfo {author} {\bibfnamefont {R.~G.}\ \bibnamefont
  {Brown}},\ }\href@noop {} {\bibinfo {title} {Dieharder: A random number test
  suite}},\ \bibinfo {howpublished}
  {\url{https://webhome.phy.duke.edu/~rgb/General/dieharder.php}}\BibitemShut
  {NoStop}%
\bibitem [{\citenamefont {L'Ecuyer}\ and\ \citenamefont
  {Simard}(2007)}]{TestU01}%
  \BibitemOpen
  \bibfield  {author} {\bibinfo {author} {\bibfnamefont {P.}~\bibnamefont
  {L'Ecuyer}}\ and\ \bibinfo {author} {\bibfnamefont {R.}~\bibnamefont
  {Simard}},\ }\bibfield  {title} {\bibinfo {title} {Testu01: A c library for
  empirical testing of random number generators},\ }\bibfield  {journal}
  {\bibinfo  {journal} {ACM Trans. Math. Softw.}\ }\textbf {\bibinfo {volume}
  {33}},\ \href {https://doi.org/10.1145/1268776.1268777}
  {10.1145/1268776.1268777} (\bibinfo {year} {2007})\BibitemShut {NoStop}%
\bibitem [{\citenamefont {Scarani}\ \emph {et~al.}(2009)\citenamefont
  {Scarani}, \citenamefont {Bechmann-Pasquinucci}, \citenamefont {Cerf},
  \citenamefont {Du\ifmmode~\check{s}\else \v{s}\fi{}ek}, \citenamefont
  {L\"utkenhaus},\ and\ \citenamefont {Peev}}]{Scarani09}%
  \BibitemOpen
  \bibfield  {author} {\bibinfo {author} {\bibfnamefont {V.}~\bibnamefont
  {Scarani}}, \bibinfo {author} {\bibfnamefont {H.}~\bibnamefont
  {Bechmann-Pasquinucci}}, \bibinfo {author} {\bibfnamefont {N.~J.}\
  \bibnamefont {Cerf}}, \bibinfo {author} {\bibfnamefont {M.}~\bibnamefont
  {Du\ifmmode~\check{s}\else \v{s}\fi{}ek}}, \bibinfo {author} {\bibfnamefont
  {N.}~\bibnamefont {L\"utkenhaus}},\ and\ \bibinfo {author} {\bibfnamefont
  {M.}~\bibnamefont {Peev}},\ }\bibfield  {title} {\bibinfo {title} {{The
  security of practical quantum key distribution}},\ }\href
  {https://doi.org/10.1103/RevModPhys.81.1301} {\bibfield  {journal} {\bibinfo
  {journal} {Rev. Mod. Phys.}\ }\textbf {\bibinfo {volume} {81}},\ \bibinfo
  {pages} {1301} (\bibinfo {year} {2009})}\BibitemShut {NoStop}%
\bibitem [{Note1()}]{Note1}%
  \BibitemOpen
  \bibinfo {note} {In principle, authentication is possible with a weakly
  correlated partially secret shared bit string~\cite {Renner03}.}\BibitemShut
  {Stop}%
\bibitem [{\citenamefont {Katz}\ and\ \citenamefont {Lindell}(2014)}]{Katz14}%
  \BibitemOpen
  \bibfield  {author} {\bibinfo {author} {\bibfnamefont {J.}~\bibnamefont
  {Katz}}\ and\ \bibinfo {author} {\bibfnamefont {Y.}~\bibnamefont {Lindell}},\
  }\href@noop {} {\emph {\bibinfo {title} {Introduction to Modern Cryptography,
  Second Edition}}},\ \bibinfo {edition} {2nd}\ ed.\ (\bibinfo  {publisher}
  {Chapman \& Hall/CRC},\ \bibinfo {year} {2014})\BibitemShut {NoStop}%
\bibitem [{\citenamefont {Bennett}\ and\ \citenamefont
  {Brassard}(1984)}]{Bennett84}%
  \BibitemOpen
  \bibfield  {author} {\bibinfo {author} {\bibfnamefont {C.~H.}\ \bibnamefont
  {Bennett}}\ and\ \bibinfo {author} {\bibfnamefont {G.}~\bibnamefont
  {Brassard}},\ }\bibfield  {title} {\bibinfo {title} {Quantum cryptography:
  public key distribution and coin tossing.},\ }\href@noop {} {\bibfield
  {journal} {\bibinfo  {journal} {Theor. Comput. Sci.}\ }\textbf {\bibinfo
  {volume} {560}},\ \bibinfo {pages} {7} (\bibinfo {year} {1984})}\BibitemShut
  {NoStop}%
\bibitem [{\citenamefont {Zhao}\ \emph {et~al.}(2008)\citenamefont {Zhao},
  \citenamefont {Fung}, \citenamefont {Qi}, \citenamefont {Chen},\ and\
  \citenamefont {Lo}}]{Zhao08}%
  \BibitemOpen
  \bibfield  {author} {\bibinfo {author} {\bibfnamefont {Y.}~\bibnamefont
  {Zhao}}, \bibinfo {author} {\bibfnamefont {C.-H.~F.}\ \bibnamefont {Fung}},
  \bibinfo {author} {\bibfnamefont {B.}~\bibnamefont {Qi}}, \bibinfo {author}
  {\bibfnamefont {C.}~\bibnamefont {Chen}},\ and\ \bibinfo {author}
  {\bibfnamefont {H.-K.}\ \bibnamefont {Lo}},\ }\bibfield  {title} {\bibinfo
  {title} {{Quantum hacking: Experimental demonstration of time-shift attack
  against practical quantum-key-distribution systems}},\ }\href
  {https://doi.org/10.1103/PhysRevA.78.042333} {\bibfield  {journal} {\bibinfo
  {journal} {Phys. Rev. A}\ }\textbf {\bibinfo {volume} {78}},\ \bibinfo
  {pages} {042333} (\bibinfo {year} {2008})}\BibitemShut {NoStop}%
\bibitem [{\citenamefont {Lydersen}\ \emph {et~al.}(2010)\citenamefont
  {Lydersen}, \citenamefont {Wiechers}, \citenamefont {Wittmann}, \citenamefont
  {Elser}, \citenamefont {Skaar},\ and\ \citenamefont {Makarov}}]{Lydersen10}%
  \BibitemOpen
  \bibfield  {author} {\bibinfo {author} {\bibfnamefont {L.}~\bibnamefont
  {Lydersen}}, \bibinfo {author} {\bibfnamefont {C.}~\bibnamefont {Wiechers}},
  \bibinfo {author} {\bibfnamefont {C.}~\bibnamefont {Wittmann}}, \bibinfo
  {author} {\bibfnamefont {D.}~\bibnamefont {Elser}}, \bibinfo {author}
  {\bibfnamefont {J.}~\bibnamefont {Skaar}},\ and\ \bibinfo {author}
  {\bibfnamefont {V.}~\bibnamefont {Makarov}},\ }\bibfield  {title} {\bibinfo
  {title} {{Hacking commercial quantum cryptography systems by tailored bright
  illumination}},\ }\href {https://doi.org/10.1038/nphoton.2010.214} {\bibfield
   {journal} {\bibinfo  {journal} {Nat. Photonics}\ }\textbf {\bibinfo {volume}
  {4}},\ \bibinfo {pages} {686} (\bibinfo {year} {2010})}\BibitemShut {NoStop}%
\bibitem [{\citenamefont {Gerhardt}\ \emph {et~al.}(2011)\citenamefont
  {Gerhardt}, \citenamefont {Liu}, \citenamefont {Lamas-Linares}, \citenamefont
  {Skaar}, \citenamefont {Kurtsiefer},\ and\ \citenamefont
  {Makarov}}]{Gerhardt11}%
  \BibitemOpen
  \bibfield  {author} {\bibinfo {author} {\bibfnamefont {I.}~\bibnamefont
  {Gerhardt}}, \bibinfo {author} {\bibfnamefont {Q.}~\bibnamefont {Liu}},
  \bibinfo {author} {\bibfnamefont {A.}~\bibnamefont {Lamas-Linares}}, \bibinfo
  {author} {\bibfnamefont {J.}~\bibnamefont {Skaar}}, \bibinfo {author}
  {\bibfnamefont {C.}~\bibnamefont {Kurtsiefer}},\ and\ \bibinfo {author}
  {\bibfnamefont {V.}~\bibnamefont {Makarov}},\ }\bibfield  {title} {\bibinfo
  {title} {Full-field implementation of a perfect eavesdropper on a quantum
  cryptography system},\ }\bibfield  {journal} {\bibinfo  {journal} {Nat.
  Commun.}\ }\textbf {\bibinfo {volume} {2}},\ \href
  {https://doi.org/10.1038/ncomms1348} {10.1038/ncomms1348} (\bibinfo {year}
  {2011})\BibitemShut {NoStop}%
\bibitem [{\citenamefont {Weier}\ \emph {et~al.}(2011)\citenamefont {Weier},
  \citenamefont {Krauss}, \citenamefont {Rau}, \citenamefont {F{\"u}rst},
  \citenamefont {Nauerth},\ and\ \citenamefont {Weinfurter}}]{Weier11}%
  \BibitemOpen
  \bibfield  {author} {\bibinfo {author} {\bibfnamefont {H.}~\bibnamefont
  {Weier}}, \bibinfo {author} {\bibfnamefont {H.}~\bibnamefont {Krauss}},
  \bibinfo {author} {\bibfnamefont {M.}~\bibnamefont {Rau}}, \bibinfo {author}
  {\bibfnamefont {M.}~\bibnamefont {F{\"u}rst}}, \bibinfo {author}
  {\bibfnamefont {S.}~\bibnamefont {Nauerth}},\ and\ \bibinfo {author}
  {\bibfnamefont {H.}~\bibnamefont {Weinfurter}},\ }\bibfield  {title}
  {\bibinfo {title} {Quantum eavesdropping without interception: an attack
  exploiting the dead time of single-photon detectors},\ }\href@noop {}
  {\bibfield  {journal} {\bibinfo  {journal} {New J. Phys.}\ }\textbf {\bibinfo
  {volume} {13}},\ \bibinfo {pages} {073024} (\bibinfo {year}
  {2011})}\BibitemShut {NoStop}%
\bibitem [{\citenamefont {Pirandola}\ \emph {et~al.}(2020)\citenamefont
  {Pirandola}, \citenamefont {Andersen}, \citenamefont {Banchi}, \citenamefont
  {Berta}, \citenamefont {Bunandar}, \citenamefont {Colbeck}, \citenamefont
  {Englund}, \citenamefont {Gehring}, \citenamefont {Lupo}, \citenamefont
  {Ottaviani}, \citenamefont {Pereira}, \citenamefont {Razavi}, \citenamefont
  {Shaari}, \citenamefont {Tomamichel}, \citenamefont {Usenko}, \citenamefont
  {Vallone}, \citenamefont {Villoresi},\ and\ \citenamefont
  {Wallden}}]{Pirandola20}%
  \BibitemOpen
  \bibfield  {author} {\bibinfo {author} {\bibfnamefont {S.}~\bibnamefont
  {Pirandola}}, \bibinfo {author} {\bibfnamefont {U.~L.}\ \bibnamefont
  {Andersen}}, \bibinfo {author} {\bibfnamefont {L.}~\bibnamefont {Banchi}},
  \bibinfo {author} {\bibfnamefont {M.}~\bibnamefont {Berta}}, \bibinfo
  {author} {\bibfnamefont {D.}~\bibnamefont {Bunandar}}, \bibinfo {author}
  {\bibfnamefont {R.}~\bibnamefont {Colbeck}}, \bibinfo {author} {\bibfnamefont
  {D.}~\bibnamefont {Englund}}, \bibinfo {author} {\bibfnamefont
  {T.}~\bibnamefont {Gehring}}, \bibinfo {author} {\bibfnamefont
  {C.}~\bibnamefont {Lupo}}, \bibinfo {author} {\bibfnamefont {C.}~\bibnamefont
  {Ottaviani}}, \bibinfo {author} {\bibfnamefont {J.~L.}\ \bibnamefont
  {Pereira}}, \bibinfo {author} {\bibfnamefont {M.}~\bibnamefont {Razavi}},
  \bibinfo {author} {\bibfnamefont {J.~S.}\ \bibnamefont {Shaari}}, \bibinfo
  {author} {\bibfnamefont {M.}~\bibnamefont {Tomamichel}}, \bibinfo {author}
  {\bibfnamefont {V.~C.}\ \bibnamefont {Usenko}}, \bibinfo {author}
  {\bibfnamefont {G.}~\bibnamefont {Vallone}}, \bibinfo {author} {\bibfnamefont
  {P.}~\bibnamefont {Villoresi}},\ and\ \bibinfo {author} {\bibfnamefont
  {P.}~\bibnamefont {Wallden}},\ }\bibfield  {title} {\bibinfo {title}
  {Advances in quantum cryptography},\ }\href
  {https://doi.org/10.1364/AOP.361502} {\bibfield  {journal} {\bibinfo
  {journal} {Adv. Opt. Photon.}\ }\textbf {\bibinfo {volume} {12}},\ \bibinfo
  {pages} {1012} (\bibinfo {year} {2020})}\BibitemShut {NoStop}%
\bibitem [{\citenamefont {Pironio}\ \emph {et~al.}(2009)\citenamefont
  {Pironio}, \citenamefont {Acin}, \citenamefont {Brunner}, \citenamefont
  {Gisin}, \citenamefont {Massar},\ and\ \citenamefont {Scarani}}]{Pironio09}%
  \BibitemOpen
  \bibfield  {author} {\bibinfo {author} {\bibfnamefont {S.}~\bibnamefont
  {Pironio}}, \bibinfo {author} {\bibfnamefont {A.}~\bibnamefont {Acin}},
  \bibinfo {author} {\bibfnamefont {N.}~\bibnamefont {Brunner}}, \bibinfo
  {author} {\bibfnamefont {N.}~\bibnamefont {Gisin}}, \bibinfo {author}
  {\bibfnamefont {S.}~\bibnamefont {Massar}},\ and\ \bibinfo {author}
  {\bibfnamefont {V.}~\bibnamefont {Scarani}},\ }\bibfield  {title} {\bibinfo
  {title} {Device-independent quantum key distribution secure against
  collective attacks},\ }\href@noop {} {\bibfield  {journal} {\bibinfo
  {journal} {New J. Phys.}\ }\textbf {\bibinfo {volume} {11}},\ \bibinfo
  {pages} {045021} (\bibinfo {year} {2009})}\BibitemShut {NoStop}%
\bibitem [{\citenamefont {Brunner}\ \emph {et~al.}(2014)\citenamefont
  {Brunner}, \citenamefont {Cavalcanti}, \citenamefont {Pironio}, \citenamefont
  {Scarani},\ and\ \citenamefont {Wehner}}]{Brunner14}%
  \BibitemOpen
  \bibfield  {author} {\bibinfo {author} {\bibfnamefont {N.}~\bibnamefont
  {Brunner}}, \bibinfo {author} {\bibfnamefont {D.}~\bibnamefont {Cavalcanti}},
  \bibinfo {author} {\bibfnamefont {S.}~\bibnamefont {Pironio}}, \bibinfo
  {author} {\bibfnamefont {V.}~\bibnamefont {Scarani}},\ and\ \bibinfo {author}
  {\bibfnamefont {S.}~\bibnamefont {Wehner}},\ }\bibfield  {title} {\bibinfo
  {title} {{Bell nonlocality}},\ }\href
  {https://doi.org/10.1103/RevModPhys.86.419} {\bibfield  {journal} {\bibinfo
  {journal} {Rev. Mod. Phys.}\ }\textbf {\bibinfo {volume} {86}},\ \bibinfo
  {pages} {419} (\bibinfo {year} {2014})}\BibitemShut {NoStop}%
\bibitem [{\citenamefont {Pironio}\ \emph {et~al.}(2010)\citenamefont
  {Pironio}, \citenamefont {Acín}, \citenamefont {Massar}, \citenamefont
  {Boyer de~la Giroday}, \citenamefont {Matsukevich}, \citenamefont {Maunz},
  \citenamefont {Olmschenk}, \citenamefont {Hayes}, \citenamefont {Luo},
  \citenamefont {Manning},\ and\ \citenamefont {Monroe}}]{Pironio10}%
  \BibitemOpen
  \bibfield  {author} {\bibinfo {author} {\bibfnamefont {S.}~\bibnamefont
  {Pironio}}, \bibinfo {author} {\bibfnamefont {A.}~\bibnamefont {Acín}},
  \bibinfo {author} {\bibfnamefont {S.}~\bibnamefont {Massar}}, \bibinfo
  {author} {\bibfnamefont {A.}~\bibnamefont {Boyer de~la Giroday}}, \bibinfo
  {author} {\bibfnamefont {D.~N.}\ \bibnamefont {Matsukevich}}, \bibinfo
  {author} {\bibfnamefont {P.}~\bibnamefont {Maunz}}, \bibinfo {author}
  {\bibfnamefont {S.}~\bibnamefont {Olmschenk}}, \bibinfo {author}
  {\bibfnamefont {D.}~\bibnamefont {Hayes}}, \bibinfo {author} {\bibfnamefont
  {L.}~\bibnamefont {Luo}}, \bibinfo {author} {\bibfnamefont {T.~A.}\
  \bibnamefont {Manning}},\ and\ \bibinfo {author} {\bibfnamefont
  {C.}~\bibnamefont {Monroe}},\ }\bibfield  {title} {\bibinfo {title} {{Random
  numbers certified by Bell’s theorem}},\ }\href
  {https://doi.org/10.1038/nature09008} {\bibfield  {journal} {\bibinfo
  {journal} {Nature}\ }\textbf {\bibinfo {volume} {464}},\ \bibinfo {pages}
  {1021} (\bibinfo {year} {2010})}\BibitemShut {NoStop}%
\bibitem [{\citenamefont {Kessler}\ and\ \citenamefont
  {Arnon-Friedman}(2020)}]{Kessler20}%
  \BibitemOpen
  \bibfield  {author} {\bibinfo {author} {\bibfnamefont {M.}~\bibnamefont
  {Kessler}}\ and\ \bibinfo {author} {\bibfnamefont {R.}~\bibnamefont
  {Arnon-Friedman}},\ }\bibfield  {title} {\bibinfo {title} {Device-independent
  randomness amplification and privatization},\ }\href@noop {} {\bibfield
  {journal} {\bibinfo  {journal} {IEEE Journal on Selected Areas in Information
  Theory}\ }\textbf {\bibinfo {volume} {1}},\ \bibinfo {pages} {568} (\bibinfo
  {year} {2020})}\BibitemShut {NoStop}%
\bibitem [{\citenamefont {Murta}\ \emph {et~al.}(2019)\citenamefont {Murta},
  \citenamefont {van Dam}, \citenamefont {Ribeiro}, \citenamefont {Hanson},\
  and\ \citenamefont {Wehner}}]{Murta19}%
  \BibitemOpen
  \bibfield  {author} {\bibinfo {author} {\bibfnamefont {G.}~\bibnamefont
  {Murta}}, \bibinfo {author} {\bibfnamefont {S.~B.}\ \bibnamefont {van Dam}},
  \bibinfo {author} {\bibfnamefont {J.}~\bibnamefont {Ribeiro}}, \bibinfo
  {author} {\bibfnamefont {R.}~\bibnamefont {Hanson}},\ and\ \bibinfo {author}
  {\bibfnamefont {S.}~\bibnamefont {Wehner}},\ }\bibfield  {title} {\bibinfo
  {title} {{Towards a realization of device-independent quantum key
  distribution}},\ }\href@noop {} {\bibfield  {journal} {\bibinfo  {journal}
  {Quantum Sci. Technol.}\ }\textbf {\bibinfo {volume} {4}},\ \bibinfo {pages}
  {035011} (\bibinfo {year} {2019})}\BibitemShut {NoStop}%
\bibitem [{\citenamefont {Hensen}\ \emph {et~al.}(2015)\citenamefont {Hensen},
  \citenamefont {Bernien}, \citenamefont {Dr{\'e}au}, \citenamefont {Reiserer},
  \citenamefont {Kalb}, \citenamefont {Blok}, \citenamefont {Ruitenberg},
  \citenamefont {Vermeulen}, \citenamefont {Schouten}, \citenamefont
  {Abell{\'a}n}, \citenamefont {Amaya}, \citenamefont {Pruneri}, \citenamefont
  {Mitchell}, \citenamefont {Markham}, \citenamefont {Twitchen}, \citenamefont
  {Elkouss}, \citenamefont {Wehner}, \citenamefont {Taminiau},\ and\
  \citenamefont {Hanson}}]{Hensen2015}%
  \BibitemOpen
  \bibfield  {author} {\bibinfo {author} {\bibfnamefont {B.}~\bibnamefont
  {Hensen}}, \bibinfo {author} {\bibfnamefont {H.}~\bibnamefont {Bernien}},
  \bibinfo {author} {\bibfnamefont {A.~E.}\ \bibnamefont {Dr{\'e}au}}, \bibinfo
  {author} {\bibfnamefont {A.}~\bibnamefont {Reiserer}}, \bibinfo {author}
  {\bibfnamefont {N.}~\bibnamefont {Kalb}}, \bibinfo {author} {\bibfnamefont
  {M.~S.}\ \bibnamefont {Blok}}, \bibinfo {author} {\bibfnamefont
  {J.}~\bibnamefont {Ruitenberg}}, \bibinfo {author} {\bibfnamefont {R.~F.~L.}\
  \bibnamefont {Vermeulen}}, \bibinfo {author} {\bibfnamefont {R.~N.}\
  \bibnamefont {Schouten}}, \bibinfo {author} {\bibfnamefont {C.}~\bibnamefont
  {Abell{\'a}n}}, \bibinfo {author} {\bibfnamefont {W.}~\bibnamefont {Amaya}},
  \bibinfo {author} {\bibfnamefont {V.}~\bibnamefont {Pruneri}}, \bibinfo
  {author} {\bibfnamefont {M.~W.}\ \bibnamefont {Mitchell}}, \bibinfo {author}
  {\bibfnamefont {M.}~\bibnamefont {Markham}}, \bibinfo {author} {\bibfnamefont
  {D.~J.}\ \bibnamefont {Twitchen}}, \bibinfo {author} {\bibfnamefont
  {D.}~\bibnamefont {Elkouss}}, \bibinfo {author} {\bibfnamefont
  {S.}~\bibnamefont {Wehner}}, \bibinfo {author} {\bibfnamefont {T.~H.}\
  \bibnamefont {Taminiau}},\ and\ \bibinfo {author} {\bibfnamefont
  {R.}~\bibnamefont {Hanson}},\ }\bibfield  {title} {\bibinfo {title}
  {Loophole-free {{Bell}} inequality violation using electron spins separated
  by 1.3 kilometres},\ }\href {https://doi.org/10.1038/nature15759} {\bibfield
  {journal} {\bibinfo  {journal} {Nature}\ }\textbf {\bibinfo {volume} {526}},\
  \bibinfo {pages} {682} (\bibinfo {year} {2015})}\BibitemShut {NoStop}%
\bibitem [{\citenamefont {Shalm}\ \emph {et~al.}(2015)\citenamefont {Shalm},
  \citenamefont {Meyer-Scott}, \citenamefont {Christensen}, \citenamefont
  {Bierhorst}, \citenamefont {Wayne}, \citenamefont {Stevens}, \citenamefont
  {Gerrits}, \citenamefont {Glancy}, \citenamefont {Hamel}, \citenamefont
  {Allman} \emph {et~al.}}]{Shalm15}%
  \BibitemOpen
  \bibfield  {author} {\bibinfo {author} {\bibfnamefont {L.~K.}\ \bibnamefont
  {Shalm}}, \bibinfo {author} {\bibfnamefont {E.}~\bibnamefont {Meyer-Scott}},
  \bibinfo {author} {\bibfnamefont {B.~G.}\ \bibnamefont {Christensen}},
  \bibinfo {author} {\bibfnamefont {P.}~\bibnamefont {Bierhorst}}, \bibinfo
  {author} {\bibfnamefont {M.~A.}\ \bibnamefont {Wayne}}, \bibinfo {author}
  {\bibfnamefont {M.~J.}\ \bibnamefont {Stevens}}, \bibinfo {author}
  {\bibfnamefont {T.}~\bibnamefont {Gerrits}}, \bibinfo {author} {\bibfnamefont
  {S.}~\bibnamefont {Glancy}}, \bibinfo {author} {\bibfnamefont {D.~R.}\
  \bibnamefont {Hamel}}, \bibinfo {author} {\bibfnamefont {M.~S.}\ \bibnamefont
  {Allman}}, \emph {et~al.},\ }\bibfield  {title} {\bibinfo {title} {Strong
  loophole-free test of local realism},\ }\href@noop {} {\bibfield  {journal}
  {\bibinfo  {journal} {Phys. Rev. Lett.}\ }\textbf {\bibinfo {volume} {115}},\
  \bibinfo {pages} {250402} (\bibinfo {year} {2015})}\BibitemShut {NoStop}%
\bibitem [{\citenamefont {Giustina}\ \emph {et~al.}(2015)\citenamefont
  {Giustina}, \citenamefont {Versteegh}, \citenamefont {Wengerowsky},
  \citenamefont {Handsteiner}, \citenamefont {Hochrainer}, \citenamefont
  {Phelan}, \citenamefont {Steinlechner}, \citenamefont {Kofler}, \citenamefont
  {Larsson}, \citenamefont {Abell{\'a}n} \emph {et~al.}}]{Giustina15}%
  \BibitemOpen
  \bibfield  {author} {\bibinfo {author} {\bibfnamefont {M.}~\bibnamefont
  {Giustina}}, \bibinfo {author} {\bibfnamefont {M.~A.}\ \bibnamefont
  {Versteegh}}, \bibinfo {author} {\bibfnamefont {S.}~\bibnamefont
  {Wengerowsky}}, \bibinfo {author} {\bibfnamefont {J.}~\bibnamefont
  {Handsteiner}}, \bibinfo {author} {\bibfnamefont {A.}~\bibnamefont
  {Hochrainer}}, \bibinfo {author} {\bibfnamefont {K.}~\bibnamefont {Phelan}},
  \bibinfo {author} {\bibfnamefont {F.}~\bibnamefont {Steinlechner}}, \bibinfo
  {author} {\bibfnamefont {J.}~\bibnamefont {Kofler}}, \bibinfo {author}
  {\bibfnamefont {J.-{\AA}.}\ \bibnamefont {Larsson}}, \bibinfo {author}
  {\bibfnamefont {C.}~\bibnamefont {Abell{\'a}n}}, \emph {et~al.},\ }\bibfield
  {title} {\bibinfo {title} {Significant-loophole-free test of bell’s theorem
  with entangled photons},\ }\href@noop {} {\bibfield  {journal} {\bibinfo
  {journal} {Phys. Rev. Lett.}\ }\textbf {\bibinfo {volume} {115}},\ \bibinfo
  {pages} {250401} (\bibinfo {year} {2015})}\BibitemShut {NoStop}%
\bibitem [{\citenamefont {Rosenfeld}\ \emph {et~al.}(2017)\citenamefont
  {Rosenfeld}, \citenamefont {Burchardt}, \citenamefont {Garthoff},
  \citenamefont {Redeker}, \citenamefont {Ortegel}, \citenamefont {Rau},\ and\
  \citenamefont {Weinfurter}}]{rosenfeldEventReadyBellTest2017}%
  \BibitemOpen
  \bibfield  {author} {\bibinfo {author} {\bibfnamefont {W.}~\bibnamefont
  {Rosenfeld}}, \bibinfo {author} {\bibfnamefont {D.}~\bibnamefont
  {Burchardt}}, \bibinfo {author} {\bibfnamefont {R.}~\bibnamefont {Garthoff}},
  \bibinfo {author} {\bibfnamefont {K.}~\bibnamefont {Redeker}}, \bibinfo
  {author} {\bibfnamefont {N.}~\bibnamefont {Ortegel}}, \bibinfo {author}
  {\bibfnamefont {M.}~\bibnamefont {Rau}},\ and\ \bibinfo {author}
  {\bibfnamefont {H.}~\bibnamefont {Weinfurter}},\ }\bibfield  {title}
  {\bibinfo {title} {Event-{{Ready Bell Test Using Entangled Atoms
  Simultaneously Closing Detection}} and {{Locality Loopholes}}},\ }\href
  {https://doi.org/10.1103/PhysRevLett.119.010402} {\bibfield  {journal}
  {\bibinfo  {journal} {Phys. Rev. Lett.}\ }\textbf {\bibinfo {volume} {119}},\
  \bibinfo {pages} {010402} (\bibinfo {year} {2017})}\BibitemShut {NoStop}%
\bibitem [{\citenamefont {Heshami}\ \emph {et~al.}(2016)\citenamefont
  {Heshami}, \citenamefont {England}, \citenamefont {Humphreys}, \citenamefont
  {Bustard}, \citenamefont {Acosta}, \citenamefont {Nunn},\ and\ \citenamefont
  {Sussman}}]{Heshami16}%
  \BibitemOpen
  \bibfield  {author} {\bibinfo {author} {\bibfnamefont {K.}~\bibnamefont
  {Heshami}}, \bibinfo {author} {\bibfnamefont {D.~G.}\ \bibnamefont
  {England}}, \bibinfo {author} {\bibfnamefont {P.~C.}\ \bibnamefont
  {Humphreys}}, \bibinfo {author} {\bibfnamefont {P.~J.}\ \bibnamefont
  {Bustard}}, \bibinfo {author} {\bibfnamefont {V.~M.}\ \bibnamefont {Acosta}},
  \bibinfo {author} {\bibfnamefont {J.}~\bibnamefont {Nunn}},\ and\ \bibinfo
  {author} {\bibfnamefont {B.~J.}\ \bibnamefont {Sussman}},\ }\bibfield
  {title} {\bibinfo {title} {Quantum memories: emerging applications and recent
  advances},\ }\href {https://doi.org/10.1080/09500340.2016.1148212} {\bibfield
   {journal} {\bibinfo  {journal} {Journal of Modern Optics}\ }\textbf
  {\bibinfo {volume} {63}},\ \bibinfo {pages} {2005} (\bibinfo {year}
  {2016})}\BibitemShut {NoStop}%
\bibitem [{\citenamefont {Ekert}(1991)}]{Ekert91}%
  \BibitemOpen
  \bibfield  {author} {\bibinfo {author} {\bibfnamefont {A.~K.}\ \bibnamefont
  {Ekert}},\ }\bibfield  {title} {\bibinfo {title} {{Quantum cryptography based
  on Bell’s theorem}},\ }\href@noop {} {\bibfield  {journal} {\bibinfo
  {journal} {Phys. Rev. Lett.}\ }\textbf {\bibinfo {volume} {67}},\ \bibinfo
  {pages} {661} (\bibinfo {year} {1991})}\BibitemShut {NoStop}%
\bibitem [{\citenamefont {Arnon-Friedman}\ \emph {et~al.}(2018)\citenamefont
  {Arnon-Friedman}, \citenamefont {Dupuis}, \citenamefont {Fawzi},
  \citenamefont {Renner},\ and\ \citenamefont {Vidick}}]{ArnonFriedman18}%
  \BibitemOpen
  \bibfield  {author} {\bibinfo {author} {\bibfnamefont {R.}~\bibnamefont
  {Arnon-Friedman}}, \bibinfo {author} {\bibfnamefont {F.}~\bibnamefont
  {Dupuis}}, \bibinfo {author} {\bibfnamefont {O.}~\bibnamefont {Fawzi}},
  \bibinfo {author} {\bibfnamefont {R.}~\bibnamefont {Renner}},\ and\ \bibinfo
  {author} {\bibfnamefont {T.}~\bibnamefont {Vidick}},\ }\bibfield  {title}
  {\bibinfo {title} {Practical device-independent quantum cryptography via
  entropy accumulation},\ }\href@noop {} {\bibfield  {journal} {\bibinfo
  {journal} {Nat. Commun.}\ }\textbf {\bibinfo {volume} {9}},\ \bibinfo {pages}
  {459} (\bibinfo {year} {2018})}\BibitemShut {NoStop}%
\bibitem [{Note2()}]{Note2}%
  \BibitemOpen
  \bibinfo {note} {Alternatively, these could also be generated as the protocol
  runs}\BibitemShut {NoStop}%
\bibitem [{\citenamefont {Richardson}\ and\ \citenamefont
  {Urbanke}(2008)}]{Richardson08}%
  \BibitemOpen
  \bibfield  {author} {\bibinfo {author} {\bibfnamefont {T.}~\bibnamefont
  {Richardson}}\ and\ \bibinfo {author} {\bibfnamefont {R.}~\bibnamefont
  {Urbanke}},\ }\href {https://doi.org/10.1017/CBO9780511791338} {\emph
  {\bibinfo {title} {{Modern Coding Theory}}}}\ (\bibinfo  {publisher}
  {Cambridge: Cambridge University Press},\ \bibinfo {year} {2008})\BibitemShut
  {NoStop}%
\bibitem [{\citenamefont {Wang}\ and\ \citenamefont {Kim}(2015)}]{WangKim15}%
  \BibitemOpen
  \bibfield  {author} {\bibinfo {author} {\bibfnamefont {L.}~\bibnamefont
  {Wang}}\ and\ \bibinfo {author} {\bibfnamefont {Y.}~\bibnamefont {Kim}},\
  }\bibfield  {title} {\bibinfo {title} {{Linear code duality between channel
  coding and Slepian-Wolf coding}},\ }in\ \href
  {https://doi.org/10.1109/ALLERTON.2015.7446997} {\emph {\bibinfo {booktitle}
  {{53rd Annual Allerton Conference on Communication, Control, and Computing
  (Allerton), Monticello, IL, USA}}}}\ (\bibinfo {year} {2015})\ pp.\ \bibinfo
  {pages} {147--152}\BibitemShut {NoStop}%
\bibitem [{\citenamefont {Polyanskiy}\ \emph {et~al.}(2010)\citenamefont
  {Polyanskiy}, \citenamefont {Poor},\ and\ \citenamefont
  {Verdu}}]{Polyanskiy10}%
  \BibitemOpen
  \bibfield  {author} {\bibinfo {author} {\bibfnamefont {Y.}~\bibnamefont
  {Polyanskiy}}, \bibinfo {author} {\bibfnamefont {H.~V.}\ \bibnamefont
  {Poor}},\ and\ \bibinfo {author} {\bibfnamefont {S.}~\bibnamefont {Verdu}},\
  }\bibfield  {title} {\bibinfo {title} {{Channel Coding Rate in the Finite
  Blocklength Regime}},\ }\href@noop {} {\bibfield  {journal} {\bibinfo
  {journal} {IEEE Trans. Inf. Theory.}\ }\textbf {\bibinfo {volume} {56}},\
  \bibinfo {pages} {2307} (\bibinfo {year} {2010})}\BibitemShut {NoStop}%
\bibitem [{\citenamefont {Arikan}(2009)}]{Arikan08}%
  \BibitemOpen
  \bibfield  {author} {\bibinfo {author} {\bibfnamefont {E.}~\bibnamefont
  {Arikan}},\ }\bibfield  {title} {\bibinfo {title} {{Channel Polarization: A
  Method for Constructing Capacity-Achieving Codes for Symmetric Binary-Input
  Memoryless Channels}},\ }\href {https://doi.org/10.1109/TIT.2009.2021379}
  {\bibfield  {journal} {\bibinfo  {journal} {IEEE Trans. Inf. Theory.}\
  }\textbf {\bibinfo {volume} {55}},\ \bibinfo {pages} {3051} (\bibinfo {year}
  {2009})}\BibitemShut {NoStop}%
\bibitem [{\citenamefont {Hassani}\ \emph {et~al.}(2014)\citenamefont
  {Hassani}, \citenamefont {Alishahi},\ and\ \citenamefont
  {Urbanke}}]{Hassani14}%
  \BibitemOpen
  \bibfield  {author} {\bibinfo {author} {\bibfnamefont {S.~H.}\ \bibnamefont
  {Hassani}}, \bibinfo {author} {\bibfnamefont {K.}~\bibnamefont {Alishahi}},\
  and\ \bibinfo {author} {\bibfnamefont {R.~L.}\ \bibnamefont {Urbanke}},\
  }\bibfield  {title} {\bibinfo {title} {{Finite-Length Scaling for Polar
  Codes}},\ }\href {https://doi.org/10.1109/TIT.2014.2341919} {\bibfield
  {journal} {\bibinfo  {journal} {IEEE Trans. Inf. Theory.}\ }\textbf {\bibinfo
  {volume} {60}},\ \bibinfo {pages} {5875} (\bibinfo {year}
  {2014})}\BibitemShut {NoStop}%
\bibitem [{\citenamefont {Gallager}(1963)}]{Gallager63}%
  \BibitemOpen
  \bibfield  {author} {\bibinfo {author} {\bibfnamefont {R.~G.}\ \bibnamefont
  {Gallager}},\ }\href@noop {} {\emph {\bibinfo {title} {{Low-Density
  Parity-Check Codes}}}}\ (\bibinfo  {publisher} {Cambridge, MA: MIT Press},\
  \bibinfo {year} {1963})\BibitemShut {NoStop}%
\bibitem [{\citenamefont {Richardson}\ and\ \citenamefont
  {Urbanke}(2001)}]{Richardson01}%
  \BibitemOpen
  \bibfield  {author} {\bibinfo {author} {\bibfnamefont {T.~J.}\ \bibnamefont
  {Richardson}}\ and\ \bibinfo {author} {\bibfnamefont {R.~L.}\ \bibnamefont
  {Urbanke}},\ }\bibfield  {title} {\bibinfo {title} {{The capacity of
  low-density parity-check codes under message-passing decoding}},\ }\href
  {https://doi.org/10.1109/18.910577} {\bibfield  {journal} {\bibinfo
  {journal} {IEEE Trans. Inf. Theory.}\ }\textbf {\bibinfo {volume} {47}},\
  \bibinfo {pages} {599} (\bibinfo {year} {2001})}\BibitemShut {NoStop}%
\bibitem [{\citenamefont {Sason}\ and\ \citenamefont
  {Urbanke}(2003)}]{Sason03}%
  \BibitemOpen
  \bibfield  {author} {\bibinfo {author} {\bibfnamefont {I.}~\bibnamefont
  {Sason}}\ and\ \bibinfo {author} {\bibfnamefont {R.}~\bibnamefont
  {Urbanke}},\ }\bibfield  {title} {\bibinfo {title} {{Parity-check density
  versus performance of binary linear block codes over memoryless symmetric
  channels}},\ }\href@noop {} {\bibfield  {journal} {\bibinfo  {journal} {IEEE
  Trans. Inf. Theory}\ }\textbf {\bibinfo {volume} {49}},\ \bibinfo {pages}
  {1611} (\bibinfo {year} {2003})}\BibitemShut {NoStop}%
\bibitem [{\citenamefont {Richardson}\ \emph {et~al.}(2001)\citenamefont
  {Richardson}, \citenamefont {Shokrollahi},\ and\ \citenamefont
  {Urbanke}}]{RSU01}%
  \BibitemOpen
  \bibfield  {author} {\bibinfo {author} {\bibfnamefont {T.~J.}\ \bibnamefont
  {Richardson}}, \bibinfo {author} {\bibfnamefont {M.~A.}\ \bibnamefont
  {Shokrollahi}},\ and\ \bibinfo {author} {\bibfnamefont {R.~L.}\ \bibnamefont
  {Urbanke}},\ }\bibfield  {title} {\bibinfo {title} {{Design of
  capacity-approaching irregular low-density parity-check codes}},\ }\href
  {https://doi.org/10.1109/18.910578} {\bibfield  {journal} {\bibinfo
  {journal} {IEEE Trans. Inf. Theory.}\ }\textbf {\bibinfo {volume} {47}},\
  \bibinfo {pages} {619} (\bibinfo {year} {2001})}\BibitemShut {NoStop}%
\bibitem [{\citenamefont {Shokrollahi}(1999)}]{Shokrollahi99}%
  \BibitemOpen
  \bibfield  {author} {\bibinfo {author} {\bibfnamefont {A.}~\bibnamefont
  {Shokrollahi}},\ }\bibfield  {title} {\bibinfo {title} {{New sequences of
  linear time erasure codes approaching channel capacity}},\ }\href@noop {}
  {\bibfield  {journal} {\bibinfo  {journal} {Proc. IEEE Int. Symp. Information
  Theory and its Applications, Honolulu, HI}\ ,\ \bibinfo {pages} {65}}
  (\bibinfo {year} {1999})}\BibitemShut {NoStop}%
\bibitem [{\citenamefont {Oswald}\ and\ \citenamefont
  {Shokrollahi}(2002)}]{Oswald02}%
  \BibitemOpen
  \bibfield  {author} {\bibinfo {author} {\bibfnamefont {P.}~\bibnamefont
  {Oswald}}\ and\ \bibinfo {author} {\bibfnamefont {A.}~\bibnamefont
  {Shokrollahi}},\ }\bibfield  {title} {\bibinfo {title} {{Cappacity-achieving
  sequences for the erasure channel}},\ }\href@noop {} {\bibfield  {journal}
  {\bibinfo  {journal} {IEEE Trans. Inf. Theory}\ }\textbf {\bibinfo {volume}
  {48}},\ \bibinfo {pages} {3017–3028} (\bibinfo {year} {2002})}\BibitemShut
  {NoStop}%
\bibitem [{\citenamefont {Pfister}\ \emph {et~al.}(2005)\citenamefont
  {Pfister}, \citenamefont {Sason},\ and\ \citenamefont {Urbanke}}]{Pfister05}%
  \BibitemOpen
  \bibfield  {author} {\bibinfo {author} {\bibfnamefont {H.~D.}\ \bibnamefont
  {Pfister}}, \bibinfo {author} {\bibfnamefont {I.}~\bibnamefont {Sason}},\
  and\ \bibinfo {author} {\bibfnamefont {R.}~\bibnamefont {Urbanke}},\
  }\bibfield  {title} {\bibinfo {title} {{Capacity-achieving ensembles for the
  binary erasure channel with bounded complexity}},\ }\href@noop {} {\bibfield
  {journal} {\bibinfo  {journal} {IEEE Trans. Inf. Theory}\ }\textbf {\bibinfo
  {volume} {51}},\ \bibinfo {pages} {2352–2379} (\bibinfo {year}
  {2005})}\BibitemShut {NoStop}%
\bibitem [{\citenamefont {Amraoui}\ \emph {et~al.}(2009)\citenamefont
  {Amraoui}, \citenamefont {Montanari}, \citenamefont {Richardson},\ and\
  \citenamefont {Urbanke}}]{Amraoui08}%
  \BibitemOpen
  \bibfield  {author} {\bibinfo {author} {\bibfnamefont {A.}~\bibnamefont
  {Amraoui}}, \bibinfo {author} {\bibfnamefont {A.}~\bibnamefont {Montanari}},
  \bibinfo {author} {\bibfnamefont {T.}~\bibnamefont {Richardson}},\ and\
  \bibinfo {author} {\bibfnamefont {R.}~\bibnamefont {Urbanke}},\ }\bibfield
  {title} {\bibinfo {title} {{Finite-Length Scaling for Iteratively Decoded
  LDPC Ensembles}},\ }\href {https://doi.org/10.1109/TIT.2008.2009580}
  {\bibfield  {journal} {\bibinfo  {journal} {IEEE Trans. Inf. Theory.}\
  }\textbf {\bibinfo {volume} {55}},\ \bibinfo {pages} {473} (\bibinfo {year}
  {2009})}\BibitemShut {NoStop}%
\bibitem [{\citenamefont {Lentmaier}\ \emph {et~al.}(2010)\citenamefont
  {Lentmaier}, \citenamefont {Sridharan}, \citenamefont {Costello},\ and\
  \citenamefont {Zigangirov}}]{Lentmaier10}%
  \BibitemOpen
  \bibfield  {author} {\bibinfo {author} {\bibfnamefont {M.}~\bibnamefont
  {Lentmaier}}, \bibinfo {author} {\bibfnamefont {A.}~\bibnamefont
  {Sridharan}}, \bibinfo {author} {\bibfnamefont {D.~J.}\ \bibnamefont
  {Costello}},\ and\ \bibinfo {author} {\bibfnamefont {K.~S.}\ \bibnamefont
  {Zigangirov}},\ }\bibfield  {title} {\bibinfo {title} {{Iterative Decoding
  Threshold Analysis for LDPC Convolutional Codes}},\ }\href
  {https://doi.org/10.1109/TIT.2010.2059490} {\bibfield  {journal} {\bibinfo
  {journal} {IEEE Trans. Inf. Theory.}\ }\textbf {\bibinfo {volume} {56}},\
  \bibinfo {pages} {5274} (\bibinfo {year} {2010})}\BibitemShut {NoStop}%
\bibitem [{\citenamefont {Kudekar}\ \emph {et~al.}(2013)\citenamefont
  {Kudekar}, \citenamefont {Richardson},\ and\ \citenamefont
  {Urbanke}}]{Kudekar13}%
  \BibitemOpen
  \bibfield  {author} {\bibinfo {author} {\bibfnamefont {S.}~\bibnamefont
  {Kudekar}}, \bibinfo {author} {\bibfnamefont {T.}~\bibnamefont
  {Richardson}},\ and\ \bibinfo {author} {\bibfnamefont {R.~L.}\ \bibnamefont
  {Urbanke}},\ }\bibfield  {title} {\bibinfo {title} {{Spatially Coupled
  Ensembles Universally Achieve Capacity Under Belief Propagation}},\ }\href
  {https://doi.org/10.1109/TIT.2013.2280915} {\bibfield  {journal} {\bibinfo
  {journal} {IEEE Trans. Inf. Theory.}\ }\textbf {\bibinfo {volume} {59}},\
  \bibinfo {pages} {7761} (\bibinfo {year} {2013})}\BibitemShut {NoStop}%
\bibitem [{\citenamefont {Mitchell}\ \emph {et~al.}(2015)\citenamefont
  {Mitchell}, \citenamefont {Lentmaier},\ and\ \citenamefont
  {Costello}}]{Mitchell15}%
  \BibitemOpen
  \bibfield  {author} {\bibinfo {author} {\bibfnamefont {D.~G.~M.}\
  \bibnamefont {Mitchell}}, \bibinfo {author} {\bibfnamefont {M.}~\bibnamefont
  {Lentmaier}},\ and\ \bibinfo {author} {\bibfnamefont {D.~J.}\ \bibnamefont
  {Costello}},\ }\bibfield  {title} {\bibinfo {title} {{Spatially Coupled LDPC
  Codes Constructed From Protographs}},\ }\href
  {https://doi.org/10.1109/TIT.2015.2453267} {\bibfield  {journal} {\bibinfo
  {journal} {IEEE Trans. Inform. Theory}\ }\textbf {\bibinfo {volume} {61}},\
  \bibinfo {pages} {4866–4889} (\bibinfo {year} {2015})}\BibitemShut
  {NoStop}%
\bibitem [{\citenamefont {Chen}\ \emph {et~al.}(2006)\citenamefont {Chen},
  \citenamefont {He},\ and\ \citenamefont {Jagmohan}}]{Chen06}%
  \BibitemOpen
  \bibfield  {author} {\bibinfo {author} {\bibfnamefont {J.}~\bibnamefont
  {Chen}}, \bibinfo {author} {\bibfnamefont {D.}~\bibnamefont {He}},\ and\
  \bibinfo {author} {\bibfnamefont {A.}~\bibnamefont {Jagmohan}},\ }\bibfield
  {title} {\bibinfo {title} {Slepian-wolf code design via source-channel
  correspondence},\ }in\ \href {https://doi.org/10.1109/ISIT.2006.262025}
  {\emph {\bibinfo {booktitle} {2006 IEEE International Symposium on
  Information Theory}}}\ (\bibinfo {year} {2006})\ pp.\ \bibinfo {pages}
  {2433--2437}\BibitemShut {NoStop}%
\bibitem [{\citenamefont {Zhang}\ and\ \citenamefont
  {Fossorier}(2005)}]{Zhang05}%
  \BibitemOpen
  \bibfield  {author} {\bibinfo {author} {\bibfnamefont {J.}~\bibnamefont
  {Zhang}}\ and\ \bibinfo {author} {\bibfnamefont {M.}~\bibnamefont
  {Fossorier}},\ }\bibfield  {title} {\bibinfo {title} {{Shuffled iterative
  decoding}},\ }\href@noop {} {\bibfield  {journal} {\bibinfo  {journal} {IEEE
  Trans. Commun.}\ }\textbf {\bibinfo {volume} {53}},\ \bibinfo {pages} {209}
  (\bibinfo {year} {2005})}\BibitemShut {NoStop}%
\bibitem [{\citenamefont {Hocevar}(2004)}]{Hocevar04}%
  \BibitemOpen
  \bibfield  {author} {\bibinfo {author} {\bibfnamefont {D.~E.}\ \bibnamefont
  {Hocevar}},\ }\bibfield  {title} {\bibinfo {title} {{A reduced complexity
  decoder architecture via layered decoding of LDPC code}},\ }\href@noop {}
  {\bibfield  {journal} {\bibinfo  {journal} {Proc. IEEE Workshop Signal
  processing and Systems (SIPS. 04), Austin, TX}\ ,\ \bibinfo {pages} {107}}
  (\bibinfo {year} {2004})}\BibitemShut {NoStop}%
\bibitem [{\citenamefont {Lentmaier}\ \emph {et~al.}(2011)\citenamefont
  {Lentmaier}, \citenamefont {Prenda},\ and\ \citenamefont
  {Fettweis}}]{Lentmaier11}%
  \BibitemOpen
  \bibfield  {author} {\bibinfo {author} {\bibfnamefont {M.}~\bibnamefont
  {Lentmaier}}, \bibinfo {author} {\bibfnamefont {M.~M.}\ \bibnamefont
  {Prenda}},\ and\ \bibinfo {author} {\bibfnamefont {G.~P.}\ \bibnamefont
  {Fettweis}},\ }\bibfield  {title} {\bibinfo {title} {{Efficient message
  passing scheduling for terminated LDPC convolutional codes}},\ }\href@noop {}
  {\bibfield  {journal} {\bibinfo  {journal} {Proc. IEEE Int. Symp. Inf.
  Theory, St. Petersburg, Russia}\ ,\ \bibinfo {pages} {1826–1830}} (\bibinfo
  {year} {2011})}\BibitemShut {NoStop}%
\bibitem [{\citenamefont {Iyengar}\ \emph {et~al.}(2012)\citenamefont
  {Iyengar}, \citenamefont {Papaleo}, \citenamefont {Siegel}, \citenamefont
  {Wolf}, \citenamefont {Vanelli-Coralli},\ and\ \citenamefont
  {Corazza}}]{Iyengar12}%
  \BibitemOpen
  \bibfield  {author} {\bibinfo {author} {\bibfnamefont {A.}~\bibnamefont
  {Iyengar}}, \bibinfo {author} {\bibfnamefont {M.}~\bibnamefont {Papaleo}},
  \bibinfo {author} {\bibfnamefont {P.}~\bibnamefont {Siegel}}, \bibinfo
  {author} {\bibfnamefont {J.}~\bibnamefont {Wolf}}, \bibinfo {author}
  {\bibfnamefont {A.}~\bibnamefont {Vanelli-Coralli}},\ and\ \bibinfo {author}
  {\bibfnamefont {G.}~\bibnamefont {Corazza}},\ }\bibfield  {title} {\bibinfo
  {title} {{Windowed decoding of protograph-based LDPC convolutional codes over
  erasure channels}},\ }\href@noop {} {\bibfield  {journal} {\bibinfo
  {journal} {IEEE Trans. Inf. Theory}\ }\textbf {\bibinfo {volume} {58}},\
  \bibinfo {pages} {2303–2320} (\bibinfo {year} {2012})}\BibitemShut
  {NoStop}%
\bibitem [{\citenamefont {Arnon-Friedman}\ \emph {et~al.}(2019)\citenamefont
  {Arnon-Friedman}, \citenamefont {Renner},\ and\ \citenamefont
  {Vidick}}]{ArnonFriedman19}%
  \BibitemOpen
  \bibfield  {author} {\bibinfo {author} {\bibfnamefont {R.}~\bibnamefont
  {Arnon-Friedman}}, \bibinfo {author} {\bibfnamefont {R.}~\bibnamefont
  {Renner}},\ and\ \bibinfo {author} {\bibfnamefont {T.}~\bibnamefont
  {Vidick}},\ }\bibfield  {title} {\bibinfo {title} {{Simple and Tight
  Device-Independent Security Proofs}},\ }\href@noop {} {\bibfield  {journal}
  {\bibinfo  {journal} {SIAM J. Comput.}\ }\textbf {\bibinfo {volume} {48}},\
  \bibinfo {pages} {181} (\bibinfo {year} {2019})}\BibitemShut {NoStop}%
\bibitem [{\citenamefont {Tan}\ \emph {et~al.}(2020)\citenamefont {Tan},
  \citenamefont {Sekatski}, \citenamefont {Bancal}, \citenamefont {Schwonnek},
  \citenamefont {Renner}, \citenamefont {Sangouard},\ and\ \citenamefont
  {Lim}}]{Tan20}%
  \BibitemOpen
  \bibfield  {author} {\bibinfo {author} {\bibfnamefont {E.~Y.-Z.}\
  \bibnamefont {Tan}}, \bibinfo {author} {\bibfnamefont {P.}~\bibnamefont
  {Sekatski}}, \bibinfo {author} {\bibfnamefont {J.-D.}\ \bibnamefont
  {Bancal}}, \bibinfo {author} {\bibfnamefont {R.}~\bibnamefont {Schwonnek}},
  \bibinfo {author} {\bibfnamefont {R.}~\bibnamefont {Renner}}, \bibinfo
  {author} {\bibfnamefont {N.}~\bibnamefont {Sangouard}},\ and\ \bibinfo
  {author} {\bibfnamefont {C.~C.-W.}\ \bibnamefont {Lim}},\ }\bibfield  {title}
  {\bibinfo {title} {{Improved DIQKD protocols with finite-size analysis}},\
  }\href@noop {} {\bibfield  {journal} {\bibinfo  {journal} {arXiv:2012.08714}\
  } (\bibinfo {year} {2020})}\BibitemShut {NoStop}%
\bibitem [{Note3()}]{Note3}%
  \BibitemOpen
  \bibinfo {note} {In more detail: since here we have $n$ independent trials of
  a Bernoulli variable with parameter $q$, the number of $W_i =1$ trials is
  binomially distributed with standard deviation $\protect \sqrt {n{q(1-q)}}$.
  Hence the \protect \emph {fraction} of such trials is a random variable with
  standard deviation $\protect \sqrt {{q(1-q)}/{n}}$, and is approximately
  normally distributed for large $n$. This implies that choosing the
  (one-sided) accept threshold 3 standard deviations away from $q$ ensures that
  the accept probability is well over 99\%, as long as the devices perform as
  expected.}\BibitemShut {Stop}%
\bibitem [{\citenamefont {Portmann}\ and\ \citenamefont
  {Renner}(2021)}]{Portmann21}%
  \BibitemOpen
  \bibfield  {author} {\bibinfo {author} {\bibfnamefont {C.}~\bibnamefont
  {Portmann}}\ and\ \bibinfo {author} {\bibfnamefont {R.}~\bibnamefont
  {Renner}},\ }\bibfield  {title} {\bibinfo {title} {{Security in Quantum
  Cryptography}},\ }\href {http://arxiv.org/abs/2102.00021} {\bibfield
  {journal} {\bibinfo  {journal} {arXiv:2102.00021}\ } (\bibinfo {year}
  {2021})}\BibitemShut {NoStop}%
\bibitem [{\citenamefont {Tomamichel}(2016)}]{Tomamichel16}%
  \BibitemOpen
  \bibfield  {author} {\bibinfo {author} {\bibfnamefont {M.}~\bibnamefont
  {Tomamichel}},\ }\bibfield  {title} {\bibinfo {title} {{Quantum Information
  Processing with Finite Resources}},\ }\href {arXiv:1504.00233v5} {\bibfield
  {journal} {\bibinfo  {journal} {Springer International Publishing}\ }
  (\bibinfo {year} {2016})}\BibitemShut {NoStop}%
\bibitem [{Note4()}]{Note4}%
  \BibitemOpen
  \bibinfo {note} {This follows straightforwardly from the fact that in either
  optimization, for any feasible $\sigma _B$, the normalized state $\protect
  \hat {\sigma }_B = \sigma _B/\protect \textrm {Tr}\left (\sigma _B\right )$
  is another feasible point, and it yields an objective value at least as large
  as that for $\sigma _B$}\BibitemShut {NoStop}%
\bibitem [{\citenamefont {Tomamichel}\ \emph {et~al.}(2010)\citenamefont
  {Tomamichel}, \citenamefont {Colbeck},\ and\ \citenamefont
  {Renner}}]{Tomamichel09}%
  \BibitemOpen
  \bibfield  {author} {\bibinfo {author} {\bibfnamefont {M.}~\bibnamefont
  {Tomamichel}}, \bibinfo {author} {\bibfnamefont {R.}~\bibnamefont
  {Colbeck}},\ and\ \bibinfo {author} {\bibfnamefont {R.}~\bibnamefont
  {Renner}},\ }\bibfield  {title} {\bibinfo {title} {{Duality Between Smooth
  Min- and Max-Entropies}},\ }\href@noop {} {\bibfield  {journal} {\bibinfo
  {journal} {IEEE Trans. on Inf. Theory}\ }\textbf {\bibinfo {volume} {56}},\
  \bibinfo {pages} {4674} (\bibinfo {year} {2010})}\BibitemShut {NoStop}%
\bibitem [{\citenamefont {Stinson}(1994)}]{Stinson94}%
  \BibitemOpen
  \bibfield  {author} {\bibinfo {author} {\bibfnamefont {D.~R.}\ \bibnamefont
  {Stinson}},\ }\bibfield  {title} {\bibinfo {title} {{Universal hashing and
  authentication codes}},\ }\href@noop {} {\bibfield  {journal} {\bibinfo
  {journal} {Designs, Codes and Cryptography}\ }\textbf {\bibinfo {volume}
  {4}},\ \bibinfo {pages} {369} (\bibinfo {year} {1994})}\BibitemShut {NoStop}%
\bibitem [{\citenamefont {Krawczyk}(1994)}]{Krawczyk94}%
  \BibitemOpen
  \bibfield  {author} {\bibinfo {author} {\bibfnamefont {H.}~\bibnamefont
  {Krawczyk}},\ }\bibfield  {title} {\bibinfo {title} {{LFSR-based hashing and
  authentication}},\ }in\ \href@noop {} {\emph {\bibinfo {booktitle} {{Advances
  in Cryptology - CRYPTO '94 in Lecture Notes in Computer Science}}}},\ Vol.\
  \bibinfo {volume} {839}\ (\bibinfo  {publisher} {Springer},\ \bibinfo {year}
  {1994})\ pp.\ \bibinfo {pages} {129--139}\BibitemShut {NoStop}%
\bibitem [{\citenamefont {Dai}\ and\ \citenamefont {Krovetz}()}]{Dai07}%
  \BibitemOpen
  \bibfield  {author} {\bibinfo {author} {\bibfnamefont {W.}~\bibnamefont
  {Dai}}\ and\ \bibinfo {author} {\bibfnamefont {T.}~\bibnamefont {Krovetz}},\
  }\bibfield  {title} {\bibinfo {title} {{VHASH Security}},\ }\href@noop {}
  {\bibinfo  {journal} {Cryptology ePrint Archive,
  http://eprint.iacr.org/2007/338}\ }\BibitemShut {NoStop}%
\bibitem [{\citenamefont {Wegman}\ and\ \citenamefont
  {Carter}(1981)}]{Wegman81}%
  \BibitemOpen
\bibfield  {journal} {  }\bibfield  {author} {\bibinfo {author} {\bibfnamefont
  {M.~N.}\ \bibnamefont {Wegman}}\ and\ \bibinfo {author} {\bibfnamefont
  {L.}~\bibnamefont {Carter}},\ }\bibfield  {title} {\bibinfo {title} {{New
  hash functions and their use in authentication and set equality}},\
  }\href@noop {} {\bibfield  {journal} {\bibinfo  {journal} {J. Comput. Syst.
  Sci.}\ }\textbf {\bibinfo {volume} {22}},\ \bibinfo {pages} {265} (\bibinfo
  {year} {1981})}\BibitemShut {NoStop}%
\bibitem [{\citenamefont {Portmann}(2014)}]{Portmann14}%
  \BibitemOpen
  \bibfield  {author} {\bibinfo {author} {\bibfnamefont {C.}~\bibnamefont
  {Portmann}},\ }\bibfield  {title} {\bibinfo {title} {{Key recycling in
  authentication}},\ }\href@noop {} {\bibfield  {journal} {\bibinfo  {journal}
  {IEEE Trans. Inf. Th.}\ }\textbf {\bibinfo {volume} {60}},\ \bibinfo {pages}
  {4383} (\bibinfo {year} {2014})}\BibitemShut {NoStop}%
\bibitem [{\citenamefont {Bancal}()}]{UVMACcode}%
  \BibitemOpen
  \bibfield  {author} {\bibinfo {author} {\bibfnamefont {J.-D.}\ \bibnamefont
  {Bancal}},\ }\href@noop {} {\bibinfo {title} {{UVMAC source code}}},\
  \bibinfo {howpublished} {\url{https://github.com/jdbancal/uvmac}}\BibitemShut
  {NoStop}%
\bibitem [{\citenamefont {Krovetz}()}]{VMACcode}%
  \BibitemOpen
  \bibfield  {author} {\bibinfo {author} {\bibfnamefont {T.}~\bibnamefont
  {Krovetz}},\ }\href@noop {} {\bibinfo {title} {{VMAC} source code}},\
  \bibinfo {howpublished} {\url{https://www.fastcrypto.org/vmac}}\BibitemShut
  {NoStop}%
\bibitem [{\citenamefont {De}\ \emph {et~al.}(2012)\citenamefont {De},
  \citenamefont {Portmann}, \citenamefont {Vidick},\ and\ \citenamefont
  {Renner}}]{De12}%
  \BibitemOpen
  \bibfield  {author} {\bibinfo {author} {\bibfnamefont {A.}~\bibnamefont
  {De}}, \bibinfo {author} {\bibfnamefont {C.}~\bibnamefont {Portmann}},
  \bibinfo {author} {\bibfnamefont {T.}~\bibnamefont {Vidick}},\ and\ \bibinfo
  {author} {\bibfnamefont {R.}~\bibnamefont {Renner}},\ }\bibfield  {title}
  {\bibinfo {title} {{Trevisan's Extractor in the Presence of Quantum Side
  Information}},\ }\href {https://doi.org/10.1137/100813683} {\bibfield
  {journal} {\bibinfo  {journal} {SIAM J. Comput.}\ }\textbf {\bibinfo {volume}
  {41}},\ \bibinfo {pages} {915} (\bibinfo {year} {2012})}\BibitemShut
  {NoStop}%
\bibitem [{\citenamefont {Mauerer}\ \emph {et~al.}(2012)\citenamefont
  {Mauerer}, \citenamefont {Portmann},\ and\ \citenamefont
  {Scholz}}]{Mauerer12}%
  \BibitemOpen
  \bibfield  {author} {\bibinfo {author} {\bibfnamefont {W.}~\bibnamefont
  {Mauerer}}, \bibinfo {author} {\bibfnamefont {C.}~\bibnamefont {Portmann}},\
  and\ \bibinfo {author} {\bibfnamefont {V.~B.}\ \bibnamefont {Scholz}},\
  }\bibfield  {title} {\bibinfo {title} {{A modular framework for randomness
  extraction based on Trevisian's construction}},\ }\bibfield  {journal}
  {\bibinfo  {journal} {arXiv:1212.0520}\ }\href
  {https://doi.org/10.48550/arXiv.1212.0520} {10.48550/arXiv.1212.0520}
  (\bibinfo {year} {2012})\BibitemShut {NoStop}%
\bibitem [{\citenamefont {Ma}\ \emph {et~al.}(2011)\citenamefont {Ma},
  \citenamefont {Zhang},\ and\ \citenamefont {Tan}}]{Ma11}%
  \BibitemOpen
  \bibfield  {author} {\bibinfo {author} {\bibfnamefont {X.}~\bibnamefont
  {Ma}}, \bibinfo {author} {\bibfnamefont {Z.}~\bibnamefont {Zhang}},\ and\
  \bibinfo {author} {\bibfnamefont {X.}~\bibnamefont {Tan}},\ }\bibfield
  {title} {\bibinfo {title} {{Explicit combinatorial design}},\ }\href@noop {}
  {\bibfield  {journal} {\bibinfo  {journal} {arXiv:1109.6147}\ } (\bibinfo
  {year} {2011})}\BibitemShut {NoStop}%
\bibitem [{\citenamefont {Bierhorst}\ \emph {et~al.}(2018)\citenamefont
  {Bierhorst}, \citenamefont {Knill}, \citenamefont {Glancy}, \citenamefont
  {Zhang}, \citenamefont {Mink}, \citenamefont {Jordan}, \citenamefont
  {Rommal}, \citenamefont {Liu}, \citenamefont {Christensen}, \citenamefont
  {Nam}, \citenamefont {Stevens},\ and\ \citenamefont {Shalm}}]{Bierhorst18}%
  \BibitemOpen
  \bibfield  {author} {\bibinfo {author} {\bibfnamefont {P.}~\bibnamefont
  {Bierhorst}}, \bibinfo {author} {\bibfnamefont {E.}~\bibnamefont {Knill}},
  \bibinfo {author} {\bibfnamefont {S.}~\bibnamefont {Glancy}}, \bibinfo
  {author} {\bibfnamefont {Y.}~\bibnamefont {Zhang}}, \bibinfo {author}
  {\bibfnamefont {A.}~\bibnamefont {Mink}}, \bibinfo {author} {\bibfnamefont
  {S.}~\bibnamefont {Jordan}}, \bibinfo {author} {\bibfnamefont
  {A.}~\bibnamefont {Rommal}}, \bibinfo {author} {\bibfnamefont {Y.-K.}\
  \bibnamefont {Liu}}, \bibinfo {author} {\bibfnamefont {B.}~\bibnamefont
  {Christensen}}, \bibinfo {author} {\bibfnamefont {S.~W.}\ \bibnamefont
  {Nam}}, \bibinfo {author} {\bibfnamefont {M.~J.}\ \bibnamefont {Stevens}},\
  and\ \bibinfo {author} {\bibfnamefont {L.~K.}\ \bibnamefont {Shalm}},\
  }\bibfield  {title} {\bibinfo {title} {{Experimentally Generated Randomness
  Certified by the Impossibility of Superluminal Signals}},\ }\href@noop {}
  {\bibfield  {journal} {\bibinfo  {journal} {Nature}\ }\textbf {\bibinfo
  {volume} {556}},\ \bibinfo {pages} {223} (\bibinfo {year}
  {2018})}\BibitemShut {NoStop}%
\bibitem [{\citenamefont {Mauerer}()}]{Mauerer12code}%
  \BibitemOpen
  \bibfield  {author} {\bibinfo {author} {\bibfnamefont {W.}~\bibnamefont
  {Mauerer}},\ }\href@noop {} {\bibinfo {title} {Trevisan's extractor source
  code}},\ \bibinfo {howpublished}
  {\url{https://github.com/wolfgangmauerer/libtrevisan}}\BibitemShut {NoStop}%
\bibitem [{Note5()}]{Note5}%
  \BibitemOpen
  \bibinfo {note} {In principle, we can pick any positive ``threshold'' value
  $\mu $ to split the two cases: this would give the final result that if $\ell
  \leq \Upsilon _b\left ( H_\protect \mathrm {min}^{{\varepsilon
  _s}}(C|Q)_{\rho } - 6 - 4\protect \qopname \relax o{log}({1}/{\varepsilon
  _\protect \mathrm {PA}}) - \protect \qopname \relax o{log}({1}/{\mu }) \right
  )$, then implementing the extractor with $\varepsilon _1= {\varepsilon
  _\protect \mathrm {PA}}/\ell $ yields a bound of $\protect \qopname \relax
  m{max}\left (\mu , \varepsilon _\protect \mathrm {PA}\right ) + 2{\varepsilon
  _s}$ in Eq.~\protect \textup {\hbox {\mathsurround \z@ \protect \normalfont
  (\ignorespaces \ref {eq:TrevNorm}\unskip \@@italiccorr )}}. However, observe
  that for a fixed value of $\protect \qopname \relax m{max}\left (\mu ,
  \varepsilon _\protect \mathrm {PA}\right )$, the optimal key length given by
  this expression is always obtained by setting $\mu $ and $\varepsilon
  _\protect \mathrm {PA}$ to be equal, hence the choice in this
  proof.}\BibitemShut {Stop}%
\bibitem [{\citenamefont {Tomamichel}\ and\ \citenamefont
  {Leverrier}(2017)}]{Tomamichel17}%
  \BibitemOpen
  \bibfield  {author} {\bibinfo {author} {\bibfnamefont {M.}~\bibnamefont
  {Tomamichel}}\ and\ \bibinfo {author} {\bibfnamefont {A.}~\bibnamefont
  {Leverrier}},\ }\bibfield  {title} {\bibinfo {title} {{A largely
  self-contained and complete security proof for quantum key distribution}},\
  }\href@noop {} {\bibfield  {journal} {\bibinfo  {journal} {Quantum}\ }\textbf
  {\bibinfo {volume} {1}},\ \bibinfo {pages} {14} (\bibinfo {year}
  {2017})}\BibitemShut {NoStop}%
\bibitem [{\citenamefont {Winkler}\ \emph {et~al.}(2011)\citenamefont
  {Winkler}, \citenamefont {Tomamichel}, \citenamefont {Hengl},\ and\
  \citenamefont {Renner}}]{Winkler11}%
  \BibitemOpen
  \bibfield  {author} {\bibinfo {author} {\bibfnamefont {S.}~\bibnamefont
  {Winkler}}, \bibinfo {author} {\bibfnamefont {M.}~\bibnamefont {Tomamichel}},
  \bibinfo {author} {\bibfnamefont {S.}~\bibnamefont {Hengl}},\ and\ \bibinfo
  {author} {\bibfnamefont {R.}~\bibnamefont {Renner}},\ }\bibfield  {title}
  {\bibinfo {title} {{Impossibility of Growing Quantum Bit Commitments}},\
  }\href@noop {} {\bibfield  {journal} {\bibinfo  {journal} {Phys. Rev. Lett.}\
  }\textbf {\bibinfo {volume} {107}},\ \bibinfo {pages} {090502} (\bibinfo
  {year} {2011})}\BibitemShut {NoStop}%
\bibitem [{Note6()}]{Note6}%
  \BibitemOpen
  \bibinfo {note} {This condition is slightly stricter than necessary, as the
  EAT only requires the Markov conditions to hold for the specific state it is
  applied to~\cite {Dupuis20,Dupuis19}. However, for brevity we incorporate
  this into the definition of EAT channels, as was done in~\cite
  {ArnonFriedman19,Liu21}.}\BibitemShut {Stop}%
\bibitem [{\citenamefont {Liu}\ \emph {et~al.}(2021)\citenamefont {Liu},
  \citenamefont {Li}, \citenamefont {Ragy}, \citenamefont {Zhao}, \citenamefont
  {Bai}, \citenamefont {Liu}, \citenamefont {Brown}, \citenamefont {Zhang},
  \citenamefont {Colbeck}, \citenamefont {Fan}, \citenamefont {Zhang},\ and\
  \citenamefont {Pan}}]{Liu21}%
  \BibitemOpen
  \bibfield  {author} {\bibinfo {author} {\bibfnamefont {W.-Z.}\ \bibnamefont
  {Liu}}, \bibinfo {author} {\bibfnamefont {M.-H.}\ \bibnamefont {Li}},
  \bibinfo {author} {\bibfnamefont {S.}~\bibnamefont {Ragy}}, \bibinfo {author}
  {\bibfnamefont {S.-R.}\ \bibnamefont {Zhao}}, \bibinfo {author}
  {\bibfnamefont {B.}~\bibnamefont {Bai}}, \bibinfo {author} {\bibfnamefont
  {Y.}~\bibnamefont {Liu}}, \bibinfo {author} {\bibfnamefont {P.~J.}\
  \bibnamefont {Brown}}, \bibinfo {author} {\bibfnamefont {J.}~\bibnamefont
  {Zhang}}, \bibinfo {author} {\bibfnamefont {R.}~\bibnamefont {Colbeck}},
  \bibinfo {author} {\bibfnamefont {J.}~\bibnamefont {Fan}}, \bibinfo {author}
  {\bibfnamefont {Q.}~\bibnamefont {Zhang}},\ and\ \bibinfo {author}
  {\bibfnamefont {J.-W.}\ \bibnamefont {Pan}},\ }\bibfield  {title} {\bibinfo
  {title} {{Device-independent randomness expansion against quantum side
  information}},\ }\href@noop {} {\bibfield  {journal} {\bibinfo  {journal}
  {Nat. Phys.}\ }\textbf {\bibinfo {volume} {17}},\ \bibinfo {pages} {448}
  (\bibinfo {year} {2021})}\BibitemShut {NoStop}%
\bibitem [{\citenamefont {Dupuis}\ and\ \citenamefont
  {Fawzi}(2019)}]{Dupuis19}%
  \BibitemOpen
  \bibfield  {author} {\bibinfo {author} {\bibfnamefont {F.}~\bibnamefont
  {Dupuis}}\ and\ \bibinfo {author} {\bibfnamefont {O.}~\bibnamefont {Fawzi}},\
  }\bibfield  {title} {\bibinfo {title} {{Entropy Accumulation With Improved
  Second-Order Term}},\ }\href@noop {} {\bibfield  {journal} {\bibinfo
  {journal} {IEEE Trans. Inf. Theory}\ }\textbf {\bibinfo {volume} {65}},\
  \bibinfo {pages} {7596} (\bibinfo {year} {2019})}\BibitemShut {NoStop}%
\bibitem [{Note7()}]{Note7}%
  \BibitemOpen
  \bibinfo {note} {Note that the expressions for $V(f_t,p) $ and $K_{\alpha
  '}(f_t)$ here differ slightly from those appearing in the Supplementary
  Information of Ref.~\cite {Liu21}. Here, following~\cite {Dupuis19}, we took
  the expressions given in Eqs. (16) and (17) of the Supplementary Information
  of~\cite {Liu21} but replace $d_A$ by 4 (the dimension of the registers
  $A_iB'_i$). Also, we avoided applying the inequality $\alpha
  '>1$.}\BibitemShut {Stop}%
\bibitem [{\citenamefont {Dupuis}\ \emph {et~al.}(2020)\citenamefont {Dupuis},
  \citenamefont {Fawzi},\ and\ \citenamefont {Renner}}]{Dupuis20}%
  \BibitemOpen
  \bibfield  {author} {\bibinfo {author} {\bibfnamefont {F.}~\bibnamefont
  {Dupuis}}, \bibinfo {author} {\bibfnamefont {O.}~\bibnamefont {Fawzi}},\ and\
  \bibinfo {author} {\bibfnamefont {R.}~\bibnamefont {Renner}},\ }\bibfield
  {title} {\bibinfo {title} {{Entropy Accumulation}},\ }\href@noop {}
  {\bibfield  {journal} {\bibinfo  {journal} {Commun. Math. Phys.}\ }\textbf
  {\bibinfo {volume} {379}},\ \bibinfo {pages} {867} (\bibinfo {year}
  {2020})}\BibitemShut {NoStop}%
\bibitem [{Note8()}]{Note8}%
  \BibitemOpen
  \bibinfo {note} {This is valid because Lemma~\ref {lem:EATlem2} does not
  require that the conditioning event is defined entirely by the registers in
  the state being considered --- it only requires that the state has a
  decomposition of the form $\rho = \DOTSB \sum@ \slimits@ _{x} p_x \rho
  _{|x}$. In this case, we indeed have such a decomposition, $\rho _{\protect
  \mathbf {B}'\protect \mathbf {X}\protect \mathbf {Y}E} = P(\Omega _\protect
  \mathrm {PE'})(\rho _{|\Omega _\protect \mathrm {PE'}})_{\protect \mathbf
  {B}'\protect \mathbf {X}\protect \mathbf {Y}E} + P(\Omega _\protect \mathrm
  {PE'}^c)(\rho _{|\Omega _\protect \mathrm {PE'}^c})_{\protect \mathbf
  {B}'\protect \mathbf {X}\protect \mathbf {Y}E}$, and can thus apply the
  lemma.}\BibitemShut {Stop}%
\bibitem [{\citenamefont {Renner}\ and\ \citenamefont {Wolf}(2003)}]{Renner03}%
  \BibitemOpen
  \bibfield  {author} {\bibinfo {author} {\bibfnamefont {R.}~\bibnamefont
  {Renner}}\ and\ \bibinfo {author} {\bibfnamefont {S.}~\bibnamefont {Wolf}},\
  }\bibfield  {title} {\bibinfo {title} {{Unconditional authenticity and
  privacy from an arbitrarily weak secret}},\ }in\ \href@noop {} {\emph
  {\bibinfo {booktitle} {{Advances in Cryptology - CRYPTO '03 in Lecture Notes
  in Computer Science}}}}\ (\bibinfo  {publisher} {Springer},\ \bibinfo {year}
  {2003})\ pp.\ \bibinfo {pages} {78--95}\BibitemShut {NoStop}%
\end{thebibliography}%

\clearpage

\section*{Acknowledgements}

JDB and EYZT would like to thank Rotem Arnon-Friedman for discussions.
We thank Sandia National Laboratories for supplying HOA2 ion traps.
This work was supported by the UK EPSRC Hub in Quantum Computing and Simulation (EP/T001062/1), the E.U. Quantum Technology Flagship Project AQTION (No. 820495), and CJB's UKRI Fellowship (MR/S03238X/1). EYZT and RR acknowledge funding by the Swiss National Science Foundation (SNSF), through the National Centers for Competence in Research QSIT and SwissMAP, and by the Air Force Office of Scientific Research (AFOSR) through grant FA9550-19-1-0202. JDB and NS acknowledge funding by the Institut de Physique Théorique (IPhT), Commissariat à l'Énergie Atomique et aux Energies Alternatives (CEA) and the Region Île-de-France in the framework of DIM SIRTEQ.

\section*{Author information}
\subsection*{Contributions}
DPN, PD, BCN, GA, DM, and RS built and operated the experimental apparatus.
DPN and PD led the collection of the experimental data and performed the data analysis.
JDB and DPN extracted the key from the raw data.
KI, RLU and JDB designed the error correction code.
JDB, EYZT, NS, PS and RR established the detailed protocol steps and derived the corresponding security proof.
NS, JDB, DPN and CJB wrote the manuscript.
CJB and DML supervised the experimental work, JDB and NS the theoretical work. NS and JDB initiated the project.
All authors contributed to the discussion and interpretation of results, and contributed to the manuscript.

\subsection*{Competing interests}

CJB is a director of Oxford Ionics.
The remaining authors declare no competing interests.

\subsection*{Corresponding authors}

Correspondence regarding experimental work should be addressed to DPN or CJB, that concerning theoretical aspects to JDB or NS.

\section*{Additional information}

Supplementary Information is available for this paper.

\vfill

\end{document}


\title{Supplementary information for\\ ``Experimental quantum key distribution certified by Bell's theorem"}
\author{D.~P.~Nadlinger}
\author{P.~Drmota}
\author{B.~C.~Nichol}
\author{G.~Araneda}
\author{D.~Main}
\author{R.~Srinivas}
\author{D.~M.~Lucas}
\author{C.~J.~Ballance}
\affiliation{Department of Physics, University of Oxford, Clarendon Laboratory, Parks Road, Oxford OX1 3PU, U.K.}

\author{K.~Ivanov}
\affiliation{School of Computer and Communication Sciences, EPFL, 1015 Lausanne, Switzerland}

\author{E.~Y-Z.~Tan}
\affiliation{Institute for Theoretical Physics, ETH Zürich, 8093 Zürich, Switzerland}

\author{P.~Sekatski}
\affiliation{Department of Applied Physics, University of Geneva, Rue de l’École-de-Médecine, 1211 Geneva, Switzerland}

\author{R.~L.~Urbanke}
\affiliation{School of Computer and Communication Sciences, EPFL, 1015 Lausanne, Switzerland}

\author{R.~Renner}
\affiliation{Institute for Theoretical Physics, ETH Zürich, 8093 Zürich, Switzerland}

\author{N.~Sangouard}
\affiliation{Universit\'e Paris-Saclay, CEA, CNRS, Institut de physique th\'eorique, 91191, Gif-sur-Yvette, France}

\author{J-D.~Bancal}
\affiliation{Universit\'e Paris-Saclay, CEA, CNRS, Institut de physique th\'eorique, 91191, Gif-sur-Yvette, France}

\date{September 29, 2021; this revision as accepted to Nature: May 10, 2022}

\maketitle

\vspace{-1cm}

\tableofcontents

\renewcommand{\figurename}{Fig.}
\crefname{figure}{Fig.}{Figs.}

\section{Notation}

\begin{center}
\begin{table}[!htbp]
\begin{tabular}{l|l}
  Symbol \, & \, Description \\
  \hline
  QKD  & \, Quantum key distribution \\
  DIQKD  & \, Device-independent quantum key distribution \\
  CPTP  & \, Completely positive and trace preserving \\
  CP  & \, Completely positive \\
  CQ & \, Classical-quantum \\
  BSC & \, Binary symmetric channel \\
  LDPC & \, Low-density parity-check \\
  SC-LDPC & \, Spatially-coupled low-density parity-check \\
  ML & \, Maximum likelihood \\
  BP & \, Belief propagation \\
  EAT & \, Entropy accumulation theorem \\
  cq-state & \, classical-quantum state\\
  $\chi(U_i=0)$ \, & \, Indicator function returning $1$ if $U_i=0$ is true, $0$ otherwise \\
  $\S_{=}(\mathcal{H})$ & \, The set of normalized quantum states  \\
  $\S_{\leq}(\mathcal{H})$ & \, The set of sub-normalized quantum states \\
  $\log$ & \, Binary logarithm \\
  $\ln$ & \, Natural logarithm \\
  $\rho$, $\tau$, $\sigma$ & \, Density matrices \\
  $D(\rho, \tau)$ & \, Trace distance between $\rho$ and $\tau$ \\
  $\mathcal{P}(\rho, \tau)$ & \, Purified distance between $\rho$ and $\tau$ \\
  $F(\rho, \tau)$ & \, Generalized fidelity between $\rho$ and $\tau$ \\
  $\lvert\lvert \mathcal{O} \rvert\rvert_1$ & \, Trace norm of the operator $\mathcal{O}$ \\
  $P(\Omega)$ & \, Probability for the event $\Omega$ to be true \\
  $\rho_{|\Omega}$ & \, Normalized state conditioned on the event $\Omega$ being true \\
  $\rho_{\land\Omega}$ & \, Subnormalized state conditioned on the event $\Omega$ being true ($\rho_{\land\Omega} = P(\Omega) \rho_{|\Omega}$) \\
  $A_i$ & \, A variable indexed by an integer $i$\\
  $\bA$ & \, A string of variables, i.e.~$\bA=(A_1,A_2,\ldots,A_n)$
  \end{tabular}
\caption{List of commonly used acronyms, symbols and expressions.}
\end{table}
\end{center}


\section{Experimental details}

Alice and Bob each control one of a pair of nominally identical room-temperature trapped-ion nodes, as first described in ref.~\cite{stephensonHighRateHighFidelityEntanglement2020}.
We have made extensive technical improvements to the experiment since, leading to higher-fidelity, more stable remote entanglement (e.g.~through beam power and magnetic field stabilisation), and improved capabilities for autonomous calibration and coordination between the nodes.
In this section, we give a brief description of the setup before discussing the entanglement performance and some details relevant for the DIQKD implementation not described in the Methods. We finally discuss some considerations regarding the isolation assumption made in the DIQKD setting.

Compared to ref.~\cite{stephensonHighRateHighFidelityEntanglement2020}, the system names (Alice/Bob) have been swapped throughout this manuscript in order for the node roles (set by an arbitrary choice in the real-time coordination logic) to match the protocol analysis.

\subsection{The two-node ion trap network}

\begin{figure*}
    \centering
    \includegraphics{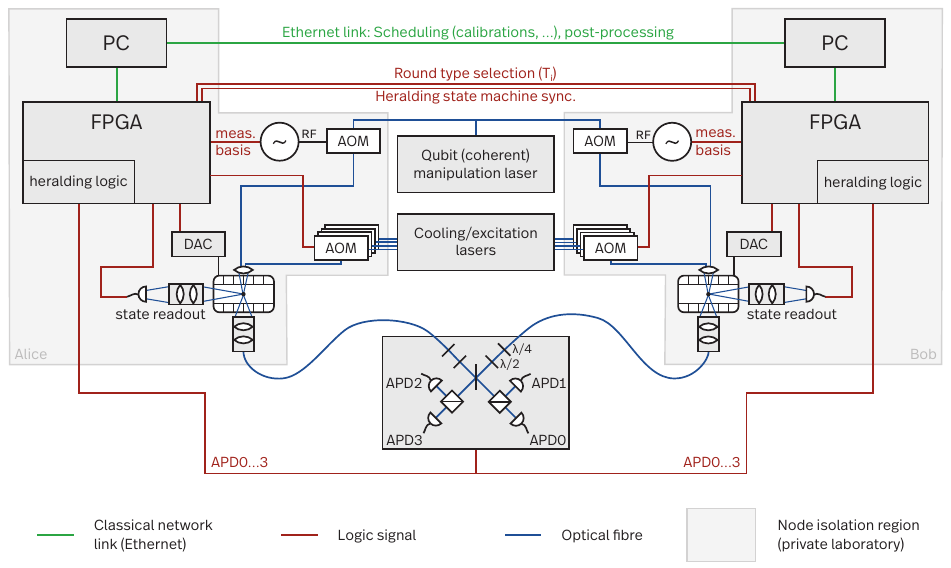}
    \vspace{-6pt}
    \caption{High-level sketch of the experimental apparatus. The points of connection between Alice's and Bob's nodes (indicated by the shaded areas) as well as the central heralding station are shown explicitly. The laser beams are derived from a common set of sources, but switched separately for each node using acousto-optical modulators (AOMs). The agile radio-frequency (RF) sources used for this, which are programmed on the fly from the FPGA-based real-time control systems, are shown explicitly for the \SI{674}{\nano\metre} qubit manipulation laser to illustrate how the measurement bases choices are physically implemented. The digital-to-analogue converters (DACs) set the trap electrode voltages and are used to disconnect the photonic link by changing the ion position.
    }
    \label{fig:block-diagram-details}
\end{figure*}

Each node is centred around a \enquote{High Optical Access 2} surface-electrode Paul trap (Sandia National Laboratories~\cite{maunzHighOpticalAccess2016}), which confines one \srplus{} ion each. We use standard techniques to implement dissipative and coherent operations using resonant laser radiation. Lasers at \SI{422}{\nano\metre} and \SI{1092}{\nano\metre} address the $5s\, {S}_{1/2} \leftrightarrow 5p\, {P}_{1/2} \leftrightarrow 4d\, {D}_{3/2}$ transition cycle for Doppler cooling, state preparation and readout; a static magnetic field of \SI{0.500}{\milli\tesla} lifts state degeneracies. For readout shelving and coherent operations, a narrow-linewidth \SI{674}{\nano\metre} laser addresses pairs of $5s\, {S}_{1/2}  \leftrightarrow 3d\, {D}_{5/2}$ Zeeman sublevels, implementing arbitrary rotations of the form $\Rxy{\theta}{\varphi} = \exp(-\ii \frac{\theta}{2}(\cos \varphi\, X + \sin \varphi\, Y)
)$.

\begin{figure}
    \input{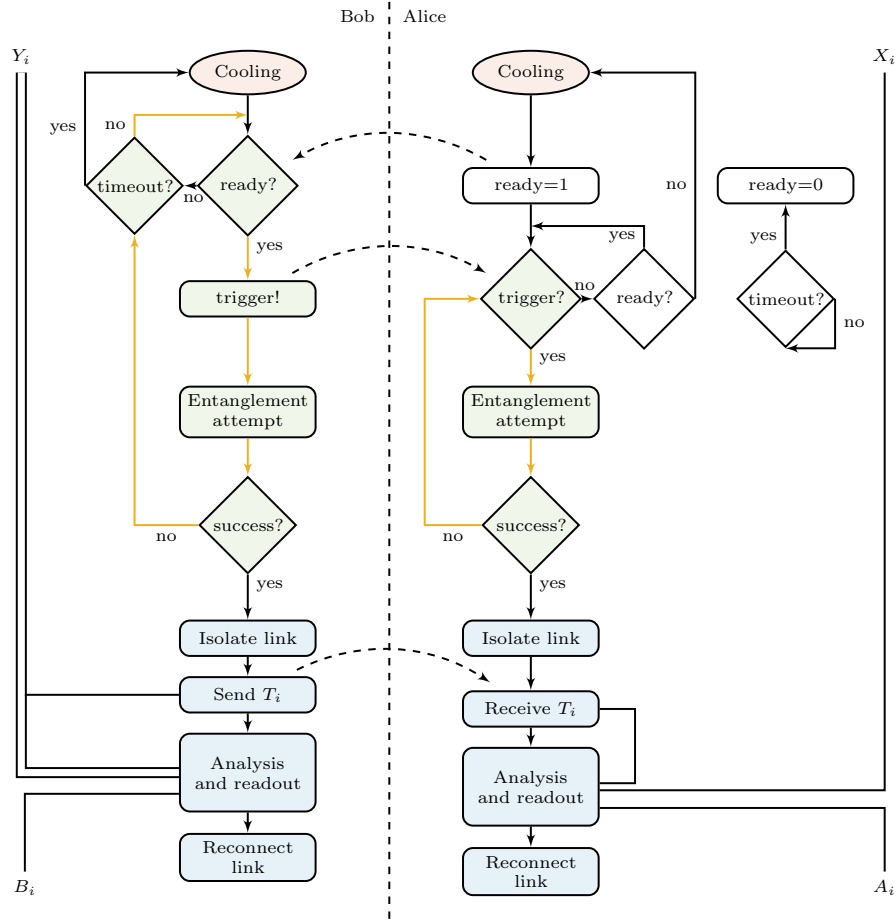}
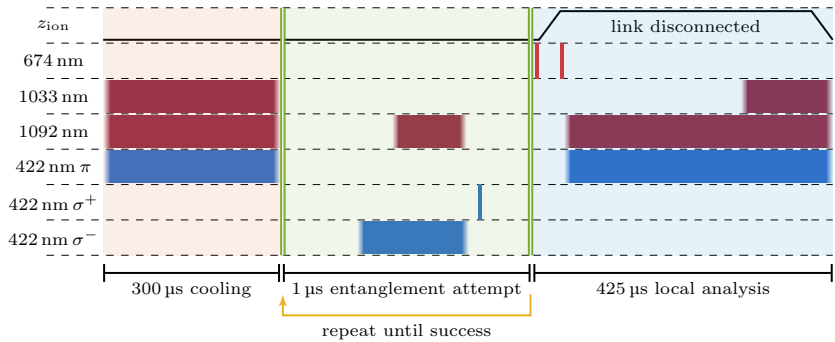
    \begin{subfigure}{\linewidth}
        \centering
        \begin{tikzpicture}[node distance = 1.6cm, auto, font=\scriptsize, thick]
    \draw [dashed] (0,1cm) -- (0,-12cm); 

	\node [align=right] at (-0.45cm, 0.8cm) {Bob};
	\node [align=left] at (0.5cm, 0.8cm) {Alice};
        
	\node [cloud, coolingStage] at (-2cm, 0) (coolingL) {Cooling};
	\node [decision, cycleStage, below of=coolingL] (ready1L) {ready?};
	\node [decision, cycleStage, left of=ready1L] (timeout1L) {timeout?};
	\node [block, cycleStage, below of=ready1L] (sendTrigger) {trigger!};
	\node [block, cycleStage, below of=sendTrigger] (cycleL) {Entanglement attempt};
	\node [decision, cycleStage, below of=cycleL] (successL) {success?};
	\node [block, analysisStage, below of=successL] (isolationL) {Isolate link};
	\node [block, analysisStage, below of=isolationL, node distance=0.8cm] (sendIskey) {Send $T_i$};
	\node [block, analysisStage, below of=sendIskey, node distance=1.1cm, minimum height=1.1cm] (analysisL) {Analysis\\and readout};
    \node [block, analysisStage, below of=analysisL, node distance=1.2cm] (reconnectL) {Reconnect link};
	
	\path [line] (coolingL) -- (ready1L);
	\path [line] (ready1L) -- node[decision answer] {no}  (timeout1L);
	\path [line, mlorange] (ready1L) -- node[decision answer] {yes} (sendTrigger);
	\path [line, mlorange] (timeout1L) |- node[decision answer] {no} ($(coolingL.south)!0.5!(ready1L.north)$);
	\path [line] (timeout1L.west) |- node[decision answer] {yes} (coolingL);
	\path [line, mlorange] (sendTrigger) -- (cycleL);
	\path [line, mlorange] (cycleL) -- (successL);
	\path [line] (successL) -- node[decision answer] {yes} (isolationL);
	\path [line, mlorange] (successL) -| node[decision answer] {no} (timeout1L);
    \path [line] (isolationL) -- (sendIskey);
    \path [line] (sendIskey) -- (analysisL);
    \path [line] (analysisL) -- (reconnectL);
	
	\node [cloud, coolingStage] at (2cm, 0) (coolingF) {Cooling};
	\node [block, below of=coolingF] (setReady) {ready=1};
	\node [decision, cycleStage, below of=setReady] (recvTrigger) {trigger?};
	\node [decision, right of=recvTrigger] (ready1F) {ready?};
	\node [block, cycleStage, below of=recvTrigger] (cycleF) {Entanglement attempt};
	\node [decision, right of=ready1F, xshift=0.4cm] (timeoutF) {timeout?};
	\node [block, above of=timeoutF] (unsetReady) {ready=0};
	\node [decision, cycleStage, below of=cycleF] (successF) {success?};
	\node [block, analysisStage, below of=successF] (isolationF) {Isolate link};
	\node [block, analysisStage, below of=isolationF, node distance=1cm] (recvIskey) {Receive $T_i$};
    \node [block, analysisStage, below of=recvIskey, node distance=1.1cm, minimum height=1.1cm] (analysisF) {Analysis\\and readout};
    \node [block, analysisStage, below of=analysisF, node distance=1.2cm] (reconnectF) {Reconnect link};
	
	\path [line] (coolingF) -- (setReady);
	\path [line] (setReady) -- (recvTrigger);
	\path [line] (recvTrigger) -- node[decision answer] {no} (ready1F);
	\path [line, mlorange] (recvTrigger) -- node[decision answer] {yes} (cycleF);
	\path [line] (ready1F) |- node[decision answer] {yes} ($(setReady.south)!0.5!(recvTrigger.north)$);
	\path [line] (ready1F.east) |- node[decision answer] {no} (coolingF);
	\path [line] (timeoutF.east) |- node[decision answer] {no} (timeoutF.south);
	\path [line] (timeoutF) -- node[decision answer] {yes} (unsetReady);
	\path [line, mlorange] (cycleF) -- (successF);
	\path [line, mlorange] (successF) -| node[decision answer] {no} ($(recvTrigger) - (1.5cm, 0cm)$) -- (recvTrigger);
	\path [line] (successF) -- node[decision answer] {yes} (isolationF);    
    \path [line] (isolationF) -- (recvIskey);
    \path [line] (recvIskey) -- (analysisF);
    \path [line] (analysisF) -- (reconnectF);
	
    \path[comm] (sendTrigger) edge[bend left] node[left] {} (recvTrigger);
    \path[comm] (sendIskey) edge[bend left] node[left] {} (recvIskey);
    \path[comm] (setReady) edge[bend right] node[left] {} (ready1L);
    
    \draw [solid, double, double distance=0.1cm] (-5.2cm, 0) node[xshift=0.05cm](nodeX) {} node[above, rotate=0] {$Y_i$} |- (analysisL);
    \draw [solid] (nodeX |- 99, 99 |- sendIskey) -- (sendIskey);
    \draw [solid] ($(analysisL.west) - (0, 0.3cm)$) -| (nodeX |- 99, 99 |- reconnectF) node[below]{$B_i$};
    
    \draw [solid] (7cm, 0) node[above, rotate=0] (nodeY) {$X_i$} |- ($(analysisF.east) - (0, 0.05cm)$);
    \draw [solid] (recvIskey.east) -- ++(0.5cm, 0) |- ($(analysisF.east) + (0, 0.05cm)$);
    \draw [solid] ($(analysisF.east) - (0, 0.3cm)$) -| (nodeY |- 99, 99 |- reconnectF) node[below]{$A_i$};
      
\end{tikzpicture}
        \caption{Flow diagram of the experimental sequence (left Bob, right Alice; separated by the dashed line). Execution starts on both systems with private random strings $X_i$ and $Y_i$. After a period of cooling, the state machines synchronise using an acknowledged trigger pulse. Subsequently, the attempt cycle is repeated until either a successful herald is received from the external heralding station, or a set attempt duration limit has elapsed after which the ions are re-cooled. When a successful herald is received, both systems disconnect from the link by shuttling the ion away from the focus of the high-NA lens. Only then, Bob announces whether he is going to perform a key generation round or a test round and executes the corresponding local analysis. Alice follows suit upon receipt of the key-round signal. Both parties store the outcome of their measurement, $B_i$ and $A_i$, locally.}
        \label{fig:exp-structogram}
    \end{subfigure}
    \begin{subfigure}{\linewidth}
        \centering
        \begin{tikzpicture}[font=\scriptsize, thick]
    
    %
    %
    
    \def\pulsestripX{-5cm}
    \def\pulsestripY{-1.cm}
    \def\fourtwotwosigma{\pulsestripX - 4.5cm}
    \def\fourtwotwops{\pulsestripX - 4cm}
    \def\fourtwotwopi{\pulsestripX - 3.5cm}
    \def\tenninetytwo{\pulsestripX - 3cm}
    \def\tenthirtythree{\pulsestripX - 2.5cm}
    \def\sixsevenfour{\pulsestripX - 2cm}
    \def\shuttledist{\pulsestripX - 1.5cm}
    
    \path (\pulsestripY - 0.7cm, \fourtwotwosigma)
    node {\SI{422}{\nano\meter}$\,\sigma^{-}$} -- ++(0, 0.5cm)
    node {\SI{422}{\nano\meter}$\,\sigma^{+}$} -- ++(0, 0.5cm)
    node {\SI{422}{\nano\meter}$\,\pi$} -- ++(0, 0.5cm)
    node {\SI{1092}{\nano\meter}} -- ++(0, 0.5cm)
    node {\SI{1033}{\nano\meter}} -- ++(0, 0.5cm)
    node {\SI{674}{\nano\meter}} -- ++(0, 0.5cm)
    node {$z_\mathrm{ion}$} -- ++(0, 0.5cm);

    \def\coolingheight{2.50}
    
    \def\cycleheight{3.5}
    \def\cyclestart{\pulsestripY + \coolingheight cm + 0.04cm} 
    
    \def\shuttleheight{0.3}
    \def\isolationstart{\cyclestart + \cycleheight cm + 0.04cm} 
    \def\isolationheight{0.38}
    \def\analysisstart{\isolationstart + \isolationheight cm}
    \def\analysisheight{2.55}
    \def\reconnectstart{\analysisstart + \analysisheight cm}
    \def\deshelveheight{1.0}
    \def\reconnectheight{1.3}
    \def\recoolheight{3.8}
    
    \def\timeline{\fourtwotwosigma-0.5cm}
    \draw [|-|] (\pulsestripY, \timeline) -- node[midway, below] {\SI{300}{\micro\second} cooling} ++(\coolingheight cm, 0);
    \draw [|-|] (\cyclestart, \timeline) -- node[midway, below] {\SI{1}{\micro\second} entanglement attempt} ++(\cycleheight cm, 0);
    \draw [|-|] (\isolationstart, \timeline) -- node[midway, below] {\SI{425}{\micro\second} local analysis} ++(\isolationheight cm + \analysisheight cm + \deshelveheight cm + \shuttleheight cm, 0);
    \foreach \num in {0,...,7} {
        \draw [thin, dashed] (\pulsestripY-0.8cm, \pulsestripX-\num * 0.5cm - 1.25cm) -- (\cyclestart, \pulsestripX-\num * 0.5cm - 1.25cm);
        \draw [thin, dashed] (\cyclestart, \pulsestripX-\num * 0.5cm - 1.25cm) -- ++(\cycleheight cm, 0);
        \draw [thin, dashed] (\isolationstart, \pulsestripX-\num * 0.5cm - 1.25cm) -- ++(\isolationheight cm, 0);
        \draw [thin, dashed] (\analysisstart, \pulsestripX-\num * 0.5cm - 1.25cm) -- ++(\analysisheight cm, 0);
        \draw [thin, dashed] (\reconnectstart, \pulsestripX-\num * 0.5cm - 1.25cm) -- ++(\deshelveheight cm, 0);
        \draw [thin, dashed] (\reconnectstart + \deshelveheight cm, \pulsestripX-\num * 0.5cm - 1.25cm) -- ++(\shuttleheight cm, 0);
    }
    
    \def\laserstripwidth{0.45}
    \pic at (\pulsestripY, \fourtwotwopi) { laserpulse={wldoppler}{\laserstripwidth}{\coolingheight} };
    \pic at (\pulsestripY, \tenninetytwo) { laserpulse={wlrepump}{\laserstripwidth}{\coolingheight} };
    \pic at (\pulsestripY, \tenthirtythree) { laserpulse={wldeshelve}{\laserstripwidth}{\coolingheight} };
    
    \pic at (\cyclestart + 0.299cm * \cycleheight, \fourtwotwosigma) { laserpulse={wldoppler}{\laserstripwidth}{0.45 * \cycleheight} };
    \pic at (\cyclestart + 0.440cm * \cycleheight, \tenninetytwo) { laserpulse={wlrepump}{\laserstripwidth}{0.3 * \cycleheight} };
    \fill[wldoppler] (\cyclestart + 0.790cm * \cycleheight, \fourtwotwops - 0.25cm) rectangle ++(0.05cm, 0.5cm);

    \fill[wlquadrupole] (\isolationstart + 0.03cm, \sixsevenfour - 0.25cm) rectangle ++(0.05cm, 0.5cm);
    
    \fill[wlquadrupole] (\analysisstart, \sixsevenfour - 0.25cm) rectangle ++(0.05cm, 0.5cm);
    \pic at (\analysisstart + 0.05cm, \fourtwotwopi) { laserpulse={wldoppler}{\laserstripwidth}{\recoolheight} };
    \pic at (\analysisstart + 0.05cm, \tenninetytwo) { laserpulse={wlrepump}{\laserstripwidth}{\recoolheight} };
    
    \pic at (\reconnectstart, \tenthirtythree) { laserpulse={wldeshelve}{\laserstripwidth}{\reconnectheight} };
    
    \draw (\pulsestripY, \shuttledist - \laserstripwidth cm * 0.45) -- (\cyclestart, \shuttledist - \laserstripwidth cm  * 0.45) -- ++(\cycleheight cm, 0);
    \draw (\isolationstart, \shuttledist-\laserstripwidth cm  * 0.45) -- ++(0.08cm, 0) -- ++(\shuttleheight, 0.9 * \laserstripwidth) -- ++(\analysisheight cm + \deshelveheight cm, 0) node[midway, below] {link disconnected} -- ++(\shuttleheight cm, -0.9 * \laserstripwidth);
    
    \draw [double, double distance=0.02cm, mlgreen] (\cyclestart, \fourtwotwosigma - 0.25cm) -- ++(0, 3.5cm);
    \draw [double, double distance=0.02cm, mlgreen] (\cyclestart + \cycleheight cm, \fourtwotwosigma - 0.25cm) -- ++(0, 3.5cm);
    \path [line, mlorange] (\cyclestart + \cycleheight cm, \timeline - 0.3cm) -- ++(0, -0.3cm) -- ++(-\cycleheight cm, 0) node[midway, below, color=black] {repeat until success} -- ++(0, 0.3cm);
    
    \fill [coolingStage] (\pulsestripY, \shuttledist + 0.25cm) rectangle ++(\coolingheight cm, -3.5cm);
    \fill [cycleStage] (\cyclestart, \shuttledist + 0.25cm) rectangle ++(\cycleheight cm, -3.5cm);
    \fill [analysisStage] (\isolationstart, \shuttledist + 0.25cm) rectangle ++(\isolationheight cm + \analysisheight cm + \reconnectheight cm, -3.5cm);  
\end{tikzpicture}
        \caption{Timing diagram of ion position $z_{\mathrm{ion}}$ and the experimental pulse sequence (not to scale). The ions are Doppler-cooled prior to the entanglement attempt loop, during which tightly timed state preparation and pulsed excitation pulses create single photons to be probabilistically collected by the photonic link. If a coincidence event heralds the successful creation of a remote entangled state, the local analysis sequence is executed locally on each system.}
        \label{fig:exp-timing}
    \end{subfigure}
    \caption{Experimental sequence. The flow diagram \cref{fig:exp-structogram} shows the basic logic implemented in the state machine driving the remote entanglement generation scheme. The timing diagram \cref{fig:exp-timing} complements this picture with the concrete laser pulses issued at different stages of the protocol.}
    \label{fig:timing}
\end{figure}

Ion-photon entanglement is generated in each node by coherent excitation to the short-lived $\ket{P_{1/2}, m_J=1/2}$ state using a $\ish \SI{5}{\pico\second}$ $\SI{422}{\nano\metre}$ laser pulse. Photons emitted into free space during spontaneous decay, entangled in polarisation with the resulting $S_{1/2}$ Zeeman state according to the dipole selection rules, are collected by lens objectives (numerical aperture 0.6) perpendicular to the magnetic field. The trap nodes are connected to a central heralding station using $2\times \SI{1.75}{\metre}$ fused silica single-mode fibres, which also avoids polarisation mixing errors. There, a $50:50$ beam splitter and two polarising beam splitters implement a partial Bell-basis measurement. The orientation of a pair of quarter- and half-wave plates at each input is calibrated daily to compensate birefringence in the input fibres. The output of four avalanche photodiode single-photon detectors is connected to heralding logic implemented in the node-local FPGA control systems. Entanglement generation is repeatedly attempted at a rate of \SI{1}{\mega\hertz} until coincident detection of two photons of opposite polarisation heralds the creation of an entangled state (see \cref{fig:timing} for the detailed sequence during DIQKD experiments).

A high-level block diagram of the relevant components is shown in \cref{fig:block-diagram-details}. Alice's and Bob's nodes each have their own, independent copy of the required experimental control system. The connections between the nodes are explicitly indicated in the figure: A classical network link (Ethernet) between the node PCs is used for high-level coordination, e.g.~to handle periodic recalibrations or interruptions due to ion-loss events, as well as to exchange the messages required for the DIQKD post-processing steps. The PCs, together with the FPGA-based real-time control systems also connected via Ethernet, make up the trusted classical processing systems available to Alice and Bob as per DIQKD assumption (iv).

For this demonstration, the two nodes are located in the same room; it is thus convenient to derive all the required laser beams from the same set of sources. As indicated in \cref{fig:block-diagram-details}, the actual switching and routing optics are kept separate and local to the nodes. A separate set of lasers could easily be used for each (e.g.~in an application with a larger distance between nodes), as we do not rely on any optical phase coherence between the nodes. The only synchronisation requirement is for the entanglement-carrying photons to be well-matched in frequency and arrival time at the heralding station compared to their \SI{7}{\nano\second} duration. In particular, we do not require the phases of the \SI{422}{\nano\metre} picosecond pulses or the absolute phase of the coherent \SI{674}{\nano\metre} pulses to be stable; the relative phase between projective measurements at each node is determined solely by the \SI{14}{\mega\hertz} RF beat note between the two \SI{674}{\nano\metre} pulses.

The \SI{674}{\nano\metre} Ti:Sapphire laser is stabilised to a high-finesse cavity using a multi-stage closed-loop feedback controller, which occasionally loses lock (\ish once per day). To enable autonomous operation of long-running experiments, a real-time signal is forwarded to the control systems of Alice and Bob if this occurs, who then pause protocol execution until the lock is established again. The nodes also negotiate an interruption if any of a number of rare issues with the experimental hardware is detected (e.g.~insufficient power in one of the laser beams); such an issue, which required manual intervention, was responsible for the $\ish\SI{4}{\hour}$ interruption during the final DIQKD data run presented in the main text. Since the total number of rounds is fixed before the start of the protocol and the timing of the round is free, pauses between the rounds have no impact on the final security.

\subsection{Entanglement characterisation}

Remote entanglement is produced in the ground state ${S}_{1/2}$ manifold. For laser-based manipulation, it is mapped to an \enquote{optical} qubit using a $\pi$ pulse on the \SI{674}{\nano\metre} $\ket{S_{1/2}, m_J = 1/2} \rightarrow \ket{D_{5/2}, m_J = -3/2}$ transition, followed by coherent manipulation on the $\ket{S_{1/2}, m_J = -1/2} \leftrightarrow \ket{D_{5/2}, m_J = -3/2}$ transition to select the measurement basis, or, for measurements in the computational basis, an additional $\ket{S_{1/2}, m_J = 1/2} \rightarrow \ket{D_{5/2}, m_J = 3/2}$ $\pi$ pulse to reduce shelving errors.

To characterise the experimental platform in detail, we perform quantum state tomography on the joint ion-ion state observed after each of the four heralding detector coincidence patterns. For this, we optionally apply a $\pi / 2$ pulse, $\Rxy{\pi / 2}{\phi}$ for $\phi \in \{0, \pi / 2, \pi, 3 \pi / 2\}$, after a successful herald and apply fluorescence detection, such that ion bright/dark results correspond to a projective measurement onto the $+1/-1$ eigenstates of the Pauli operators $\{\pm X, \pm Y, Z\}$. This over-complete set of measurement bases is chosen to mitigate any systematic bias in the $xy$ plane due to imperfect analysis pulses. The joint outcomes for all combinations are recorded for a number of repetitions.

\begin{figure*}
    \centering
    \includegraphics{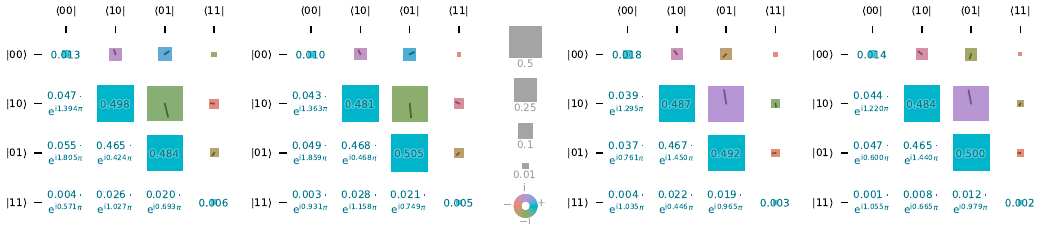}
    \includegraphics{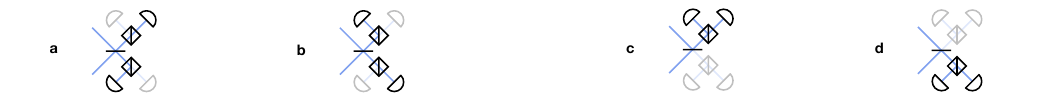}
    \caption{
        Maximum-likelihood density matrix estimates for the joint ion-ion state after each of the heralding detector click patterns (a-d). The fully entangled fractions, i.e.~fidelities to the closest maximally entangled state, are \SI{96.1(3)}{\percent}, \SI{96.4(1)}{\percent}, \SI{95.8(2)}{\percent} and \SI{95.8(2)}{\percent}, respectively, for an average fidelity of \SI{96.0(1)}{\percent} ($1\sigma$ statistical errors from parametric bootstrapping). For the \num{180000} successful heralds in this dataset a total of \num{1133526525} entanglement attempts were made, corresponding to an average success probability of \num{1.6e-4}. With an attempt rate of \SI{1}{\MHz}, interrupted by \SI{300}{\micro\second} of cooling every \SI{500}{\micro\second}, this results in a heralding rate of \SI{100}{\second^{-1}}.
    }
    \label{fig:ion-ion-mle}
\end{figure*}

\begin{table}
    \begin{subfigure}[b]{0.48\textwidth}
      \centering
      \begin{tabular}{cc|cccc}
        $O_A$ & $O_B$ & $+_A+_B$ & $+_A-_B$ & $-_A+_B$ & $-_A-_B$ \\
        \hline
        X & X & 229 & 332 & 375 & 284 \\
        X & Y & 27 & 526 & 650 & 15 \\
        X & -X & 317 & 184 & 310 & 374 \\
        X & -Y & 532 & 23 & 24 & 580 \\
        X & Z & 541 & 558 & 650 & 609 \\
        Y & X & 548 & 16 & 20 & 579 \\
        Y & Y & 284 & 301 & 389 & 241 \\
        Y & -X & 25 & 509 & 618 & 27 \\
        Y & -Y & 286 & 253 & 260 & 379 \\
        Y & Z & 551 & 626 & 715 & 594 \\
        -X & X & 356 & 296 & 237 & 347 \\
        -X & Y & 618 & 19 & 27 & 500 \\
        -X & -X & 271 & 362 & 321 & 211 \\
        -X & -Y & 20 & 633 & 563 & 29 \\
        -X & Z & 684 & 671 & 539 & 514 \\
        -Y & X & 23 & 621 & 527 & 17 \\
        -Y & Y & 400 & 231 & 286 & 292 \\
        -Y & -X & 617 & 33 & 23 & 516 \\
        -Y & -Y & 251 & 378 & 308 & 277 \\
        -Y & Z & 717 & 594 & 544 & 561 \\
        Z & X & 525 & 693 & 620 & 527 \\
        Z & Y & 655 & 562 & 634 & 502 \\
        Z & -X & 661 & 579 & 635 & 589 \\
        Z & -Y & 543 & 674 & 521 & 575 \\
        Z & Z & 82 & 2415 & 2378 & 17 \\
      \end{tabular}
      \subcaption{Number of observations per Pauli eigenstate for detector pattern 0011, for a total of $43182$ clicks.}
    \end{subfigure}%
    \hfill
    \begin{subfigure}[b]{0.48\textwidth}
      \centering
      \begin{tabular}{cc|cccc}
        $O_A$ & $O_B$ & $+_A+_B$ & $+_A-_B$ & $-_A+_B$ & $-_A-_B$ \\
        \hline
        X & X & 337 & 201 & 296 & 369 \\
        X & Y & 571 & 40 & 22 & 596 \\
        X & -X & 195 & 373 & 343 & 285 \\
        X & -Y & 35 & 548 & 633 & 23 \\
        X & Z & 564 & 575 & 664 & 638 \\
        Y & X & 31 & 562 & 589 & 32 \\
        Y & Y & 341 & 257 & 208 & 418 \\
        Y & -X & 556 & 29 & 50 & 575 \\
        Y & -Y & 230 & 356 & 396 & 254 \\
        Y & Z & 464 & 662 & 681 & 540 \\
        -X & X & 242 & 356 & 365 & 192 \\
        -X & Y & 23 & 608 & 525 & 29 \\
        -X & -X & 365 & 248 & 204 & 387 \\
        -X & -Y & 618 & 19 & 23 & 532 \\
        -X & Z & 622 & 622 & 556 & 555 \\
        -Y & X & 618 & 30 & 22 & 561 \\
        -Y & Y & 246 & 410 & 330 & 229 \\
        -Y & -X & 41 & 579 & 541 & 19 \\
        -Y & -Y & 370 & 236 & 224 & 335 \\
        -Y & Z & 762 & 537 & 560 & 580 \\
        Z & X & 708 & 515 & 583 & 666 \\
        Z & Y & 526 & 704 & 588 & 647 \\
        Z & -X & 481 & 736 & 691 & 529 \\
        Z & -Y & 706 & 545 & 597 & 612 \\
        Z & Z & 60 & 2284 & 2459 & 27 \\
      \end{tabular}
      \subcaption{Number of observations per Pauli eigenstate for detector pattern 0101, for a total of $43524$ clicks.}
    \end{subfigure}
    
    \vskip2\baselineskip
    
    \begin{subfigure}[b]{0.48\textwidth}
      \centering
      \begin{tabular}{cc|cccc}
        $O_A$ & $O_B$ & $+_A+_B$ & $+_A-_B$ & $-_A+_B$ & $-_A-_B$ \\
        \hline
        X & X & 357 & 242 & 310 & 368 \\
        X & Y & 588 & 21 & 14 & 658 \\
        X & -X & 247 & 338 & 346 & 357 \\
        X & -Y & 19 & 571 & 658 & 29 \\
        X & Z & 596 & 608 & 683 & 659 \\
        Y & X & 19 & 603 & 680 & 26 \\
        Y & Y & 284 & 267 & 277 & 375 \\
        Y & -X & 604 & 18 & 28 & 643 \\
        Y & -Y & 289 & 284 & 408 & 273 \\
        Y & Z & 506 & 716 & 695 & 658 \\
        -X & X & 345 & 359 & 329 & 251 \\
        -X & Y & 20 & 674 & 565 & 27 \\
        -X & -X & 352 & 340 & 292 & 325 \\
        -X & -Y & 637 & 17 & 26 & 573 \\
        -X & Z & 694 & 668 & 562 & 562 \\
        -Y & X & 685 & 20 & 28 & 560 \\
        -Y & Y & 237 & 383 & 335 & 300 \\
        -Y & -X & 28 & 623 & 604 & 15 \\
        -Y & -Y & 372 & 298 & 287 & 343 \\
        -Y & Z & 723 & 629 & 534 & 735 \\
        Z & X & 797 & 573 & 556 & 659 \\
        Z & Y & 608 & 727 & 629 & 642 \\
        Z & -X & 548 & 785 & 697 & 524 \\
        Z & -Y & 667 & 623 & 657 & 569 \\
        Z & Z & 50 & 2489 & 2513 & 24 \\
      \end{tabular}
      \subcaption{Number of observations per Pauli eigenstate for detector pattern 1010, for a total of $46016$ clicks.}
    \end{subfigure}
    \hfill
    \begin{subfigure}[b]{0.48\textwidth}
      \centering
      \begin{tabular}{cc|cccc}
        $O_A$ & $O_B$ & $+_A+_B$ & $+_A-_B$ & $-_A+_B$ & $-_A-_B$ \\
        \hline
        X & X & 229 & 358 & 408 & 305 \\
        X & Y & 36 & 553 & 667 & 16 \\
        X & -X & 363 & 229 & 322 & 417 \\
        X & -Y & 558 & 35 & 34 & 698 \\
        X & Z & 581 & 631 & 774 & 669 \\
        Y & X & 589 & 28 & 32 & 646 \\
        Y & Y & 286 & 343 & 470 & 259 \\
        Y & -X & 20 & 593 & 679 & 26 \\
        Y & -Y & 346 & 284 & 255 & 447 \\
        Y & Z & 560 & 681 & 731 & 620 \\
        -X & X & 409 & 327 & 235 & 354 \\
        -X & Y & 759 & 26 & 29 & 551 \\
        -X & -X & 283 & 433 & 370 & 236 \\
        -X & -Y & 23 & 662 & 594 & 31 \\
        -X & Z & 773 & 755 & 580 & 643 \\
        -Y & X & 39 & 645 & 582 & 22 \\
        -Y & Y & 411 & 271 & 288 & 351 \\
        -Y & -X & 691 & 33 & 28 & 609 \\
        -Y & -Y & 277 & 461 & 301 & 282 \\
        -Y & Z & 707 & 674 & 556 & 587 \\
        Z & X & 640 & 713 & 620 & 605 \\
        Z & Y & 735 & 568 & 668 & 605 \\
        Z & -X & 679 & 653 & 609 & 604 \\
        Z & -Y & 550 & 828 & 671 & 662 \\
        Z & Z & 69 & 2603 & 2517 & 13 \\
      \end{tabular}
      \subcaption{Number of observations per Pauli eigenstate for detector pattern 1100, for a total of $47278$ clicks.}
    \end{subfigure}
  
    \caption{Number of observations of each eigenstate for tensor product combinations $O_A \otimes O_B$ of Pauli operators in the Alice-Bob state tomography experiment, for each of the four detector click patterns.}
    \label{tab:tomography-counts}
\end{table}

A dataset collected one day prior to the DIQKD link characterisation and key generation experiments is given in \cref{tab:tomography-counts}. These outcome counts represent the complete set of input data for a diluted fixed-point iteration procedure~\cite{Rehacek2007} to obtain the ML estimate; no additional corrections (such as for readout errors) are applied. The resulting density matrices are shown in \cref{fig:ion-ion-mle}. The average state fidelity to the closest maximally entangled state was \SI{96.0(2)}{\percent} with a heralding probability of \num{1.6e-4}, equivalent to an effective rate of $\SI{100}{\second^{-1}}$ including periodic recooling. (The average rate quoted in main text \refmaintextfiglinkperformance{} is significantly lower than this as it also includes the time necessary for frequent reloading of the ion trap in Alice due to a vacuum leak at the time the experiment was performed.)

The fidelity estimates are roughly consistent with the results from ion-photon tomography, but a detailed model accounting for the error sources has yet to be established. We currently suspect an uncharacterised source of polarisation mixing, for instance in the high-NA free-space optics, to be the primary contribution. Errors from local ion qubit state manipulation are not significant at the current fidelity level; randomised benchmarking indicates gate fidelities of $\ish \num{5e-4}$ per Clifford group element on either of the \SI{674}{\nano\metre} transitions involved, and the readout state discrimination error is $<\num{e-3}$.

To stabilise the magnetic fields at the ion location, we use high-precision feedback controllers acting onto the quantisation coil currents (including active feed-forward to compensate for ambient fields generated by \SI{50}{\hertz} power mains), resulting in magnetic field noise amplitudes on the \si{\nano\tesla} level. Typical coherence times observed on the \SI{674}{\nano\metre} qubit transitions without any dynamical decoupling techniques are quoted in \cref{tab:coherence-times}, showing that ion coherence times will not significantly limit the experiment performance even for longer speed-of-light delays between the nodes (not before \SI{422}{\nano\metre} fibre loss becomes significant).

\begin{table}[ht]
    \begin{tabular}{c@{\hspace{10pt}}c@{\hspace{10pt}}c@{\hspace{15pt}}c}
        \toprule
        \multirow{2}{*}[-2pt]{Lower qubit state} & \multirow{2}{*}[-3pt]{\shortstack[c]{Transition magnetic\\ field sensitivity}} & \multicolumn{2}{c}{Coherence time} \\
        \cmidrule{3-4}
        & & Alice & Bob \\ \midrule
        $\ket{{S}_{1/2},m_J=-1/2}$ & $\SI{-11.2}{\mega\hertz/ \milli\tesla}$ & \SI{4.2}{\milli\second} & \SI{7.5}{\milli\second} \\
        $\ket{{S}_{1/2},m_J=+1/2}$ & $\SI{-39.2}{\mega\hertz/ \milli\tesla}$ & \SI{1.0}{\milli\second} & \SI{1.7}{\milli\second} \\
        \bottomrule
    \end{tabular}
    \caption{Measured $1 / e$ coherence times on optical transitions with $\ket{{D}_{5/2},m_J=-3/2}$ as the upper qubit state, extracted from a Gaussian fit to the contrast decay in a Ramsey experiment with varying wait duration, without any dynamical decoupling pulses. The $\ket{{S}_{1/2},m_J=-1/2} \leftrightarrow \ket{{D}_{5/2},m_J=-3/2}$ qubit is used to store the remote entanglement between heralding and measurement basis choice.}
    \label{tab:coherence-times}
\end{table}

\subsection{DIQKD primitives}\label{sec:DIQKDPrimitives}

\begin{table*}
    \begin{tabular}{c|ccc}
        \diagbox{$X_i$}{$Y_i$} & 0 & 1 & 2 \\
        \hline
        0
        &
        $\begin{pmatrix}
        0.415(1) & 0.0688(5) \\
        0.0961(6) & 0.420(1) \\
        \end{pmatrix}$
        &
        $\begin{pmatrix}
        0.439(1) & 0.0805(6) \\
        0.0735(5) & 0.4068(10) \\
        \end{pmatrix}$
        &
        $\begin{pmatrix}
        0.5017(5) & 0.00339(6) \\
        0.0110(1) & 0.4839(5) \\
        \end{pmatrix}$
        \\
        1
        &
        $\begin{pmatrix}
        0.3928(10) & 0.0916(6) \\
        0.0851(6) & 0.431(1) \\
        \end{pmatrix}$
        &
        $\begin{pmatrix}
        0.0820(6) & 0.437(1) \\
        0.3970(10) & 0.0841(6) \\
        \end{pmatrix}$
        &
        —
    \end{tabular}
    \caption{
        Empirical probabilities to observe classical outcome $(A_i$, $B_i)$ for measurements settings $(X_i$, $Y_i)$. Matrix rows represent Alice's outcomes, columns that of Bob. The probabilities are estimated from the \num{1920000} total characterisation rounds also shown in main text \refmaintextfiglinkperformance{} and \cref{fig:link-performance-poststamp}, with the multinomial standard errors given in parentheses. Alice's classical outcomes were inverted (such that $0$ corresponds to finding the ion in the $S_{1/2}$ manifold after shelving), giving maximum correlations for the key generation settings $X = 0, Y = 2$, and maximising the probability that $A_i \oplus B_i = X_i \cdot Y_i$ in Bell test rounds.
    }
    \label{tab:box-statistics}
\end{table*}

\begin{figure*}
    \includegraphics[width=\linewidth]{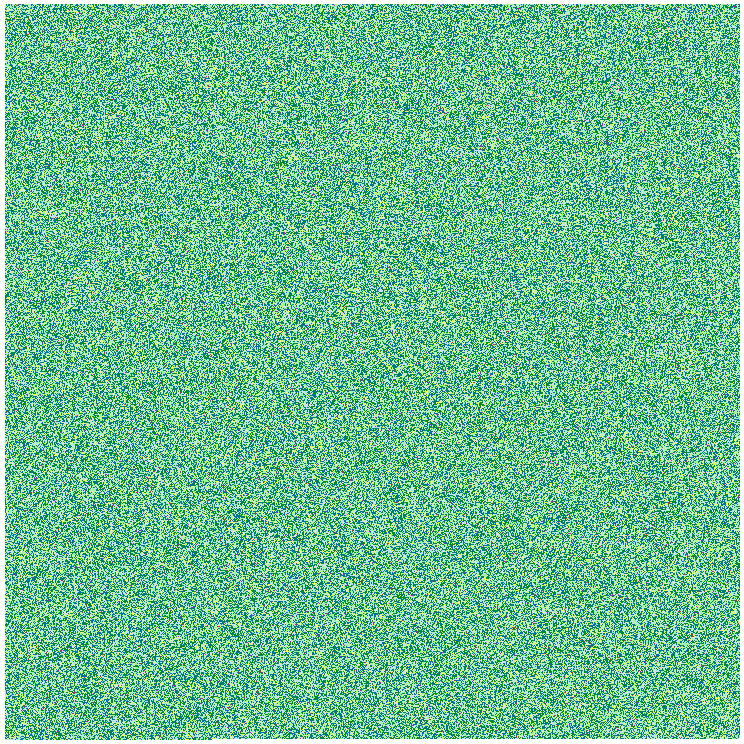}
    \vspace{2pt}
    \includegraphics[width=\linewidth]{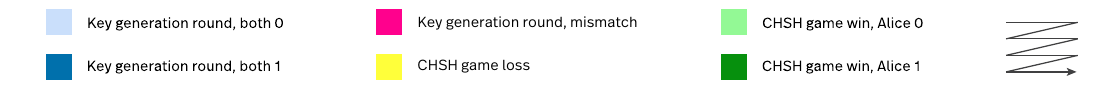}
    \caption{Per-round outcomes for the link characterisation run summarised in main text \refmaintextfiglinkperformance{}. Each square shows the round type ($\gamma \approx 0.5$) and outcome for one of \num{1920000} rounds in the dataset, arranged row by row from left to right and top to bottom as indicated by the arrow.}
    \label{fig:link-performance-poststamp}
\end{figure*}

To obtain a high CHSH winning probability, a set of measurement angles widely known to be optimal for pure Bell states is chosen: Alice measures in the computational basis for $X = 0$ and applies $\Rxy{\pi / 2}{0}$ prior to measuring for $X = 1$. Bob applies $\Rxy{\pm \pi / 4}{\phi}$ for $Y = 0$ and $Y = 1$, respectively, with the phase $\phi$ calibrated independently for each heralding click pattern to absorb the differing phases between the $\ket{01}$ and $\ket{10}$ components. For the key generation rounds, where $Y = 2$, both nodes measure in the computational basis. Alice inverts her classical outcome throughout to match the usual theoretical convention where the Bell state is taken to be maximally correlated along the $Z$ axis.

To characterise the quality of the resulting correlations, we perform \num{1920000} DIQKD data acquisition rounds, but immediately recombine the inputs $\bX, \bY$ and outputs $\bA, \bB$ for analysis. The resulting empirical probabilities $P(A_i B_i | X_i Y_i)$ are given in \cref{tab:box-statistics}. \Cref{fig:link-performance-poststamp} shows the outcomes of each individual round.

Numerical optimisation of the measurement angles based on maximum-likelihood estimates for the joint ion-ion state, such as the data in \cref{fig:ion-ion-mle}, predicts a statistically significant improvement in quantum bit error rate for a slightly tilted choice of measurement axes (and similarly for the CHSH score). Thus far, we were neither able to observe these improvements in practice, nor to conclusively trace the prediction back to any tomography bias.

The required random strings for the setting choices, $\bX$ and $\bY$, (as well as the initial shared secret key $\bK_0$) directly originate from a commercial random number generator (model: ID Quantique Quantis QRNG 4M USB). A cursory inspection of $2^{32} - 1$ bytes ($\ish\num{3e10}$ bits) using two widely used suites of randomness tests (\emph{dieharder} version 3.31.1 \cite{DieharderCode}, and the \enquote{Rabbit} and \enquote{Alphabit} batteries from \emph{TestU01} version 1.2.3 \cite{TestU01}) revealed no obvious defects in the quality of the outputs. We choose a power-of-two fraction for the test round probability ($\gamma = 13 / 2^8$) and implement the round type choice in a straightforward fashion by consuming eight random source bits to avoid any unintended bias between experiment and security analysis.

\subsection{Shuttling-based link isolation}\label{sec:ShuttlingBasedLinkIsolation}

As discussed in the main text and Methods, private data on the nodes is assumed to be isolated from any adversary in DIQKD, which is intrinsically in tension with the requirement for the qubits to be well-coupled to the optical fibre link during the heralded entanglement generation step for any matter-based remote entanglement experiment. For the isolation assumption to be valid, the link thus needs to be disconnected every round while the node systems could leak information about the private inputs or outputs. In a real deployment, this isolation could be established to an effectively arbitrary degree through a variety of technical means, e.g.~through free-space shutters, fast micro-electromechanical fibre switches, shuttling of the ions from the imaging system focus by a macroscopic distance, or transfer of the entanglement to another ion species in combination with spectral filtering.

In this demonstration, we isolate the ions by displacing them relative to the focus of the high-NA objective coupling the lens to the fibre through modulation of the trapping potential. While the axial potential minimum can be moved in our traps by \ish\SI{1}{\milli\metre} through modulation of many electrodes, here, we only modify the potentials of the four second-nearest-neighbour electrodes (Q15, Q16, Q23 and Q24) due to technical limitations in the DAC system. In the disconnected configuration, the ion is moved \SI{3}{\micro\meter} from the centre. We keep the nearest-neighbour electrodes fixed to maintain the motional mode structure of the trap; resolved-sideband measurements indicate no shuttle-induced heating within the measurement uncertainty $\Delta\bar{n} \approx 0.1$.

\begin{figure}[h]
    \begin{tikzpicture}[font=\footnotesize]
    \def\width{7cm}
    \def\height{5cm}
    \def\colsep{1.5cm}
    \begin{axis}[name=shuttleFollower,
        ymode=log,
        width=\width, height=\height,
        xlabel={Time after DAC step / \si{\micro\second}}, xlabel near ticks,
        ylabel={Link isolation}, ylabel near ticks,
        ymin=1e-6, ymax=1, 
        try min ticks log=5,
        xmin=-5, xmax=105,
        extra x ticks={30},
        ymajorgrids
        ]
        \def\srcfile{data/shuttle-extinction/dynamic_bob.txt};
        \addplot[mark=o, only marks, mark size=1pt] table[x=x, y=y] {\srcfile};
        \addplot[mark=none, thick, dashed] coordinates {(30, 1e-7) (30, 10)};
        \addplot[mark=none, draw=mlblue, name path=yMin] table[x=x, y=ymin] {\srcfile};
        \addplot[mark=none, draw=mlblue, name path=yMax] table[x=x, y=ymax] {\srcfile};
        \addplot[fill opacity=0.3, fill=mlblue] fill between[of=yMax and yMin];
    \end{axis}
    \begin{axis}[name=shuttleLeader, at={($(shuttleFollower.south) + (0cm, -1.5cm)$)}, anchor=north,
        ymode=log,
        width=\width, height=\height,
        xlabel={Time after DAC step / \si{\micro\second}}, xlabel near ticks,
        ylabel={Link isolation}, ylabel near ticks,
        ymin=1e-6, ymax=1, 
        try min ticks log=5,
        xmin=-5, xmax=105,
        ymajorgrids
        ]
        \def\srcfile{data/shuttle-extinction/dynamic_alice.txt};
        \addplot[mark=o, only marks, mark size=1pt] table[x=x, y=y] {\srcfile};
        \addplot[mark=none, thick, dashed] coordinates {(55, 1e-7) (55, 10)};
        \addplot[mark=none, draw=mlred, name path=yMin] table[x=x, y=ymin] {\srcfile};
        \addplot[mark=none, draw=mlred, name path=yMax] table[x=x, y=ymax] {\srcfile};
        \addplot[fill opacity=0.3, fill=mlred] fill between[of=yMax and yMin];
    \end{axis}
    \begin{axis}[name=repumpFollower, at={($(shuttleFollower.east) + (\colsep, 0cm)$)}, anchor=west,
        width=\width, height=\height,
        xlabel={\SI{1033}{\nano\meter} deshelving pulse duration / \si{\micro\second}}, xlabel near ticks,
        ylabel={Bright fraction}, ylabel near ticks,
        xmin=-5, xmax=105,
        ymin=-0.07, ymax=1.07,
        ]
        \def\srcfile{data/deshelving-dynamics/deshelve_bob.txt};
        \addplot[mark=*, only marks, mark size=1pt] table[x=x, y=y] {\srcfile};
        \addplot[mark=none, color=mlblue, domain=0:100, samples=500] {1-exp(-x / 3.88)};
        \draw[|-|] (axis cs: 0, 0.632) -- (axis cs: 3.88, 0.632) node[right] {\SI{3.88(2)}{\micro\second}};
    \end{axis}
    \begin{axis}[name=repumpLeader, at={($(shuttleLeader.east) + (\colsep, 0cm)$)}, anchor=west,
        width=\width, height=\height,
        xlabel={\SI{1033}{\nano\meter} deshelving pulse duration / \si{\micro\second}}, xlabel near ticks,
        ylabel={Bright fraction}, ylabel near ticks,
        xmin=-5, xmax=105,
        ymin=-0.07, ymax=1.07,
        ]
        \def\srcfile{data/deshelving-dynamics/deshelve_alice.txt};
        \addplot[mark=*, only marks, mark size=1pt] table[x=x, y=y] {\srcfile};
        \addplot[mark=none, color=mlred, domain=0:100, samples=500] {1 - exp(-x / 6.7)};
        \draw[|-|] (axis cs: 0, 0.632) -- (axis cs: 6.7, 0.632) node[right] {\SI{6.70(7)}{\micro\second}};
    \end{axis}
\end{tikzpicture}
    \caption{Link isolation and state scrambling dynamics; top Alice, bottom Bob.
        Left: Suppression of photon collection via the remote fibre link while shuttling the ion away from the focus of the high-NA objective. The dashed line at \SI{30}{\micro\second} (\SI{55}{\micro\second}) marks the start of the analysis pulse and subsequent readout. Circles mark the mode of the background-subtracted posterior distribution of the signal and shaded areas span the \SI{95}{\percent} confidence interval. Right: Population in the ${S}_{1/2}$ ground level after deshelving population initially prepared in the $\ket{{D}_{5/2}, m_J=-3/2}$ state vs.~duration. An exponential fit and the resulting time constant is shown. The link is reconnected by shuttling the ion back after deshelving for \SI{100}{\micro\second}.
    }
\label{fig:LinkIsolation}
\end{figure}
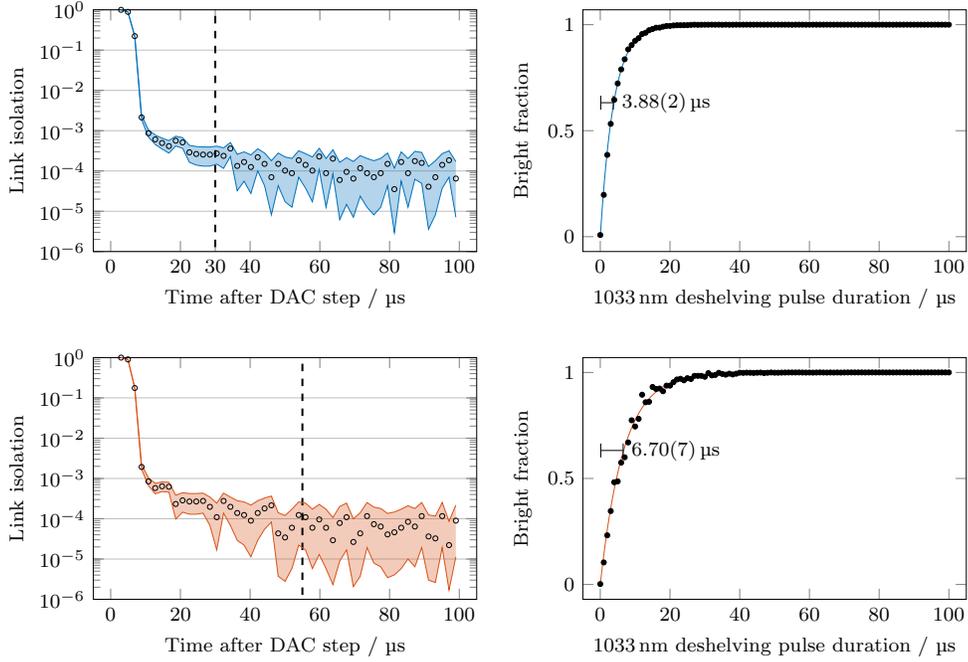

The precise timing sequence is shown in \cref{fig:timing}: After the heralding logic signals the successful creation of remote entanglement, both systems first map the Zeeman qubit into the less magnetic field-sensitive optical qubit using a $\ket{{S}_{1/2}, m_J=+1/2} \rightarrow \ket{{D}_{5/2}, m_J=-3/2}$ $\pi$ pulse, independent of the later measurement basis. Then, the DAC voltages are switched and a delay of \SI{30}{\micro\second} ($\sim$\,time constant of electrode low-pass filter) ensures the link is fully disconnected before Bob sends the round type signal ($T_i$) to Alice and continues with local analysis and readout. Alice waits to receive $T_i$ from Bob before proceeding with her local operations, incurring an additional \SI{25}{\micro\second} delay due to processing latencies. Due to the non-demolition nature of the ion qubit measurements, the state also needs to be scrambled before reconnecting the link. For this, the \SI{1033}{\nano\meter} deshelving laser is applied together with the cooling lasers for \SI{100}{\micro\second} before the ion is shuttled back into the focus of the high-NA objective. The typical dynamics of these processes are shown in \cref{fig:LinkIsolation}.

We stress that isolation is an assumption in DIQKD that cannot be verified using device-independent techniques, and should thus be unambiguously established through independent technical means. This is the reason for the active measures described above that we took to limit possible information leakage. Although essential for cryptography, the isolation assumption can be partially relaxed: if the amount of information possibly leaked to the adversary can be upper bounded, the key can remain secure at the expense of shortening its length accordingly. In our case, some amount of information may still possibly leak despite the measures taken, but not to the point of compromising the extraction of a secret key. To see this, we monitor the detectors at the central heralding station during qubit readout in the DIQKD link characterisation run; the count totals listed in \cref{tab:link-leakage}. From this, we estimate the probability of a photon leaking from Alice during a bright measurement (classical outcome $1$) to \num{6e-4}, that from Bob to \num{4.4e-3}. (Here, we have factored in the $\sim\SI{25}{\percent}$ coupling loss and $\sim\SI{70}{\percent}$ quantum efficiency of the single-photon detectors used, as well as their dark count rate, which accounts for most of the background.) Combined with the marginal probability $1/2$ for a bright measurement on one side, the probability of a photon leaking from either system during a measurement is \num{2.5e-3}. The information leaked by this channel is thus upper bounded by the binary entropy $h(\num{2.5e-3}) = \num{0.025}$, which is significantly lower than the key rate of $0.064$ obtained in the DIQKD demonstration run. We also emphasize that this bound could be made arbitrarily small in a real deployment by the various means mentioned in the beginning of this section.
\begin{center}
    \begin{table}[htbp]
        \begin{tabular}{cc@{\hspace{10pt}}ccccc@{\hspace{10pt}}c}
            \toprule
            \multicolumn{2}{c}{Ion state} & \multirow{2}{*}[-2pt]{$k=0$} & \multirow{2}{*}[-2pt]{$k=1$} & \multirow{2}{*}[-2pt]{$k=2$} & \multirow{2}{*}[-2pt]{$k=3$} & \multirow{2}{*}[-2pt]{$k > 3$} & \multirow{2}{*}[-2pt]{$\bar{k}$} \\
            \cmidrule(r{6pt}){1-2}
            Alice & Bob & & & & & & \\
            \midrule
            dark & dark & 141506 & 3864 & 80 & 2 & 0 & 0.0277 \\
            dark & bright & 656728 & 19770 & 378 & 6 & 0 & 0.0304 \\
            bright & dark & 670039 & 18685 & 322 & 7 & 0 & 0.0281 \\
            bright & bright & 141707 & 4347 & 76 & 3 & 0 & 0.0308 \\
            \bottomrule
        \end{tabular}
        \caption{Leakage after shuttling-based link isolation during the DIQKD link characterisation run. The columns give the numbers of rounds during which $k$ photons were registered at the central heralding station detectors and the respective means $\bar{k}$, conditioned on the classical outcomes measured on both sides. The background counts measured for the dark classical outcome $00$ are, to within the statistical uncertainty, entirely due to detector dark counts.}
        \label{tab:link-leakage}
    \end{table}
\end{center}


\section{Definition and assumptions of DIQKD}\label{sec:assumptions}

\subsection{General setting of QKD}
QKD is a cryptographic protocol which allows (upon success) two separated users sharing a classical and a quantum channel to expand a secret key, i.e.~a bit string unknown to an adversary. These legitimate parties are commonly referred to as Alice and Bob. In addition to knowing the protocol, the adversary may have complete knowledge of the functioning of Alice and Bob's devices, it may have unlimited computational power and it may fully control the channels outside of Alice and Bob's labs. The information-theoretic security of the key, i.e.~the mathematical claim that the adversary gains virtually no information about the key produced by the QKD protocol, relies on the following physical assumptions:

\medskip

(i) \textit{Quantum theory:} The operations performed by the users and by the adversary admit a description in terms of quantum states, operations and measurements. In particular, the statistics observed by the legitimate parties admit a quantum description.

\medskip

(ii) \textit{Isolation:} The parties can prevent information leaking to the adversary. The users' systems can be shielded so that classical and quantum information only leave Alice's (Bob's) lab under her (his) control.

\medskip

(iii) \textit{Input randomness:} Alice and Bob's input choices are truly random, i.e.~they are not correlated with their other devices nor the adversary.

\medskip

(iv) \textit{Trusted information processing:} The computers processing Alice and Bob's classical information perform the expected computations.

\medskip

(v) \textit{Trusted quantum operations:} Alice and Bob's quantum devices perform the expected quantum operations. In particular, quantum devices are accurately characterized and they maintain perfect calibration: sources of quantum states produce the expected quantum states and quantum measurements are performed according to their specifications.

\medskip

QKD is often presented with the additional assumption that the classical channel between Alice and Bob is authenticated, i.e.~so that the adversary cannot alter the content of messages transiting on it~\cite{Scarani09}. QKD is then often considered a key generation protocol rather than a key expansion one. However, fulfilling this assumption still requires the consumption of an initial secret of some sort~\footnote{In principle, authentication is possible with a weakly correlated partially secret shared bit string~\cite{Renner03}.}. In order to demonstrate clearly that our experiment produces more secrecy than what is required to authenticate its classical communication (it would be trivial to produce a key shorter than the initial secret seed), we include an explicit authentication step into our protocol. Authentication is then not an assumption anymore.

As a cryptographic protocol, QKD is subject to the specific assumption of isolation (ii), thereby differentiating QKD from other quantum tasks which do not take place in an adversarial setting. It is indeed a fundamental assumption of cryptography that information can be kept from exiting the user's labs: without this assumption, no cryptography is possible. This assumption allows the outcomes of each round, as well as the settings and initial shared key $\bK_0$, to be safely stored locally. Since no law of physics forbids information from flowing in space a priori, this assumption deserves special attention in any practical context and should be clearly stated~\cite{Katz14}.

Another critical assumption of conventional QKD is that of trusted quantum operations (v). In the BB84 protocol~\cite{Bennett84} for instance, this assumption requires Alice's device to prepare at each round one of the following four possible qubit states and to send it to Bob: $|\pm  z\rangle$ or $|\pm  x\rangle$. It then requires Bob to perform a measurement that is a projection onto either $|\pm  z\rangle$ or $|\pm  x\rangle$. Unfortunately, numerous attacks have shown that it is difficult to gain confidence that this assumption holds in practice. For instance, attacks on the protocol may be successful because the calibration of the quantum devices is imperfect~\cite{Zhao08}, or because the devices can be brought to operate outside their intended regime, thereby drastically changing their response~\cite{Lydersen10, Gerhardt11, Weier11}. The aim of device-independent QKD is to improve the security of QKD by lifting this last, often problematic, assumption.

\subsection{DIQKD assumptions}
Device-independent QKD takes place in the same context as QKD. It is also a cryptographic protocol, with the same purpose as QKD, but with an improved security obtained by relying only on the first four assumptions above~\cite{Pirandola20}. In other words, security of DIQKD is guaranteed when the protocol succeeds even if assumption (v) is not satisfied. DIQKD can thus be understood as a QKD protocol with partially characterized and partially trusted devices. Quantum devices producing prescribed states and performing prescribed measurements shall be used to ensure that the protocol succeeds, but the security of the key produced in the case of success is provable without trusting that quantum operations were performed as expected (in this sense, they could even have been chosen by the adversary). Below, we describe more precisely the part of the devices which does not need to be characterized. Clearly, relying on fewer assumptions means that DIQKD is both more secure than conventional QKD and more demanding to realize experimentally.

It is worth noting that DIQKD \emph{only} relaxes assumption (v) above. In particular, the scope of DIQKD is not to relax the isolation assumption (ii). This assumption, although often tricky to guarantee in practice, is treated identically in DIQKD as it is in standard QKD. In particular, this means that DIQKD remains susceptible to attacks breaking this assumption, such as if Alice or Bob's labs happen to contain an emitter broadcasting secret information to the adversary through a means that is not properly shielded. Ultimately, it is thus the responsibility of the users to gain confidence about the proper isolation of their systems by independent means than those provided by the protocol. We comment below on possible ways to improve confidence on the proper isolation of devices in the context of DIQKD.

\subsection{From assumptions to security}
Assumptions (ii) and (iv) set the stage for a security claim: parties are well-identified and we can trust that the computational steps of the protocol happen as expected. Security of the key produced by the DIQKD protocol then reduces to the existence of a security proof. Our security proof, given in Sec.~\ref{sec:proof}, is a mathematical consequence of two conditions. As we show now, these two conditions themselves are direct implications of the DIQKD assumptions.

The first condition is the Markov chain condition needed in the application of the entropy accumulation theorem (see Eq.~\eqref{eq:markov} and below). This condition is directly granted by assumption (iii).

The second condition plays a central role in device-independent security proofs. We derive it now from assumptions (i)-(iii). Assumptions (i) together with the isolation assumption (ii) imply that the Hilbert space describing the two devices factorizes as a tensor product $\mathcal{H}=\mathcal{H}_A\otimes\mathcal{H}_B$ (the same condition is also a direct consequence of assumption (i) together with the simple fact that Alice and Bob are separated parties). The fact that the measurement settings are uncorrelated with the setup (assumption (iii)) implies that the quantum state measured in the $i^\text{th}$ round of the protocol does not depend on the inputs $X_i$ and $Y_i$: it can be written simply $\rho_i\in\mathcal{H}$. The same condition (iii) together with the isolation assumption (ii) directly imply that Alice's (Bob's) measurement in the $i^\text{th}$ round of the protocol does not depend on $Y_i$ ($X_i$): it can be written as a set of POVM elements $M_{A_i|X_i}$ ($M_{B_i|Y_i})$. Therefore, the probability of observing the outcomes $(A_i, B_i)$ can be written, through (i) again, following Born's rule as
\begin{equation}\label{eq:locality}
P(A_i,B_i|X_i,Y_i) = \tr{\rho_i\, M_{A_i|X_i}\otimes M_{B_i|Y_i}}.
\end{equation}
This condition is used in the security proof when deriving the entropy bound $\eta(\omega)$ as a function of the winning probability $\omega$ of the CHSH game~\cite{Pironio09}.

Note that $\rho_i$ in Eq.~\eqref{eq:locality} is the state measured at round $i$ and in practice this state could be different from the one distributed by the users' devices (e.g.~if the adversary tampers with it). It is precisely one of the advantages offered by the device-independent approach that the security proof holds as long as Eq.~\eqref{eq:locality} is correct for \emph{some} state $\rho_i$ and \emph{some} measurements $M_{A_i|X_i}$, $M_{B_i|Y_i}$, i.e.~even if these are not the intended state and measurements.

Condition~\eqref{eq:locality} identifies precisely the parts of Alice and Bob's devices which need not be trusted (in contrast with the parts requiring independent assessement to guarantee the validity of the remaining assumptions, such as the proper isolation of the systems). Since both the state and measurements are untrusted, one often refers to this part of the parties' devices as ``black boxes'' receiving classical inputs $X_i$ and $Y_i$, and producing outcomes $A_i$ and $B_i$~\cite{Brunner14}. These black boxes are described by Eq.~\eqref{eq:locality}, i.e.~they are quantum, but otherwise fully untrusted. The fact that the security is independent of the description of this part of the users' devices is what the term \emph{device-independent} refers to~\cite{Pironio10}.

\subsection{The DIQKD assumptions in the experiment}
We see no reason to doubt the validity of assumptions (i) and (iv) in the present experiment and therefore take them for granted. Similarly, we have no reason to believe that the QRNG we use is correlated to the rest of the setup. We checked that the numbers produced by this generator pass common randomness test suites (see Sec.~\ref{sec:DIQKDPrimitives}). Therefore, we take assumption (iii) for granted. Note that if needed, established measurement dependence methods could in principle allow to relax assumption (iii) to deal with deviations from the ideal case~\cite{Kessler20}. As mentioned before, the isolation assumption (ii) requires more appreciation.

In our case, Alice and Bob's ``labs'' are physically separated, well identified in space, and connected to a central station by two $\SI{1.75}{\metre}$-long optical fibres. Therefore, it is clear that their Hilbert space factorizes into a tensor product. Each ``lab'' includes a trapped ion, individual laser modes, optics, electronics (FPGA) and computer (see Fig.~\ref{fig:block-diagram-details}). The PCs allow for trusted storage and processing of classical information locally, including the initial key $\bK_0$, the settings $\bX$ and $\bY$ and the outcomes $\bA$, $\bB$. The lasers are used to cool the ions, excite and manipulate them, and detect their internal state.

As the user's lasers are obtained from common laser sources, it is worth emphasizing that each user only has access to its own laser mode which contain just half of the initial laser light. These beams can thus be seen as two independent laser beams, possibly only correlated in frequency and intensity (a laser beam impinging on a beam splitter gives rise to two independent laser beams). Therefore, laser light essentially plays here the role of a power supply, similarly to the electric power grid to which all equipment is connected. Since no fluctuations or actual signal on these power supplies may depend on the choice of settings $X_i$, $Y_i$ of a given round -- by assumption (iii) -- their effect on the actual measured quantum state and measurement operators can always be cast into the form of Eq.~\eqref{eq:locality} by including in $\rho_i$ the state of the lasers. Shared lasers and power supplies are thus entirely compatible with the device-independent framework of DIQKD. It is in fact a genuinely device-independent guarantee that any fluctuations in shared equipment do not compromise the security of the resulting key.

In the current experiment, each lab has a \enquote{hatch} to the outside that allows photons produced upon excitation of the ions to exit the labs and be combined at a central station, resulting ideally in a shared entangled ion-ion state in a heralded way. This hatch is then closed by moving the position of the ions while preserving entanglement. Specifically, the ions are displaced relative to the focus of the high-NA objective coupling the lens to the fibre through modulation of the trapping potential. The qubit state left after readout is also scrambled by applying deshelving and cooling lasers before reconnecting the link. Moving ions while preserving entanglement and scrambling the ion state are efficient procedures and well developed features of the trapped ion platform, which we use here as a novel method of isolation.

\subsection{Possible improvements}
Assumption (ii) could be further enforced by additional measures. The use of independent laser sources for example could facilitate the shielding of the devices as it would become sufficient to guarantee that no unwanted optical signal exits the devices through the fibre link. Currently, we did not take any special measure to avoid information from leaking through the laser modes. The labs have been designed to favour the coupling of the ions with the fibres rather than with the laser modes, but ultimately other couplings should not be neglected. Alternatively, the incoming laser could be passed through an optical isolator with appropriately characterised extinction, or the qubits could be manipulated via RF instead of laser light.

Similarly, one could arbitrarily reduce the amount of information leaked through the fibre link by shuttling the ion a further distance from the focus of the fibre interface, or by adding a separate, macroscopic mechanism to disconnect the optical fibre, e.g.~a fast shutter or micro-electromechanical fibre switch. As discussed in Sec.~\ref{sec:ShuttlingBasedLinkIsolation}, the extinction ratio of the fibre \enquote{hatches} is currently limited to $\sim 10^{-4}$. Importantly, the correct operation of such additional isolation devices applied to the fibre could be verified completely independently of the ion trap hardware.

In the context of improving isolation, space-like separation between Alice's and Bob's measurements could also have a positive impact. For instance, it could rule out certain attacks without further isolation considerations. The impact of space-like separation for DIQKD is however fundamentally limited since space-like separation only forbids information from reaching certain points in space during very short intervals -- at most for about \SI{40}{\milli\second} for any two locations on the earth's surface. Compared to other possible improvements mentioned here, space-like separation is also considerably more difficult to achieve. Additionally, space-like separation does not help to solve concrete isolation issues such as the isolation of the ion-fibre link: a leakage of photons through a fibre is entirely compatible with the constraints imposed by the speed of light. In fact, if all isolation issues are properly taken care of, then space-like separation becomes unnecessary~\cite{Pironio10,Pironio09,Murta19,Pirandola20}. In other words, all attacks that can be ruled out by invoking space-like separation are also ruled out by isolation. It is also worth noting that the notion of space-like separation that should be used here fundamentally differs from that used in the context of loophole-free Bell tests: in a QKD experiment all settings can be stored in advance locally because classical information can be held locally (otherwise there is no point in doing cryptography in the first place). In contrast, storing of measurement settings in advance is generally not accepted in the context of loophole-free Bell tests, where space-like separation tends to be defined between the measurement event of one party and the settings generation event of the other one, imposing that settings should be freshly generated on the fly~\cite{Hensen2015,Shalm15,Giustina15,rosenfeldEventReadyBellTest2017}. DIQKD experiments are thus clearly distinct from loophole-free Bell tests (see also Methods).

We envision that a significantly stronger confidence in the devices' isolation, i.e.~in assumption (ii), will be reached when quantum memories are able to store reliably $\sim10^6$ qubits~\cite{Heshami16}. This will allow for all entangled systems to be distributed before the DIQKD protocol begins. Contrary to all existing setups which require opening Alice and Bob's devices on a regular basis to distribute new entangled resources, no quantum \enquote{hatch} will be required, thus allowing for strong shielding to be applied on both parties without interruption throughout the whole protocol and improving the confidence in the absence of information leakage: the protocol will only require classical communication exchange. Ultimately, such a demonstration will however still rely on the same four assumptions defined above.


\section{DIQKD protocol}\label{sec:protocol}
\begin{figure}
\includegraphics[width=0.7\textwidth]{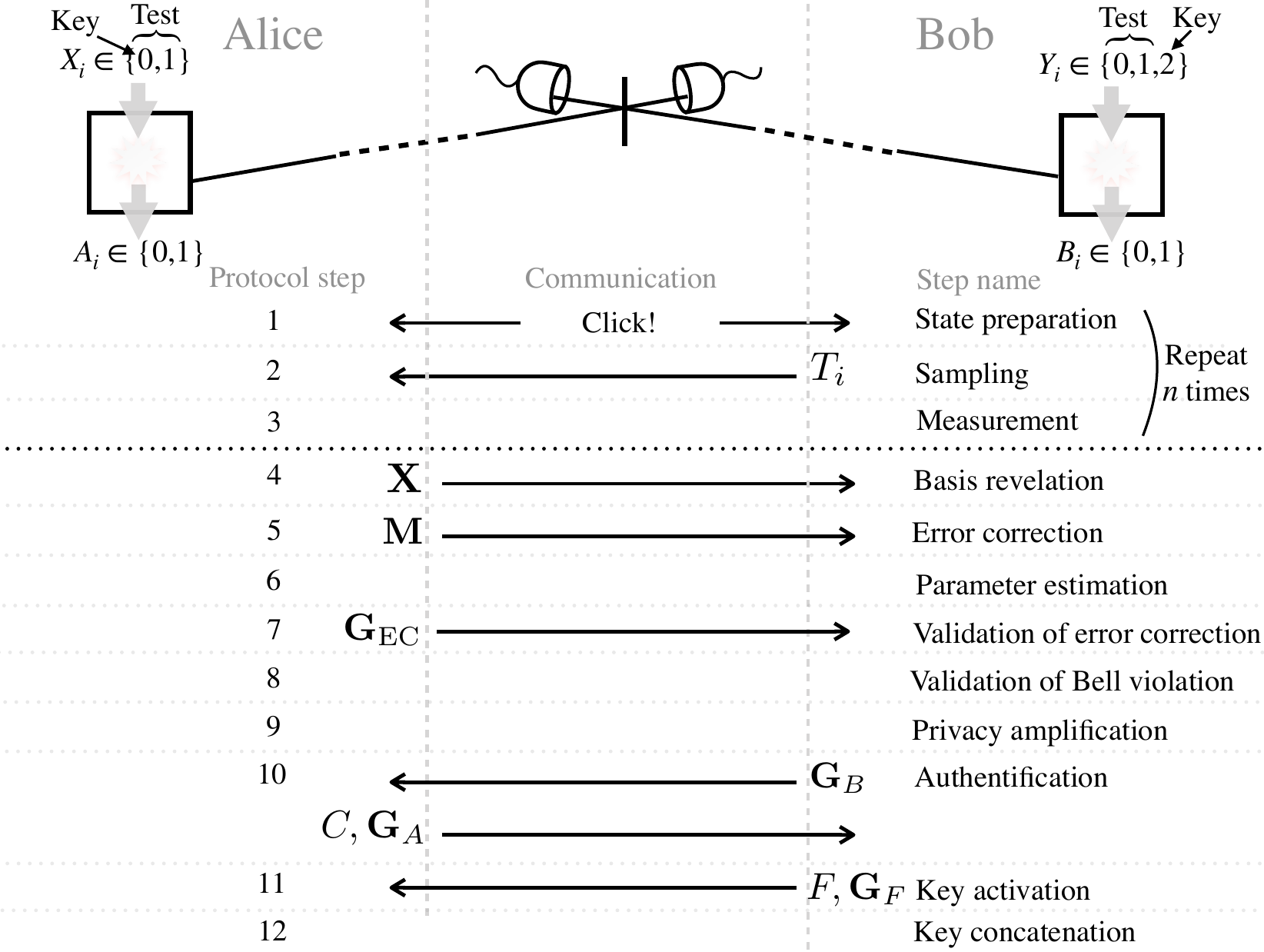}
\caption{Schematic representation of the key distribution setup and protocol. Alice's input $X_i=0$ is chosen both for the key and test generation rounds while Bob's input $Y_i=2$ is exclusively used for the key generation rounds. Classical communications happening during the DIQKD protocol are represented by arrows.}
\label{fig:classicalComm}
\end{figure}

Here we describe our DIQKD protocol in detail. It follows a similar general structure as most DIQKD protocols proposed thus far~\cite{Ekert91,Pironio09,ArnonFriedman18}. The protocol involves two distant parties, Alice and Bob, who initially share a secret key $\bK_0$. The aim of the protocol is to create a shared secret key $\bK_1$ that is longer than $\bK_0$ by performing $n\in\mathbb{N}$ successive measurements on distributed quantum systems followed by post-processing steps, see Fig.~\ref{fig:classicalComm}. The protocol is successful when both parties are confident that their copy of $\bK_1$ is sound.

\begin{figure}
\includegraphics[width=0.65\textwidth]{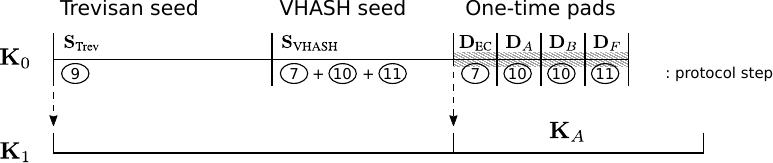}
\caption{Secret key expansion (not to scale) : A secret key $\mathbf{K}_0$ is used to create a longer secret key $\mathbf{K}_1$. The part of the initial key used for one-time pads during the protocol needs to be discarded. The rest of the initial key can be reused. If the length of the key $\mathbf{K}_A$ is larger than the part of $\bK_0$ consumed for one-time pad, the new key is longer than the initial one.}
\label{fig:K0K1}
\end{figure}

\ \\
\paragraph{Parameters.}
The protocol depends on the following parameters: the testing probability $\gamma\in(0,1)$, the total number of rounds $n\in \mathbb{N}$, the threshold CHSH winning probability $\omega_\thresh\in\left(\frac{3}{4},\frac{1+1/\sqrt{2}}{2}\right]$, and the syndrome length $m$. These parameters are fixed before the protocol starts.

\ \\
\paragraph{Preparation.}
At the beginning of the protocol, both parties share a secret key $\bK_0$. This key contains the variables $\bS_\mathrm{Trev}$, $\bS_\mathrm{VHASH}$, $\bD_\mathrm{EC}$, $\bD_{A}$, $\bD_{B}$ and $\bD_{F}$, see Fig.~\ref{fig:K0K1}. Each party also holds a set of (non-shared) secret random variables~\footnote{alternatively, these could also be generated as the protocol runs}. Namely, Alice holds a string $\bX=(X_1,\ldots,X_n)$ with $X_i\in\{0,1\}$ of measurement settings chosen uniformly, i.e.~with distribution
\begin{equation}
\label{distribuion_Aliceinputs}
P(X_i=0)=P(X_i=1)=1/2.
\end{equation}
Bob holds a string $\bY=(Y_1,\ldots,Y_n)$ with $Y_i\in\{0,1,2\}$ that are sampled from the following distribution:
\begin{equation}
\label{distribuion_Bobinputs}
P(Y_i=0)=P(Y_i=1)=\gamma/2\ ,\ \  P(Y_i=2)=1-\gamma.
\end{equation}
From this string, Bob also defines the test variables $T_i=\chi(Y_i\neq 2)$, $\chi$ being the indicator function.

\ \\
\paragraph{Protocol steps.}
The quantum steps of the protocol are repeated $n$ times; each round is indexed by $i=1,\ldots,n$.
\begin{enumerate}
\item[] Repeat $n$ times:
\begin{enumerate}

\item[1.] [\textit{State preparation}] The parties wait for the heralding logic signal announcing that their respective ions have been prepared in an entangled state.

\item[2.] [\textit{Sampling}] Bob sends the value of $T_i$ to Alice. If $T_i=0$, Alice overrides $X_i$: she sets $X_i=0$. 

\item[3.] [\textit{Measurement}] Alice and Bob measure their quantum systems in the bases defined by $X_i$ and $Y_i$ respectively, and store their results in the variable $A_i$ and $B_i$ respectively ($A_i, B_i \in \{0,1\}$).

\end{enumerate}
\end{enumerate}

Once these steps are performed, the parties proceed with the following classical post-processing steps.

\begin{enumerate}
\item[]
\begin{enumerate}
\item[4.] [\textit{Basis revelation}] Alice publicly announces her measurement settings $\bX$.
\item[5.] \label{step:EC} [\textit{Error correction}] Alice computes a $m$-bit long syndrome $\bM \in \{0,1\}^m$ for her string of outcomes $\bA=(A_1,\ldots,A_n)$ and sends it to Bob. Bob reconstructs a guess $\bAt$ of $\bA$ from $\bX,\bY,\bB$ and $\bM$.
\item[6.] [\textit{Parameter estimation}] Bob computes the score
\begin{equation}
U_i = \begin{cases}
\chi(\tilde{A}_i \oplus B_i = X_i \cdot Y_i) & \textup{if } T_i=1\\
\perp & \textup{if } T_i=0\\
\end{cases}
\end{equation}
for each round $i=1,\ldots,n$.
\item[7] \label{step:ValidationEC} [\textit{Validation of error correction}] Alice computes a short encrypted almost-universal hash $\bG_\mathrm{EC}$ of $\bA$ with seed $\bS_\mathrm{VHASH}$ and one-time pad $\bD_\mathrm{EC}$ and sends it to Bob. Bob computes the short hash $\tilde\bG_\mathrm{EC}$ of $\bAt$ with seed $\bS_\mathrm{VHASH}$ and one-time pad $\bD_\mathrm{EC}$ and checks that it matches with Alice's hash $\bG_\mathrm{EC}$. If $\tilde\bG_\mathrm{EC}\neq \bG_\mathrm{EC}$, the protocol aborts. 
\item[8] \label{step:ValidationCHSH} [\textit{Validation of Bell violation}] 
Bob checks the condition
\begin{equation}\label{eq:validation}
\sum_i \chi(U_i=0) \leq n\gamma\left(1 - \omega_\thresh\right).
\end{equation}
If the condition is violated, the protocol aborts.
\item[9.] \label{step:PA} [\textit{Privacy amplification}] Using seed $\bS_\mathrm{Trev}$, Alice applies a strong extractor to $\bA$ to obtain $\bKA$. Bob uses the same seed and extractor to create $\bKB$ from $\bAt$.
\item[10.] \label{step:Authentication} [\textit{Authentication}] Using the seed $\bS_\mathrm{VHASH}$ and one-time pad $\bD_{B}$, and an unconditionally-secure message authentication code, Bob computes a tag $\bG_B$ for $\bT$. He sends the tag to Alice. Alice checks whether the tag matches the one she computes with the same part of $\bK_0$ and the communication she received from Bob earlier. If it is the case, she defines $C=1$, otherwise she sets $C=0$. Alice sends $C$ to Bob. Alice then similarly computes a tag $\bG_A$ for $(\bX,\bG_\mathrm{EC},C)$, using the seed $\bS_\mathrm{VHASH}$ and a one-time pad $\bD_{A}$, and sends it to Bob. Bob checks that the tag matches the one he computes from the corresponding information he received from Alice. If the tags don't match, or if $C=0$, Bob defines $F=0$. Otherwise he sets $F=1$.
\item[11.] [\textit{Key activation}] Bob sends $F$ to Alice together with a tag $\bG_F$ computed for $F$ using the seed $\bS_\mathrm{VHASH}$ and a one-time pad $\bD_F$. Alice checks that the tag matches the one she computes from the information she received from Bob. If the tags don't match or if $F=0$, the protocol aborts.
\item[12.] [\textit{Key concatenation}] Define $\bK_0'$ as $\bK_0$ with $\bD_\mathrm{EC}$, $\bD_{A}$, $\bD_{B}$ and $\bD_{F}$ removed. Alice defines her new shared secret key $\bKA'$ as the concatenation of $\bK_0'$ and $\bKA$, see Fig.~\ref{fig:K0K1}. Similarly, Bob defines his new shared secret $\bKB'$ as the concatenation of $\bK_0'$ together with $\bKB$.
\end{enumerate}
\end{enumerate}

At the end of the protocol, the keys $\bKA'$ and $\bKB'$ are identical and secret with high probability, in the sense defined below. Alice sets $\bK_1=\bK_A'$. If $|\bK_1|>|\bK_0|$, the protocol extended a secret key between Alice and Bob.

\section{Model of the statistics}\label{sec:model}

To define the expected honest behaviour of the system, e.g.~when tuning the protocol parameters to match our experimental setup, we condense the observed statistics from \cref{sec:DIQKDPrimitives} into a simplified model with two parameters $S$ and $Q$, respectively corresponding to the CHSH Bell violation and quantum bit error rate (QBER). The security of our demonstration does not depend on the validity of this model.

More precisely, we assume that the statistics in the test rounds satisfy
\begin{equation}\label{eq:modelP1}
P(A,B|0,0) = P(A,B|0,1) = P(A,B|1,0) = P(A,1-B|1,1) = \frac{1 + (-1)^{A\oplus B}S/4}{4}
\end{equation}
and that the statistics in the key rounds are sampled according to
\begin{equation}\label{eq:modelP2}
P(A,B|0,2) = \frac{\delta_{A,B}-(-1)^{A\oplus B}Q}2
\end{equation}
We conservatively set $S=2.64$ and $Q=0.018$, since we expect the setup to hold these parameters for a time sufficient for the experiment to run (cf.~\refmaintextfiglinkperformance{} of the main text). Note that this model does not assume that the underlying state is a Werner state -- that would impose the relation
\begin{equation}
S = 2\sqrt{2}(1-2Q).
\end{equation}

\section{Error correction}\label{sec:EC}

In a DIQKD protocol, the violation of a Bell inequality generally guarantees the secrecy of the outcomes that Alice observes during the key rounds. These outcomes can then be considered as a raw key. But this key is not shared with Bob unless Bob holds a copy of it. Unfortunately, Bob's measurements generally do not create an exact copy of Alice's outcomes. In particular, Bob's outcomes observed during the test rounds are not expected to be strongly correlated with Alice's. Some amount of communication is thus needed to allow Bob to reconstruct Alice's outcomes in the key and test rounds.

Here, we consider the simple case in which this reconstruction is realized by having Alice send a single message $\bM$ of length $m$ to Bob. While a message of arbitrary length is always sufficient in this context (Alice could simply set $\bM=\bA$), any information revealed in this way reduces the secrecy of Alice's key: all information revealed between the parties is public. Therefore, the length $m$ of the communicated message should be as small as possible to reduce information leakage. In fact, the ability to correct errors with minimal message length is crucial to the feasibility of DIQKD: larger messages must be compensated by larger Bell violations and additional measurement rounds. This calls for error correction codes operating as close to the Shannon limit in the finite size regime as possible.

However, the fact that the error correction scheme should succeed with a small message length $m$ is not the only constraint it should fulfill. It is equally desirable that the error correction code would be relatively insensitive to noise specificities, and that its algorithm would admit an efficient implementation. Here we describe an error correction code fulfilling all of these requirements.

\begin{figure}
\centering
\includegraphics[width=0.35\textwidth]{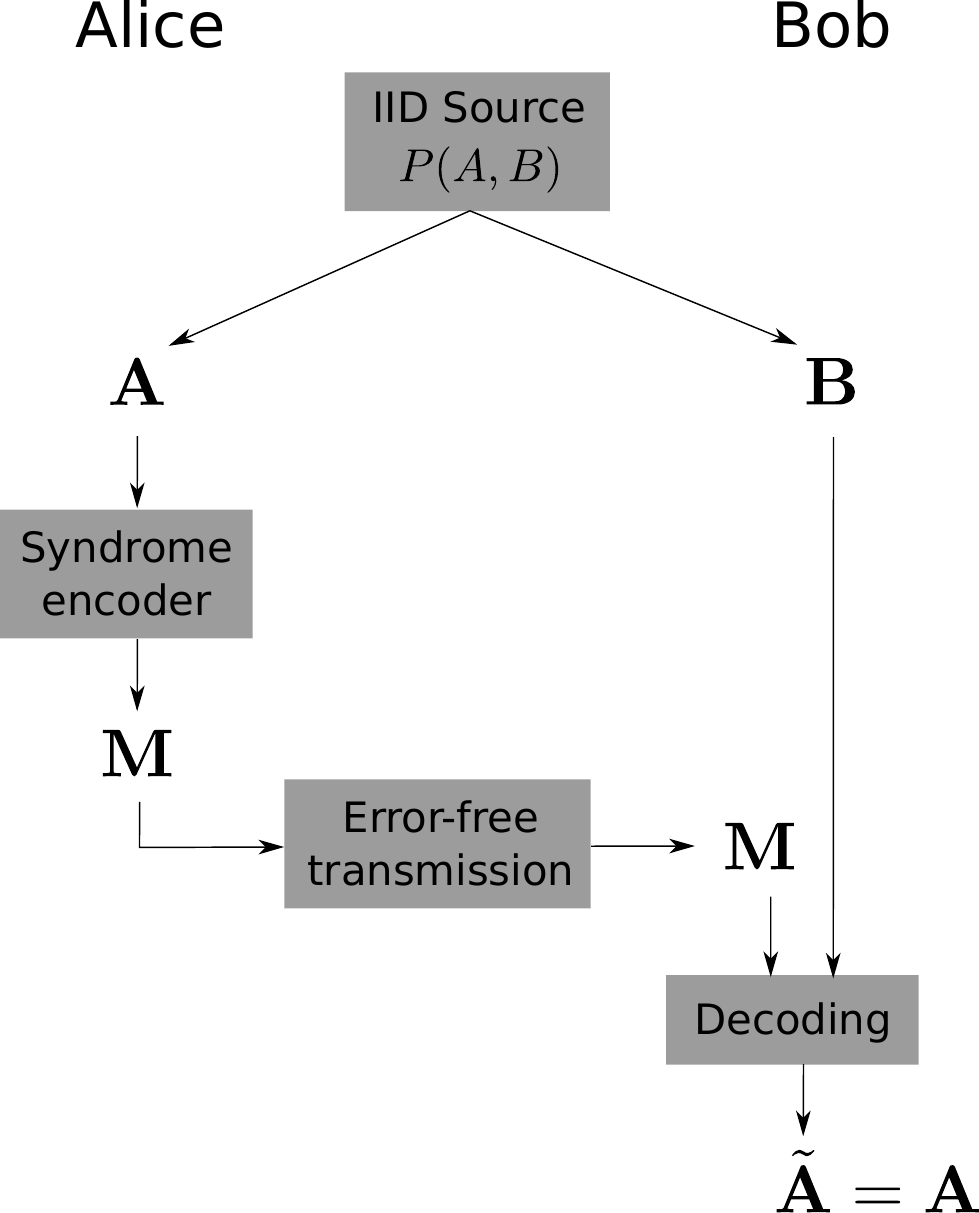}
\caption{Overall error correction setting: asymmetric Slepian-Wolf coding. Strings $\bA$ and $\bB$ are jointly sampled from a distribution $P(A,B)$ and given to two distinct parties Alice and Bob. The aim is for Bob to end up with a copy of Alice's string $\bA$, while only exchanging a `short' message $\bM$.}
\label{fig:setupEC}
\end{figure}

In general, the setting for error correction is equivalent to a Slepian-Wolf scheme with asymmetric coding, as depicted in Fig.~\ref{fig:setupEC}. In this setting, a correlated source described by the joint probability distribution $P(A,B)$ is sampled $n$ times. Alice computes a syndrome $\bM$ from her variable $\bA$ and sends it to Bob over a noiseless channel. Bob then reconstructs a copy $\tilde \bA$ of Alice's outcomes $\bA$ from $\bB$ and $\bM$. In this setting, the \textit{overhead} $\eta=m/n$ can be asymptotically as small as $H(A|B)$ with the following scheme~\cite{Richardson08}. Consider a noisy channel transforming $A$ into $B$ described by the conditional distribution $P(B|A)=\frac{P(A,B)}{P(A)}$. Consider an optimal (capacity-achieving) binary linear code of length $n$ with code rate $R=1-m/n$ for this channel. Take the $m\times n$ parity-check matrix $H$ of this code and compute the \textit{syndrome} $\bM=H\bA$ (mod $2$). Then $\bA$ can be recovered from $(\bB,\bM)$. While the standard formulation corresponds to the symmetric channel with uniform input distribution $P(A)$, it was recently shown \cite{WangKim15} that the $H(B|A)$ limit is also achievable in a general setting where the input distribution is not uniform.

\subsection{Minimum syndrome length for a binary symmetric channel}
In the case where $\bA$ and $\bB$ have the same statistics as the input and output of a Binary Symmetric Channel (BSC) with error parameter $\delta$, i.e.~of a channel which flips each input bit with probability $\delta$, we can compute the smallest possible syndrome length $m$ in presence of a finite number of samples $n$ and assuming that the error correction scheme fails with probability $\eps$. Following Eq. (289) of~\cite{Polyanskiy10}, the effective finite-size channel capacity is given by $C(n,\delta)$ with
\begin{equation}\label{eq:CN}
\begin{split}
n\cdot C(n,\delta) = &n(1-h(\delta)) - \sqrt{n\delta(1-\delta)}\log_2\frac{1-\delta}{\delta}Q^{-1}(\eps)\\
&+\frac{1}{2}\log_2(n) + O(1),
\end{split}
\end{equation}
where $n$ is the length of the ``transmitted'' bit string (which in our model is $\bB$), $h(x)$ is the binary entropy function, $\eps$ is the error correction failure probability and $Q(x)=\int_x^\infty \frac{1}{\sqrt{2\pi}}e^{-t^2/2}dt=\frac{1}{2}\left(1-\text{erf}(x/\sqrt{2})\right)$. For example, when considering that the error correction scheme might fail with a probability of $\eps=10^{-3}$, we have $Q^{-1}(\eps)\simeq 3$. The minimum length of the syndrome is then
\begin{equation}\label{eq:syn_length}
m_\mathrm{BSC}(n,\delta)=n(1-C(n,\delta)).
\end{equation}

\subsection{Concrete setting}
As discussed in Sec.~\ref{sec:model}, the honest model for our setup includes 5 sources of outcomes for Alice and Bob, corresponding to each choice of measurement setting. The corresponding distributions $P(A,B|X,Y)$ are parametrised in terms of the Bell value $S$ and the QBER $Q$ only. More precisely, the first four distributions correspond to a BSC with error $\delta_1=\delta_2=\delta_3=\delta_4=\frac{4-S}8=\delta'$ (with a bit flip in the fourth case). These first sources are on average sampled $n_1=n_2=n_3=n_4=\gamma n/4$ times each. The last source is also a BSC but with a different error: $\delta_5=Q=\delta''$, which is sampled $(1-\gamma)n$ times on average. Therefore, we need to deal here with the transmission of $\gamma n$ samples of a BSC with bit flip probability $\delta'$ and $(1-\gamma)n$ samples of a BSC with bit flip probability $\delta''$.

Note that the true statistics differ from this model (see Sec.~\ref{sec:DIQKDPrimitives}). We rely on the universality of our code to deal with these deviations.

Assuming an error-correcting code performing close to the finite-length limits, we have in principle two options: encode each string on its own or encode all of them in one block. From \eqref{eq:CN} it follows that the larger the block length the smaller the required overhead for error-free performance. Hence it is preferable to encode the data from all channels together. In this case, we also need a joint decoding procedure.

\subsection{Practical coding approaches}
For practical application, we need an error-correcting code which is capacity-achieving under low-complexity decoding algorithm. Since we are operating in the finite-length regime, we would also like to get the finite-length performance close to theoretical limits. Simplicity of the code construction is an additional advantage due to the possible changes in the experimental setting. There are three reasonable choices that we can use: low-density parity-check (LDPC) codes, spatially-coupled LDPC (SC-LDPC) codes and polar codes. 

Polar codes \cite{Arikan08} have low-complexity encoding and decoding procedures but have several issues. Perhaps most importantly, the code is channel-dependent, i.e.~any change in $\gamma, Q$ or $S$ will require a separate optimization step, which in case of BSC can only be done through costly Monte-Carlo simulations. Another problem is that whereas the optimal code has gap to the asymptotic value that scales as $\Theta(\frac{1}{\sqrt n})$ (it follows from \eqref{eq:CN}), for polar codes it is of order $\Theta(\frac{1}{n^{1/3.579}})$ \cite{Hassani14}, which is significantly worse and implies that their finite-length performance is rather far from theoretical limits. 

LDPC codes were introduced in the early 1960s \cite{Gallager63}. They are described as dual spaces of some sparse matrices, which typically have a simple structure. Sparsity of this matrix gives rise to low-complexity iterative belief propagation (BP) decoding algorithm. It is possible to compute the asymptotic performance of LDPC codes under BP decoding \cite{Richardson01}. Although LDPC codes can achieve capacity of BSC under computationally unfeasible maximum likelihood (ML) decoding \cite{Sason03}, their \textit{BP thresholds}, i.e., the largest channel error probability that can be corrected by BP decoder with infinite number of iterations, are typically not capacity-achieving. There exist some code constructions that come close to the asymptotic limits \cite{RSU01} but the optimization for certain channels (such as mixture of two BSCs in our setting) might be nontrivial. In case of the binary erasure channel (the receiver either gets the transmitted bit correctly with probability $1-\eps$ or receives an \textit{erasure symbol} "$?$" with probability $\eps$) LDPC codes achieve capacity with linear decoding complexity \cite{Shokrollahi99, Oswald02, Pfister05}. The optimal scaling behavior of LDPC codes is established for the erasure channel and conjectured for the general channels \cite{Amraoui08}.

SC-LDPC codes are constructed as a chain of LDPC codes that are coupled together. It was observed numerically in \cite{Lentmaier10} and proved in \cite{Kudekar13} that BP threshold of SC-LDPC codes converges to the ML threshold of the underlying LDPC code. Moreover, they achieve capacity universally, i.e., a code of rate $R$ allows the error-free transmission through any channel with capacity greater than $R$ when the code length grows to infinity. Another advantage of SC-LDPC codes is the simplicity of the code construction since it is much easier to find LDPC codes which are capacity achieving under ML decoding than to optimize LDPC codes so that their BP threshold is close to capacity. Since SC-LDPC codes build on LDPC codes, in what follows, we start by describing LDPC codes and then proceed to describe the SC-LDPC codes we use.

\subsubsection{LDPC codes}
An LDPC code of length $n$ and dimension $k$ is defined as a dual space of $(n-k)\times n$ full-rank parity-check matrix $H$, which needs to be sparse. A code is called $(d_v,d_c)$-\textit{regular} if each row of $H$ contains exactly $d_c$ ones and each column exactly $d_v$ ones and irregular otherwise. The matrix $H$ is typically represented as a Tanner graph, which is a bipartite graph that consists of \textit{check} and \textit{variable} nodes, which correspond to the rows and columns of $H$. If $H_{i,j}=1$, then check node $i$ and variable node $j$ are connected by an edge. In Tanner graph representation, $d_v$ and $d_c$ become the degrees of variable and check nodes.

One way to construct good LDPC codes of variable length is to use a protograph construction \cite{Mitchell15}. A protograph with \textit{design rate} $R=1-n_c/n_v$ is a small bipartite graph with partitions of size $n_c$ and $n_v$ that correspond to check and variable nodes, respectively. A code of length $n=Mn_v$ is obtained by applying a \textit{lifting} procedure with \textit{lifting factor} $M$. This procedure can be described as follows. Take the $n_c\times n_v$ biadjacency matrix $B$ of the protograph and replace its zeros with $M\times M$ all-zero matrices and ones with random $M\times M$ permutation matrices $\Pi_{i,j}$. The resulting code has length $Mn_v$ and rate $R\ge 1-n_c/n_v$. In practice, for random matrices $\Pi_{i,j}$ there is almost always an equality \cite{Mitchell15}. The advantage of protograph-based LDPC codes is that their properties, such as BP decoding threshold, can be derived directly from protographs \cite{Mitchell15}.

\begin{example}
Consider the $(2,3)$-regular LDPC code. The corresponding protograph is given at Figure \ref{fig:proto1} and its biadjacency matrix can be written as 
$
B=\begin{pmatrix}
1 & 1 & 1\\
1 & 1 & 1
\end{pmatrix}
$. The lifting process for $M=3$ is given at figures \ref{fig:proto2}-\ref{fig:proto3}.
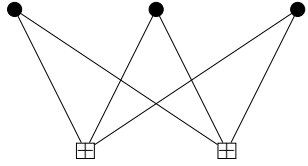
\begin{figure}
\centering
\begin{tikzpicture}
    \node (0) at (1,0) [draw, inner sep=0pt] {$+$};
    \node (1) at (3,0) [draw, inner sep=0pt] {$+$};
    \coordinate (3) at (0,2);
    \draw[fill=black] (3) circle (0.1);
    \coordinate (4) at (2,2);
    \draw[fill=black] (4) circle (0.1);
    \coordinate (5) at (4,2);
    \draw[fill=black] (5) circle (0.1);
    
    \begin{scope}[on background layer]
    \draw (0) edge (3);
    \draw (0) edge (4);
    \draw (0) edge (5);
    \draw (1) edge (3);
    \draw (1) edge (4);
    \draw (1) edge (5);
    \end{scope}
\end{tikzpicture}
\caption{(2,3)-regular protograph. Three vertices are connected to two check nodes.}
\label{fig:proto1}
\end{figure}

\begin{figure}
\begin{subfigure}{\linewidth}
    \centering
    \begin{tikzpicture}
   
    \draw[fill=white] (1.75,0) ellipse (1cm and 0.4cm);
    \node (10) at (1.25,0) [draw, inner sep=0pt] {$+$};
    \node (20) at (1.75,0) [draw, inner sep=0pt] {$+$};
    \node (30) at (2.25,0) [draw, inner sep=0pt] {$+$};

    \draw[fill=white] (4.25,0) ellipse (1cm and 0.4cm);
    \node (11) at (3.75,0) [draw, inner sep=0pt] {$+$};
    \node (21) at (4.25,0) [draw, inner sep=0pt] {$+$};
    \node (31) at (4.75,0) [draw, inner sep=0pt] {$+$};

    \draw[fill=white] (0.5,2) ellipse (1cm and 0.4cm);
    \node (40) at (0,2) [circle,fill=black] {};
    \node (50) at (0.5,2) [circle,fill=black] {};
    \node (60) at (1,2) [circle,fill=black] {};

    \draw[fill=white] (3,2) ellipse (1cm and 0.4cm);
    \node (41) at (2.5,2) [circle,fill=black] {};
    \node (51) at (3,2) [circle,fill=black] {};
    \node (61) at (3.5,2) [circle,fill=black] {};
    
    \draw[fill=white] (5.5,2) ellipse (1cm and 0.4cm);
    \node (42) at (5,2) [circle,fill=black] {};
    \node (52) at (5.5,2) [circle,fill=black] {};
    \node (62) at (6,2) [circle,fill=black] {};
    
    \begin{scope}[on background layer]
    \draw (10) edge (40);
    \draw (20) edge (50);
    \draw (30) edge (60); 
    \draw (11) edge (40);
    \draw (21) edge (50);
    \draw (31) edge (60);

    \draw (10) edge (41);
    \draw (20) edge (51);
    \draw (30) edge (61); 
    \draw (11) edge (41);
    \draw (21) edge (51);
    \draw (31) edge (61); 
    
    \draw (10) edge (42);
    \draw (20) edge (52);
    \draw (30) edge (62); 
    \draw (11) edge (42);
    \draw (21) edge (52);
    \draw (31) edge (62); 
    \end{scope}
\end{tikzpicture}
    \caption{Lifting with $M=3$: vertices and check nodes are split into three copies, with connections following the original topology.}
    \label{fig:proto2}
\end{subfigure}
\begin{subfigure}{\linewidth}
    \centering
   \begin{tikzpicture}
   
    \node (10) at (1.25,0) [draw, inner sep=0pt] {$+$};
    \node (20) at (1.75,0) [draw, inner sep=0pt] {$+$};
    \node (30) at (2.25,0) [draw, inner sep=0pt] {$+$};

    \node (11) at (3.75,0) [draw, inner sep=0pt] {$+$};
    \node (21) at (4.25,0) [draw, inner sep=0pt] {$+$};
    \node (31) at (4.75,0) [draw, inner sep=0pt] {$+$};

    \node (40) at (0,2) [circle,fill=black] {};
    \node (50) at (0.5,2) [circle,fill=black] {};
    \node (60) at (1,2) [circle,fill=black] {};

    \node (41) at (2.5,2) [circle,fill=black] {};
    \node (51) at (3,2) [circle,fill=black] {};
    \node (61) at (3.5,2) [circle,fill=black] {};
    
    \node (42) at (5,2) [circle,fill=black] {};
    \node (52) at (5.5,2) [circle,fill=black] {};
    \node (62) at (6,2) [circle,fill=black] {};
    
    \begin{scope}[on background layer]
    \draw (10) edge (40);
    \draw (20) edge (60);
    \draw (30) edge (50); 
    \draw (11) edge (60);
    \draw (21) edge (50);
    \draw (31) edge (40);

    \draw (10) edge (51);
    \draw (20) edge (41);
    \draw (30) edge (61); 
    \draw (11) edge (61);
    \draw (21) edge (51);
    \draw (31) edge (41); 
    
    \draw (10) edge (42);
    \draw (20) edge (62);
    \draw (30) edge (52); 
    \draw (11) edge (62);
    \draw (21) edge (42);
    \draw (31) edge (52); 
    \end{scope}
\end{tikzpicture}
    \caption{Tanner graph of lifted $(2,3)$-regular LDPC code. The mapping between check and variable nodes, within connections compatible with the protograph, is random.}
    \label{fig:proto3}
\end{subfigure}
\end{figure}
\end{example}

\subsubsection{SC-LDPC codes}

\textit{A spatially coupled} protograph is obtained by chaining together $L$ copies of LDPC protographs, where $L$ is called the \textit{coupling factor}. We assign time index $t$ to each of these copies. Suppose a check node $c_i$ and a variable node $v_j$ in a protograph are connected with $B_{i,j}$ edges (with $B_{i,j}>1$ possibly). Then we spread these edges forward, i.e., we connect $B_{i,j}$ edges from node $v_j$ at time $t$ to check nodes $c_j$ at time $t,t+1,\dots,t+w$, where $w$ is called \textit{coupling width}. For a regular protograph it is reasonable to take $w=d_v-1$ (typically, $d_v << L$). 
Figures \ref{fig:sc1} and \ref{fig:sc2} illustrate the process for a (3,6)-regular protograph. A SC-LDPC code is then obtained by applying the lifting procedure described earlier to a spatially-coupled protograph.

Let us now compute the rate of the obtained SC-LDPC code. We take $L$ copies of the original protograph and chain them together. The number of variable nodes is now fixed and therefore the code length is $Ln_v$. A coupling width $w$ means that variable nodes of the protograph at position $t$ are also connected to the check nodes of protographs at positions $t+1,\dots,t+w$ and therefore we need to add extra $wn_c$ check nodes in addition to $Ln_c$ that we already have. Therefore, the resulting SC-LDPC code has rate
\begin{equation}
    R=1-\frac{(L+w)n_c}{Ln_v}=1-\frac{(1+(d_v-1)/L)n_c}{n_v}.
\end{equation}

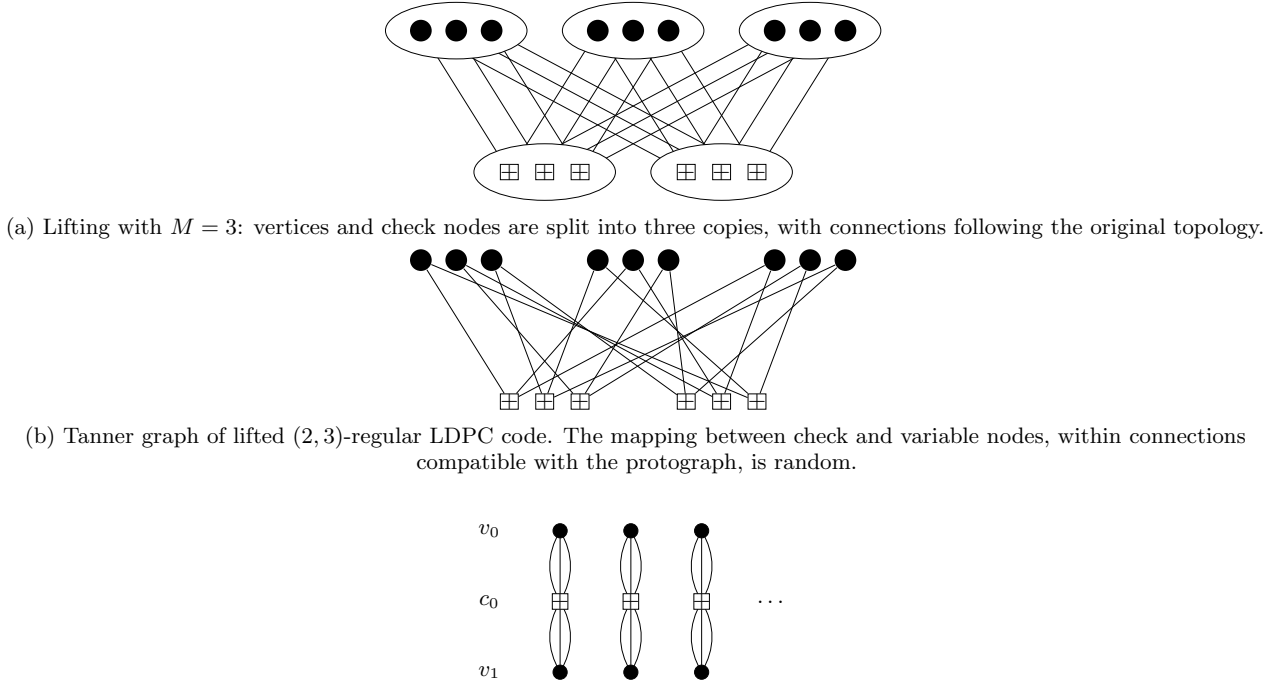
\begin{figure}
    \centering
\begin{tikzpicture}
    \draw (-1,0) node {$c_0$};
    \draw (-1,1) node {$v_0$};
    \draw (-1,-1) node {$v_1$};

    \coordinate (0) at (0,0);
    \draw[fill=white] (-0.11,-0.11) rectangle (0.11,0.11);
    \draw[fill=white] (0) node {$+$};
    \coordinate (1) at (0,1);
    \draw[fill=black] (1) circle (0.1);
    \coordinate (2) at (0,-1);
    \draw[fill=black] (2) circle (0.1);
    \begin{scope}[on background layer]
    \draw (0) edge (1);
    \draw (0) edge[bend left] (1);
    \draw (0) edge[bend right] (1);
    \draw (0) edge (2);
    \draw (0) edge[bend left] (2);
    \draw (0) edge[bend right] (2);
    \end{scope}
    
    \coordinate (0) at (1,0);
    \draw[fill=white] (0.89,-0.11) rectangle (1.11,0.11);
    \draw[fill=white] (0) node {$+$};
    \coordinate (1) at (1,1);
    \draw[fill=black] (1) circle (0.1);
    \coordinate (2) at (1,-1);
    \draw[fill=black] (2) circle (0.1);
    \begin{scope}[on background layer]
    \draw (0) edge (1);
    \draw (0) edge[bend left] (1);
    \draw (0) edge[bend right] (1);
    \draw (0) edge (2);
    \draw (0) edge[bend left] (2);
    \draw (0) edge[bend right] (2);
    \end{scope}
    
    \coordinate (0) at (2,0);
    \draw[fill=white] (1.89,-0.11) rectangle (2.11,0.11);
    \draw[fill=white] (0) node {$+$};
    \coordinate (1) at (2,1);
    \draw[fill=black] (1) circle (0.1);
    \coordinate (2) at (2,-1);
    \draw[fill=black] (2) circle (0.1);
    \begin{scope}[on background layer]
    \draw (0) edge (1);
    \draw (0) edge[bend left] (1);
    \draw (0) edge[bend right] (1);
    \draw (0) edge (2);
    \draw (0) edge[bend left] (2);
    \draw (0) edge[bend right] (2);
    \end{scope}
    
    \draw (3,0) node {\dots};

\end{tikzpicture}
    \caption{Uncoupled (3,6)-regular protographs. Each variable node has degree 3 and each check node has degree 6. Although there are duplicate edges in the protograph, there will be none in the code after the coupling and lifting steps.}
    \label{fig:sc1}
\end{figure}
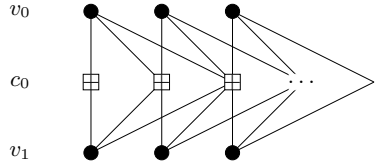
\begin{figure}
    \centering
\begin{tikzpicture}
    \draw (-1,0) node {$c_0$};
    \draw (-1,1) node {$v_0$};
    \draw (-1,-1) node {$v_1$};

    \coordinate (0) at (0,0);
    \draw[fill=white] (-0.11,-0.11) rectangle (0.11,0.11);
    \draw[fill=white] (0) node {$+$};
    \coordinate (1) at (0,1);
    \draw[fill=black] (1) circle (0.1);
    \coordinate (2) at (0,-1);
    \draw[fill=black] (2) circle (0.1);
    
    \coordinate (3) at (1,0);
    \draw[fill=white] (0.89,-0.11) rectangle (1.11,0.11);
    \draw[fill=white] (3) node {$+$};
    \coordinate (4) at (1,1);
    \draw[fill=black] (4) circle (0.1);
    \coordinate (5) at (1,-1);
    \draw[fill=black] (5) circle (0.1);
    
    \coordinate (6) at (2,0);
    \draw[fill=white] (1.89,-0.11) rectangle (2.11,0.11);
    \draw[fill=white] (6) node {$+$};
    \coordinate (7) at (2,1);
    \draw[fill=black] (7) circle (0.1);
    \coordinate (8) at (2,-1);
    \draw[fill=black] (8) circle (0.1);
    
    \coordinate (9) at (3,0);
    \coordinate (10) at (4,0);
    
    \draw[white,fill=white] (2.8,-0.11) rectangle (3.11,0.11);
    \draw (3,0) node {\dots};
    
    \begin{scope}[on background layer]
    \draw (1) edge (0);
    \draw (1) edge (3);
    \draw (1) edge (6);
    \draw (2) edge (0);
    \draw (2) edge (3);
    \draw (2) edge (6);
    \draw (4) edge (3);
    \draw (4) edge (6);
    \draw (4) edge (9);
    \draw (5) edge (3);
    \draw (5) edge (6);
    \draw (5) edge (9);
    \draw (7) edge (6);
    \draw (7) edge (9);
    \draw (7) edge (10);
    \draw (8) edge (6);
    \draw (8) edge (9);
    \draw (8) edge (10);
    \end{scope}
\end{tikzpicture}
    \caption{Spatially coupled (3,6)-regular LDPC ensemble, $w=2$}
    \label{fig:sc2}
\end{figure}

\subsubsection{Belief propagation decoding}
Assume that we have a parity check matrix $H$ of some LDPC code $C$. In our setting, Bob has the string $\bB=(B_1,\dots,B_n)$ and wants to reconstruct the string $\bA=(A_1,\dots,A_n)$ from the knowledge of $\bB$, the syndrome $\bM=(M_1,\dots,M_m)=H\bA$ and the joint distribution $P(A,B)$. Therefore, the goal of the decoder is to find
\begin{equation}
\tilde \bA=\arg \max_{\hat \bA:\ H\hat \bA=\bM} \prod_{i=1}^n P(A=\hat A_i,B=B_i|X_i,Y_i).
\end{equation}

The straightforward procedure is to check all possible codewords of $C$ and return the most probable. However, the number of codewords grows as $2^{k}$, where $k$ is the code dimension, which quickly makes such procedure unfeasible, hence the need for low-complexity algorithms.

Belief propagation is an iterative algorithm which makes use of the sparsity of the parity-check matrix $H$ and for constant $(d_v,d_c)$ has running time $O(N_\mathrm{it}\cdot n)$, where $N_\mathrm{it}$ is a number of decoding iterations performed. Let
\begin{equation}
    \bl=(l_1,\dots,l_n),\quad l_i=\ln\frac{P(A_i=0,B_i)}{P(A_i=1,B_i)}
\end{equation}
be the logarithmic reliability ratios vector corresponding to the vector $\bB$. The prior knowledge of which bits have error probability $\delta'$ and which have error probability $\delta''$ is incorporated in $l_i$ by appropriately setting the corresponding probabilities $P(A_i=0,B_i)$ and $P(A_i=1,B_i)$. The recovery of $\tilde \bA$ can be performed by an iterative exchange of messages between variable and check nodes of the Tanner graph of the code. The messages at iteration $i$ are defined as follows \cite{Richardson08, Chen06}:
\begin{align}\label{eq:bpmsg}
    \mu_{v\to c}^{(i)} = \begin{cases}
    l_v,\ i=0\\
    l_v+\sum_{c'\in C_v\setminus\{c\}}\mu_{c'\to v}^{(i)},\ i>0
    \end{cases}\\
    \mu_{c\to v}^{(i)}=2\tanh^{-1}\left((-1)^{M_c}\prod_{v'\in V_c\setminus\{v\}}\tanh(\mu_{v'\to c}^{(i-1)}/2)\right),
\end{align}
where the set $C_v$ contains indices of check nodes, incident to variable node $v$, and $V_c$ contains indices of variable nodes, incident to check node $c$. After $N_\mathrm{it}$ iterations we can obtain the string $\tilde \bA=(\tilde A_1,\dots,\tilde A_n)$ as follows:
\begin{equation}
\tilde A_v=\frac{1-x_v}{2}, x_v=\sgn(l_v+\sum_{c\in C_v}\mu_{c\to v}^{(N_{it})}),
\end{equation}
where $x_v$ are the hard-decision values obtained by converting the corresponding log-likelihood ratios into $\{-1,+1\}$.
For SC-LDPC codes with underlying $(d_v,d_c)$-regular LDPC codes the error probability is expected to decrease doubly exponentially with decoding iterations if $d_v\ge 3$ \cite{Mitchell15}. There exists various improvements of this algorithm in the literature for general LDPC codes \cite{Zhang05, Hocevar04} as well as for SC-LDPC codes \cite{Lentmaier11, Iyengar12}. However, in the literature there is no analysis of the behaviour of SC-LDPC codes under such decoders for non-erasure channels, hence the use of the standard BP algorithm, which is provably good, for our experiments. The similar lack of theoretical analysis exists with early termination schemes, which seek to find a criterion when the decoding process can be stopped.

\subsection{Simulations}
Although the existing theory predicts good asymptotic behaviour of SC-LDPC codes, their finite-length performance is well studied only for the case of the erasure channel. Furthermore, our setting with the transmission through two BSCs makes the analysis even more complicated. In the asymptotic setting with infinite block length, the smallest achievable overhead is
\begin{equation}\label{eq:asympt}
\tilde \eta^\infty = \gamma H(A'|B')+(1-\gamma)H(A''|B''),
\end{equation}
where $P(A',B')$ and $P(A'',B'')$ are the probability distributions from Eq.~\eqref{eq:modelP1} and Eq.~\eqref{eq:modelP2}, determined by $S$ and $Q$ respectively. For the finite-length setting, to get an estimate of the best-possible overhead, we can assume the case of independent decoding of BSC($\delta'$) and BSC($\delta''$) and use Eq.~\eqref{eq:syn_length} to get the estimated syndrome length
\begin{equation}
\tilde m(n) = m_\mathrm{BSC}(\gamma n,\delta')+m_\mathrm{BSC}((1-\gamma)n,\delta'').
\end{equation}
The estimated overhead is then naturally
\begin{equation}\label{eq:bestGuess}
\tilde \eta(n) = \frac{\tilde m(n)}{n}.
\end{equation}
For the mixture of two BSCs the threshold might be better but we don't know a priori where it is exactly. Also there might be some loss due to the length and overhead adaptation of the code construction. Therefore, we resort to numeric simulations to locate the threshold.

The simulation setting for the error correction part can be described as follows. Assume that the length $n$ and the error correction overhead $\eta=m/n$ are fixed. Then we choose the base protograph with the design rate close to $1-\eta$, which is then followed by the coupling and lifting process, where the coupling factor is $L=80$ (SC-LDPC codes are optimal for $L\to \infty$, but in practice we just pick a sufficiently big number). Note that the resulting code has slightly larger length $n'$ (since it is a multiple of the lifting factor) and overhead $\eta'$ (due to coupling). Hence, to finish the code construction we need to remove some variable and check nodes from the Tanner graph in a way that keeps the spatial coupling advantage (which comes from low-degree check nodes at the sides of the chain). The simple and rather efficient strategy can be described as follows:
\begin{enumerate}
    \item Remove $n'-n$ variable nodes with the smallest degrees (the node is less reliable if it has smaller degree)
    \item Pick two sets of $(\eta'-\eta)n$ checks with the largest degrees and perform pairwise merging, i.e.~replace every pair of nodes $c$, $c'$ with a new node $c''$ s.t. $\adj(c'')=\adj(c)\cup\adj(c')$.
\end{enumerate}

Before the decoding process, the string is shuffled so that noisier bits become more evenly distributed in the codeword, which helps the decoding process. Note that the shuffling pattern is known to the decoder so it can correctly assign the reliabilities.

\begin{figure}
    \centering
    \includegraphics[width=0.8\linewidth]{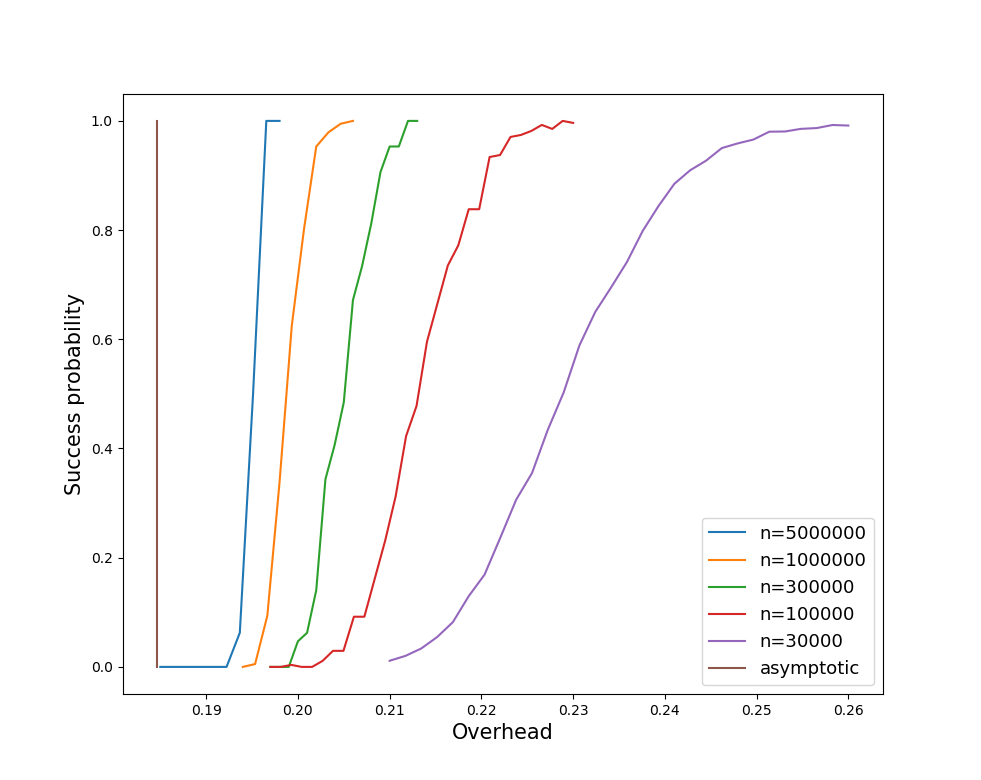}
    \caption{Performance of the error-correction scheme: probability of a successful error correction as a function of the ratio between the syndrome and block length for various block length. Here we have $\gamma=13/256, S=2.6507$ and $Q=0.0239$. The threshold for error-free correction approaches the asymptotic bound as the block length $n$ increases.}
    \label{fig:perf}
\end{figure}

Figure \ref{fig:perf} demonstrates how the probability of successful string recovery changes with coding overhead for different values of block length $n$. In this setting, $\gamma=13/256, S=2.6507, Q=0.0239$. It can be seen that for larger lengths the curves become smoother and sharper. Furthermore, for $n=5e6$ the corresponding curve is almost a vertical line and it follows that the string is always recovered for overheads larger than $\eta^*=0.196$, which is only 6\% larger than the asymptotic value $\tilde {\eta}^\infty\approx 0.1847$ given by Eq.~\eqref{eq:asympt} and 3.7\% larger than $\tilde {\eta}(5e6)\approx0.189$ given by  Eq.~\eqref{eq:bestGuess}. Note that if we only assume the global symmetric error $\hat \delta=\gamma\delta'+(1-\gamma)\delta''\approx0.032$, the corresponding asymptotic overhead is $h(\hat \delta)\approx0.204$. From  Fig.~\ref{fig:perf} it follows that the knowledge of exact bit reliabilities allows an error-free transmission with this overhead in the finite size regime for $n\ge 1e6$.

Let us now demonstrate the threshold performance of the considered code construction, i.e., how much ``extra'' overhead we need in addition to the asymptotic value for the string reconstruction for finite block length $n$ and how it decreases with $n$. Since we are in the finite length setting, we define the threshold as the smallest overhead $\eta$ s.t. the probability of successful recovery is larger than some fixed value $p^*$. Let us denote the gap function as
\begin{equation}
g(\eta) = \eta - \tilde\eta^\infty.
\end{equation}

\begin{figure}
    \centering
    \includegraphics[width=0.8\linewidth]{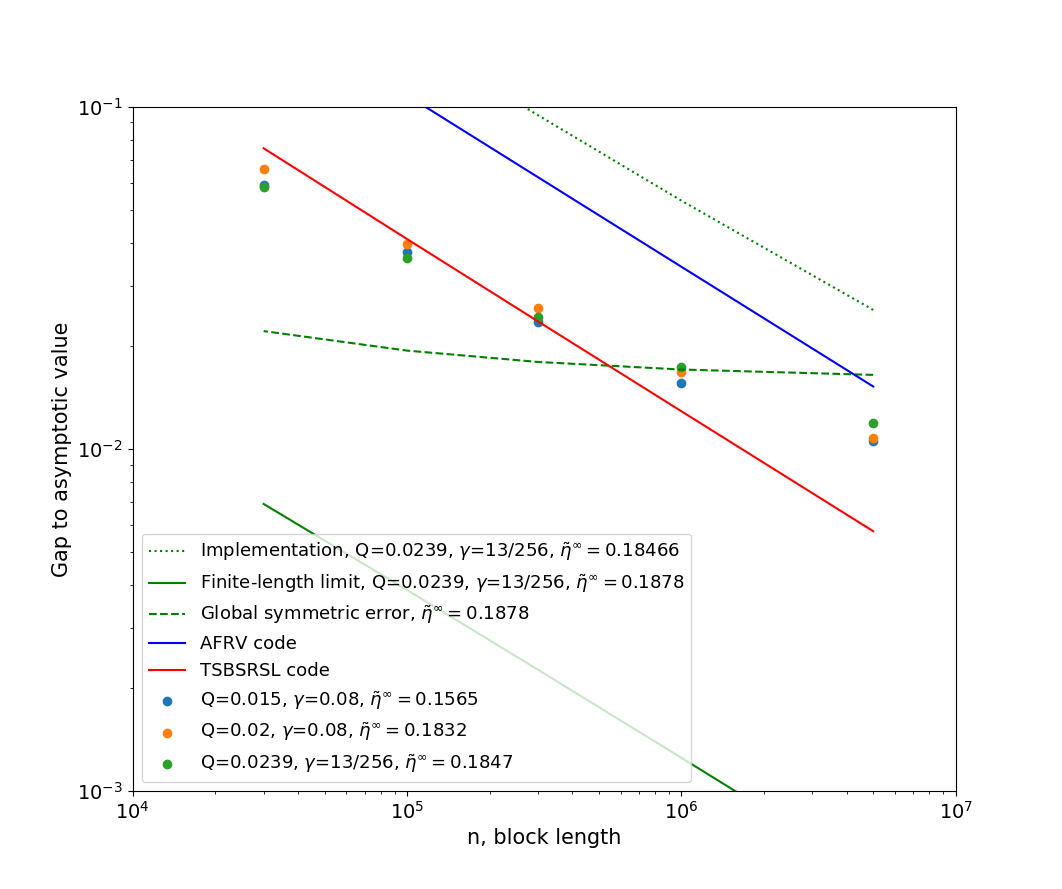}
    \caption{Convergence of the code performance: gap to the asymptotic limit as a function of the block length. The success probability here is fixed to be $p^*=0.9$. Blue, orange and green dots are the result of simulations. The green full line corresponds to the ideal convergence. The dashed green curve correspond to the best convergence possible when treating the channel as a single BSC. For comparison, the convergence of two error-correction schemes previously considered for DIQKD which do not admit a known efficient decoding algorithm are given by the red and blue curves. The choice of the syndrome length for our implementation of DIQKD is represented by the dotted green line. It is chosen above the simulation results curve on purpose to allow for a decoding success probability larger than 90\%.}
    \label{fig:gap9}
\end{figure}

The performance of our scheme for $p^*=0.9$ is demonstrated at Fig.~\ref{fig:gap9}. Two of the experimental plots correspond to $\gamma=0.08, S=2.6192$ and $Q\in\{0.015,0.02\}$ and the third to $\gamma=13/256$ and the probabilities defined by Tab.~\ref{tab:priors}, which roughly correspond to $S=2.6507, Q=0.0239$. All regimes demonstrate a similar convergence, which confirms the universality of the code construction. Curves corresponding to Eq.~\eqref{eq:bestGuess} and Eq.~\eqref{eq:syndromeLength} for the setting $\gamma=13/256, S=2.6507, Q=0.0239$, are also present. They essentially provide lower and upper bounds.

The curve corresponding to using only global symmetric error $\hat \delta$ is also provided as a reference in Fig.~\ref{fig:gap9}. For the sake of comparison with Eq.~\eqref{eq:bestGuess}, we take the same value of $\tilde\eta^\infty$. In this case, the convergence to the asymptotic value is faster since we don't have a small block anymore. However, the best-possible overhead is much higher than when we are allowed to distinguish different bit reliabilities, and in fact, bit reliabilities allow to lower the threshold for any finite size $n$.

In Fig.~\ref{fig:gap9}, we also compare our construction to some error-correction schemes previously considered for DIQKD which do not admit a known efficient decoding algorithm and are thus not suited for practical purpose. Namely, we consider the bound from~\cite{ArnonFriedman19,Murta19}:
\begin{equation}
m_{AFRV}=n\cdot \tilde\eta^\infty + \underset{0\leq\eps'\leq\eps}\min\ 4\log(2\sqrt{2}+1)\sqrt{2n\log\left(\frac{8}{\eps'^2}\right)}+\log\left(\frac{8}{\eps'^2}+\frac{2}{2-\eps'}\right)+\log\left(\frac{1}{\eps-\eps'}\right)
\end{equation}
and the one from~\cite{Tan20}:
\begin{equation}
m_{TSBSRSL}=n\cdot \tilde\eta^\infty + \underset{0\leq\eps'\leq\eps}\min\ 2\log(5)\sqrt{n\log\left(\frac{2}{\eps'^2}\right)} + 2\log\left(\frac{1}{\eps-\eps'}\right)+4.
\end{equation}
We notice that our construction achieves smaller overhead than the AFRV bound, and a comparable overhead compared to the TSBSRSL bound.

Finally, we note that the results of our simulations in Fig.~\ref{fig:gap9} yield a critical threshold that is systematically shifted from the bound given in Eq.~\eqref{eq:bestGuess}. This can be attributed to the fact that scaling depends more on the lifting factor $M$ than on the total block length $n$.

One straightforward benefit from using one big block of data compared to treating separately more and less noisy bits is that we only need to design one code instead of two. Another advantage becomes clear assuming the similar behaviour of the gap function for the case of single BSC (which is confirmed by experiments). Joint decoding gives us $g(\tilde\eta(n))$, which is smaller compared to $\gamma \cdot g(\tilde\eta(\gamma n))+(1-\gamma) \cdot g(\tilde\eta((1-\gamma)n))$ for the independent decoding.

\subsection{Practical implementation of the coding scheme}\label{sec:ECtuning}
Given the result of our numerical experiments, we choose the syndrome length as a function of the parameters $S$, $Q$ $\gamma$ and $n$ with the following formula:
\begin{equation}\label{eq:syndromeLength}
m = n((1-\gamma) h(Q) + \gamma h((4-S)/8)) + 50\sqrt{n}.
\end{equation}
This choice guarantees a proper error correction in all simulations (see Fig.~\ref{fig:gap9}). It also includes some margin to guarantee successful decoding with a probability larger than $90\%$, as well as in the case where the experimental statistics would differ from the expected ones (our construction would work seamlessly in such a case, but only provided that sufficient information is available in the syndrome).

Finally, we need to define the prior probabilities for the decoder. Given our knowledge of the setup ahead of the DIQKD experiment (see Sec.~\ref{sec:DIQKDPrimitives}), we choose decoding prior probabilities according to Tab.~\ref{tab:priors}.

\begin{table}
    \begin{equation*}
        P(A',B')=
        \begin{pmatrix}
            0.4210 & 0.0807 \\
            0.0847 & 0.4136
        \end{pmatrix}        
        \quad\quad
        P(A'',B'')=
        \begin{pmatrix}
            0.5017 & 0.0034 \\
            0.0110 & 0.4839 \\
        \end{pmatrix}
    \end{equation*}
    \caption{Channel statistics assumed for the test rounds (left) and key generating rounds (right) for error correction: these probabilities define the decoder priors. The matrices give the expected correlations between each bit in Alice's and Bob's strings $(A_i)_i$, $(B_i)_i$ (after flipping Bob's result for the $X_i = Y_i = 1$ case), with rows corresponding to Alice's outcomes and columns to Bob's.}
    \label{tab:priors}
\end{table}

In the case of the statistics given by Tab.~\ref{tab:priors} with $\gamma=13/256$ and $n=\num{1.5e6}$, \cref{eq:syndromeLength} gives an overhead of $\eta=m/n\simeq 0.229$, for which we expect a high success probability according to \cref{fig:perf}. This is confirmed by Monte-Carlo simulations: we observe 5 failures out of \num{15000} samples for this case when $m=\num{337500}$. Similarly, in the case $S=2.64$, $Q=1.8\%$, $\gamma=13/256$, $n=\num{1.5e6}$, which corresponds to the parameters reported in the main text, 
we observe 3 failure out of \num{15000} samples when $m=\num{296517}$. Therefore, in both cases the choice of syndrome length given by \cref{eq:syndromeLength} yields a success probability $p^*\geq 99.9\%$.

For the actual DIQKD demonstration, we run a BP decoder on Bob's node using an Intel Core i7-9700 CPU (\SI{3.00}{\giga\hertz}), which took \SI{250}{\second} to converge for $n = \num{1.5e6}$.

\section{Protocol feasibility and completeness}

For a DIQKD protocol to be practical, we would want it to accept with high probability when the setup behaves as expected. To achieve this, the protocol parameters should be chosen according to details of the experimental setup. We discuss this point here and how it leads to specific choices of parameters.

The condition that the expected behaviour is accepted with high probability is usually formalized as follows:

\begin{definition}
A protocol is said to be $\ecom$\textbf{-complete} if the honest implementation aborts with probability at most $\ecom$.
\end{definition}

Upon examination of our protocol, we find that the abort probability $\ecom$ of the expected behaviour can be made arbitrarily small by choosing appropriate protocol parameters and having enough rounds. More specifically, we do so by considering the three points where the protocol may abort, and ensuring that the expected behaviour is accepted with high probability at each step.

First, we need to consider the error-correction step, which accepts if the hashes $\bG_\mathrm{EC}$ and $\tilde\bG_\mathrm{EC}$ match.
These hashes always match in a honest implementation if the reconstructed string $\tilde{\textbf{A}}$ is equal to $\textbf{A}$ (see Sec.~\ref{sec:hashing}), so it suffices to ensure that this occurs with high probability. Following the analysis presented in Sec.~\ref{sec:EC}, we deduce that a safe boundary for our regime of parameters ensuring proper error correction is given by the syndrome length given in Eq.~\eqref{eq:syndromeLength}. Our simulations show that the contribution of the error correction to the error probability is then smaller than $\leq 0.1\%$.

Second, the authentication and activation steps accept if the authentication tags match. For the honest implementation, this is trivially satisfied (i.e.~as long as all messages are received without errors, which is ensured by the reliability of classical communication).

Finally, we need to ensure that the number of test rounds in which the CHSH game is lost (i.e.~cases for which $\tilde{A}_i \oplus B_i \neq X_i\cdot Y_i$) should pass the Bell-test verification: one must have $\sum_i \chi(U_i=0) \leq n\gamma (1-\omega_\thresh)$ with high probability. To ensure this, we consider the random variable $W_i=\chi(U_i=0)$ that corresponds to the $i^\text{th}$ round being a test round losing the CHSH game. For the honest implementation, each of these is an independent Bernoulli variable with parameter $q=P(W_i=1)=\gamma (1-\omega)$, where $\omega=\frac{4+S}{8}$ is the CHSH winning probability we expect to achieve. We can thus expect the condition to be verified with high probability by accepting whenever the fraction of rounds with $W_i =1$ is at most
\begin{equation}\label{qest}
q_\thresh = q + k \sqrt{\frac{q(1-q)}{n}},
\end{equation}
for some suitable $k$ (this describes how much ``allowance'' we give between the expected behaviour and the threshold the protocol accepts).
For definiteness, we choose here $k=3$, which essentially corresponds to allowing 3 standard deviations of tolerance, contributing to an error probability well below 1\%~\footnote{In more detail: since here we have $n$ independent trials of a Bernoulli variable with parameter $q$, the number of $W_i =1$ trials is binomially distributed with standard deviation $\sqrt{n{q(1-q)}}$. Hence the \emph{fraction} of such trials is a random variable with standard deviation $\sqrt{{q(1-q)}/{n}}$, and is approximately normally distributed for large $n$. This implies that choosing the (one-sided) accept threshold 3 standard deviations away from $q$ ensures that the accept probability is well over 99\%, as long as the devices perform as expected.}. Altogether, this guarantees that the protocol is complete with $\ecom \leq 0.01$.

In terms of the protocol parameter $\omega_\thresh$, Eq.~\eqref{qest} means taking
\begin{equation}
\label{west}
\omega_\thresh = 1-\frac{q_\thresh}{\gamma}.
\end{equation}
Eq.~\eqref{west} sets a relation between $\omega_\thresh$ and $\gamma$ which guarantees that the inequality~\eqref{eq:validation} of the validation step can be satisfied with high probability. We then choose the value of the remaining parameters, such as $\gamma$, in order to maximize the key rate.

\section{Basic definitions and notions of security}
This section presents the notions of distances, entropies and security on which this work relies.

\subsection{Trace distance}
To establish the security of the reported protocol, we need to compare the state obtained at the end of the protocol with a uniformly random state, i.e.~with the state expected in an ideal scenario in which Eve has no information about the final key. The relevant distance in this case is the trace distance~\cite{Portmann21}, which we discuss quickly in this subsection.

We recall first that the trace norm of an operator $\mathcal{O}$ is defined as
\begin{equation}
\lvert\lvert \mathcal{O} \rvert\rvert_1 = \tr{\sqrt{\mathcal{O}^\dag \mathcal{O}}}.
\end{equation}
In addition to being non-negative definite $\lvert\lvert\mathcal{O}\rvert\rvert_1 \geq 0,$ the trace norm satisfies homogeneity $\lvert\lvert c \mathcal{O} \rvert\rvert_1 = \lvert c \rvert \lvert\lvert \mathcal{O} \rvert\rvert_1$ for any complex number $c$ and the triangle inequality $\lvert\lvert \mathcal{O}+\mathcal{O}' \rvert\rvert_1 \leq \lvert\lvert \mathcal{O} \rvert\rvert_1 + \lvert\lvert \mathcal{O}' \rvert\rvert_1$.

Since we need to consider sub-normalized states, we define the corresponding notation. Let $\S_\leq(\mathcal{H})$ denote the set of `sub-normalized' quantum states on $\mathcal{H}$, i.e.~the set of all linear operators $\rho$ on $\mathcal{H}$ satisfying $\rho\geq 0$ and $\tr{\rho}\leq 1$. This set includes in particular the set of normalized quantum states, which we refer to as $S_=(\mathcal{H})$.

\begin{definition}
Let $\rho,\tau\in \S_{\leq}(\mathcal{H})$. The trace distance between $\rho$ and $\tau$ is given by (see~\cite{Tomamichel16}, section 3.2)
\begin{equation}
\label{deftracedistance}
D(\rho,\tau) = \frac{1}{2} \lvert\lvert \rho-\tau \rvert\rvert_1 +\frac{1}{2}\lvert \tr{\rho-\tau}\rvert.
\end{equation}
\end{definition}
Note that for normalized states, the trace distance reduces to the first term in the right hand side of Eq.~\eqref{deftracedistance}.

\subsection{Smooth min and max entropies}
The quantum conditional min and max entropies of sub-normalized states are defined below, following the definitions in~\cite{Tomamichel16}. 
\begin{definition}
Let $\rho_{AB}\in\S_\leq(\mathcal{H})$. The min-entropy of $A$ conditioned on $B$ of $\rho_{AB}$ is defined as
\begin{equation}\label{eq:Hmindef}
H_{\min}(A|B)_{\rho}=\underset{\sigma_B\in S_{\leq}(\mathcal{H}_B)}{\sup} \sup \left\{\lambda\in\mathbb{R} : \rho_{AB} \leq 2^{-\lambda} \openone_A \otimes\sigma_B\right\},
\end{equation}
where the operator inequality here is to be understood in semi-definite terms.
\end{definition}
\begin{definition}
Let $\rho_{AB}\in\S_\leq(\mathcal{H})$. The max-entropy of $A$ conditioned on $B$ of $\rho_{AB}$ is defined as
\begin{equation}\label{eq:Hmaxdef}
H_{\max}(A|B)_{\rho}=\underset{\sigma_B\in S_{\leq}(\mathcal{H}_B)}{\max} \log || \sqrt{\rho_{AB}}\sqrt{\openone_A\otimes\sigma_B} ||_1^2.
\end{equation}
\end{definition}
\noindent Without loss of generality, the optimization over $\sigma_B$ in either of the above definitions can be restricted to normalized states~\footnote{This follows straightforwardly from the fact that in either optimization, for any feasible $\sigma_B$, the normalized state $\hat{\sigma}_B = \sigma_B/\tr{\sigma_B}$ is another feasible point, and it yields an objective value at least as large as that for $\sigma_B$}.

In order to discuss min and max entropies in the neighborhood of a quantum state, we need to rely on a metric for quantum states. Following~\cite{Tomamichel09,Tomamichel16}, we do this with the help of the purified distance, whose definition we recall below. 

\begin{definition}
Let $\rho,\tau\in \S_\leq(\mathcal{H})$. The generalized fidelity between $\rho$ and $\tau$ is given by
\begin{equation}
F(\rho,\tau) = \underset{\bar{\mathcal{H}}\supseteq \mathcal{H}}{\sup}\\ \underset{\substack{\bar\rho, \bar\tau\in S_{=(\bar{\mathcal{H}})} \\ \rho = \Pi\bar\rho\Pi, \tau=\Pi\bar\tau\Pi}}{\sup} \lvert \lvert \sqrt{\bar \rho}\sqrt{\bar \sigma} \rvert \rvert_1,
\end{equation}
where the first supremum runs over all embeddings of $\mathcal{H}$ into larger spaces $\bar{\mathcal{H}}$, and $\Pi$ is the projection from $\bar{\mathcal{H}}$ to $\mathcal{H}$. 
\end{definition}
It can be shown that the generalized fidelity can be equivalently computed as
\begin{equation}
F(\rho,\tau) = \tr{\sqrt{\sqrt{\rho}\,\tau\,\sqrt{\rho}}} + \sqrt{(1-\tr{\rho})(1-\tr{\tau})}.
\end{equation}
\begin{definition}
Let $\rho,\tau\in \S_\leq(\mathcal{H})$. The purified distance between $\rho$ and $\tau$ is given by (see~\cite{Tomamichel16}, section 3.4)
\begin{equation}
\label{linkpurifiedgeneralized}
\mathcal{P}(\rho,\tau) = \sqrt{1-F(\rho,\tau)^2}.
\end{equation}
\end{definition}
Note that the purified distance is contractive with respect to trace non-increasing CP maps (see~\cite{Tomamichel09}, lemma 7), i.e.~for
$\rho,\tau\in \S_\leq(\mathcal{H})$ and $\mathcal{E}$ a trace non-increasing CP map, 
\begin{equation}
\label{contractiveP}
\mathcal{P}(\mathcal{E}(\rho),\mathcal{E}(\tau)) \leq \mathcal{P}(\rho,\tau).
\end{equation}
Note also the relationship between trace distance and purified distance (see~\cite{Tomamichel09}, lemma 6):
\begin{equation}
\label{DvsP}
D(\rho, \tau) \leq \mathcal{P}(\rho,\tau)
\end{equation}
where $\rho,\tau\in \S_\leq(\mathcal{H})$. Finally, The purified distance is a metric on the set of sub-normalized states (see~\cite{Tomamichel09}, lemma 5). This allows one to define balls in the neighborhood of a sub-normalized states.
\begin{definition}
Let $\rho\in \S_\leq(\mathcal{H})$ and $0\leq\eps<\sqrt{\tr{\rho}}$. The $\eps$-ball in $\mathcal{H}$ around $\rho$ is given by
\begin{equation}
\mathcal{B}^\eps(\rho)=\{\tau \in \S_\leq(\mathcal{H}): \mathcal{P}(\rho, \tau)\leq\eps\}.
\end{equation}
\end{definition}
Note that from Ineqs.~\eqref{contractiveP},~\eqref{DvsP} and the definition of the trace distance in Eq.~\eqref{deftracedistance}, the following inequalities hold for any state $\tau \in \S_\leq(\mathcal{H})$ in the $\eps$-ball around $\rho \in \S_\leq(\mathcal{H})$
\begin{align}
\label{ineqepstracenorm}
\eps \geq & \, \mathcal{P}(\rho,\tau)  
\geq \mathcal{P}(\mathcal{E}(\rho),\mathcal{E}(\tau)) 
\geq D(\mathcal{E}(\rho),\mathcal{E}(\tau)) \geq \frac{1}{2}||\mathcal{E}(\rho)-\mathcal{E}(\tau))||_1
\end{align}
where $\mathcal{E}$ is a trace non-increasing CP map.\

We can now define the smooth conditional min and max entropies respectively as the maximum min entropy and the minimum max entropy within the neighborhood of a state.
\begin{definition}
Let $\rho_{AB}\in\S_\leq(\mathcal{H})$ and $0\leq\eps<\sqrt{\tr{\rho_{AB}}}$. The $\eps$-smooth min- and max-entropy of $A$ conditioned on $B$ of $\rho_{AB}$ are given by
\begin{equation}\label{eq:smoothHmindef}
H_{\min}^\eps(A|B)_{\rho} = \underset{\tilde\rho_{AB}\in\mathcal{B}^\eps(\rho_{AB})}{\max} H_{\min}(A|B)_{\tilde \rho}.
\end{equation}
\begin{equation}\label{eq:smoothHmaxdef}
H_{\max}^\eps(A|B)_{\rho} = \underset{\tilde\rho_{AB}\in\mathcal{B}^\eps(\rho_{AB})}{\min} H_{\max}(A|B)_{\tilde \rho}.
\end{equation}
\end{definition}

\subsection{Security definitions}
In a DIQKD protocol, the two parties Alice and Bob either accept the data produced during the protocol, or abort the protocol. We denote the event in which Alice and Bob accept the result of the protocol by $\Omega$. In case the parties accept, which happens with some a priori unknown probability $P(\Omega)$, keys $\bKA$ and $\bKB$ are produced for Alice and Bob respectively.
Following~\cite{ArnonFriedman19}, we say that our protocol is $\ecorr${\bf -correct} if the keys $\bKA$ and $\bKB$ are produced and are different with probability at most $\ecorr$, i.e.
\begin{equation}\label{define:correct}
P(\bKA\neq\bKB \land \Omega) \leq \ecorr.
\end{equation}
We say that our protocol is $\esec${\bf -secret} if
\begin{equation}\label{define:secret}
P(\Omega) \frac12 || (\rho_{|\Omega})_{\bK_A, \overline E} - \mathbb{U}_{\bK_A}\otimes (\rho_{|\Omega})_{\overline E} ||_1\leq \esec,
\end{equation}
where $\rho$ is the state at the end of the protocol, and the conditioning in $\rho_{|\Omega}$ indicates that the state is normalized, conditioned on the protocol accepting. The subscripts $\bK_A$ and $\overline E$ indicate the subsystems which are not traced out.
$\mathbb{U}_{\bK_A}$ is the fully mixed state on register $\bK_A$ and $\overline E$ is Eve's full side information. Finally, we say that our protocol is $\esnd${\bf -sound} if it is both $\ecorr$-correct and $\esec$-secret with
\begin{equation}
\ecorr + \esec \leq \esnd
\end{equation}
for all possible implementations of the devices.

\bigskip

We establish the correctness of our key through the choice of an $\eps$-almost-universal hashing function, and its secrecy through Trevisan privacy amplification as described below. This together guarantees the soundness of the key.

\section{Classical cryptographic algorithms}

Several classical processing algorithms are necessary to run a DIQKD protocol. Sec.~\ref{sec:EC} describes our choice of error correction algorithm. Here, we define and describe the cryptographic algorithms which play a role in the security of our final key.

\subsection{Strongly universal hashing}\label{sec:hashing}
We consider families of functions $H$ with $h\in H$, $h:\{0,1\}^n \to \{0,1\}^l$ for $n,l\in\mathbb{N}$. Denote $\underset{h\in H}{P}$  the probability of an event when the function $h$ is chosen randomly from the family $H$.

\begin{definition}[Epsilon-almost-universal (AU) hashing~\cite{Stinson94}]
$H$ is an $\eh$-almost-universal family of hash functions if for all two strings $\bA\neq\tilde \bA$ we have the guarantee that 
\begin{equation}\label{eq:EAUhashing}
\underset{h\in H}{P}(h(\bA) = h({\tilde \bA})) \leq \eh.
\end{equation}
\end{definition}

\begin{definition}[Epsilon-almost-$\Delta$-universal (A$\Delta$U) hashing~\cite{Krawczyk94}]
$H$ is an $\eh$-almost-$\Delta$-universal family of hash functions if for all two strings $\bA\neq\tilde \bA$ and for all $t\in\{0,1\}^l$ we have the guarantee that 
\begin{equation}\label{eq:EDUhashing}
\underset{h\in H}{P}(h(\bA) \oplus h({\tilde \bA}) = t) \leq \eh.
\end{equation}
\end{definition}

\begin{definition}[Epsilon-almost-strongly-universal (ASU) hashing~\cite{Stinson94}]
$H$ is an $\eh$-almost-strongly-universal family of hash functions if for all two strings $\bA\neq\tilde \bA$ and all $t,\tilde t\in \{0,1\}^l$ we have the guarantee that 
\begin{equation}\label{eq:EASUhashing}
\underset{h\in H}{P}(h(\bA) = t, h({\tilde \bA}) = \tilde t) \leq \frac{\eh}{2^l}.
\end{equation}
\end{definition}

In this work, we use the VHASH hashing family $H_\mathrm{VHASH}$ with an output of size $l=64$ bits~\cite{Dai07}. For messages of length up to $2^{62} = 4,194,304$ TiB, this is a $\eh$-A$\Delta$U family of hash functions with a seed of length $\sim 1280$ and a collision probability $\eh=2^{-61}\simeq 4.3\cdot 10^{-19}$~\cite{Dai07}. This implies in particular that $H_\mathrm{VHASH}$ is a $\eh$-AU family of hash functions.

Following Wegman and Carter~\cite{Wegman81}, we protect the VHASH seed by systematically encrypting hashes with a fresh one-time-pad. This way, for every exchange of a hash between the parties, the information revealed to the adversary reduces to that which can be inferred in any case from knowning whether a message is accepted or not~\cite{Portmann14}. Effectively, this defines a new family of hash functions G:
\begin{equation}
g(\bA) = h(\bA) \oplus k_\mathrm{otp},
\end{equation}
where $h$ is chosen from $H_\mathrm{VHASH}$ and $k_\mathrm{otp}$ is chosen uniformly in $\{0,1\}^l$.

Since $k_\mathrm{otp}$ is chosen uniformly, it is clear that the tags created by $G$ are uniform:
\begin{equation}
\underset{g\in G}{P}(g(\bA)=t)=\frac{1}{2^l}.
\end{equation}
Therefore, $G$ is a $\eh$-ASU family of hash functions:
\begin{equation}
\begin{split}
\underset{g\in G}{P}(g(\bA)=t, g(\tilde \bA) = \tilde t) &=\underset{g\in G}{P}(g(\bA)=t, g(\bA)\oplus g(\tilde \bA) = t \oplus \tilde t)\\
&=\underset{g\in G}{P}(g(\bA)=t)\underset{g\in G}{P}(g(\bA)\oplus g(\tilde \bA) = t \oplus \tilde t|g(\bA)=t)\\
&\leq \frac{\eh}{2^l}
\end{split}
\end{equation}
for all two strings $\bA\neq\tilde \bA$ and all tags $t, \tilde t$. In particular, this also implies the following relation for $\bA\neq\tilde \bA$:
\begin{equation}\label{eq:conditionalHashing}
\underset{g\in G}{P}(g(\bA)=t | g(\tilde \bA) = \tilde t) \leq \eh.
\end{equation}

We use this family of hash functions at several stages of our protocol. In steps~5 and~7 of the protocol, we use it to ensure the correctness of the error correction, see Sec.~\ref{sec:correctness}. What is crucial here, is that VHASH is a AU family of hash functions. We then encrypt the hash with a one-time-pad to protect the VHASH seed, allowing us to reuse the same seed later. Effectively, this is done by simply sampling from the $G$ family directly. In step~10 of the protocol, we then use the same family $G$ to authenticate classical communication. This time, the crucial property is that $G$ is a ASU hashing family, see~\ref{sec:secrecy}. UVMAC, our implementation of the $G$ hashing family~\cite{UVMACcode}, is based on the reference implementation of the VHASH family~\cite{VMACcode}.

\subsection{Privacy amplification}\label{sec:PA}

The reduction of the eavesdropper's information about the key is obtained by mean of a quantum-proof strong $(k,\eps)$-extractor that we define now. An extractor Ext: $\{0,1\}^n\times\{0,1\}^s \to \{0,1\}^\ell$ is a function which takes a weak source of $n$ random bits and a uniformly random, short $s$-bit long seed, and produces some output made with $\ell$ bits, which is almost uniform. An extractor is said to be strong if the output is approximately independent of the seed. In our case, the randomness of the source is measured relative to a quantum side information. We denote the classical-quantum state (cq-state) describing the source and adversary's registers by $\rho_{CQ}$. In such a state, $C$ is a classical register and $Q$ a quantum register. The state of the $n$-bit classical register of the source alone is given by $\rho_C=\text{Tr}_{Q}({\rho_{CQ}})$ and the state of the quantum side information by $\rho_Q=\text{Tr}_{C}({\rho_{CQ}})$.

\bigskip

\begin{definition}[Quantum-proof strong extractor] A function $\mathrm{Ext}$ : $\{0,1\}^n\times\{0,1\}^s \to \{0,1\}^\ell$ is a quantum-proof $(k, \eps)$-strong extractor, if for all cq-states $\rho_{CQ}$ with $H_{\min}(C|Q)_\rho \geq k$, we have
\begin{equation}
\frac{1}{2}|| \mathcal{E}_\mathrm{Ext}(\rho_{CQ}\otimes\mathbb{U}_S) - \mathbb{U}_K \otimes \rho_Q \otimes \mathbb{U}_S ||_1 \leq \eps.
\end{equation}
\label{def_strongextractor}
\end{definition}
Here, $\mathcal{E}_\mathrm{Ext}$ is the map associated to Ext, $\mathbb{U}_K$ is the fully mixed state on register $K$, i.e.~for a system of dimension $2^\ell$, and $\mathbb{U}_S$ is a fully mixed state on register $S$, i.e.~for a system of dimension $2^s$.

\bigskip

We consider Trevisan's construction of randomness extractors. Trevisan extractors produce large outputs by concatenating the output of several small extractors. A weak design is used to define the seed of each individual extractor from an initial larger seed in such a way as to minimize mutual overlap between seeds. Thanks to this mechanism, Trevisan extractors only require a remarkably short initial seed. Moreover, when constructed with 1-bit extractors, the resulting Trevisan construction is quantum-proof~\cite{De12}. Let us formally introduce Trevisan's construction.

\begin{definition}[Weak design] A family of sets $\bar{S}_1,\hdots,\bar{S}_\ell \subset [s]$ is a weak $(\ell,t,r,s)$-design if 
\begin{itemize}
    \item For all $i$, $|\bar{S}_i|=t$
    \item For all $i$, $2^{|\bar{S}_j \cap \bar{S}_i|} \leq r\ell.$
\end{itemize}
\end{definition}
$\ell$ is thus the number of elements in the family of sets $\{\bar{S}_i\}_{i=1}^\ell$, $t$ is the number of elements in each set and $r$ bounds the overlap between the sets. Let us consider a long seed $\bS \in \{0,1\}^s$ and $ \bS_{\bar{S}_i}$ the string formed by the bits of $\bS$ at the position speficied by the elements of $\bar{S}_i$ -- a string of length $t$.

\bigskip

\begin{definition}[Trevisan extractor] Given a weak $(\ell,t,r,s)$-design $\{\bar{S}_i\}_{i=1}^\ell$ and a one-bit extractor $\mathrm{Ext}_1$: $\{0,1\}^n \times \{0,1\}^t \to \{0,1\}$, Trevisan's construction is defined as $\mathrm{Ext}(\bC,\bS)\coloneqq (\mathrm{Ext}_1(\bC,\bS_{\bar{S}_1}), \hdots, \mathrm{Ext}_1(\bC,\bS_{\bar{S}_\ell}))$.
\end{definition}

As an example, Trevisan's construction applied onto a quantum-proof $(k,\eps)$-strong one-bit extractor and a weak $(\ell,t,r,s)$-design gives a quantum proof $(k+r\ell,\ell\eps)$-strong extractor~\cite{Mauerer12}.

\bigskip

In our case, we use the block weak design construction proposed in~\cite{Mauerer12}, section III B 2, together with the polynomial hashing one-bit extractor presented in~\cite{Mauerer12}, section III C 3, which is based on a composition of two hash functions -- a Reed-Salomon code and a Hadamard code. The block weak design has $r=1$. This leads to a quantum-proof $\left(\ell + 6 + 4\log \frac{1}{\eps_1}, \ell \eps_1\right)$-strong extractor with a seed of length
\begin{equation}\label{eq:seedlength}
s = \max\left(2, \left\lceil \frac{ \log(\ell - e)- \log(t_+ - e)  }{\log(e)-\log(e-1)}\right\rceil+1\right) t_+^2
\end{equation}
with $t_+$ the first prime number larger or equal to
\begin{equation}
t=2\left\lceil\log n + 2 \log \frac{2}{\eps_1}\right\rceil.
\end{equation}
Here, $\eOne$ characterizes the distance to a random bit of the output of the one-bit extractor (it is the security parameter of the underlying Reed-Solomon-Hadamard code) and $\ell$ is the length of the extracted string. Here, we take into account the bound on the overlap of the block weak design given in~\cite{Ma11}, see also~\cite{Bierhorst18}. An open-source code implementation of this Trevisan extractor is available in~\cite{Mauerer12code}.

\begin{proposition}[Trevisan extraction with block weak design and polynomial hashing] \label{prop:trevisan}
Let $\rho_{CQ}\in \S_\leq(\mathbb{C}^{2^n}\otimes\mathcal{H})$ be a (possibly subnormalized) cq-state and let $S$ be an $s$-bit classical register. We write $\mathcal{E}_\mathrm{Trev}:\{0,1\}^n\times\{0,1\}^s \to \{0,1\}^\ell$ the Trevisan extractor with block weak design constructed on top of 1-bit polynomial hashing as defined in~\cite{Mauerer12}. Define the function
\begin{equation}
\Upsilon_b(x)=b\, W\left( \frac{e^{\frac{x}{b}}}{b}\right)
\end{equation}
where $W$ is the principal branch of the Lambert function and $b=\frac{4}{\ln{2}} \simeq 5.77$. 
Then for any $\es,\ePA>0$, if
\begin{equation}\label{eq:Trevlength}
1 \leq \ell \leq \Upsilon_b\left( H_\mathrm{min}^{\es}(C|Q)_{\rho} - 6 - 5\log\left(\frac{1}{\ePA}\right) \right),
\end{equation}
then implementing the extractor with $\eOne = {\ePA}/ \ell$ yields
\begin{equation}\label{eq:TrevNorm}
\frac{1}{2}|| \mathcal{E}_\mathrm{Trev}\left(\rho_{CQ} \otimes \mathbb{U}_S\right) - \mathbb{U}_K \otimes \rho_{Q} \otimes \mathbb{U}_S ||_1 \leq \ePA + 2\es.
\end{equation}
Moreover, the seed length is given by Eq.~\eqref{eq:seedlength}.
\end{proposition}

Here, $\mathcal{E}_\mathrm{Trev}\left(\rho_{CQ} \otimes \mathbb{U}_S\right)$ is the state in $\S_\leq(\mathbb{C}^{2^\ell}\otimes\mathcal{H}\otimes\mathbb{C}^{2^s})$ obtained after applying the Trevisan extractor onto the register $C$ of $\rho_{CQ}$ with a uniformly random seed $S$. Note that this map does not act on register $Q$, and it does not modify the register $S$: it only affects the $C$ register.

In essence, this proposition states that secrecy can be guaranteed simply by quantifying the conditional smooth min-entropy $H_\mathrm{min}^{\es}(C|Q)$ on the state left at the end of the protocol. Bounding this smooth min-entropy then becomes the main focus of the security analysis.

We emphasize that the extractor considered here is a quantum-proof strong randomness extractor. 
Furthermore, note that $\Upsilon_b(x)$ for $b>0$ can be equivalently defined as the inverse of the function $f(y) = y + b\ln(y)$ (i.e.~$\Upsilon_b(x)$ is the unique value $y$ satisfying $x = y + b\ln(y)$), and it is strictly increasing because $f$ is strictly increasing. 
This implies that for $x\geq 1$ we have
\begin{equation}
x - b\ln(x) \leq \Upsilon_b(x) \leq x,
\end{equation}
because for the value $y =\Upsilon_b(x)$, we have $y\geq1$ as well (since $\Upsilon_b$ is strictly increasing and $\Upsilon_b(1)=1$) and thus $x=y+b\ln(y) \geq y = \Upsilon_b(x)$. This in turn implies 
$b\ln(x)\geq b\ln(y)$ and thus $\Upsilon_b(x)=y 
= x - b\ln(y) \geq x - b\ln(x)$.
Hence, this extractor extracts the whole entropy of the string up to a logarithmic correction (and the function $\Upsilon_b(x)$ can be safely replaced by $x- b \ln(x)= x-4\log(x)$ in the key rate evaluation). Finally, we note that the seed requirement for this extractor only grows polylogarithmically with $n$, as shown by Eq.~\eqref{eq:seedlength}.

\begin{proof} 
Since the state $\rho_{CQ}$ has smooth min-entropy $H_\mathrm{min}^{\es}(C|Q)_\rho$, there exist a neighboring state $\overline \rho_{CQ}\in \mathcal{B}^{\es}(\rho_{CQ})$ at a purified distance $\mathcal{P}(\rho_{CQ},\overline\rho_{CQ})$ bounded by $\es$ with identical min-entropy: $H_\mathrm{min}(C|Q)_{\overline\rho}=H_\mathrm{min}^{\es}(C|Q)_\rho$. 
Using the triangle inequality twice, we have:
\begin{align}
\frac{1}{2}|| \mathcal{E}_\mathrm{Trev}(\rho_{CQ} \otimes \mathbb{U}_S) - \mathbb{U}_K \otimes \rho_{Q} \otimes \mathbb{U}_S ||_1 &\leq \frac{1}{2}||  \mathcal{E}_\mathrm{Trev}(\rho_{CQ} \otimes \mathbb{U}_S) -  \mathcal{E}_\mathrm{Trev}(\overline \rho_{CQ} \otimes \mathbb{U}_S) ||_1 \label{Ineq_extractror}\\
&\ \ \ \ + \frac{1}{2}|| \mathcal{E}_\mathrm{Trev}(\overline \rho_{CQ} \otimes \mathbb{U}_S) - \mathbb{U}_K \otimes \overline\rho_{Q} \otimes \mathbb{U}_S ||_1 \nonumber \\
&\ \ \ \ + \frac{1}{2}|| \mathbb{U}_K\otimes\overline\rho_{Q}\otimes \mathbb{U}_S - \mathbb{U}_K \otimes \rho_{Q}\otimes \mathbb{U}_S ||_1. \nonumber
\end{align}
Applying Ineq.~\eqref{ineqepstracenorm} to the first term in the right hand side of the above inequality by considering a map $\mathcal{E}$ taking states $\tau_{QC}$ and returning $\mathcal{E}_\mathrm{Trev}(\tau_{QC} \otimes \mathbb{U}_S),$ we conclude that this first term is bounded by $\es.$
Applying the same argument to the third term by considering a map $\mathcal{E}$ taking states $\tau_{QC}$ and returning $\mathbb{U}_K \otimes \tau_{Q}\otimes \mathbb{U}_S,$ we deduce that
\begin{align}
\frac{1}{2}|| \mathcal{E}_\mathrm{Trev}\left(\rho_{CQ} \otimes \mathbb{U}_S\right) - \mathbb{U}_K \otimes \rho_{Q} \otimes \mathbb{U}_S ||_1 &\leq
\frac{1}{2}|| \mathcal{E}_\mathrm{Trev}\left(\overline \rho_{CQ} \otimes \mathbb{U}_S\right) - \mathbb{U}_K \otimes \overline\rho_{Q}\otimes\mathbb{U}_S ||_1 + 2\es.
\end{align}

In order to bound the first term on the RHS of the above expression, we consider two cases~\footnote{In principle, we can pick any positive ``threshold'' value $\mu$ to split the two cases: this would give the final result that if $\ell \leq \Upsilon_b\left( H_\mathrm{min}^{\es}(C|Q)_{\rho} - 6 - 4\log({1}/{\ePA}) - \log({1}/{\mu}) \right)$, then implementing the extractor with $\eOne = {\ePA}/\ell$ yields a bound of $\max\left(\mu, \ePA\right) + 2\es$ in Eq.~\eqref{eq:TrevNorm}. However, observe that for a fixed value of $\max\left(\mu, \ePA\right)$, the optimal key length given by this expression is always obtained by setting $\mu$ and $\ePA$ to be equal, hence the choice in this proof.}: $\tr{\overline{\rho}_{CQ}} \leq \ePA$ and $\tr{\overline{\rho}_{CQ}} > \ePA$. For the first case, that term is immediately upper bounded by $\ePA$, and we are done. For the second case, note that since $\Upsilon_b^{-1}(\ell) = \ell + b\ln(\ell)$ is a strictly increasing function, the upper inequality in condition~\eqref{eq:Trevlength} is equivalent to
\begin{equation}
\ell + 4\log(\ell) \leq H_\mathrm{min}^{\es}(C|Q)_\rho - 6 - 5\log\frac{1}{\ePA}.
\end{equation}
Let $\hat{\rho}_{CQ}$ be the normalized version of $\overline{\rho}_{CQ}$, i.e.~so we have $\overline{\rho}_{CQ} = \tr{\overline{\rho}_{CQ}}\hat{\rho}_{CQ}$.
Since $H_\mathrm{min}(C|Q)_{\hat{\rho}} = H_\mathrm{min}(C|Q)_{\overline{\rho}} - \log (1/\tr{\overline{\rho}_{CQ}})$ (this follows straightforwardly from the definition of the min-entropy; see Eq.~\eqref{eq:Hmindef}) and keeping in mind that $H_\mathrm{min}(C|Q)_{\overline\rho} = H_\mathrm{min}^{\es}(C|Q)_\rho$, we have
\begin{align}
H_\mathrm{min}(C|Q)_{\hat{\rho}} &= H_\mathrm{min}^{\es}(C|Q)_\rho - \log \frac{1}{\tr{\overline{\rho}_{CQ}}} \nonumber\\
&\geq \ell + 4\log(\ell) + 6 + 5\log\frac{1}{\ePA} - \log \frac{1}{\tr{\overline{\rho}_{CQ}}}
\nonumber\\
&\geq 4\log\frac{\ell}{\ePA} + 6 + \ell, \quad \text{ since } \tr{\overline{\rho}_{CQ}} > \ePA.
\label{eq:TrevHmin}
\end{align}
From~\cite{Mauerer12}, we know that the Trevisan extractor $\mathcal{E}_\mathrm{Trev}$ with block weak design and polynomial hashing is a quantum-proof $(4\log\frac{1}{\eOne}+6+\ell, \ell\eOne)$-strong extractor in the sense of Definition \ref{def_strongextractor}.
Hence the bound~\eqref{eq:TrevHmin} implies that implementing the extractor with $\eOne = {\ePA}/\ell$ gives
\begin{equation}
\frac{1}{2}|| \mathcal{E}_\mathrm{Trev}(\hat{\rho}_{CQ}\otimes\mathbb{U}_S) - \mathbb{U}_K \otimes \hat{\rho}_Q \otimes \mathbb{U}_S ||_1 \leq \ell\eOne = \ePA ,
\end{equation}
which yields the desired bound (recalling that $\overline{\rho} = \tr{\overline{\rho}}\hat{\rho}$ and $\tr{\overline{\rho}}\leq 1$).
\end{proof}

\section{Security statement}

Here, we give an expression for the secure key length produced by our DIQKD protocol. For this, we define a family of linear lower bounds
\begin{equation}\label{eq:defineg}
g_t(\omega) = \eta(t) + (\omega-t) \eta'(t)
\end{equation}
on the CHSH-based entropy bound~\cite{Pironio09}
\begin{equation}\label{eq:entropyCHSH}
\eta(\omega) = \begin{cases}
0 & \textup{ if } \frac14\leq \omega \leq \frac34\\
1-h\left(\frac{1 + \sqrt{16\omega(\omega-1)+3}}2\right) & \textup{ elseif } \omega\in \tilde{Q}\\
\textup{undefined} & \textup{ else,}
\end{cases}
\end{equation}
where $\tilde{Q} = [\omega_\mathrm{min}, \omega_\mathrm{max}]$, and $\omega_\mathrm{min}=\frac{1-1/\sqrt{2}}2$, $\omega_\mathrm{max}=\frac{1+1/\sqrt{2}}2$ are the minimum and maximum quantum winning probabilities for the CHSH game.

We also define the functions $f_t$ on ternary probability distributions $p:\{0,1,\perp\}\to\mathbb{R}$ as
\begin{equation}\label{eq:definef}
f_t(p) = \sum_{u=0,1,\perp} p(u) f_t(\delta_u),
\end{equation}
i.e.~as the affine extension of
\begin{equation}\label{eq:definef2}
f_t(\delta_u)=\begin{cases}
\frac{1}\gamma g_t(0) + \left(1 - \frac{1}\gamma\right) g_t(1)& \textup{ if } u=0\\
g_t(1) & \textup{ if } u=1 \textup { or } u = \perp,
\end{cases}
\end{equation}
where $\delta_u$ are the three delta distributions over $\{0,1,\perp\}$. It is convenient to denote $p$ by the triple $p=(p(0),p(1),p(\perp))$. Then, $\delta_u = (\chi(u=0),\chi(u=1),\chi(u=\perp))$.

Recall that the variance of $f_t$ over a distributions $p$ is given by
\begin{equation}\label{eq:defineVar}
\mathrm{Var}_p(f_t) = \sum_{u=0,1,\perp} p(u) f_t(\delta_u)^2 - f_t(p)^2.
\end{equation}

Finally, we define:
\begin{align}
\vartheta_\eps &= \log\frac{1}{1-\sqrt{1-\eps^2}} \leq \log\frac{2}{\eps^2}\\
q(\omega) &= (\gamma(1-\omega), \gamma \omega, 1-\gamma)\\
\Delta(f_t, \omega ) &= \eta(\omega) - f_t(q(\omega))\\
V(f_t,p) &= \frac{\ln{2}}{2}\left(\log{33} + \sqrt{2 + \mathrm{Var}_p(f_t)}\right)^2\\
K_{\alpha'}(f_t) &= \frac{1}{6(2-\alpha')^3 \ln 2} 2^{(\alpha'-1)(2 + g_t(1) - g_t(\omega_\mathrm{min}))} \ln^3\left( 2^{2 +g_t(1) - g_t(\omega_\mathrm{min})} + e^2\right).
\end{align}

We can now formulate the secure key length of our protocol.

\begin{proposition}[Secure key length]\label{prop:keyrate}
Define the following parameters
\begin{equation}
\begin{split}
n \in \mathbb{N} &  \quad \textup{the number of measurement rounds}\\
\gamma \in (0,1) & \quad \textup{the testing probability}\\
\omega_\thresh & \quad \textup{the threshold CHSH winning probability}\\
m & \quad \textup{the length of the error correction syndrome} \\
t\in\left(\frac{3}{4}, \frac{1+1\sqrt{2}}{2}\right] & \quad \textup{a CHSH winning probability}\\
\eh > 0 & \quad \textup{the hashing collision probability}\\
\ePA > 0 & \quad \textup{the privacy amplification parameter}\\
\eEA > 0 & \quad \textup{the entropy accumulation parameter}\\
\alpha' \in (1,2) & \quad\textup{a Renyi parameter}\\
\alpha'' \in \left(1,1+\frac{1}{\log5}\right) & \quad \textup{another Renyi parameter}\\
\end{split}
\end{equation}
Define also three smoothing parameters $\es, \es', \es'' > 0$ satisfying the condition
\begin{eqnarray}
\es' + 2\es'' < \es,
\end{eqnarray}
and let $b=\frac{4}{\ln{2}}$. A key $\bK_A$ of length
\begin{equation}
\begin{split}
\ell = \Upsilon_{b}\Bigg[&
n\, g_t\left(\omega_\thresh\right) + n \inf_{\omega\in \tilde{Q}} \Delta\left(f_t, \omega\right) - (\alpha'-1)V\left(f_t,q(\omega)\right) \\
& - n(\alpha'-1)^2 K_{\alpha'}(f_t) - n\, \gamma - n(\alpha''-1) \log^2(5)
\\
&  
- \frac{1}{\alpha'-1}\left(\vartheta_{\es'} + \alpha' \log\left(\frac{1}{\eEA}\right)\right)
- \frac{1}{\alpha''-1}\left(\vartheta_{\es''} + \alpha'' \log\left(\frac{1}{\eEA}\right)\right)\\
& - 3 \vartheta_{\es-\es'-2\es''} 
- 5\log\left(\frac{1}{\ePA}\right)  - m - 264
\Bigg]
\end{split}
\end{equation}
produced by the protocol described in Sec.~\ref{sec:protocol} is $(\max(\eEA, \ePA + 2\es) + 4 \eh)$-sound.
\end{proposition}

\begin{figure*}
    \centering
    \includegraphics[width=\textwidth]{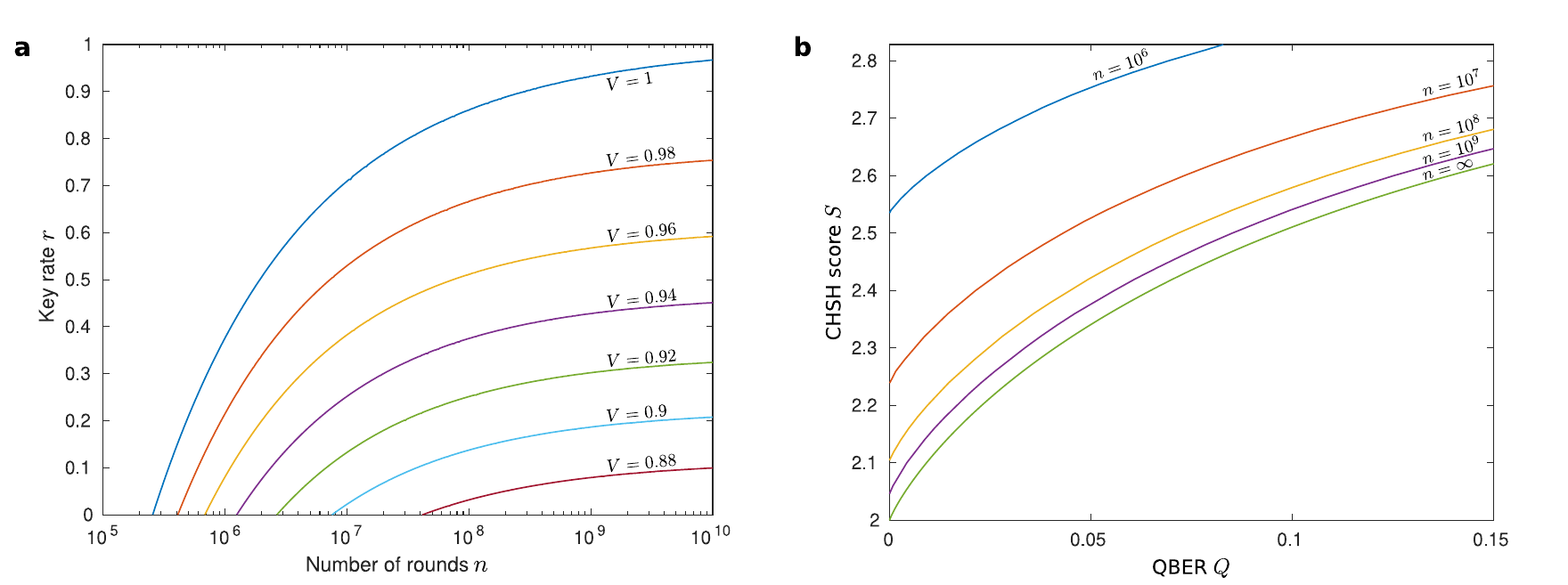}
    \caption{\textbf{Secure key for various physical models.}  Here, the soundness is fixed to $\esnd=10^{-10}$, the completeness error is estimated to be $\ecom \leq 0.01$, and all free parameters are numerically optimized over. \textbf{a,}~Key rate $r=\ell/n$ in the depolarizing model as a function of the visibility $V$, i.e.~with $S=2\sqrt{2}V$ and $Q=(1-V)/2$. A positive key rate is found starting from $n\simeq \num{2.6e5}$ rounds. \textbf{b,}~Minimal number of round $n$ needed to obtain a positive key length as a function of $S$ and $Q$. All points lying above a line present a positive key length for the corresponding number of rounds $n$.}
    \label{fig:noise_SQ}
\end{figure*}

\begin{figure*}
    \centering
    \includegraphics[width=\textwidth]{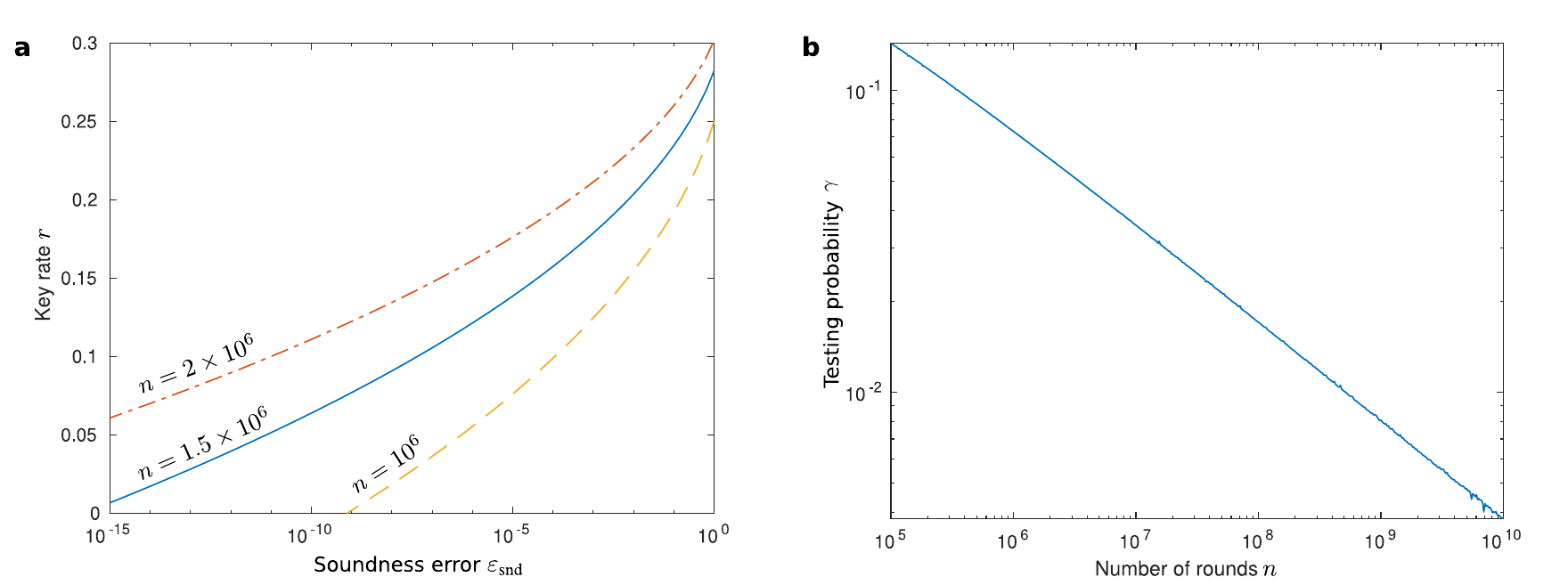}
    \caption{\textbf{Parameter dependence of the key length.} Here we focus on the case presented in the main text with $S=2.64$, $Q=0.018$ and $\ecom \leq 0.01$. \textbf{a,}~Dependency of the key rate on the soundness parameter when $\gamma=13/256$. \textbf{b,}~Optimal testing probability $\gamma$ as a function of the number of rounds for $\esnd=10^{-10}$.}
    \label{fig:parametersDependency}
\end{figure*}

The parameters $\ell, n, \gamma, \omega_\thresh$ and $m$ are explicit parameters in the DIQKD protocol, while the parameters $t, \eh, \ePA, \eEA, \alpha', \alpha'', \es, \es'$ and $ \es''$ are more abstract quantities that appear in the derivation of the theorem. For a given choice of $n$, $\gamma$, $m$ and $\omega_\thresh$, the remaining parameters can be optimized in order to maximize the final key length $\ell$ --- in general, we would simply perform this optimization heuristically. In our case, we express $m$ and $\omega_\thresh$ through \Cref{eq:syndromeLength,west,qest}, fix $n$ and $\gamma$, and optimize all other parameters, except $\eh=2^{-61}$ which is fixed by our use of the VHASH algorithm. Proposition~\ref{prop:keyrate} is proven in the next Section. General properties of the key length certified by Proposition~\ref{prop:keyrate} are illustrated in Fig~\ref{fig:noise_SQ} and~\ref{fig:parametersDependency}.

\section{Security analysis -- Proof of Proposition~\ref{prop:keyrate}}\label{sec:proof}

In order to assess the security of our DIQKD protocol, we consider a modified protocol including additional quantities and which slightly rearranges some of the steps. Importantly, while the additional quantities that are defined in this new protocol are not present in the original protocol, all variables of the original protocol are still defined in this new protocol and their statistics are strictly identical (e.g.~they are not affected by the step rearrangement). Hence, security in this new protocol implies security in the original protocol. We thus focus from now on this modified protocol, whose security is easier to analyze thanks to its additional variables. Our analysis borrows several steps from earlier security proofs and in particular~\cite{Tan20}. To our knowledge, this is the first proof of DIQKD directly applicable to a Trevisan extractor.

\subsection{Modified protocol}
The starting point of the modified protocol is similar to that of the DIQKD protocol in that the two parties share a secret key $\bK_0$. The variables $\bX,\bY$ and $\bT$ are however sampled during the protocol. Also, all the quantum systems are distributed in one shot:

\begin{enumerate}
\item[1.] [\textit{State preparation}] The parties receive their shares of $n$ entangled quantum states, which can be seen as one initial global bipartite state $\rho^0$.
\end{enumerate}

The parties then proceed with the following steps repeated $n$ times; each round is indexed by $i=1,\ldots,n$.
\begin{enumerate}
\item[] Repeat $n$ times:
\begin{enumerate}
\item[2.] [\textit{Sampling}] Alice chooses $X_i$ uniformly, Bob sample $Y_i$ from~\eqref{distribuion_Aliceinputs} and computes $T_i=\chi(Y_i\neq 2)$. Bob sends the value of $T_i$ to Alice. If $T_i=0$, Alice overrides $X_i$: she sets $X_i=0$.
\item[3A] [\textit{Measurement of Alice}] Alice measures her quantum system in the basis defined by $X_i$, and stores the result in the variable $A_i$.
\item[3B] [\textit{Measurement of Bob}] Bob measures his quantum system in the basis defined by $Y_i$, and stores the result in the variable $B_i$.
\item[4.] [\textit{Basis revelation}] Alice announces her measurement settings $X_i$.
\item[$4^{+}$.] [\textit{Erasure}] Bob sets $B'_i = \begin{cases}
B_i & \textup{if } T_i=1\\
0 & \textup{if } T_i=0.\\
\end{cases}$
\item[$4^{++}$.] [\textit{Virtual parameter estimation}] Define the score
\begin{equation}
U'_i = \begin{cases}
\chi(A_i + B'_i = X_i Y_i) & \textup{if } T_i=1\\
\perp & \textup{if } T_i=0.\\
\end{cases}
\end{equation}
\end{enumerate}
\end{enumerate}

After these steps are performed, the parties proceed with the following classical post-processing steps.
\begin{enumerate}
\item[5.] [\textit{Error correction}] Alice computes a $m$-bit long syndrome $\bM\in\{0,1\}^m$ for her string of outcomes $\bA=(A_1,\ldots,A_n)$ and sends it to Bob. Bob reconstructs a guess $\bAt$ of $\bA$ from $\bX, \bY, \bB$ and $\bM$.
\item[6.] [\textit{Parameter estimation}] Bob computes the score
\begin{equation}
U_i = \begin{cases}
\chi(\tilde{A}_i \oplus B'_i = X_i \cdot Y_i) & \textup{if } T_i=1\\
\perp & \textup{if } T_i=0\\
\end{cases}
\end{equation}
for each round $i=1,\ldots,n$.
\item[7-12.] Steps 7 to 12 are unchanged compared to the initial protocol.
\end{enumerate}

In the modified protocol, variables $\bX, \bY, \bT$ and the basis revelation are created/performed round by round instead of all at once in the DIQKD protocol. On the other hand, the quantum state is distributed at once in the beginning of the protocol, instead of being generated round by round in the DIQKD protocol. The modified protocol also defines two new registers: $\bB'$ and $\bU'$. The parameter estimation is then done with $\bB'$ instead of $\bB$. Note however that this does not change the estimation statistics since $\bB'$ only differs from $\bB$ in the key-generating rounds. In the same way, one directly checks that none of these modifications affect the statistics that can be observed in the original protocol, nor the underlying states.

We write $\rho$ the state on registers $\bA \tilde\bA \bB \bB' \bX \bY \ooM \bU \bU' E$ at the end of the modified protocol. Here, $\oM=(\bM,\bG_\mathrm{EC})$, and $\ooM=(\oM,\bG_B,C,\bG_A,F,\bG_F)$ is all the classical information revealed during the protocol in addition to $\bX$ and $\bT$, see Fig.~\ref{fig:classicalComm}. The reduced state on $\bA \tilde \bA \bB \bB' \bX \bY \oM \bU E$ after step 6 of the modified protocol is identical to the state at the end of step 6 of the original protocol. Since the Validation and Privacy amplification steps happen after step 6 and they only depend on these registers, the security of the original protocol can be assessed by analysing the modified protocol only up to this step, which is what we do now.

\subsection{Events}

It is useful to define the following events on $\rho$:
\begin{equation}
\begin{split}
\Omega_g: & \quad \bA=\tilde \bA \textup{, i.e.~Bob reconstructed Alice's outcomes correctly}\\
\Omega_h: & \quad \bG_\mathrm{EC}=\tilde\bG_\mathrm{EC} \textup{, i.e.~the error correction hashes match}\\
\Omega_\mathrm{PE'}: & \quad \sum_i \chi(U'_i=0) \leq n\gamma(1-\omega_\thresh) \textup{, i.e.~the test on the virtual parameters passes}\\
\Omega_\mathrm{PE}: & \quad \sum_i \chi(U_i=0) \leq n\gamma(1-\omega_\thresh) \textup{, i.e.~the test passes}\\
\Omega_c: & \quad \textup{All communication is received without modification}\\
\Omega_a: & \quad \textup{The authentication tags $\bG_A$ $\bG_B$ and $\bG_F$ match}\\
\Omega:& \quad \Omega_a \land \Omega_h \land \Omega_\mathrm{PE} \textup{, i.e.~the protocol does not abort}\\
\hat \Omega: & \quad \Omega_c \land \Omega_a \land \Omega_g \land \Omega_h\\
\overline \Omega: & \quad \hat \Omega \land \Omega_\mathrm{PE'}\\
\end{split}
\end{equation}

We write $p(\Omega)$ the probability associated with an event $\Omega$ on the state $\rho$. $\rho_{\land\Omega}$ is the (typically subnormalized) branch of $\rho$ on which $\Omega$ is true. $\rho_{|\Omega}=\frac{\rho_{\land \Omega}}{p(\Omega)}$ is the state of the setup conditioned on the event $\Omega$ being true.

\subsection{Correctness}\label{sec:correctness}
\label{proof_correctness}
Since the keys $\bK_A$ and $\bK_B$ are obtained with identical deterministic operations applied to $\bA$ and $\tilde \bA$ respectively,
\begin{equation}
P(\bK_A \neq \bK_B \land \Omega) \leq P(\bA \neq \tilde \bA \land \Omega)
\end{equation}
and it suffices to bound the probability that $\bA$ and $\tilde \bA$ differ when the protocol does not abort. Moreover, since hashes matching is necessary for the protocol to succeed, we can further bound
\begin{equation}
\begin{split}
P(\bK_A \neq \bK_B \land \Omega) & \leq P(\bA \neq \tilde \bA \land \bG_\mathrm{EC} = \tilde\bG_\mathrm{EC})\\
 & \leq P(\bG_\mathrm{EC} = \tilde\bG_\mathrm{EC} | \bA \neq \tilde \bA)\\
 & = P(\bH_\mathrm{EC} = \tilde\bH_\mathrm{EC} | \bA \neq \tilde \bA)\\
& \leq \eh,
\end{split}
\end{equation}
where we use the fact that $\bG_\mathrm{EC}=\bH_\mathrm{EC}\oplus k_\mathrm{otp}$ for some $k_\mathrm{otp}$ and unencrypted hash $\bH_\mathrm{EC}$ (similarly for $\tilde\bG_\mathrm{EC}$ with the same $k_\mathrm{otp}$) and Eq.~\eqref{eq:EAUhashing}. Hence, comparing to Eq.~\eqref{define:correct}, we have correctness with $\ecorr=\eh$.

\subsection{Secrecy}\label{sec:secrecy}
Following Eq.~\eqref{define:secret}, our protocol is $\esec$-secret if
\begin{equation}
P(\Omega)\frac{1}{2} || \mathcal{E}_\mathrm{Trev}(\rho_{|\Omega})_{\bK_A \overline E} - \mathbb{U}_{\bK_A} \otimes (\rho_{|\Omega})_{\overline E} ||_1 = \frac{1}{2} || \mathcal{E}_\mathrm{Trev}(\rho_{\land \Omega})_{\bK_A \overline E} - \mathbb{U}_{\bK_A} \otimes (\rho_{\land\Omega})_{\overline E} ||_1 \leq \esec,
\end{equation}
where Eve's full side information consists of $\oE = (\bX,\bT,\ooM,E)$. By noticing that $\rho_{\land \Omega} = \rho_{\land \Omega \land \Omega_c} + \rho_{\land \Omega \land \Omega_c^c}$, where we denote by $\Omega^c$ the complement of event $\Omega$, and that $\mathcal{E}_\mathrm{Trev}$ is a linear operator, we can bound the LHS using the triangle inequality
\begin{equation}
\begin{split}
 \frac{1}{2}  || & \mathcal{E}_\mathrm{Trev}(\rho_{\land \Omega})_{\bK_A \overline E} - \mathbb{U}_{\bK_A} \otimes \left(\rho_{\land \Omega}\right)_{\overline E} ||_1\\
\leq & \frac{1}{2} || \mathcal{E}_\mathrm{Trev}(\rho_{\land \Omega \land \Omega_c})_{\bK_A \overline E} - \mathbb{U}_{\bK_A} \otimes \left(\rho_{\land \Omega \land \Omega_c}\right)_{\overline E} ||_1 + \frac{1}{2} || \mathcal{E}_\mathrm{Trev}(\rho_{\land \Omega \land \Omega_c^c})_{\bK_A \overline E} - \mathbb{U}_{\bK_A} \otimes \left(\rho_{\land \Omega \land \Omega_c^c}\right)_{\overline E} ||_1\\
\leq & \frac{1}{2} || \mathcal{E}_\mathrm{Trev}(\rho_{\land \Omega \land \Omega_c})_{\bK_A \overline E} - \mathbb{U}_{\bK_A} \otimes \left(\rho_{\land \Omega \land \Omega_c}\right)_{\overline E} ||_1 + 3 \eh.
\end{split}
\end{equation}
In the last line, we use the fact that the trace distance between normalized states is bounded by one and hence the trace distances between subnormalized states (with the same norm) is bounded by their norm. This allows us to bound the right-most trace distance by the norm of the state $\rho_{\land\Omega\land\Omega_c^c}$, i.e.~by the probability 
\begin{equation}
\begin{split}
P(\Omega \land \Omega_c^c) &= P(\Omega_a \land \Omega_h \land \Omega_\mathrm{PE} \land \Omega_c^c)\\
& \leq P(\Omega_a \land \Omega_c^c)\\
& = P\left(\Omega_a^{(1)} \land \Omega_a^{(2)} \land \Omega_a^{(3)} \land \left(\Omega_c^{(1)} \land \Omega_c^{(2)} \land \Omega_c^{(3)}\right)^c\right)\\
& = P\left(\Omega_a^{(1)} \land \Omega_a^{(2)} \land \Omega_a^{(3)} \land \left(\Omega_c^{(1)}\right)^c \right) + P\left(\Omega_a^{(1)} \land \Omega_a^{(2)} \land \Omega_a^{(3)} \land \left(\Omega_c^{(1)} \land \left(\Omega_c^{(2)}\right)^c\right) \right)\\
& \quad + P\left(\Omega_a^{(1)} \land \Omega_a^{(2)} \land \Omega_a^{(3)} \land \left(\Omega_c^{(1)} \land \Omega_c^{(2)} \land \left(\Omega_c^{(3)}\right)^c\right) \right)\\
& \leq P\left(\Omega_a^{(1)} \land \left(\Omega_c^{(1)}\right)^c \right) + P\left(\Omega_a^{(2)} \land \left(\Omega_c^{(2)}\right)^c \right) + P\left(\Omega_a^{(3)} \land \left(\Omega_c^{(3)}\right)^c \right)\\
& \leq P\left(\Omega_a^{(1)} | \left(\Omega_c^{(1)}\right)^c\right) + P\left(\Omega_a^{(2)} | \left(\Omega_c^{(2)}\right)^c\right) + P\left(\Omega_a^{(3)} | \left(\Omega_c^{(3)}\right)^c\right)\\
& \leq 3 \eh,
\end{split}
\end{equation}
where we splitted the events $\Omega_a=\Omega_a^{(1)} \land \Omega_a^{(2)} \land \Omega_a^{(3)}$ and $\Omega_c=\Omega_c^{(1)} \land \Omega_c^{(2)} \land \Omega_c^{(3)}$ into the relevant parts corresponding to each of the three authentication checks (the one to authenticate the communications related to $\bT$, the one related to $(\bX, \bG_\mathrm{EC}, C)$, and the one related to $F$). Concretely, $P\left(\Omega_a^{(1)} | \left(\Omega_c^{(1)}\right)^c\right)$ is the probability that a third party, knowing the message and tag pair $(\bT, t=\bG_B(\bT))$, chooses another valid pair $(\bT', \tilde t) \neq (\bT, t)$ for Alice, i.e.~such that $\bG_B(\bT') = \tilde t$. Since the new pair is valid, we must  have $\bT'\neq\bT$ (otherwise either $(\bT', \tilde t) = (\bT, t)$, or $\bG_B(\bT') \neq \tilde t$).  Hence it is enough to consider the case where $\bT \neq \bT'$, then the probability
\begin{equation}
    \begin{split}
    P\left(\Omega_a^{(1)} | \left(\Omega_c^{(1)}\right)^c\right) = P\left(\bG_B(\bT')=\tilde t| \bG_B(\bT)= t \right) \leq \varepsilon_h
    \end{split}
\end{equation}
is bounded by Eq.~\eqref{eq:conditionalHashing} since we use an authentication scheme based on a ASU family of hash functions (see Sec.~\ref{sec:hashing} for more details). 

Similarly, noting that  $\rho_{\land \Omega \land \Omega_c} = \rho_{\land \Omega \land \Omega_c \land \Omega_g} + \rho_{\land \Omega \land \Omega_c \land \Omega_g^c}$, we can further bound the quantity
\begin{equation}
\begin{split}
\frac{1}{2} || & \mathcal{E}_\mathrm{Trev}(\rho_{\land \Omega})_{\bK_A \overline E} - \mathbb{U}_{\bK_A} \otimes \left(\rho_{\land \Omega}\right)_{\overline E} ||_1\\
\leq & \frac{1}{2} || \mathcal{E}_\mathrm{Trev}(\rho_{\land \Omega \land \Omega_c \land \Omega_g})_{\bK_A \overline E} - \mathbb{U}_{\bK_A} \otimes \left(\rho_{\land \Omega \land \Omega_c \land \Omega_g}\right)_{\overline E} ||_1 \\
 & + \frac{1}{2} ||  \mathcal{E}_\mathrm{Trev}(\rho_{\land \Omega \land \Omega_c \land \Omega_g^c})_{\bK_A \overline E} - \mathbb{U}_{\bK_A} \otimes \left(\rho_{\land \Omega \land \Omega_c \land \Omega_g^c}\right)_{\overline E} ||_1 + 3\eh\\
\leq & \frac{1}{2} || \mathcal{E}_\mathrm{Trev}(\rho_{\land \Omega \land \Omega_c \land \Omega_g})_{\bK_A \overline E} - \mathbb{U}_{\bK_A} \otimes \left(\rho_{\land \Omega \land \Omega_c \land \Omega_g}\right)_{\overline E} ||_1 + 4\eh\\
\end{split}
\end{equation}
because
\begin{equation}
P(\Omega \land \Omega_c \land \Omega_g^c) = P(\Omega_a \land \Omega_h \land \Omega_\mathrm{PE} \land \Omega_c \land \Omega_g^c) \leq P(\Omega_h \land \Omega_g^c) \leq P(\Omega_h | \Omega_g^c) \leq \eh.
\end{equation}

Writing $\hat\Omega = \Omega_c \land \Omega_a \land \Omega_g \land \Omega_h$, $\overline\Omega = \hat\Omega \land \Omega_\text{PE'}$ and noticing that $\Omega_g\land\Omega_\mathrm{PE} = \Omega_g\land\Omega_\mathrm{PE'}$, we can rewrite this bound as
\begin{equation}\label{eq:normToBound}
\begin{split}
\frac{1}{2} || \mathcal{E}_\mathrm{Trev}(\rho_{\land \Omega})_{\bK_A \overline E} &- \mathbb{U}_{\bK_A} \otimes \left(\rho_{\land \Omega}\right)_{\overline E} ||_1\\
& \leq \frac{1}{2} || \mathcal{E}_\mathrm{Trev}(\rho_{\land \hat\Omega \land \Omega_\mathrm{PE'}})_{\bK_A \overline E} - \mathbb{U}_{\bK_A} \otimes \left(\rho_{\land \hat\Omega \land \Omega_\mathrm{PE'}}\right)_{\overline E} ||_1 + 4\eh\\
& = \frac{1}{2} || \mathcal{E}_\mathrm{Trev}(\rho_{\land \overline \Omega})_{\bK_A \overline E} - \mathbb{U}_{\bK_A} \otimes \left(\rho_{\land \overline \Omega}\right)_{\overline E } ||_1 + 4\eh
\end{split}
\end{equation}

It remains to bound the first term on the RHS of the above expression.
To do so, we consider critical values on the unknown probabilities $P(\hat\Omega|\Omega_\text{PE'}) $ and $P(\Omega_\text{PE'}) $, and analyse the situation differently depending on how these probabilities  compare to these critical values. Namely, consider three cases for the quantum state $\rho$:
\begin{itemize}
\item[Case 1:] $P(\hat\Omega|\Omega_\text{PE'}) \leq \es^2$
\item[Case 2:] $P(\Omega_\text{PE'}) \leq \eEA$
\item[Case 3:] Neither of the above are true
\end{itemize}
These cases may overlap, but at least one of them must always be true. It is thus sufficient to bound Eq.~\eqref{eq:normToBound} in each of the three cases individually. 

In case 1, we have
\begin{equation}
\frac{1}{2} || \mathcal{E}_\mathrm{Trev}(\rho_{\land \overline \Omega})_{\bK_A \overline E} - \mathbb{U}_{\bK_A} \otimes \left(\rho_{\land \overline \Omega}\right)_{\overline E } ||_1 \leq P(\overline\Omega) = P(\hat\Omega | \Omega_\mathrm{PE'}) P(\Omega_\mathrm{PE'}) \leq \es^2.
\end{equation}

In case 2, we have
\begin{equation}
\frac{1}{2} || \mathcal{E}_\mathrm{Trev}(\rho_{\land \overline \Omega})_{\bK_A \overline E} - \mathbb{U}_{\bK_A} \otimes \left(\rho_{\land \overline \Omega}\right)_{\overline E } ||_1 \leq P(\overline\Omega) = P(\hat\Omega | \Omega_\mathrm{PE'}) P(\Omega_\mathrm{PE'}) \leq \eEA.
\end{equation}

We now focus on case 3, for which $P(\hat\Omega|\Omega_\text{PE'}) > \es^2$ and $P(\Omega_\text{PE'}) > \eEA$. 
In this case, we use Proposition 1 to bound the trace distance via a condition on a smooth min-entropy: according to Proposition 1, as long as the length of the extracted string $\ell=|\bK_A|\geq 1$ is smaller than
\begin{equation}\label{eq:lbound}
\ell \leq \Upsilon_{{4}/{\ln 2}}\left( H_\mathrm{min}^{\es}(\bA|\overline E)_{(\rho_{|\Omega_\text{PE'}})_{\land \hat\Omega}} - 6 - 5\log\left(\frac{1}{\ePA}\right) \right),
\end{equation}
we have the guarantee that
\begin{equation}
\frac{1}{2}|| \mathcal{E}_\mathrm{Trev}((\rho_{|\Omega_\text{PE'}})_{\land \hat\Omega})_{\bK_A\overline E} - \mathbb{U}_{\bK_A}\otimes \left((\rho_{|\Omega_\text{PE'}})_{\land \hat\Omega}\right)_{\overline E} ||_1 \leq \ePA + 2 \es,
\end{equation}
which implies (since $\rho_{\land \overline \Omega} =  P(\Omega_\text{PE'}) (\rho_{|\Omega_\text{PE'}})_{\land \hat\Omega} $ and $P(\Omega_\text{PE'}) \leq 1$)
\begin{equation}
\frac{1}{2}|| \mathcal{E}_\mathrm{Trev}(\rho_{\land \overline\Omega})_{\bK_A\overline E} - \mathbb{U}_{\bK_A}\otimes \left(\rho_{\land \overline\Omega}\right)_{\overline E} ||_1 \leq \ePA + 2 \es.
\end{equation}

Under the condition~\eqref{eq:lbound}, we thus obtain secrecy for our key by grouping all three cases together and setting
\begin{equation}\label{eq:esec}
\esec=\max(\eEA, \ePA+2\es)+4\eh.
\end{equation}
This last expression follows from noting that $\es^2\leq \es \leq \ePA+2\es$ since $1 > \es, \ePA \geq 0$. We are thus left with the task of lower-bounding $H_\mathrm{min}^{\es}(\bA|\overline E)_{(\rho_{|\Omega_\text{PE'}})_{\land \hat\Omega}}$. Here, Eve's information is given by $\overline E=(\bX,\bT,\ooM,E)$, where $\bT$ is a function of $\bY$, so it suffices to bound $H_\mathrm{min}^{\es}(\bA|\bX\bY\ooM E)_{(\rho_{|\Omega_\text{PE'}})_{\land \hat\Omega}}$.\\

Since the probability of the event $\hat\Omega$ on the conditional normalized state $\rho_{|\Omega_{\mathrm{PE'}}}$ is greater than $\es^2$, we can apply Lemma 10 of~\cite{Tomamichel17} to obtain
\begin{equation}\label{eq:HAleak}
\begin{split}
H_\mathrm{min}^{\es}(\bA|\bX\bY \ooM E)_{(\rho_{|\Omega_\mathrm{PE'}})\land \hat\Omega} &\geq H_\mathrm{min}^{\es}(\bA|\bX\bY \ooM E)_{\rho_{|\Omega_\mathrm{PE'}}}\\
&\geq H_\mathrm{min}^{\es}(\bA|\bX\bY E)_{\rho_{|\Omega_\mathrm{PE'}}} - |\ooM|, 
\end{split}
\end{equation}
where in the second line we have applied a chain rule for smoothed min-entropy (see for example lemma 11 in the Supplementary Material of Ref~\cite{Winkler11}). Here, $|\ooM|$ is the total amount of classical information exchanged between Alice and Bob during the protocol, measured in bit units, including the error correction syndrome being sent by Alice to Bob in Step 5 of the protocol:
\begin{equation}\label{eq:Mleak}
\begin{split}
|\ooM| & = |\bM| + |\bG_\mathrm{EC}| + |\bG_B| + |C| + |\bG_A| + |F| + |\bG_F|\\
& = m + 64 + 64 + 1 + 64 + 1 + 64\\
&= m + 258
\end{split}
\end{equation}
We are thus left with bounding $H_\mathrm{min}^{\es}(\bA|\bX\bY E)_{\rho_{|\Omega_\mathrm{PE'}}}$. Given the registers that appear here, we can do this on the state after step $4^{++}$ of the modified protocol. From now on, we thus focus on the state $\rho_{|\Omega_{PE'}}$ at this stage, conditioned on $\Omega_\text{PE'}$ being fulfilled, and omit the subscript in the remaining of this section.

Following the approach of~\cite{ArnonFriedman19}, we use the chain rule given in Eq. (6.57) of~\cite{Tomamichel16}, to express the min entropy of Alice's outcomes in terms of a joint entropy and an entropy on Bob's outcomes only:
\begin{equation}\label{eq:HAHABHB}
\begin{split}
H_\mathrm{min}^{\es}(\bA|\bX\bY E) &\geq H_\mathrm{min}^{\es'}(\bA\bB'|\bX\bY E) - H_\mathrm{max}^{\es''}(\bB'|\bA\bX\bY E) - 3\vartheta_{\es-\es'-2\es''}\\
&\geq H_\mathrm{min}^{\es'}(\bA\bB'|\bX\bY E) - H_\mathrm{max}^{\es''}(\bB'|\bX\bY E) - 3\vartheta_{\es-\es'-2\es''}\\
\end{split}
\end{equation}
where $\es'+2\es'' < \es$ and
\begin{equation}
\vartheta_\eps = \log\frac{1}{1-\sqrt{1-\eps^2}}.
\end{equation}
We are then left with bounding two entropy terms. We do this for each term individually, each time with an application of (a different version of) the Entropy Accumulation Theorem (EAT).

\subsubsection{Lower-bounding $H_\mathrm{min}^{\es'}(\bA\bB'|\bX\bY E)$}\label{sec:HminEAT}

We first introduce the concepts of \emph{EAT channels} and \emph{tradeoff functions}, which are involved in applying the EAT.

\newcommand{\Sp}{S'}
\begin{definition}\label{def:EATchann}
A \emph{sequence of EAT channels} is a sequence 
$\mathcal{M}_1,\mathcal{M}_2,\dots,\mathcal{M}_n$
where each $\mathcal{M}_i$ is a channel from a register $R_{i-1}$ to registers $D_i S_i \Sp_i R_i$, 
which satisfies the following properties:
\begin{itemize}
\item All $D_i$ are classical registers with a common alphabet $\mathcal{D}$, and all $S_i$ have the same finite dimension.
\item 
For each $\mathcal{M}_i$, the value of $D_i$ is determined from the registers $S_i \Sp_i$ alone. Formally, this means 
$\mathcal{M}_i$ is of the form $\mathcal{P}_i \circ \mathcal{M}'_i$, where $\mathcal{M}'_i$ is a channel from $R_{i-1}$ to $S_i \Sp_i R_i$, and $\mathcal{P}_i$ is a channel from $S_i \Sp_i$ to $D_i S_i \Sp_i$ of the form
\begin{align}
\mathcal{P}_i(\rho_{S_i \Sp_i}) = \sum_{s\in\mathcal{S},s'\in\mathcal{S}'} (\Pi_{S_i,s} \otimes \Pi_{\Sp_i,s'}) \rho_{S_i \Sp_i} (\Pi_{S_i,s} \otimes \Pi_{\Sp_i,s'}) \otimes \ketbra{d(s,s')}{d(s,s')}_{D_i},
\label{eq:nodisturb}
\end{align}
where $\{\Pi_{S_i,s}\}_{s\in\mathcal{S}}$ and $\{\Pi_{\Sp_i,s'}\}_{s'\in\mathcal{S}'}$ are families of orthogonal projectors on $S_i$ and $\Sp_i$ respectively, and $d: \mathcal{S} \times \mathcal{S}' \to \mathcal{D}$ is a deterministic function.
\item
For any normalized state $\rho^0_{R_0E}$, the state
$\rho = (\mathcal{M}_n \circ\dots \circ\mathcal{M}_1 \otimes \mathcal{I}_E)\left(\rho^0_{R_0E}\right)$ satisfies the Markov conditions~\footnote{This condition is slightly stricter than necessary, as the EAT only requires the Markov conditions to hold for the specific state it is applied to~\cite{Dupuis20,Dupuis19}. However, for brevity we incorporate this into the definition of EAT channels, as was done in~\cite{ArnonFriedman19,Liu21}.}
\begin{align}
I(S_1 \dots S_{i-1} : \Sp_i | \Sp_1 \dots \Sp_{i-1} E)_\rho 
= 0 \qquad \forall i=1,2,\dots,n.
\label{eq:markov}
\end{align}
\end{itemize}
\end{definition}

\begin{definition}
Let $\fmin$ be a real-valued affine function defined on probability distributions over the alphabet $\mathcal{D}$. It is called a \emph{min-tradeoff function} for a sequence of EAT channels $\{\mathcal{M}_i\}$ if for any distribution $q$ on $\mathcal{D}$, we have 
\begin{align}
\fmin(q) \leq \inf_{\sigma \in \Sigma_i(q)} H(S_i|\Sp_iR)_\sigma \qquad \forall i=1,2,\dots,n,
\end{align}
where $\Sigma_i(q)$ denotes the set of states of the form $(\mathcal{M}_i\otimes\mathcal{I}_{R})(\omega_{R_{j-1}R})$ such that the reduced state on $D_i$ has distribution $q$. 
Analogously, a real-valued affine function $\fmax$ defined on probability distributions over $\mathcal{D}$ is called a \emph{max-tradeoff function} if 
\begin{align}
\fmax(q) \geq \sup_{\sigma \in \Sigma_i(q)} H(S_i|\Sp_iR)_\sigma \qquad \forall i=1,2,\dots,n.
\end{align}
The infimum and supremum of an empty set are defined as $+\infty$ and $-\infty$ respectively.
\end{definition}

We now follow the analysis of~\cite{Liu21}.
Here, we notice that steps 2, 3, $4^+$ and $4^{++}$ of our modified protocol correspond to a standard CHSH test protocol as studied in~\cite{Dupuis19,Liu21}. Hence, they can be modelled as an infrequent-sampling bipartite EAT channel (see~\cite{Dupuis19,Liu21}), which is to say each iteration of those steps corresponds to an EAT channel $\mathcal{M} : R_{i - 1} \to A_i B'_i X_i Y_i U_i R_i$ of the form
\begin{equation}
\mathcal{M} [\rho] = \gamma \mathcal{M}^{\mathrm{test}} [\rho] + (1 - \gamma) \mathcal{M}^{\mathrm{gen}} [\rho] \otimes | \perp \rangle \langle \perp |_{U_i},
\end{equation}
where $\mathcal{M}^{\mathrm{test}} : R_{i - 1} \to A_i B'_i X_i Y_i U_i R_i$, with $\bra{\perp}\mathcal{M}^{\mathrm{test}} [\rho]\ket{\perp}=0$ and $\mathcal{M}^{\mathrm{gen}} : R_{i - 1} \to A_i B'_i X_i Y_i R_i$. (In terms of Definition~\ref{def:EATchann} above, we are implicitly identifying $A_iB'_i$ with $S_i$ and $X_iY_i$ with $\Sp_i$. Note that these channels indeed satisfy the Markov conditions~\eqref{eq:markov} under this identification, hence they are valid EAT channels.)
As shown in~\cite{Liu21}, one can construct a min-tradeoff function for these channels using the function $g_t$ defined in~\eqref{eq:defineg}.
This yields the following bound on the conditional smooth min-entropy $H_\mathrm{min}^{\es}(\bA \bB'|\bX \bY E)_{\rho_{|\Omega_\mathrm{PE'}}}$:

\begin{theorem} [Theorem 3 of~\cite{Liu21}]\label{thm:eat}
For the quantities defined above, the following bound~\footnote{Note that the expressions for $V(f_t,p) $ and $K_{\alpha'}(f_t)$ here differ slightly from those appearing in the Supplementary Information of Ref.~\cite{Liu21}. Here, following~\cite{Dupuis19}, we took the expressions given in Eqs. (16) and (17) of the Supplementary Information of~\cite{Liu21} but replace $d_A$ by 4 (the dimension of the registers $A_iB'_i$). Also, we avoided applying the inequality $\alpha'>1$.} holds:
\begin{equation}\label{eq:EAT}
\begin{split}
H_\mathrm{min}^{\es'}(\bA\bB'|\bX\bY E)_{\rho_{|\Omega_\mathrm{PE'}}} >\ & n\, g_t\left(\omega_\thresh\right)\\
& 
- \frac{\vartheta_{\es'}}{\alpha'-1} - \frac{\alpha'}{\alpha'-1}\log\left(\frac{1}{P(\Omega_\mathrm{PE'})}\right)
\\
& + n \inf_{\omega\in \tilde{Q}} \Delta\left( f_t, \omega\right) - (\alpha'-1)V\left(f_t,q(\omega)\right)\\
& - n(\alpha'-1)^2 K_{\alpha'}(f_t)
\end{split}
\end{equation}
where
\begin{equation}
\begin{split}
q(\omega) &= (\gamma(1-\omega), \gamma \omega, 1-\gamma)\\
\tilde{Q} &= \left[\frac{1-1/\sqrt{2}}2,\frac{1+1/\sqrt{2}}2\right]\\
\Delta( f_t, \omega) &= \eta(\omega) - f_t(q(\omega))\\
V(f_t,p) &= \frac{\ln{2}}{2}\left(\log{33} + \sqrt{2 + \mathrm{Var}_p(f_t)}\right)^2\\
K_{\alpha'}(f_t) &= \frac{1}{6(2-\alpha')^3 \ln 2} 2^{(\alpha'-1)(2 + \mathrm{Max}(f_t) - \mathrm{Min}_{\tilde{Q}_\gamma}(f_t))} \ln^3\left( 2^{2+\mathrm{Max}(f_t) - \mathrm{Min}_{\tilde{Q}_\gamma} (f_t)} + e^2\right).
\end{split}
\end{equation}
\end{theorem}

Here, we used definitions \Cref{eq:defineg,eq:entropyCHSH,eq:definef,eq:definef2,eq:defineVar}. Notice that $\eta(\omega)$ is an increasing function on $\omega\in \left(\frac{3}{4},\omega_{\max}\right] $, and $g_t(\omega)$ is a lower bound on $\eta(\omega)$ tangent at $t$. Therefore, we have $g_t(\delta_1) \geq g_t(\delta_0)$ for $t\in \left(\frac34, \omega_{\max}\right]$. Focusing on this domain, we can compute the maximum of $f_t$ over all distributions:
\begin{equation}
\mathrm{Max}(f_t) = g_t(1).
\end{equation}
The minimum of $f_t$ over all distributions $\tilde Q_\gamma=\{q(\omega) | \omega\in\tilde Q\}$ compatible with the sampling probability $\gamma$ and with quantum theory is then given by
\begin{equation}
\mathrm{Min}_{\tilde{Q}_\gamma}(f_t) = g_t(\omega_\mathrm{min}),
\end{equation}
where $\omega_\mathrm{min} = \frac{1-1/\sqrt{2}}2$ is the lowest CHSH winning probability compatible with quantum theory. 

\subsubsection{Upper-bounding $H_\mathrm{max}^{\es''}(\bB'|\bX\bY E)$}\label{sec:HmaxEAT}

We define a sequence of EAT channels $\tilde{\mathcal{M}}_i$ that apply steps 2, 3B, 4 and $4^{+}$ of the modified protocol (note that this means they do not produce the registers $A_i$). As above, these maps indeed satisfy the Markov conditions~\eqref{eq:markov}, identifying $B'_i$ with $S_i$ and $X_iY_i$ with $\Sp_i$.
However, we cannot apply the EAT directly to bound $H_\mathrm{max}^{\es''}(\bB'|\bX\bY E)$, because of a technical issue that the EAT would require the conditioning event $\Omega_\mathrm{PE'}$ to be determined by the registers $\bB' \bX\bY E$ alone, which is not the case here. 
Hence we instead apply some intermediate results from~\cite{Dupuis20}, in a slightly different order as compared to the derivation of the EAT. We first list the relevant results --- in these statements, $H_{\alpha}$ denotes a particular form of quantum conditional R\'{e}nyi entropy (see e.g.~\cite{Tomamichel16,Dupuis20} for explanations of other versions), namely: 
\begin{equation}
H_{\alpha}(A|B)_\rho \coloneqq -D_{\alpha}(\rho_{AB} \Vert \mathbb{I}_A \otimes \rho_{B}), \text{ where } D_{\alpha}(\rho \Vert \sigma) \coloneqq \frac{\alpha}{\alpha-1} \log \left\lVert \rho^{\frac{1}{2}} \sigma^{\frac{1-\alpha}{2\alpha}} \right\rVert_{2\alpha}^2 .
\end{equation}

\begin{lemma}[Prop.~4.5 in~\cite{Dupuis20}]\label{lem:EATlem1}
Consider a sequence of EAT channels $\{\mathcal{M}_i\}$ and a state of the form $\rho = (\mathcal{M}_n \circ\dots \circ\mathcal{M}_1 \otimes \mathcal{I}_E)\left(\rho^0_{R_0E}\right)$ for some normalized $\rho^0_{R_0E}$.
Let $\fmax$ be a max-tradeoff function for $\{\mathcal{M}_i\}$, and consider any $h\in\mathbb{R}, \eps\in(0,1), \alpha\in(1,1+2/V')$, where $V'=2\ceil{\lvert\lvert \nabla \fmax \rvert\rvert_\infty} + 2\log(1+2d_S)$ with $d_S$ being the dimension of the systems $S_i$. 
Then for any event $\Omega$ on $\mathbf{D}$ that implies $\fmax(\freq_{\str{d}}) \leq h$ for all $\str{d}\in\Omega$, we have
\begin{equation}
H_{\frac{1}{\alpha}}(\str{S}|\str{S}'E)_{\rho_{|\Omega}} < nh + n \frac{\alpha-1}{4} V'^2 + \frac{\alpha}{\alpha-1} \log \frac{1}{P(\Omega)}.
\end{equation}
\end{lemma}

\begin{lemma}[Lemma B.6 of~\cite{Dupuis20}]\label{lem:EATlem2}
Let $\rho_{AB}$ be a normalized state of the form $\sum_{x} p_x \rho_{AB|x}$ where $\{p_x\}$ is a probability distribution over $\mathcal{X}$. Then for any $x \in \mathcal{X}$ and any $\alpha\in(1,2]$, we have
\begin{equation}
H_{\frac{1}{\alpha}}(A|B)_{\rho_{|x}} \leq H_{\frac{1}{\alpha}}(A|B)_{\rho} + \frac{\alpha}{\alpha-1} \log \frac{1}{p_x}.
\end{equation}
\end{lemma}

\begin{lemma}[Lemma B.10 of~\cite{Dupuis20}]\label{lem:EATlem3}
For any normalized state $\rho_{AB}$ and any $\eps\in(0,1), \alpha\in(1,2]$, we have
\begin{equation}
H_\mathrm{max}^{\eps}(A|B)_{\rho} \leq H_{\frac{1}{\alpha}}(A|B)_{\rho} + \frac{\vartheta_{\eps}}{\alpha-1}.
\end{equation}
\end{lemma}

To apply these lemmas, we first note that since the register $B_i'$ is set to a deterministic value whenever $Y_i=2$ (which happens with probability $1-\gamma$), the following bound always holds, without needing to consider the conditioning event $\Omega_\mathrm{PE'}$:
\begin{equation}
H(B_i'|X_i Y_i R)_{(\tilde{\mathcal{M}}\otimes \mathcal{I}_R)(\omega_{R_{i-1}R})} \leq \gamma.
\end{equation} 
This means that by choosing a max-tradeoff function that simply takes the constant value $\gamma$, we can first apply Lemma~\ref{lem:EATlem1} \emph{without} conditioning on any event, to conclude that the unconditioned state $\rho$ in our above analysis satisfies
\begin{equation}
H_{\frac{1}{\alpha''}}(\bB'|\bX\bY E)_{\rho} < n\gamma + n\frac{\alpha''-1}{4} V'^2,
\end{equation}
for any $\alpha''\in(1,1+2/V')$, with $V'=2\log{5}$ (notice that $\ceil{\lvert\lvert \nabla\fmax \rvert\rvert_\infty}=0$ since $\fmax$ is a constant function). Only now do we condition on $\Omega_\mathrm{PE'}$ by applying Lemma~\ref{lem:EATlem2}~\footnote{This is valid because Lemma~\ref{lem:EATlem2} does not require that the conditioning event is defined entirely by the registers in the state being considered --- it only requires that the state has a decomposition of the form $\rho = \sum_{x} p_x \rho_{|x}$. In this case, we indeed have such a decomposition, $\rho_{\bB'\bX\bY E} 
= P(\Omega_\mathrm{PE'})(\rho_{|\Omega_\mathrm{PE'}})_{\bB'\bX\bY E} + P(\Omega_\mathrm{PE'}^c)(\rho_{|\Omega_\mathrm{PE'}^c})_{\bB'\bX\bY E}$, and can thus apply the lemma.}, which yields 
\begin{equation}
H_{\frac{1}{\alpha''}}(\bB'|\bX\bY E)_{\rho_{|\Omega_\mathrm{PE'}}} < n\gamma + n\frac{\alpha''-1}{4} V'^2 + \frac{\alpha''}{\alpha''-1}\log\frac{1}{P(\Omega_\mathrm{PE'})}.
\end{equation}
Finally, applying Lemma~\ref{lem:EATlem3}, we get the bound
\begin{equation}\label{eq:HB}
H_\mathrm{max}^{\es''}(\bB'|\bX\bY E)_{\rho_{|\Omega_\mathrm{PE'}}} < n\gamma + n\frac{\alpha''-1}{4} V'^2 + \frac{\vartheta_{\es''}}{\alpha''-1} + \frac{\alpha''}{\alpha''-1}\log\frac{1}{P(\Omega_\mathrm{PE'})}.
\end{equation}

\subsubsection{Bringing the pieces together}
To conclude the analysis of case 3, we remember that $P(\Omega_\mathrm{PE'})>\eEA$. Using this bound in \Cref{eq:EAT,eq:HB}, and merging the result together with \Cref{eq:esec,eq:lbound,eq:HAleak,eq:Mleak,eq:HAHABHB}, we obtain Proposition~\ref{prop:keyrate} for this case, which concludes the proof.

\bibliography{references}